\documentclass{JFM-FLM_Au}

\usepackage[T1]{fontenc}
\usepackage{newtxtext,newtxmath} 
\usepackage{amsmath}
\usepackage{listings}
\usepackage{xcolor}
\usepackage{gb4e}
\usepackage{hyperref}
\usepackage{caption}
\usepackage{subcaption}

\noautomath
\usepackage{graphicx}
\usepackage{epstopdf}
\usepackage{color}
\usepackage{multirow,multicol}
\usepackage{tikz}
\usepackage{adjustbox}
\usetikzlibrary{arrows, chains, positioning, quotes, shapes.geometric}
\newcount\ndots

\def\drwln#1#2{\raise 2.5pt\vbox{\hrule width #1pt height #2pt}}

\mathchardef\scriptL="324C
\lefttitle{Shahadat, M. R. B. et al.}
\righttitle{Shahadat, M. R. B. et al.}

\title{Numerical Study of Compressibility and Velocity Parameter Effects on Spatially Evolving Supersonic Turbulent Shear Layers }

\author{Shahadat, M. R. B.\aff{1}, Li, Z.\aff{2}, Jaberi, F.A.\aff{1} \and Livescu, D.\aff{3} }

\affiliation{\aff{1} Department of Mechanical Engineering, Michigan State University, East Lansing, USA
\aff{2}College of Engineering, Texas A\&M Corpus Christi, Corpus Christi, USA
\aff{3} Computer and Artificial Intelligence Division, Los Alamos National Laboratory, Los Alamos, USA
}
\corresau{D. Livescu, livescu@lanl.gov; F. A. Jaberi, jaberi@msu.edu}

\makeatletter
\renewcommand{\@evenhead}{}
\renewcommand{\@oddhead}{}

\renewcommand{\@evenfoot}{}
\renewcommand{\@oddfoot}{}

\def\maketitle{%
  \thispagestyle{plain}%
  \vbox{\hsize\textwidth\vbox{\vskip -30pt % Pull title up to reclaim space
    \begin{center}%
      {\LARGE\bfseries\@title\par}%
      \vskip 2em%
      {\large\lineskip .5em\begin{tabular}[t]{c}\@author\end{tabular}\par}%
      \vskip 1.5em%
      {\small\@affiliation\par}% Restores university affiliations
    \end{center}%
    \vskip 1.5em%
  }}%
}

\def\fofbox#1{}

\def\fofbox{\at@end@macro}
\makeatother

\begin{document}
\maketitle

\begin{abstract}
Direct Numerical Simulations (DNS) of a spatially developing supersonic turbulent shear layer are conducted for a range of convective Mach numbers ($M_c$) and velocity parameters ($\lambda$) to examine the effects of compressibility and advection on the growth rate, self-similarity, flow statistics, asymmetry, and entrainment of the layer. At distant downstream locations, self-similarity is attained for all cases. The self-similar region is identified by the collapse of normalized mean streamwise velocity, the constant peak of normalized Reynolds stresses, and the linear growth rate of the shear layer thickness and momentum thickness. Despite significant variations in lower-order and higher-order statistics across different $M_c$ and $\lambda$ values, profiles of all turbulence quantities examined collapse within the self-similar region using our proposed self-similar scalings. The self-similar forms of continuity, momentum, and energy equations have been formulated incorporating compressibility and centerline shifts. The self-similar normalized density distribution inside the layer is used to explain the effects of compressibility on various flow statistics including the far-field cross-stream velocity. The density variation is linked to dissipation effects as revealed by our analysis of the self-similar energy equation. An approximate equation for the cross-stream velocity is developed and the profiles of cross-stream velocity obtained from this equation show good agreement with the DNS results. A geometric interpretation of the entrainment ratio is presented and the approximate equation for the cross-stream velocity is used to provide a general closed-form expression of the entrainment ratio. The entrainment ratio increases with $M_c$ and $\lambda$, favoring excess entrainment on the high-speed side.    

\end{abstract}


\section{Introduction} \label{sec: introduction}
The study of the "mixing" of two parallel streams has received considerable attention in the last few decades because of its fundamental flow nature, complex physics, and a wide range of engineering applications. The concept of shear/mixing layer is relevant to air-breathing propulsion systems (\citealt{zhang19}), supersonic ejectors (\citealt{dutton91}), missiles and projectile flows (\citealt{dutton93}), jet noise reduction (\citealt{pantano02}), boundary region of the jet, the slipstream behind a wing and the interface between a recirculation region (\citealt{dutton03}). A significant reduction in the growth rate of the shear/mixing layer with increasing compressibility is the notable disparity between the compressible and incompressible layers (\citealt{zhang19}). In this study, we are only considering single-fluid compressible supersonic shear layers, while the variable-density case will be addressed in a follow-up paper.
The important parameter that quantifies the compressibility of the shear layer is the convective Mach number, which was used by many researchers in the literature (\citealt{bogdanoff83}, \citealt{rashko88}, \citealt{kim20}, \citealt{zhang19}, \citealt{dimotakis_mc}, \citealt{dutton03}). Assuming equal specific heat ratios, if the inflow velocities of the two streams are $U_1$ and $U_2$ and the local speeds of sound in the free streams are $a_1$ and $a_2$, the convective Mach number is defined as
\begin{equation}
M_c=\frac{U_1-U_2}{a_1+a_2}
\end{equation}

In several previous studies, a similar parameter was used, which was called the relative Mach number ($M_r=\frac{\Delta U}{a} $), where $\Delta U$ is the free stream velocity difference and $a$ is the average sound speed ( \citealt{jackson90}, \citealt{soetrisno89}, \citealt{dutton91}). 

The geometrical features of large-scale structures in shear layers have been shown to be significantly influenced by the convective Mach number. When the convective Mach number falls below 0.4, the flow exhibits "low" compressibility (\citealt{zhang17,zhang19}). In scenarios where the convective Mach number ranges between 0.4 and 0.8, the flow experiences moderate compressibility. In contrast, if the convective Mach number exceeds 0.8, the flow enters a highly compressible state (\citealt{zhang19}). 

Over the past several decades, researchers have studied the incompressible shear layers analytically, experimentally, and numerically. For these types of shear layer, a significant amount of work is available on the shear layer growth rate, mean and fluctuation fields, inflow conditions, large-scale vortex structures, and scalar transport.
The first analytical study of the shear layer was conducted by \cite{gortler42}. Using Prandtl's eddy viscosity model, he derived the analytical solution of the planar shear layer. \cite{spencer70} investigated the pressure and velocity field in an incompressible shear layer. \cite{champagne76} experimentally investigated a two-dimensional incompressible shear layer and showed the connection of flow development to the initial conditions.
Dimotakis and his coworkers studied large-scale flow dynamics (\citealt{dimotakis_large}), flow instability (\citealt{dimotakis_instability}), turbulent mixing (\citealt{dimotakis_mixing}), entrainment, growth and orientation of the shear layer (\citealt{dimotakis_entrainment}), wall effects (\citealt{dimotakis_wall}) and the effects of downstream disturbance (\citealt{dimotakis_down}) on the mixing layer. \cite{dimotakis_entrainment} showed that the spatially developing shear layer entrains an unequal amount of fluid from each stream, and the ratio of entrainment on the two streams is different from unity. This effect has important consequences on mixing and reaction in mixing layers, and was studied by \cite{mungal92}, \cite{konrad1977experimental}, \cite{masutani1986structure}, \cite{karasso1996scalar}, among many others. \cite{rashko74} studied large-scale structures, coherent and pairing processes, and the growth rate of the shear layer. They observed that in an incompressible shear layer, the predominant features are large-scale two-dimensional structures. In a related study, \cite{mungal92} conducted experimental investigations and discovered that similar large-scale two-dimensional structures dominate the flow field under the condition of a low convective Mach number. As a more comprehensive understanding of the incompressible shear layers began to be acquired, the research attention started to shift towards the exploration of the compressible shear layers to understand their complex physics and flow properties (\citealt{zhang19}).
According to previous experimental and computational studies, as the compressibility increases, the growth rate and turbulent intensities of the shear layer decrease significantly (\citealt{bogdanoff83}, \citealt{pantano02}, \citealt{zhang17}, \citealt{kim20}, \citealt{shahadat2022numerical}). The large-scale structures are also influenced by the level of compressibility (\citealt{rashko74}).  \cite{mungal92} showed that as the convective Mach number increases, the large-scale two-dimensional structures are gradually oriented obliquely in the spanwise direction. At very high convective Mach numbers, these large-scale structures become disorganized and highly three-dimensional. 
\cite{rashko88, rashko882} found that compressibility effects become notably pronounced when the convective Mach number reaches 0.5. As the convective Mach number increases beyond this point, these effects become more pronounced. Beyond a convective Mach number of 1.00, the impact of compressibility seems to stabilize and level off. \cite{lele02} connected the reduced growth rate with compressibility with the pressure-strain rate correlation. \cite{kim20} made several significant observations in their study. They noted that the normalized growth rate of the shear layer decreases as the compressibility level increases. Additionally, they found that as the convective Mach number increases, both the spanwise and transverse turbulence intensities decrease, but the streamwise turbulence intensity remains unaffected by compressibility. Furthermore, \cite{kim20} demonstrated that the streamwise anisotropy component tends to increase with higher convective Mach numbers, while the transverse and spanwise components tend to decrease. \cite{dutton93} observed that with increasing compressibility, only the transverse turbulence intensity decreases, keeping the streamwise and spanwise turbulence intensities constant, and therefore the ratio of transverse to spanwise turbulence intensities decreases and the spanwise direction becomes more important with increasing compressibility. In fact, in the more idealized homogeneous shear flow configuration, analytical solutions can be obtained showing the direct connection between the pressure mode, the dilatational velocity in the transverse direction, and the reduced growth rate \citep{livescu2004small}. 

Using DNS of a temporally evolving shear layer, \cite{sandham1990compressible} observed that as compressibility increases, large-scale structures become more three-dimensional in nature. In addition, they examined the impact of compressibility on the growth rate of the shear layer, noting a decrease in the growth rate as compressibility increases. In their DNS study of a temporal mixing layer, \cite{vreman1996compressible} observed a consistent trend of reduction in the growth rate in a range of convective Mach numbers from 0.2 to 1.2 and suggested that the reduction of the turbulent pressure fluctuation term is responsible for this reduced growth rate with increasing compressibility, which aligns with the findings of \cite{lele02}. Using the DNS of temporal mixing layers, \cite{pantano02} showed that the streamwise, transverse, and spanwise turbulence intensities decrease with increasing compressibility. \cite{pantano02} also observed the drop in pressure strain with the increase of the convective Mach number, which was consistent with \cite{vreman1996compressible} and \cite{lele02}. They found a similar trend of anisotropy as in \cite{kim20}. 
%Using the DNS of temporal mixing layers, \cite{livescu20} showed the reduction in the growth rate with density Atwood number for a variable density shear layer. 
\cite{zhou12} conducted a DNS study of a spatially developing shear layer but since their simulations were for moderate compressible conditions, they could not provide insights into vortex dynamics characteristics of highly compressible shear layers. \cite{zhang19} conducted a DNS of a spatially developing shear layer at higher compressibility conditions ($M_c$=1.0) but their study focused primarily on the evolution of vortex structures. \cite{fu2006numerical} studied the spatially developing compressible shear layer for a range of convective Mach numbers but their study is also focused on  flow structures and shocklets. There seem to be limited DNS studies of spatially developing shear/mixing layers under highly compressible conditions ($M_c$ $>$ 1.0). One of the main goals of this study is to address this research gap.

There is another important parameter that affects the growth rate and statistical behavior of the shear layer called the velocity parameter ($\lambda$) which is defined as (\citealt{abramovich84}, \citealt{sabin65}, \citealt{chinzei86}, \citealt{birch73}, \citealt{dutton91}, \citealt{dutton93}):  
\begin{equation}
\lambda=\frac{U_1-U_2}{U_1+U_2}=\frac{1-r}{1+r}=\frac{\Delta U}{2U_c}
\end{equation}
Where $r$ is the ratio of low-speed to high-speed stream velocities:
\begin{equation}
r=\frac{U_2}{U_1}
\end{equation}
and $U_c$ is the convective velocity or the average velocity of the two streams: 
\begin{equation}
U_c=\frac{U_1+U_2}{2}
\end{equation}

Obviously, $\lambda$ varies between 0 and 1, with $\lambda=0$ meaning that two streams have equal velocities ($U_1$=$U_2$) (no shear condition), and $\lambda=1$ meaning that the low-speed stream $U_2$=0 (\citealt{wei22,gautam24}), resulting in a single stream shear layer. The velocity parameter is analogous to the Atwood number in Rayleigh-Taylor instability (\citealt{livescu09}, \citealt{wei22}, \citealt{glimm01}, \citealt{wei12}). There is a notable increase in the growth rate with increasing $\lambda$ (\citealt{mehta91,ragab88}).  For the incompressible case, \cite{wei22} thoroughly examined the effects of $\lambda$, emphasizing its significance on the growth rate and self-similar scalings. However, no such study has been performed for the compressible case, and another main goal of the current work is to study the effects of $\lambda$ on compressible spatially developing shear layers. 

One of the important tools for finding scaling relations among mean flow quantities in various flows has been self-similar analysis. \cite{townsend1976structure} showed that at very high Reynolds numbers and sufficiently far downstream, shear flows also become self-similar. The key indicators for self-similarity are the linear shear layer growth and the consistent collapse of mean velocity and peak values of turbulent intensities at different streamwise locations when they are properly normalized. For the compressible spatially developing shear layer, \cite{jaberi10} observed the linear growth rate of the momentum thickness and the location-wise collapse of the mean stream-wise velocity in the self-similar coordinate in their DNS and LES data. \cite{kim20} showed that the shear layer thickness grows linearly and the mean streamwise velocity and turbulent kinetic energy collapse within the self-similar region. They also noted that, while the growth rate varies for different convective Mach numbers, the mean streamwise velocity profiles exhibit a similar behavior within this self-similar region. Previous self-similar analyses of compressible shear layers have mainly focused on mean continuity and streamwise momentum equations (\citealt{livescu20,pantano02,wei22}). So far, a self-similar analysis of the transverse momentum and energy equations has not been considered. The overall scalings for mean streamwise velocity, transverse velocity, and Reynolds shear stress have been discussed in the literature (\citealt{wei22,livescu20}) but the scalings for mean density, temperature, pressure, normal Reynolds stress in the transverse direction, and dissipation are still missing. In addition, the behavior of mean transverse velocity has received limited attention in the existing literature, partially due to the lack of experimental data.

%\cite{livescu20} discussed the self-similar continuity and streamwise momentum equation in a temporal shear layer. Additionally, they suggested the proper self-similar scaling for mean streamwise velocity, transverse velocity, and Reynolds shear stress. Their self-similar analysis was similar to the analysis performed by \cite{pantano02}. \cite{wei22} discussed the self-similar continuity and streamwise momentum equation for an incompressible spatially developing shear layer. Their most important contributions were the consideration of centerline shifting in the self-similar equations and the formulation of an approximate equation for self-similar transverse velocity. They also suggested the proper self-similar scalings mean streamwise velocity, transverse velocity, and Reynolds shear stress. The self-similar analysis of \cite{wei22} was very systematic and comprehensive but it was focused on the incompressible spatially developing shear layer. \cite{bretonnet2007deflection} discussed the self-similar energy equation but the flow was laminar and they studied it under a few limiting conditions. 

To address the aforementioned shortcomings in the literature, we have formulated self-similar forms of continuity, streamwise momentum, transverse momentum, and energy equations, incorporating both compressibility and centerline shifts. We show that the convective Mach number and velocity parameter emerge naturally in the self-similar equations, and therefore we conducted separate investigations on the influences of these two parameters. We suggest self-similar scalings for various mean and turbulent quantities, including, for the first time, the density, pressure, and dissipation. We investigate density variations within the shear layer, linking these variations to dissipation effects as revealed by our analysis of the self-similar energy equation. Furthermore, we provide approximate closed-form analytical solutions for the self-similar quantities, highlighting the interface shift and the role of parameters $M_c$ and $\lambda$. In addition, we discuss the asymmetric nature of flow within spatially evolving shear layers. In this regard, we provide a geometrical interpretation for the entrainment ratio as well as an approximate closed-form analytical solution, while also examining how it changes with respect to $M_c$ and $\lambda$. We test and validate all of our analytical scalings and predictions against our high-resolution, fully converged spatially developing mixing layer DNS data.

The study is organized as follows: the governing equations are presented in Section~\ref{sec:equations} and the flow setup, as well as simulation parameters, are discussed in Section~\ref{sec:flow setup}. In Section~\ref{sec:computational approach}, we discuss the initial and boundary conditions, the numerical scheme and the numerical accuracy. Self-similarity is discussed in Section~\ref{sec:self-similarity}. We apply the self-scaling to the mean continuity, streamwise momentum, transverse momentum, and energy equations and present the proper scalings for different mean and turbulent quantities. In Section~\ref{sec:Mc}, we discuss the effects of compressibility on the growth rate and flow statistics of the shear layer and apply the self-similar scalings presented in Section~\ref{sec:self-similarity} to demonstrate the collapse of different mean and turbulent quantities. In Section ~\ref{sec:At}, we discuss the effects of $\lambda$ on the growth rate and flow statistics of the shear layer and present the collapsed profiles using our suggested self-similar scalings. In Section~\ref{sec:asym}, we discuss the effects of $M_c$ and $\lambda$ on the asymmetry and entrainment of the shear layer. Finally, Section~\ref{sec:conclusion} presents the concluding remarks.

\section{Governing Equation} \label{sec:equations}

In this work, compressible mass, momentum, energy, and scalar equations together with the equation of state are solved numerically in non-dimensional and conservative form for the spatially evolving shear layer configuration. For the equation of state, the ideal gas approximation is used. Gravity is not considered. The equations for the instantaneous variables may be written as:

\begin{equation}
    \frac{\partial{\rho}}{\partial t}+\frac{\partial (\rho u_j)}{\partial x_j}=0
\end{equation}

\begin{equation}
     \frac{\partial{(\rho u_i)}}{\partial t}+\frac{\partial (\rho u_i u_j)}{\partial x_j}=-\frac{\partial P}{\partial x_i}+\frac{\partial \tau_{ij}}{\partial x_j}
\end{equation}

\begin{equation}
     \frac{\partial{(\rho e)}}{\partial t}+\frac{\partial (\rho e u_j)}{\partial x_j}=\frac{\partial}{\partial x_j} (k\frac{\partial T}{\partial x_j})-P\frac{\partial u_j}{\partial x_j}+\tau_{ij}\frac{\partial u_i}{\partial x_j}
\end{equation}

Here $t$ represents the time variable and index $i$ represents the corresponding spatial direction. 

The viscous stress tensor is:
\begin{equation}
    \tau_{ij}=\mu [\frac{\partial u_i}{\partial x_j}+\frac{\partial u_j}{\partial x_i}-\frac{2}{3}\frac{\partial u_k}{\partial x_k}\delta_{ij}]
\end{equation}
where $\delta_{ij}$ is the Kronecker delta. Sutherland’s law for air was used to provide the viscosity dependence on temperature,  $\mu (T)$ (\cite{jaberi10}). The equation of state in non-dimensional for is written as:
\begin{equation}\label{eq:eqofstates}
    P =\frac{\rho T}{\gamma M^2}
\end{equation}

In equation (\ref{eq:eqofstates}), $P$ and $T$ denote the pressure and temperature; $\gamma$=$\frac{C_p}{C_v}$ is the specific heat ratio, and $M$ is the reference Mach number resulting from the non-dimensionalization. The value of $\gamma$ used in this study is $1.40$. 

\section{Flow Setup and Parameters}\label{sec:flow setup}
\begin{figure}
     \centering
     \begin{subfigure}[b]{1\textwidth}
         \centering
         \includegraphics[width=\textwidth]{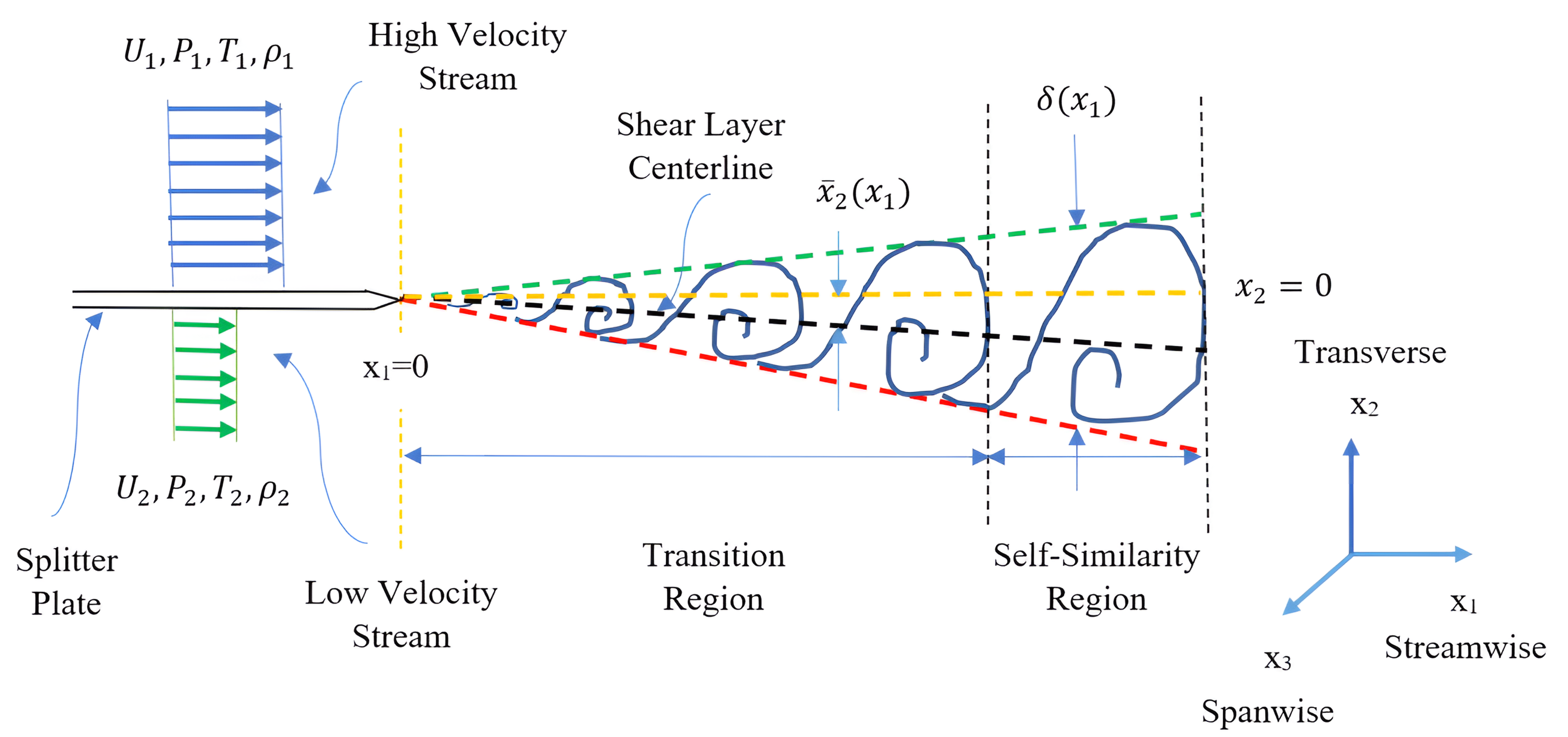}
     \end{subfigure}
     \caption{Planar view of spatially developing shear layer. The two free streams are represented by subscripts $1$ and $2$, respectively. Stream $1$ represents the high-speed side and stream $2$ represents the low-speed side. (\textcolor{green}{\textbf{- - - - -}}) line represents the upper edge and (\textcolor{red}{\textbf{- - - - -}}) line represents the lower edge of the shear layer. The difference between these two lines is the shear layer thickness ($\delta$). (\textcolor{yellow}{\textbf{- - - - -}}) line ($x_2$=0) is the geometric center of the shear layer. (\textcolor{black}{\textbf{- - - - -}}) line is the shifted centerline and $\bar x_2$ is the shifting distance from the geometric centerline.  The densities of the two streams can be the same or different. The same density represents the constant density shear layer and different densities represent the variable density shear layer. The coordinate system is chosen in such a way that $x_1$ represents the streamwise direction, $x_2$ represents the transverse or cross-stream direction, and $x_3$ represents the spanwise direction. }
     \label{fig:Screenshot 2022-10-24 145102}
     \hfill
\end{figure}

In this work, a simple spatial configuration is used with two fluid streams with different speeds flowing in the same direction. It differs from the temporal configuration in which a pair of fluid streams with equal speeds flow in opposite directions resulting in zero convective velocity (\citealt{livescu20}). In the temporal configuration, the mixing layer grows in time and not in the streamwise direction. Figure~\ref{fig:Screenshot 2022-10-24 145102} shows the schematic view of a spatially developing shear layer. The coordinate system is chosen in such a way that $x_1$ indicates the streamwise direction, $x_2$ indicates the transverse direction and $x_3$ indicates the spanwise direction (Figure~\ref{fig:Screenshot 2022-10-24 145102}). The origin of the Cartesian coordinate is at the tip of the splitter plate. Thus, two streams with essentially uniform flow velocities (i.e., $U_1$ and $U_2$), separated by the splitter plate at $x_1$ = 0, suddenly come into contact with each other at $x_1$$>$0. The two free streams are represented by subscripts 1 and 2, respectively, and may have the same or different specific heat ratios. In figure ~\ref{fig:Screenshot 2022-10-24 145102}, stream $1$ is considered to be the high-speed side, and stream $2$ is considered to be the low-speed side. The flow is shown to reach a self-similar state following the transition region. 

In the following sections, numerical values are computed from averages taken over time and in the spanwise direction. To make the profiles sufficiently "smooth," enough time averaging was performed for all the simulations. Since the flow is periodic in the spanwise ($x_3$) direction, spanwise averaging was also performed followed by time averaging. All averages presented in the paper have been verified to be converged after time averaging. In this study, temporal sampling was performed over 4 mean passover times.

For flows with density variations, it makes a difference whether different quantities are calculated based on Reynolds or Favre averaging. The Reynolds average of a quantity $q$ is interchangeably represented by $<q>$ or $\bar{q}$, with the overbar preferred for shorter expressions. The corresponding Reynolds fluctuations are expressed as $q^\prime$=q-$<q>$. The Favre average of a quantity $q$ is represented by $\tilde{q}=\frac{<\rho q>}{<\rho>} $ and the Favre average fluctuations are expressed as ${q}^{\prime\prime}=q- \tilde{q}$. $M_c$ and $\lambda$ are the two important parameters considered in this study, which we have defined in section~\ref{sec: introduction}. The local Mach number for free stream $1$ is $M_1$=$\frac{U_1}{a_1}$ and that for free stream $2$ is $M_2$=$\frac{U_2}{a_2}$, where $a_1$ and $a_2$ correspond to the sound speeds in the two streams. 

 We characterize the growth rate of the shear layer by several commonly used quantities: momentum thickness, shear layer thickness, vorticity thickness, and visual thickness (\citealt{livescu20,pantano02}). The momentum thickness $\theta$ is calculated based on the mean streamwise momentum profile and  is defined as (\citealt{livescu20,jaberi10}):
\begin{equation}
    \theta (x_1)=\frac{1}{\rho_0 (U_1-U_2)^2}\int_{-\infty}^{+\infty}{\bar{\rho} (U_1-\tilde{u_1}(x_1,x_2))(\tilde{u_1}(x_1,x_2)-U_2)\,dx_2},
\end{equation}
where $\bar{\rho}$ is the mean density and $\rho_0$ is the free stream density, which is the same for the two streams. $\tilde{u_1}(x_1,x_2)$ is the Favre average streamwise velocity which is found by time and space averaging. Based on the mean streamwise velocity profile, a similar thickness may be defined as (\citealt{livescu20}): 
\begin{equation}\label{eq:mmt}
    \theta_m (x_1)=\frac{1}{ (U_1-U_2)^2}\int_{-\infty}^{+\infty}{ (U_1-\bar{u_1}(x_1,x_2))(\bar{u_1}(x_1,x_2)-U_2)\,dx_2}
\end{equation}
The definition in equation (\ref{eq:mmt}) is based on Reynolds averaging instead of Favre averaging and is most commonly used for incompressible flows.

The definition of shear layer thickness, which is effectively the cross-stream width of the mixing layer, is based on the distance between two points, denoted as $x_{2,1}$ and $x_{2,2}$, located at the two edges of the mixing layer and defined by (\citealt{livescu20,kim20}): 
\begin{equation}
    U(x_1,x_{2,1})=(U_1-\alpha\Delta U)
\end{equation}
\begin{equation}
    U(x_1,x_{2,2})=(U_2+\alpha\Delta U)
\end{equation}
In this work, $\alpha=0.1$ is used, denoting 10\% of velocity difference ($\Delta U= U_1-U_2$).  

As mentioned above, the literature includes various other definitions of shear layer thickness that can be used, but they generally tend to exhibit more fluctuations compared to the integral thicknesses discussed here (\citealt{livescu20}). For example, the vorticity thickness is computed from the gradient of the mean velocity profiles as (\citealt{elliot90,livescu20}):
\begin{equation}
    \delta_\omega=\frac{\Delta U}{max|\frac{d\bar{u_1}}{dx_2}|}
\end{equation}

It should be noted that the vorticity thickness can occasionally lead to inaccurate predictions since it relies on a very small segment of the shear layer, where the velocity gradient is most pronounced. If significant asymmetry is present within the shear layer, it is better not to use this definition to avoid any misleading representation of layer growth (\citealt{livescu20}).

\section{Computational Approach}\label{sec:computational approach}
\subsection{Initial and Boundary Conditions} \label{sec:bc}
Direct numerical simulations have been performed to simulate the shear layer in spatial configuration. In all simulated cases, the Reynolds number, based on the inflow vorticity thickness, is 640. The molecular Prandtl and Schmidt numbers are $Pr=Sc=0.72$. The supersonic inflow is specified by a tangent hyperbolic profile for the mean streamwise velocity as:
\begin{equation}
    \bar{u}_1(x_2)=\frac{U_1+U_2}{2}+\frac{U_1-U_2}{2}tanh(\frac{2x_2}{\delta_{\omega,0}}),
\end{equation}
where the initial vorticity thickness, $\delta_{w,0}$, represents the initial thickness of the interface, and $\bar{u}_2$=0 and $\bar{u}_3$=0. The hyperbolic tangent profile is the most widely used function in shear layer simulations (\citealt{riley86,pantano02,olson11,obrien14,almagro17,jaberi10,livescu20}). 
The temperature and density are kept uniform at the inlet with a uniform pressure of $1/\gamma {M}^2$. The mean inlet velocities are perturbed by a random inflow perturbation to promote instability. The growth rates of mixing layers can be influenced in the early stages by the details of the disturbances (\citealt{fathali08}), but we used this method to avoid long-lived large-scale structures and establish self-similar growth as quickly as possible. 
%The scalar profile follows a similar trend to the mean streamwise velocity having a value of 1 in free stream $1$ and 0 in free stream $2$. {\color{red} It doesn't look you have any results involving the scalar. If that is the case, any reference to a sclaar should be removed in the journal version.}

Supersonic outflow boundary conditions are used at the outflow, while periodic boundary conditions are applied in the spanwise directions. To prevent wave reflections back into the computational domain, non-reflective boundary conditions are employed in the cross-stream direction, following the approach outlined by \cite{lele92}. Additionally, the tangential stresses $\tau_{12}$, $\tau_{13}$, and normal heat flux vectors are set to have zero spatial derivatives at cross-stream boundaries. 

\subsection{Numerical Methods} \label{sec:methods} 

In this study, sixth-order compact finite differences were utilized to compute viscous and diffusive fluxes (\citealt{lele92}), while the seventh-order MP scheme was used to evaluate inviscid fluxes (\citealt{jaberi10}). \cite{jaberi09} demonstrated that the high-order MP scheme provides sufficient accuracy to perform turbulent simulations in high-speed flows. The MP scheme involves three main steps: (i) transform variables into the local characteristics space, (ii) construct interfacial fluxes using a higher-order method (7th order in this work) and (iii) calculate interfacial fluxes in the physical coordinate system using Roe-averaged eigenvalues (\citealt{jaberi10}). Explicit integration of the governing equations was carried out using the third-order Runge-Kutta scheme in time (\citealt{shu88}) with the usual CFL conditions. %The CFL number used in the simulations was {\color{red} CFL.}

\subsection{Computational Accuracy} \label{sec: accuracy}
\subsubsection{Domain Size} \label{sec:domain}

\begin{figure}
     \centering
     \begin{subfigure}[b]{0.49\textwidth}
         \includegraphics[width=\textwidth]{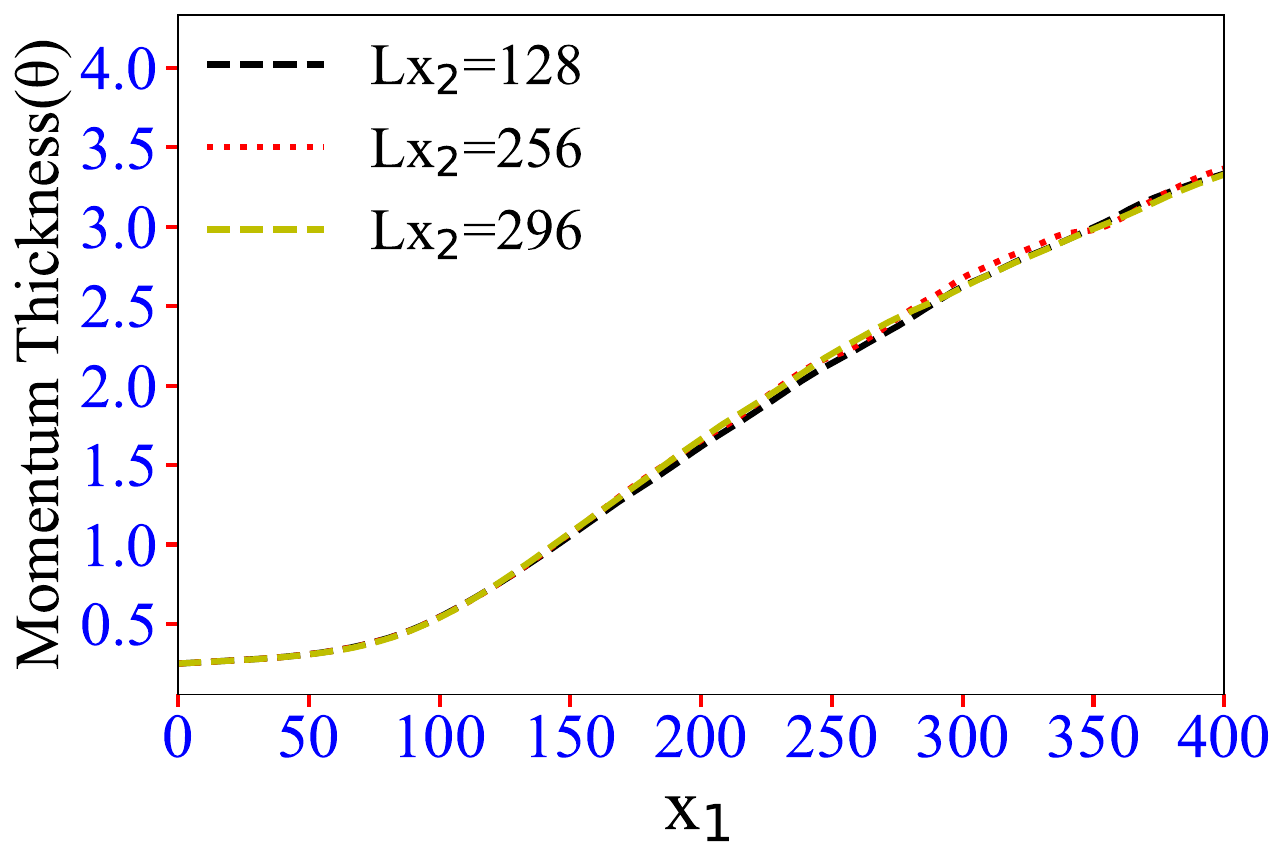}
         \subcaption{}
     \end{subfigure}
     \centering
     \begin{subfigure}[b]{0.49\textwidth}
         \centering
         \includegraphics[width=\textwidth]{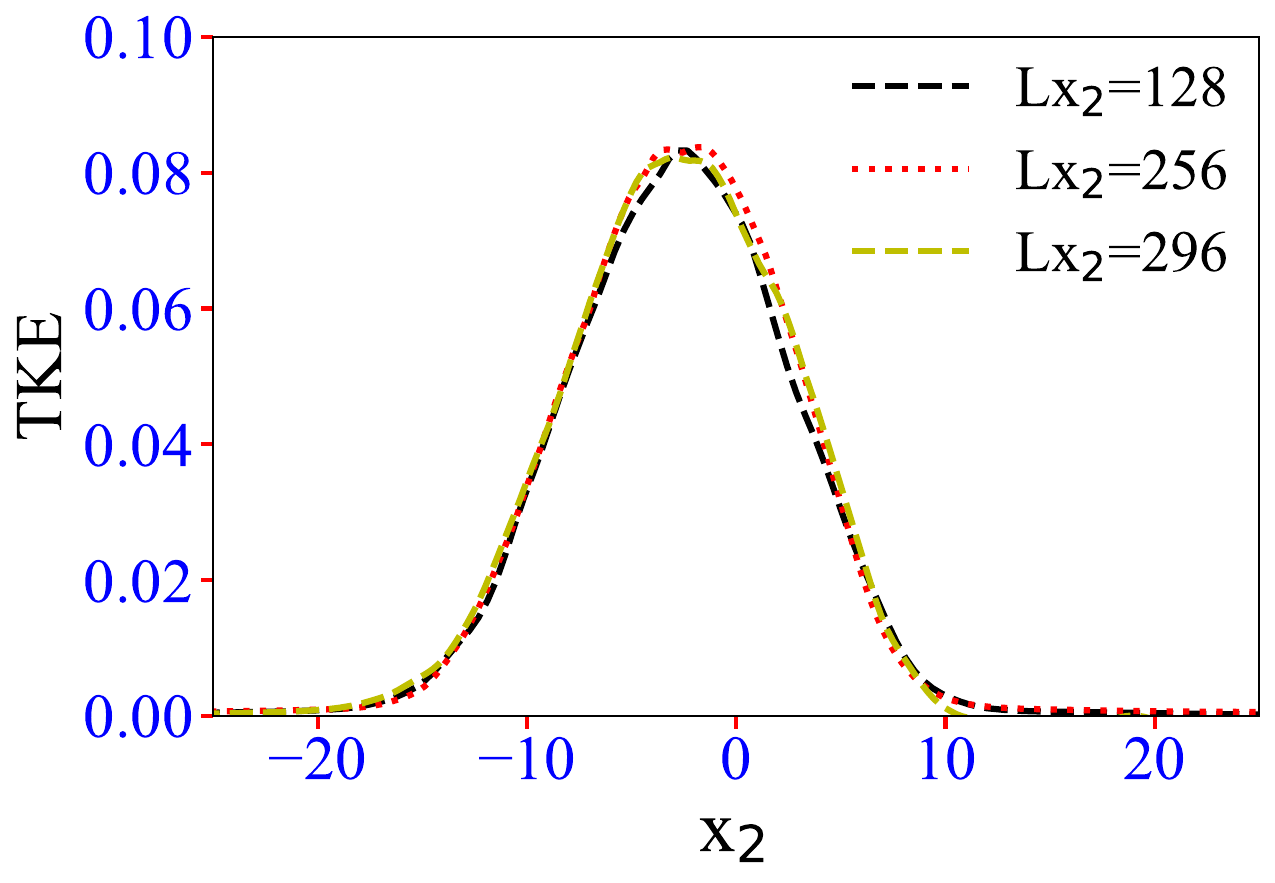}
         \subcaption{}
     \end{subfigure}
     \caption{Transverse domain size independency for (a) momentum thickness (b) TKE }
     \label{fig:domain1}
\end{figure}

The evolution of momentum thickness and turbulent kinetic energy is presented in figure~\ref{fig:domain1} for three different transverse domain sizes at $M_c$ 1.2. As shown in the figure, the results for these three domain sizes exhibit acceptable convergence. Since the results of the domain size $L_{x2}$=256 and $L_{x2}$=296 are almost identical, $L_{x2}$=256 was used in most of the simulated cases. A slightly larger $L_{x2}$=268 was used for one of the cases to reach the same convergence level as for the rest of the cases. 

\begin{figure}
     \centering
     \begin{subfigure}[b]{0.48\textwidth}
         \centering
         \includegraphics[width=\textwidth]{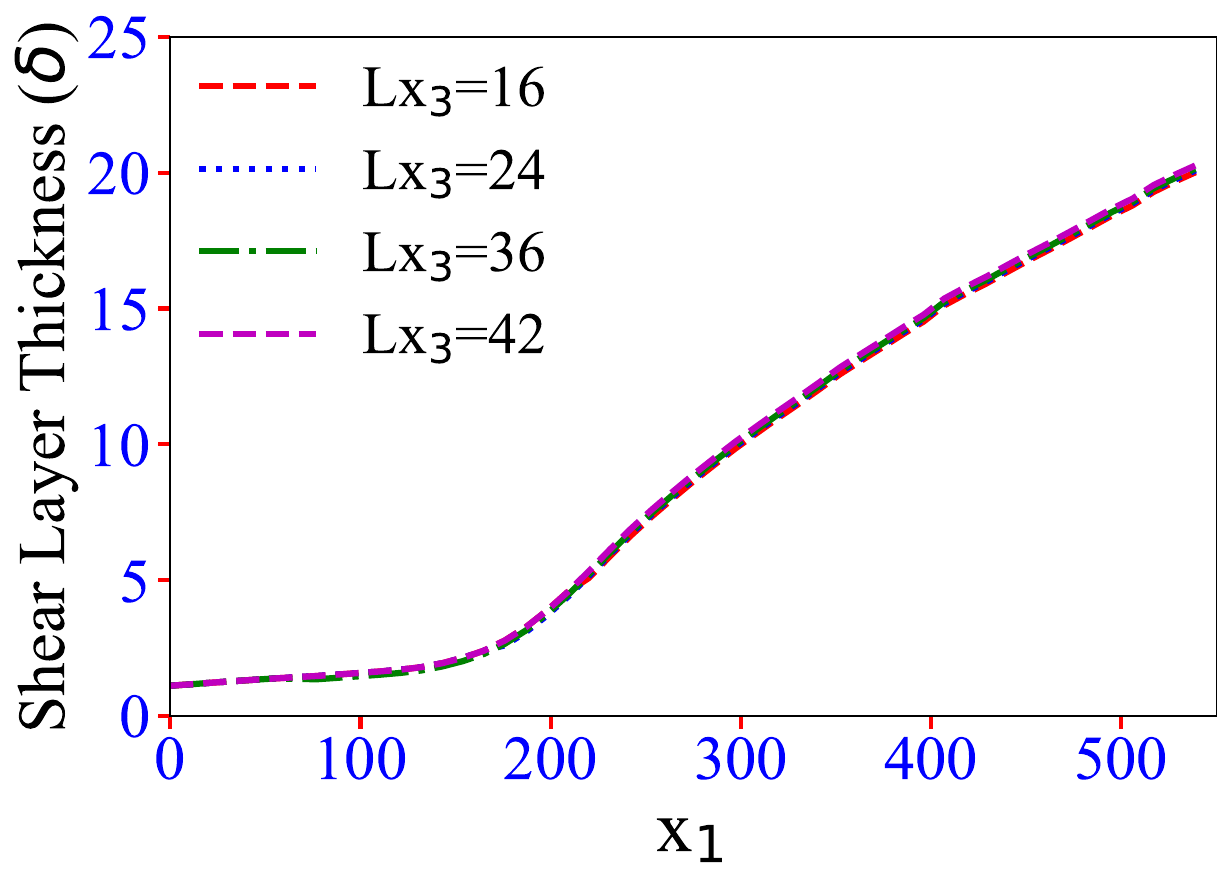}
         \subcaption{}
     \end{subfigure}
     \centering
     \begin{subfigure}[b]{0.48\textwidth}
         \centering
         \includegraphics[width=\textwidth]{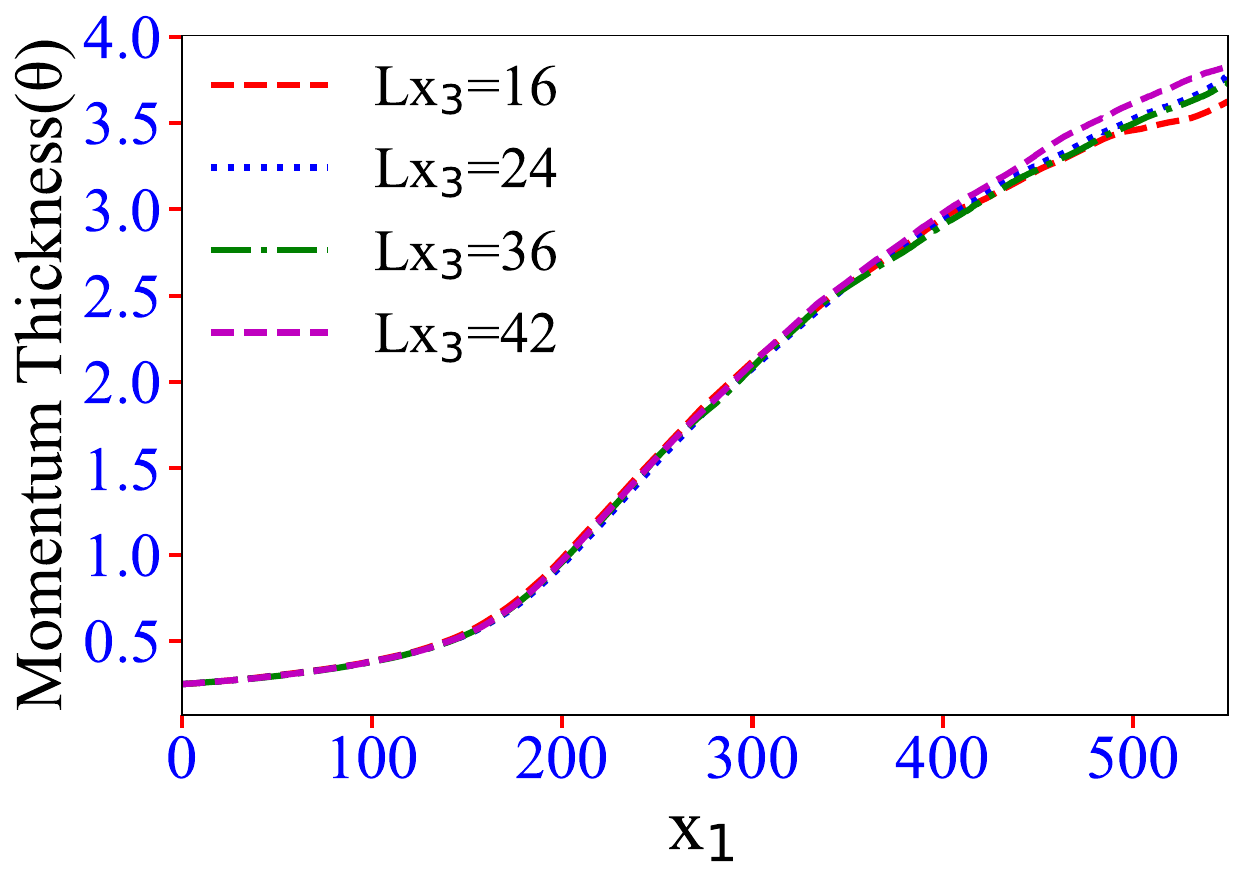}
         \subcaption{}
     \end{subfigure}
     \caption{ Spanwise domain size independency for (a) shear layer thickness (b) momentum thickness }
     \label{fig:domain2}
\end{figure}

The domain size in the spanwise direction is also important for minimizing the effect of periodicity on turbulence statistics. Four different domain sizes in the spanwise direction are compared in figures~\ref{fig:domain2} (a) and (b) for the shear layer and momentum thicknesses. Based on these comparisons, the domain sizes in the spanwise direction used in the simulations was $L_{x3}$= 42. Table~\ref{tab:simulated case} below presents all the domain size values for the simulation cases presented here. 

\subsubsection{Grid Resolution} \label{sec:grid}

\begin{figure}
     \centerline
     \centering
     \begin{subfigure}[b]{0.49\textwidth}
         \centering
         \includegraphics[width=\textwidth]{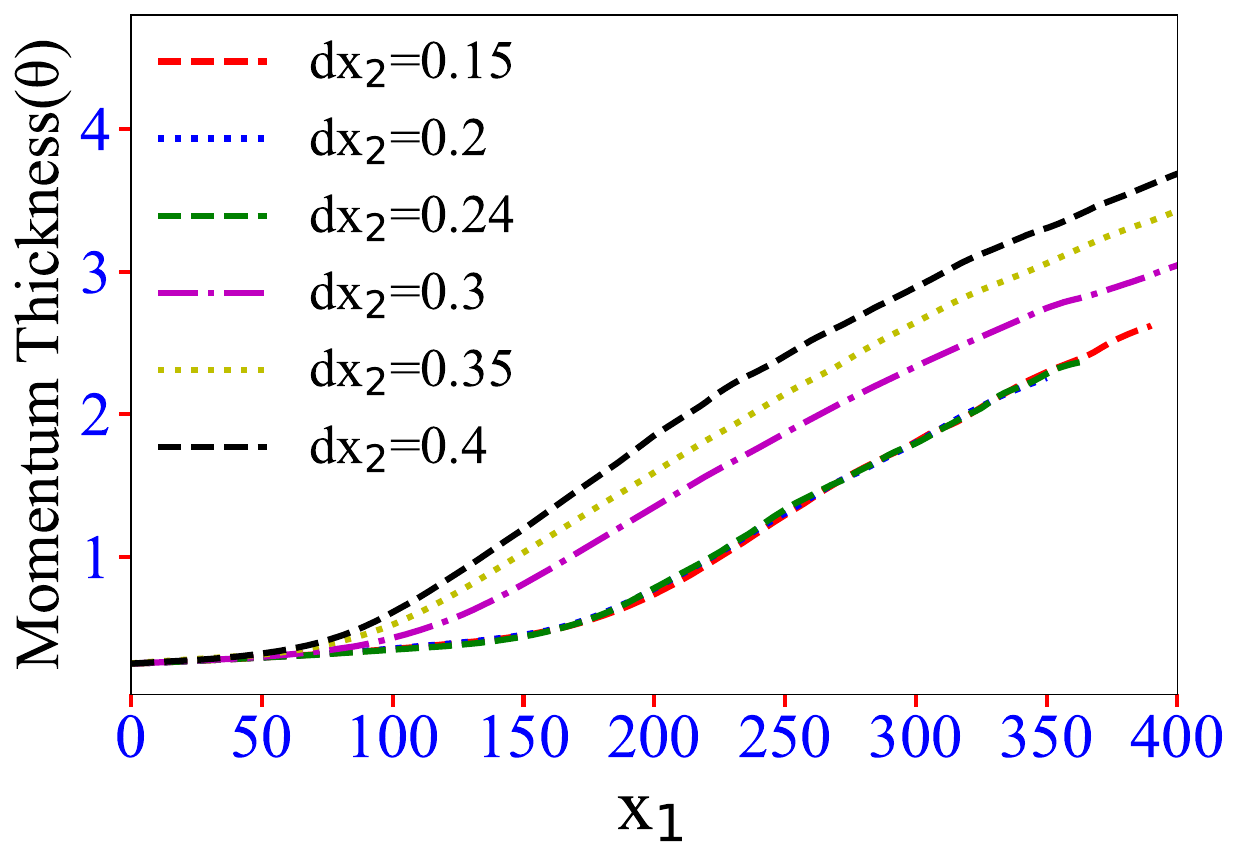}
         \subcaption{}
     \end{subfigure}
     \centering
     \begin{subfigure}[b]{0.49\textwidth}
         \centering
         \includegraphics[width=\textwidth]{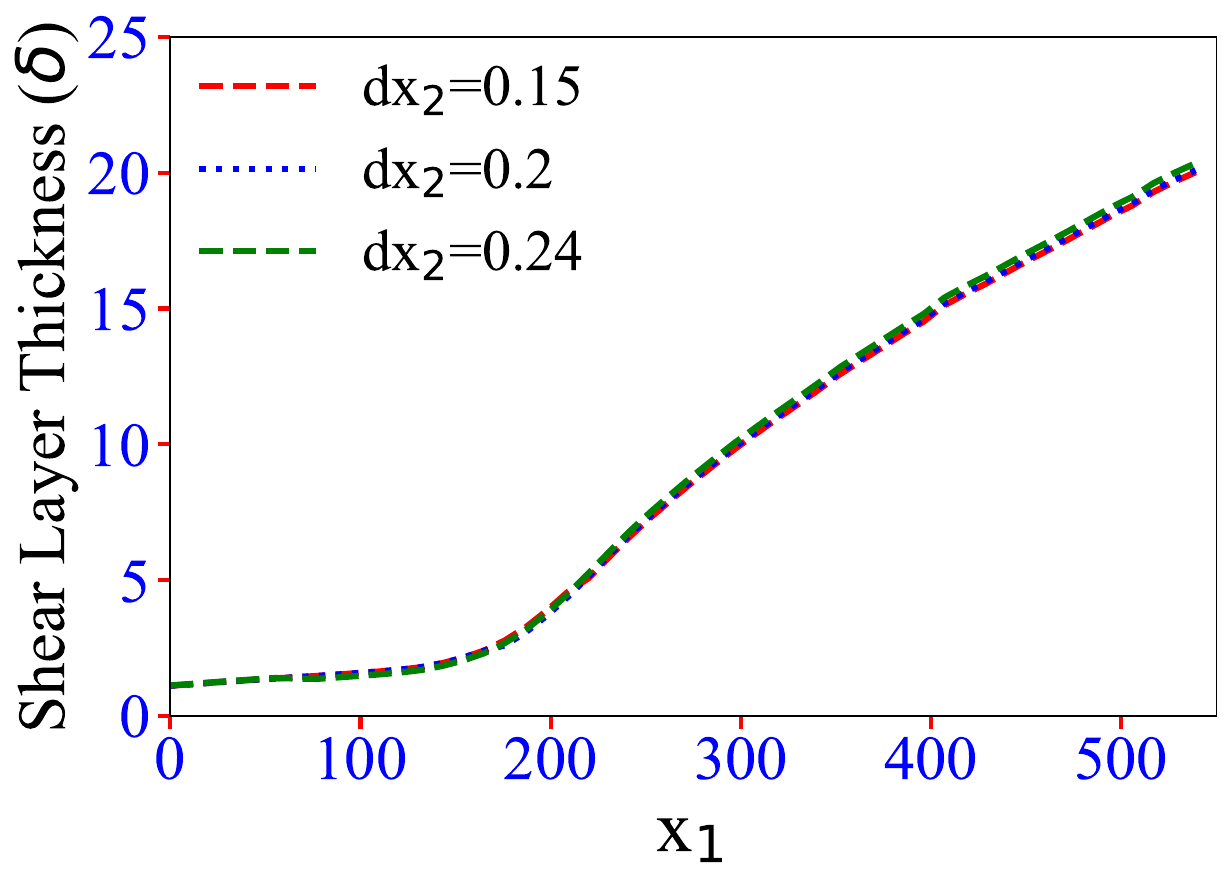}
         \subcaption{}
     \end{subfigure}
     \centering
     \begin{subfigure}[b]{0.50\textwidth}
         \centering
         \includegraphics[width=\textwidth]{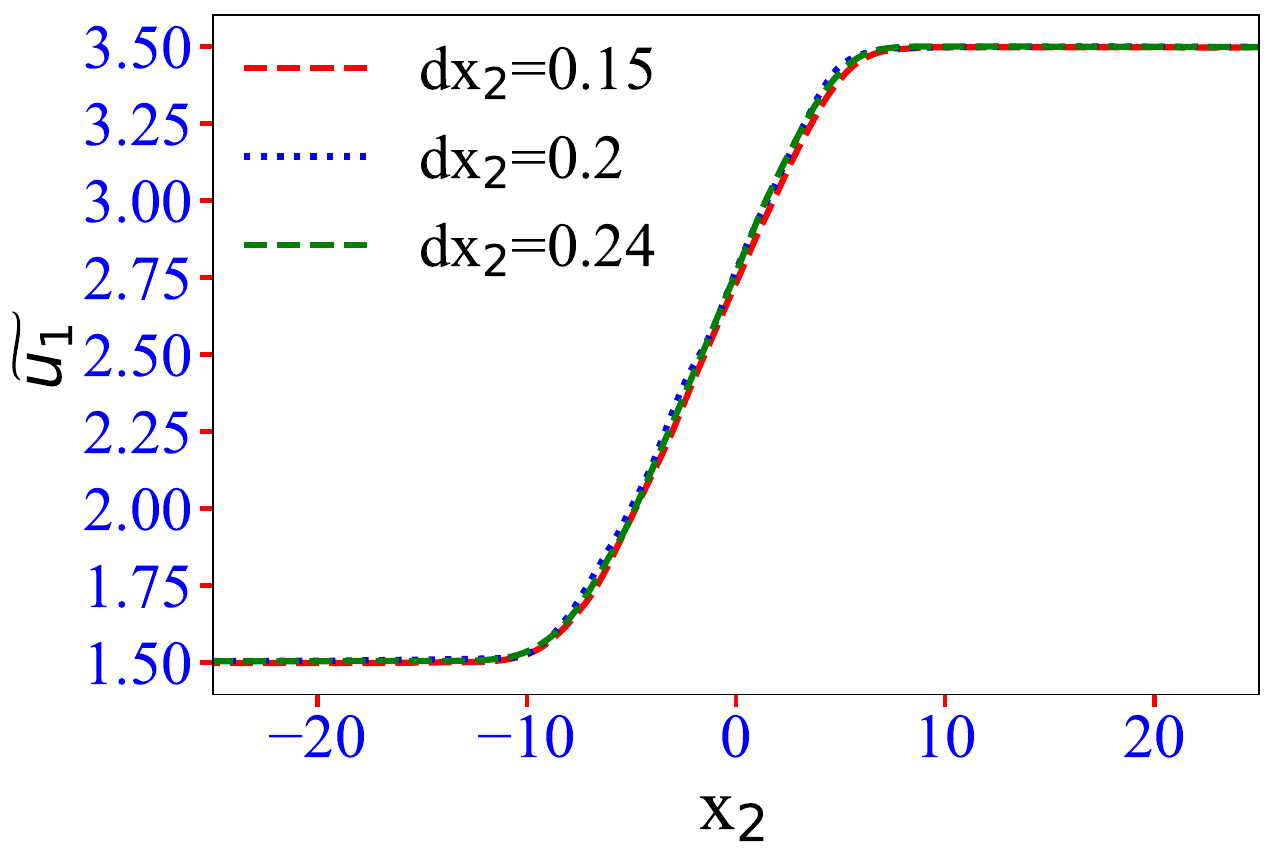}
         \subcaption{}
     \end{subfigure}
     \centering
     \begin{subfigure}[b]{0.49\textwidth}
         \centering
         \includegraphics[width=\textwidth]{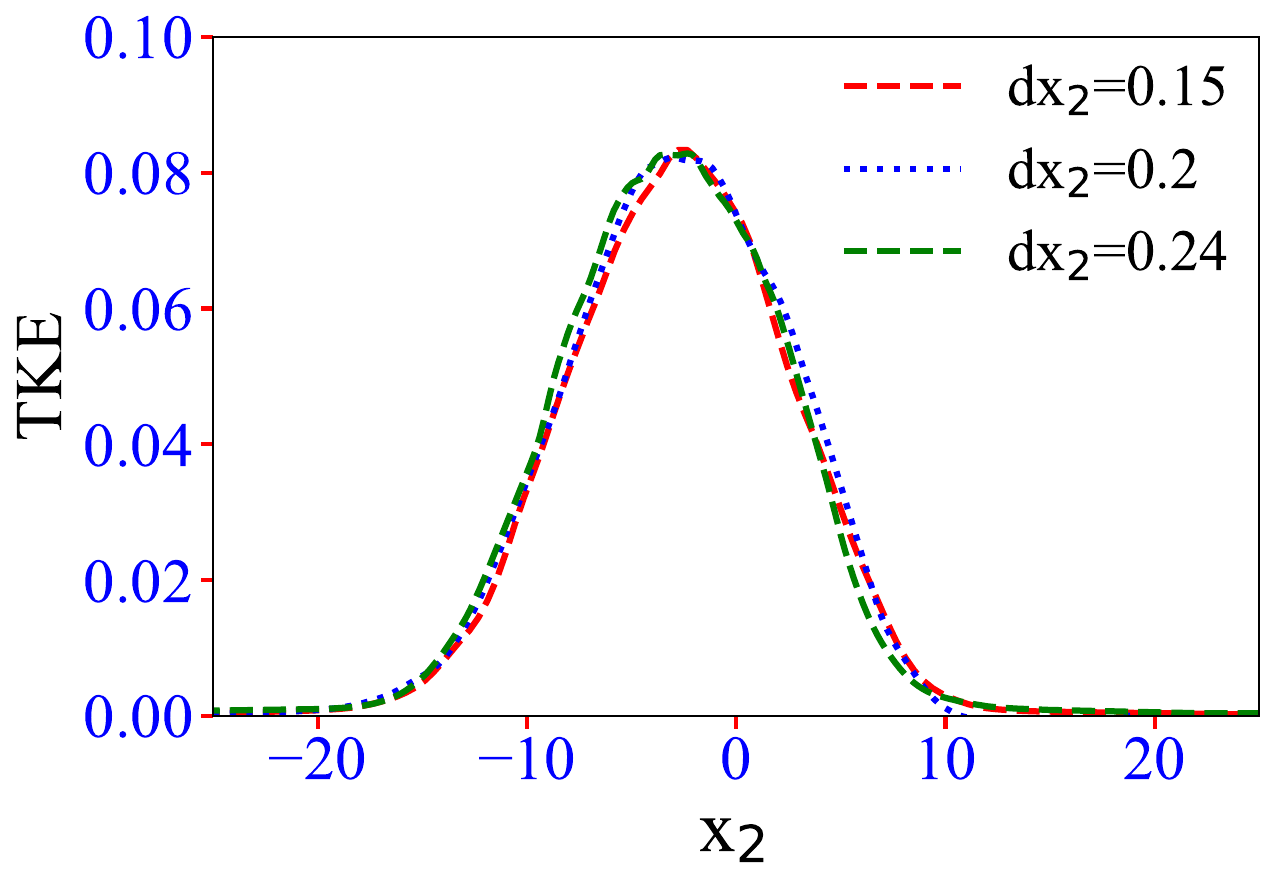}
         \subcaption{}
     \end{subfigure}
     \centering
     \begin{subfigure}[b]{0.485\textwidth}
         \centering
         \includegraphics[width=\textwidth]{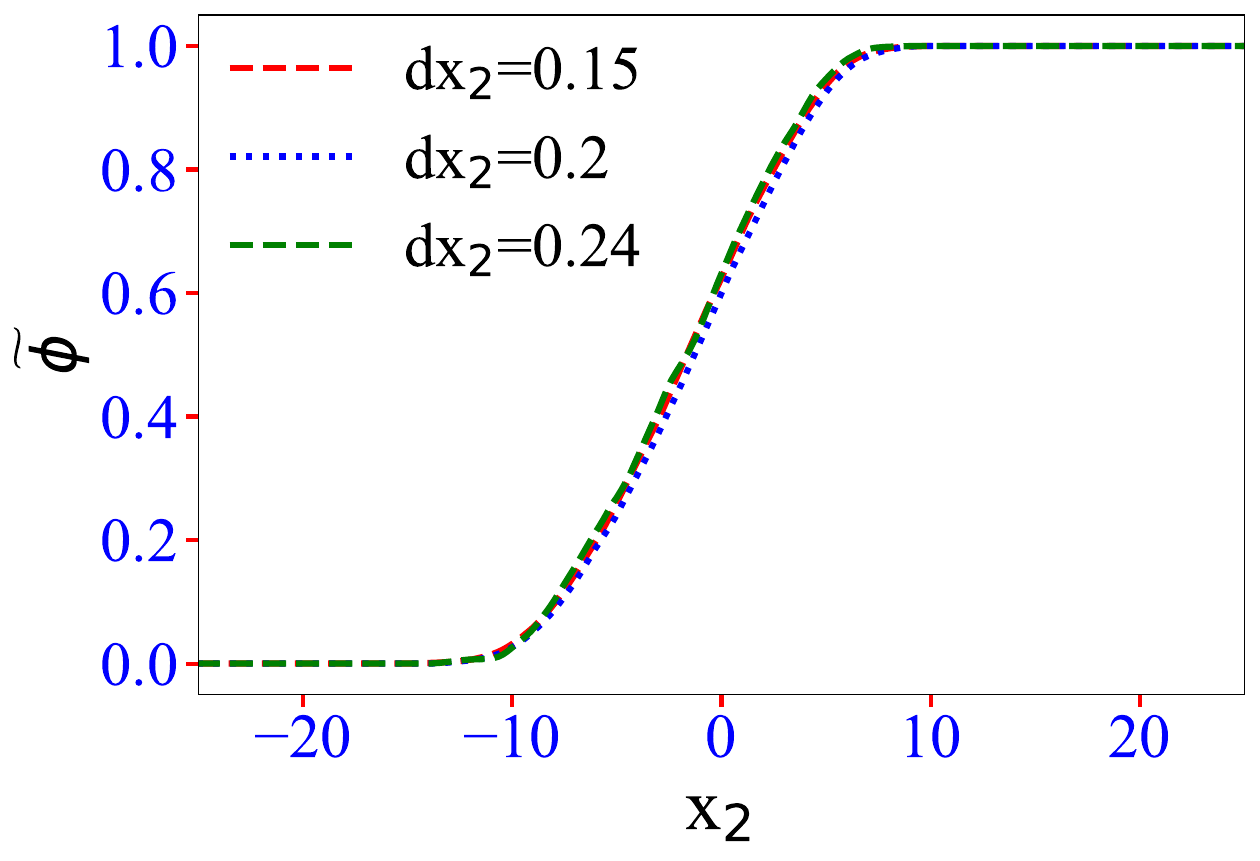}
         \subcaption{}
     \end{subfigure}
     \centering
     \begin{subfigure}[b]{0.505\textwidth}
         \centering
         \includegraphics[width=\textwidth]{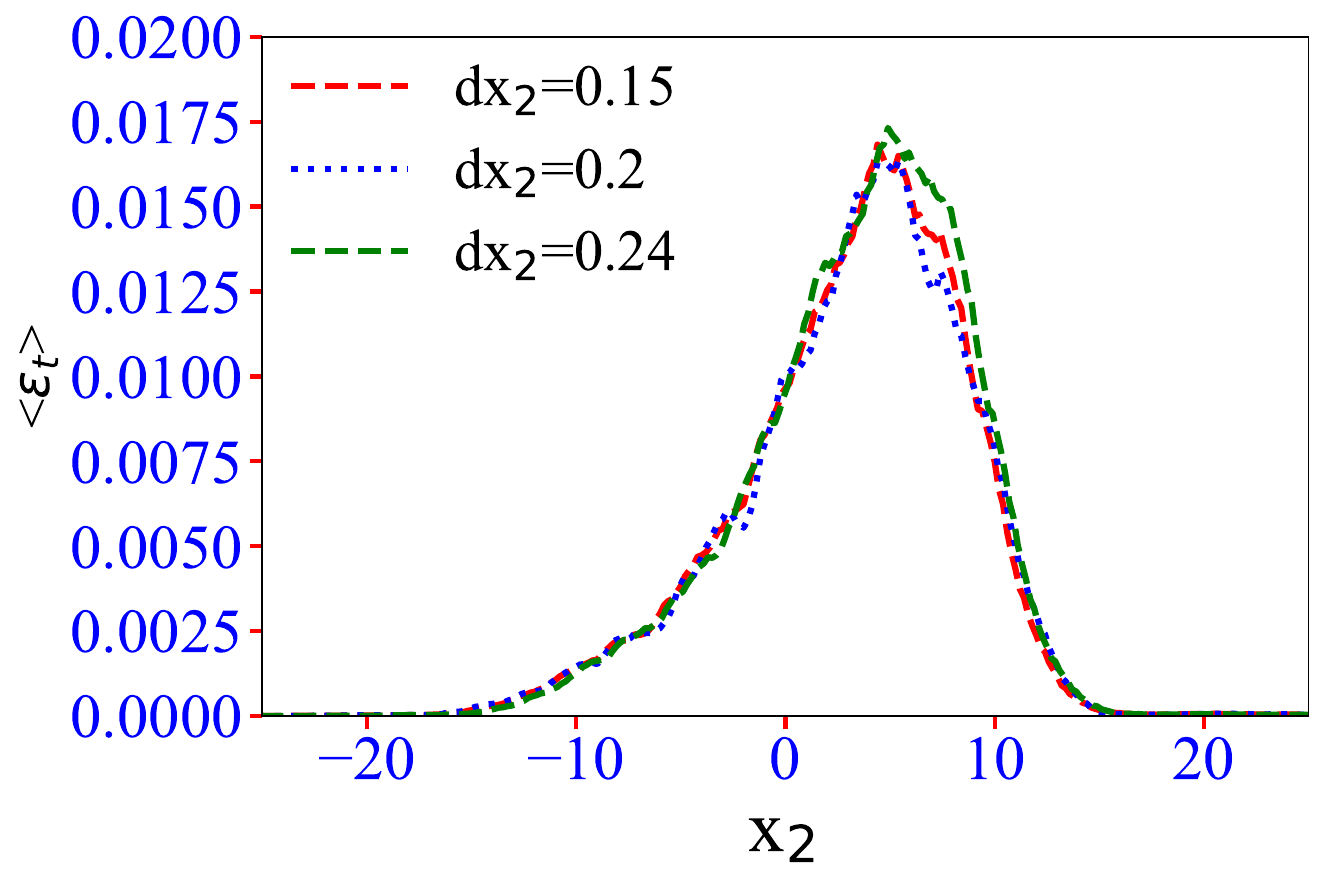}
         \subcaption{}
     \end{subfigure}
     \caption{Grid convergence for (a) Momentum thickness (b) Shear Layer thickness (c) Mean streamwise velocity (d) TKE (e) Mean scalar (f) Mean dissipation}
     \label{fig:grid}
\end{figure}

 Uniform grid resolution was used in all three directions. The variation of momentum thickness with grid resolution is presented in figure~\ref{fig:grid} (a). Momentum thickness is found to be more sensitive to grid resolution, particularly with respect to the grid size in the transverse direction. From figure~\ref{fig:grid} (a), for the finest grid sizes presented, $dx_2=0.15$ and $dx_2=0.2$, the results of the momentum thickness are very close. The shear layer thickness, mean streamwise velocity, turbulent kinetic energy, mean scalar, and mean dissipation evolutions are also very close for the transverse grid sizes of 0.24, 0.2, and 0.15. In our simulations, a grid size of 0.2 is employed in all three directions, as it is computationally affordable and, with this grid size, the statistics considered are accurately calculated. 

\subsection{Cases Simulated} \label{sec:cases}

This study systematically examines the impacts of $M_c$ and $\lambda$ by considering two distinct sets of simulations. Using the first set, we investigate the influence of $M_c$ while maintaining a constant velocity parameter of $\lambda$=0.4. Using the second set, we explore the effects of $\lambda$ while keeping the convective Mach number constant at M$_c$=1.2. The reference case for our study, denoted as case A04M12 in table \ref{tab:simulated case}, has $M_c=1.2$ and $\lambda=0.4$.

\begin{table}
    \centering
    \begin{adjustbox}{max width=\textwidth}
    \begin{tabular}{|c|c|c|c|c|c|c|c|c|c|c|c|}
    \hline
     Cases & Domain Size  & Grid Resolution & $\lambda$ & M$_c$ & U$_1$ & U$_2$ & $\Delta U$ & r & $\rho_0$ & M$_1$ & M$_2$  \\[3pt]
    \hline
    A02M12 & 800x256x42 & 4000x1280x210 & 0.2 & 1.2 & 3.0 & 2.00 & 1.0 & 0.667 & 4.0 & 7.2 & 4.8 \\
    \hline
    A03M12 & 700x256x42 & 3500x1280x210 & 0.3 & 1.2 & 3.25 & 1.75 & 1.5 & 0.538 & 1.77 & 5.2 & 2.8 \\
    \hline
    \textbf{A04M12} & \textbf{550x256x42} & \textbf{2750x1280x210} & \textbf{0.4} & \textbf{1.2} & \textbf{3.50} & \textbf{1.50} & \textbf{2.0} & \textbf{0.428} & \textbf{1.0} & \textbf{4.2} & \textbf{1.8} \\
    \hline
    A05M12 & 500x268x42 & 2500x1340x210 & 0.5 & 1.2 & 3.75 & 1.25 & 2.5 & 0.333 & 0.64 & 3.6 & 1.2 \\
    \hline
    A04M08 & 500x256x42 & 2500x1280x210 & 0.4 & 0.8 & 3.50 & 1.50 & 2.0 & 0.428 & 1.0 & 2.8 & 1.2 \\
    \hline
    A04M10 & 550x256x42 & 2750x1280x210 & 0.4 & 1.0 & 3.50 & 1.50 & 2.0 & 0.428 & 1.0 & 3.5 & 1.5 \\
    \hline
    A04M14 & 600x256x42 & 3000x1280x210 & 0.4 & 1.4 & 3.50 & 1.50 & 2.0 & 0.428 & 1.0 & 4.9 & 2.1 \\
    \hline
    A04M16 & 600x256x42 & 3000x1280x210 & 0.4 & 1.6 & 3.50 & 1.50 & 2.0 & 0.428 & 1.0 & 5.6 & 2.4 \\
    \hline
    \end{tabular}
    \end{adjustbox}
    \caption{Details of the simulated cases}
    \label{tab:simulated case}
\end{table}

\begin{figure}
     \centering
     \begin{subfigure}[b]{0.49\textwidth}
         \centering
         \includegraphics[width=\textwidth]{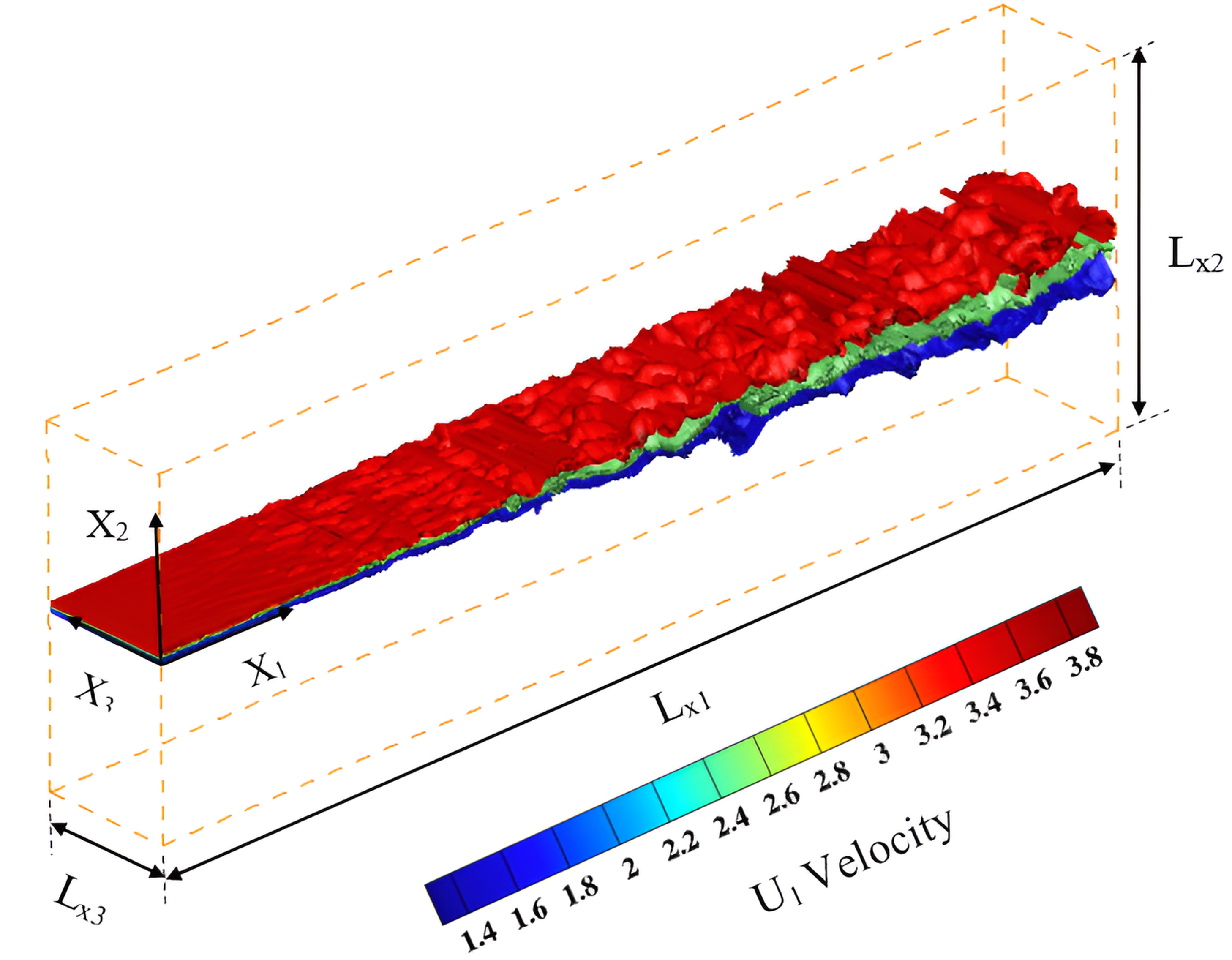}
         \caption{}
     \end{subfigure}
     \hfill
     \centering
     \begin{subfigure}[b]{0.49\textwidth}
         \centering
         \includegraphics[width=\textwidth]{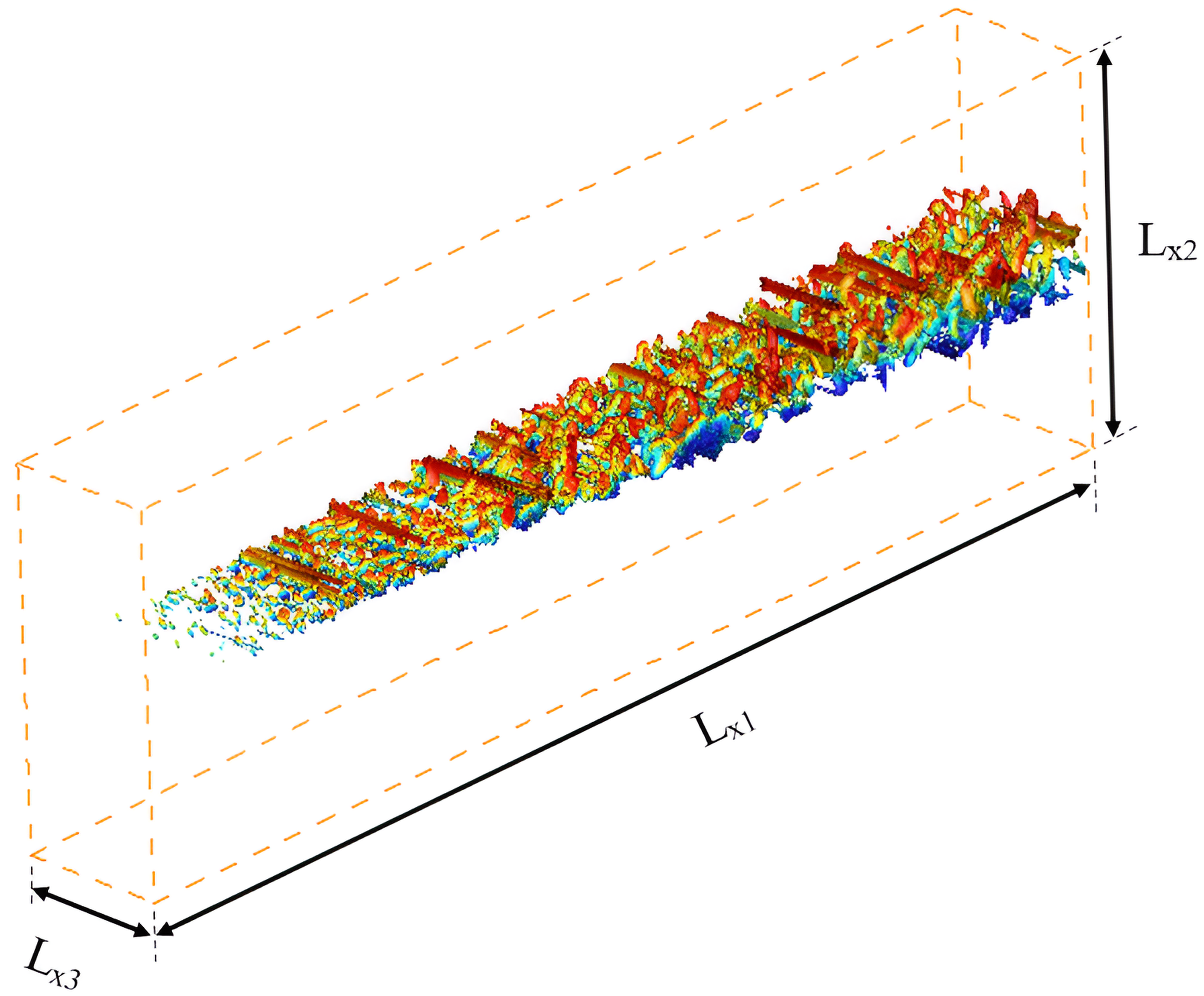}
         \caption{}
     \end{subfigure}
     \centering
     \begin{subfigure}[b]{1.0\textwidth}
         \centering
         \includegraphics[width=\textwidth]{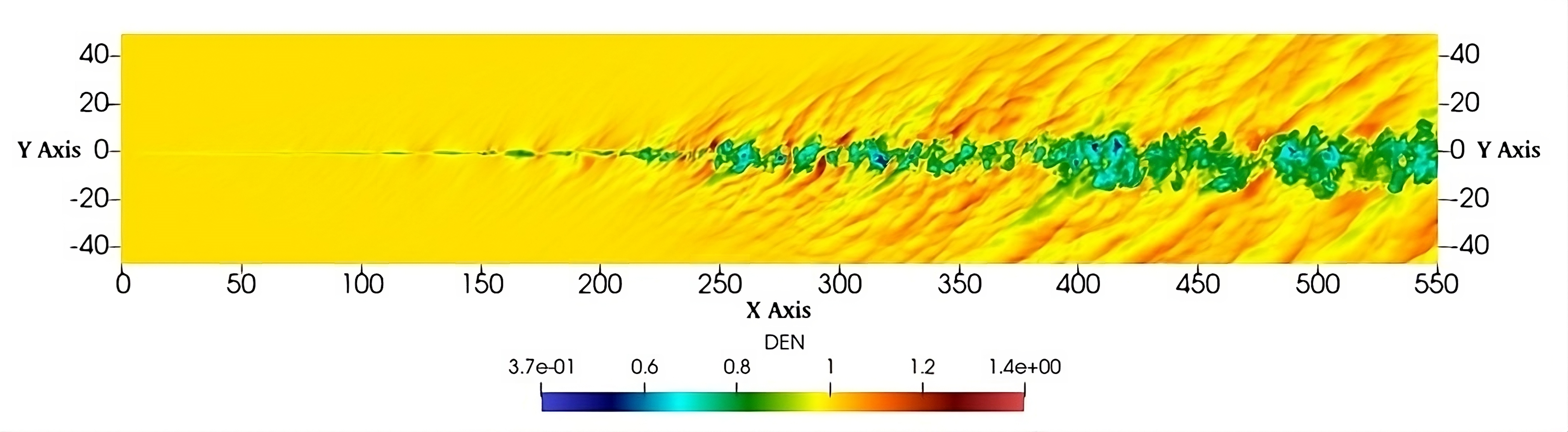}
         \caption{}
     \end{subfigure}
     \centering
     \begin{subfigure}[b]{1.0\textwidth}
         \centering
         \includegraphics[width=\textwidth]{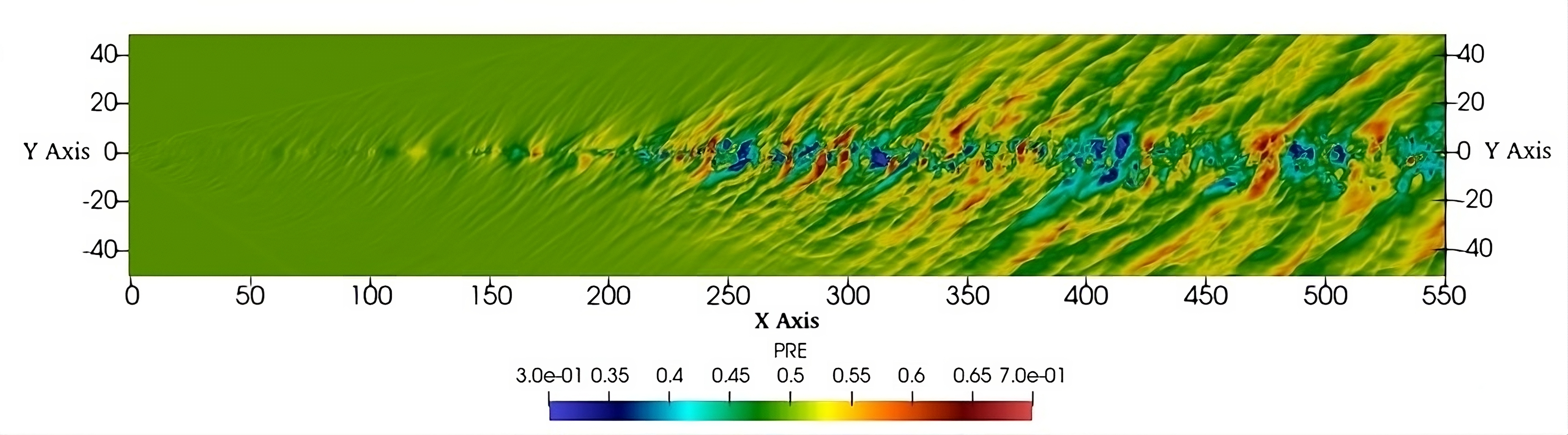}
         \caption{}
     \end{subfigure}
     \caption{3D and 2D contours of the reference case: (a) 3D iso-contours of instantaneous streamwise velocity ($u_1$) (b) 3D iso-contours of Q-criterion colored by streamwise velocity (c) 2D iso-contours of instantaneous density and (d) 2D iso-contours of instantaneous pressure. Figures (c) and (d) are magnified to provide a closer examination of the core of the shear layer. Figures show that the shear layer is confined to a small region of the overall domain and the domain sizes are large enough to avoid significant interactions with the boundaries. }
     \label{fig:3dplot}
\end{figure}

We controlled the non-dimensional speed of sound in the two streams by adjusting the uniform inlet pressure for cases with various $M_c$ values. This was done while keeping $\Delta U$ and $U_c$ constant to maintain a constant $\lambda$ value. In contrast, for cases with different $\lambda$ values, we changed the velocities of the two streams while keeping M$_c$ constant. To achieve this, we adjusted the temperature of the fluids to modify the non-dimensional speed of sound for both free streams. A comprehensive list of all simulated cases with their associated flow parameters is given in table~\ref{tab:simulated case}.

Figure \ref{fig:3dplot} shows 3D and 2D contours of streamwise velocity, density and pressure for the reference case. Figures \ref{fig:3dplot} (c) and (d) present the contours magnified around the core of the shear layer. The  figures highlight that the domain sizes in the streamwise and spanwise directions are large enough for the shear layer to grow and for waves to smoothly leave the domain.

\subsection{Streamwise evolution of mean profiles} \label{sec:mean}

\begin{figure}
     \centerline
     \centering
     \begin{subfigure}[b]{0.395\textwidth}
         \centering
         \includegraphics[width=\textwidth]{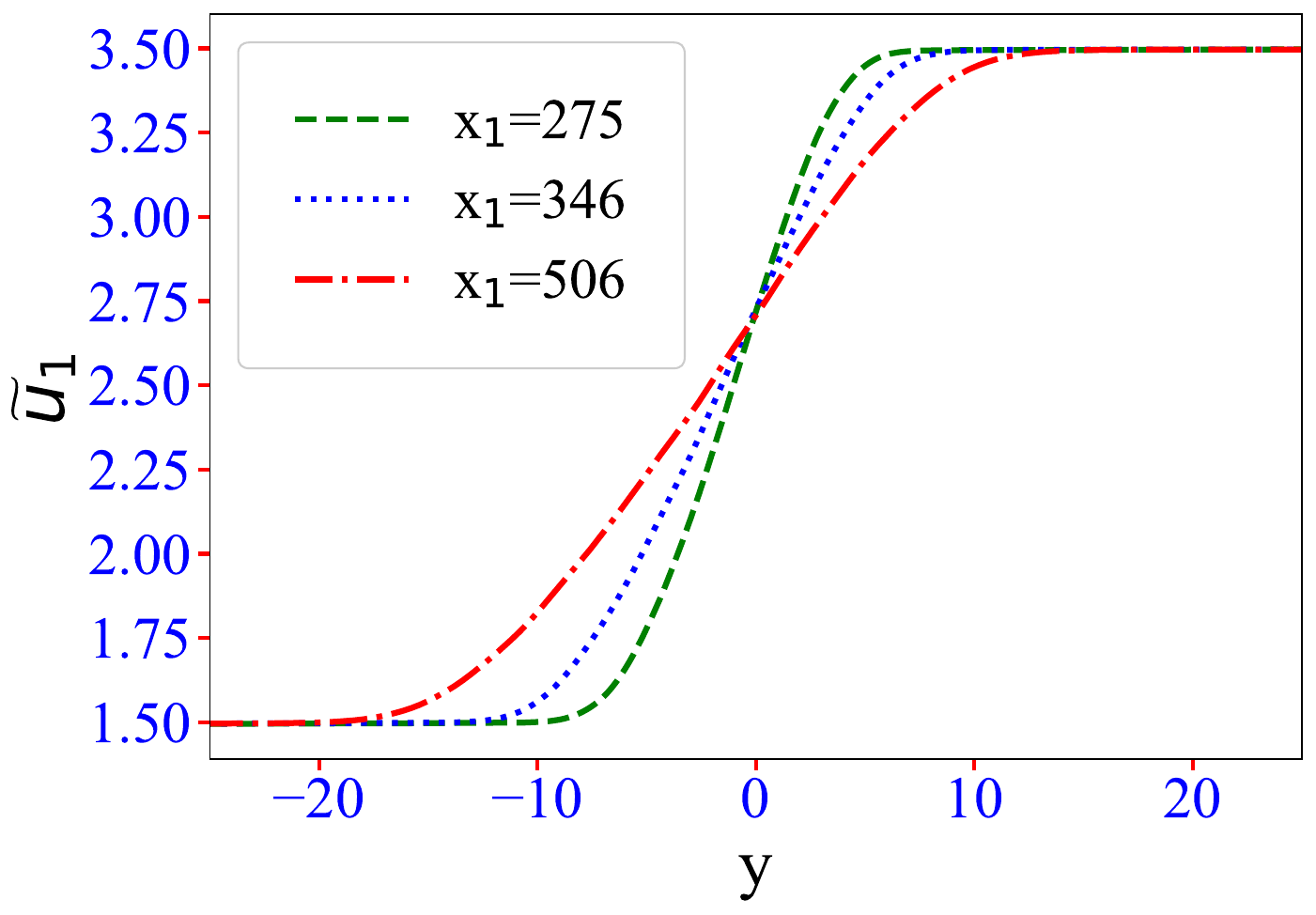}
         \subcaption{}
     \end{subfigure}
     \centering
     \begin{subfigure}[b]{0.39\textwidth}
         \centering
         \includegraphics[width=\textwidth]{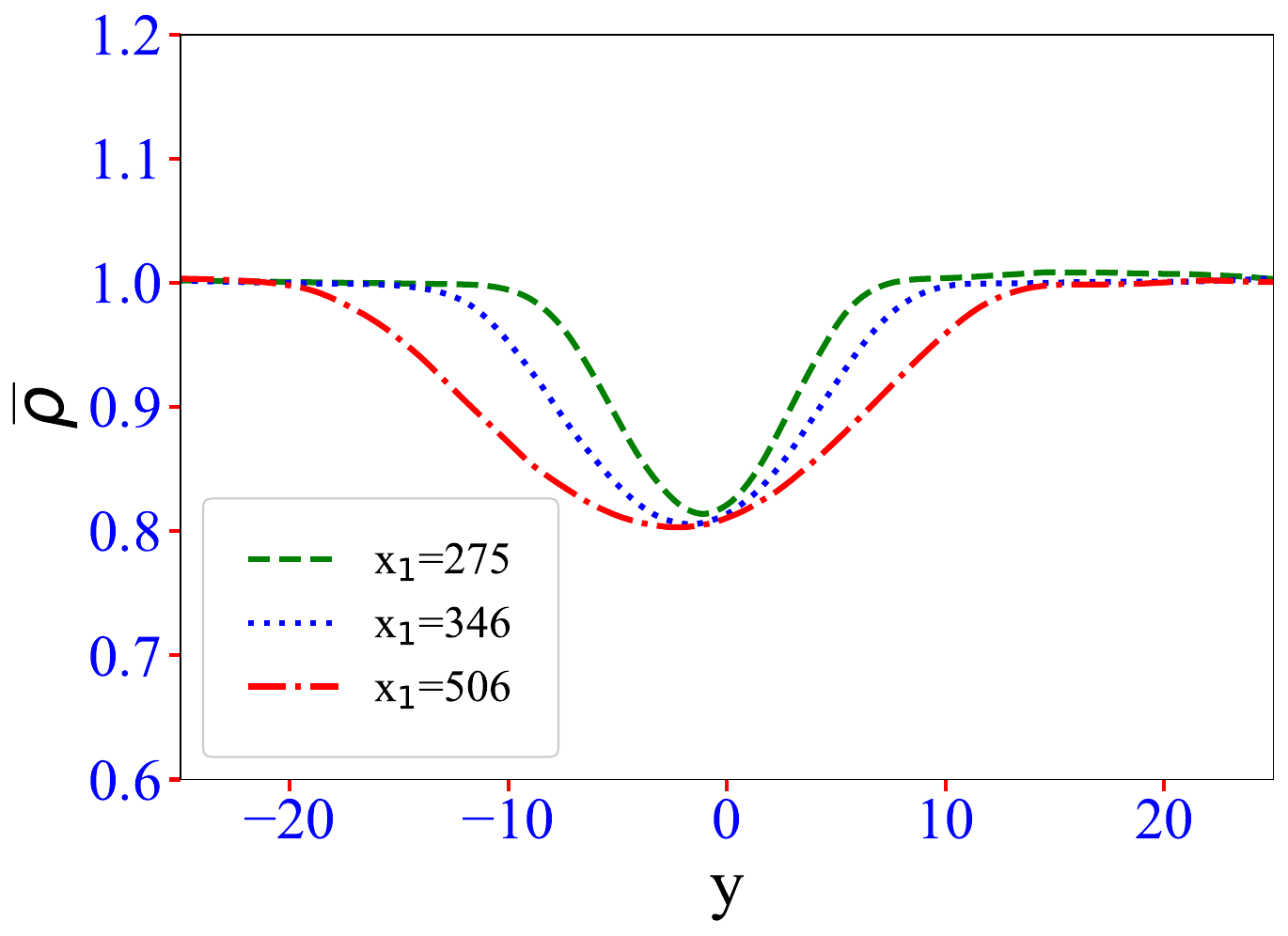}
         \subcaption{}
     \end{subfigure}
     \centering
     \begin{subfigure}[b]{0.40\textwidth}
         \centering
         \includegraphics[width=\textwidth]{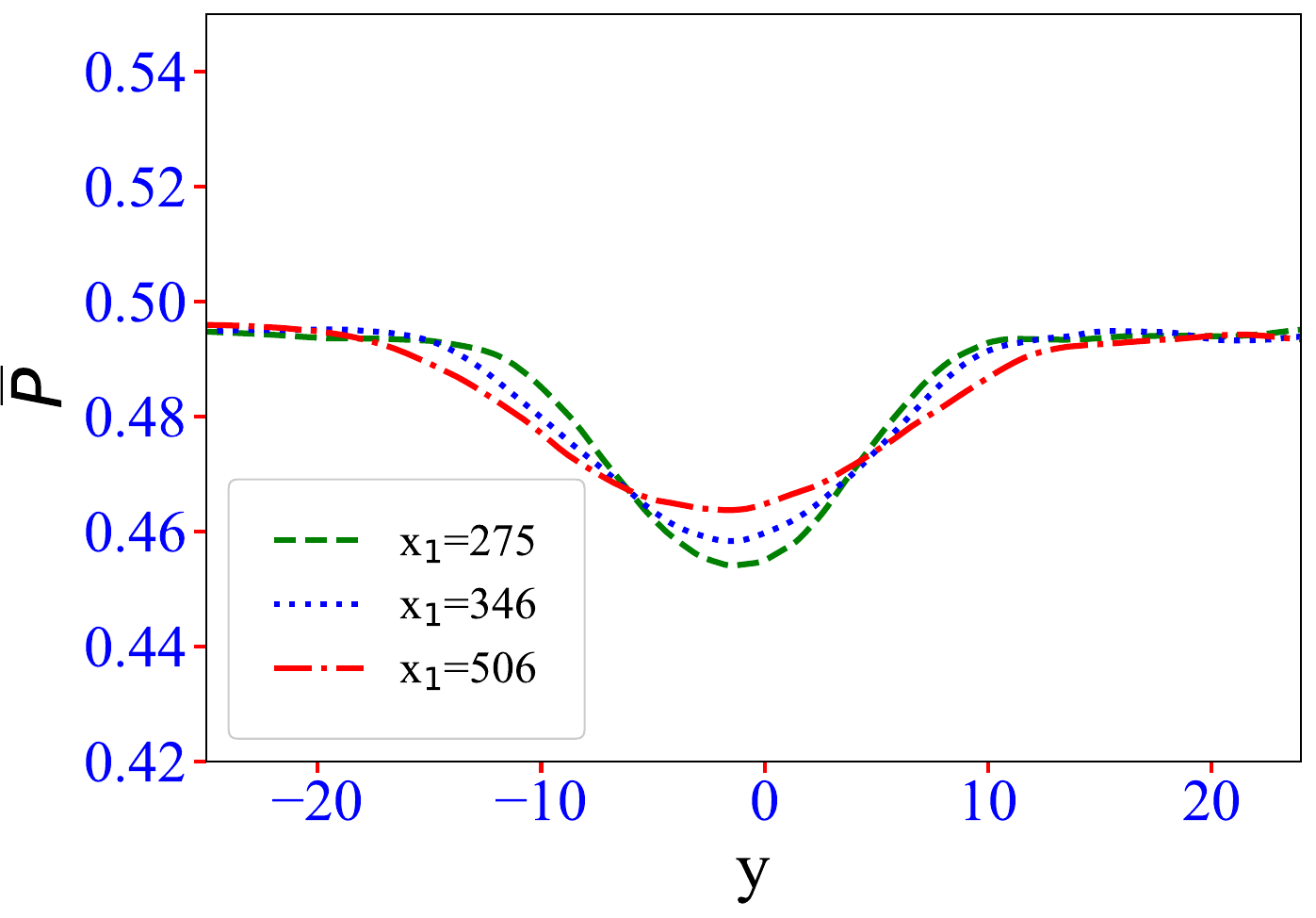}
         \subcaption{}
     \end{subfigure}
     \centering
     \begin{subfigure}[b]{0.41\textwidth}
         \centering
         \includegraphics[width=\textwidth]{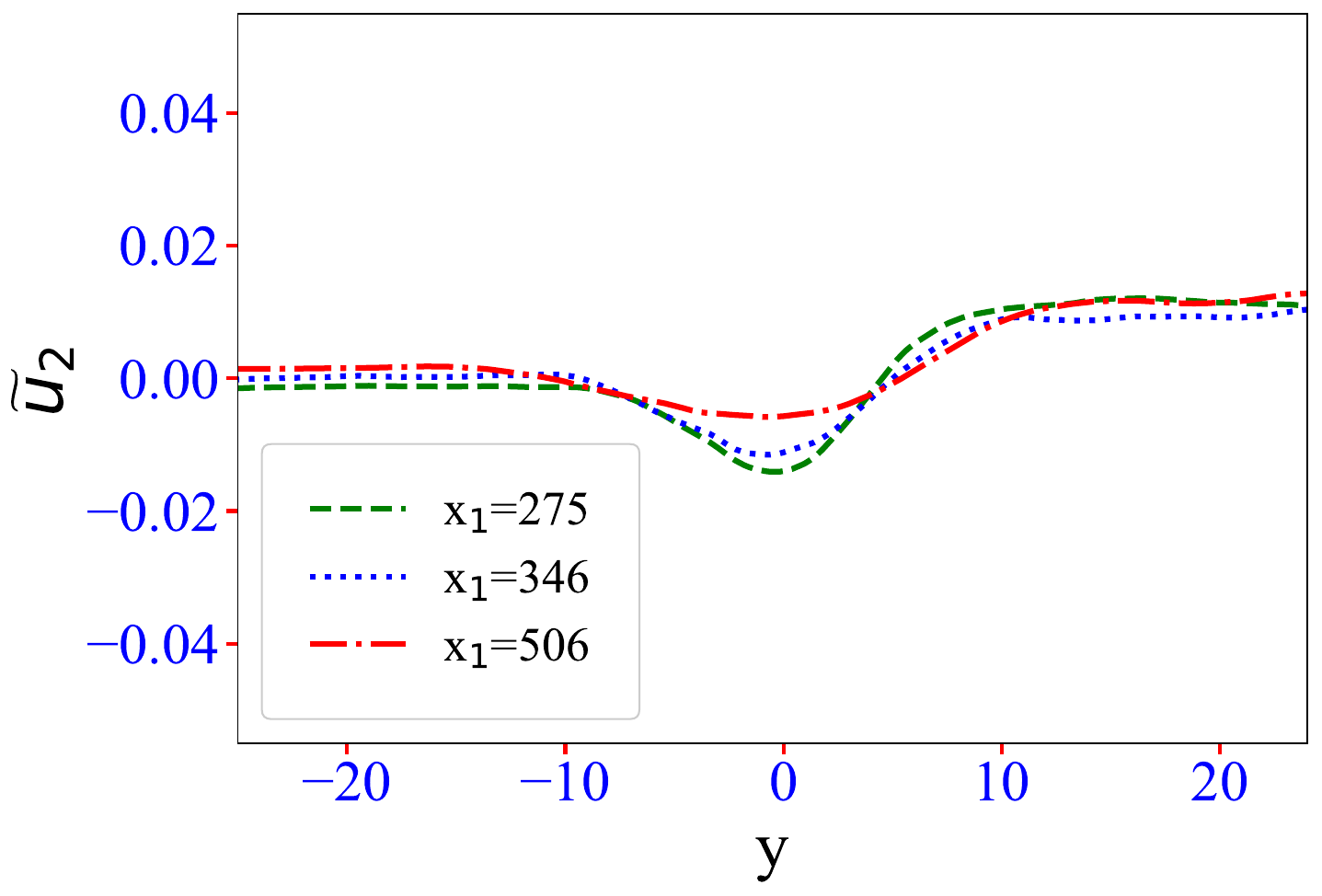}
         \subcaption{}
     \end{subfigure}
     \centering
     \begin{subfigure}[b]{0.39\textwidth}
         \centering
         \includegraphics[width=\textwidth]{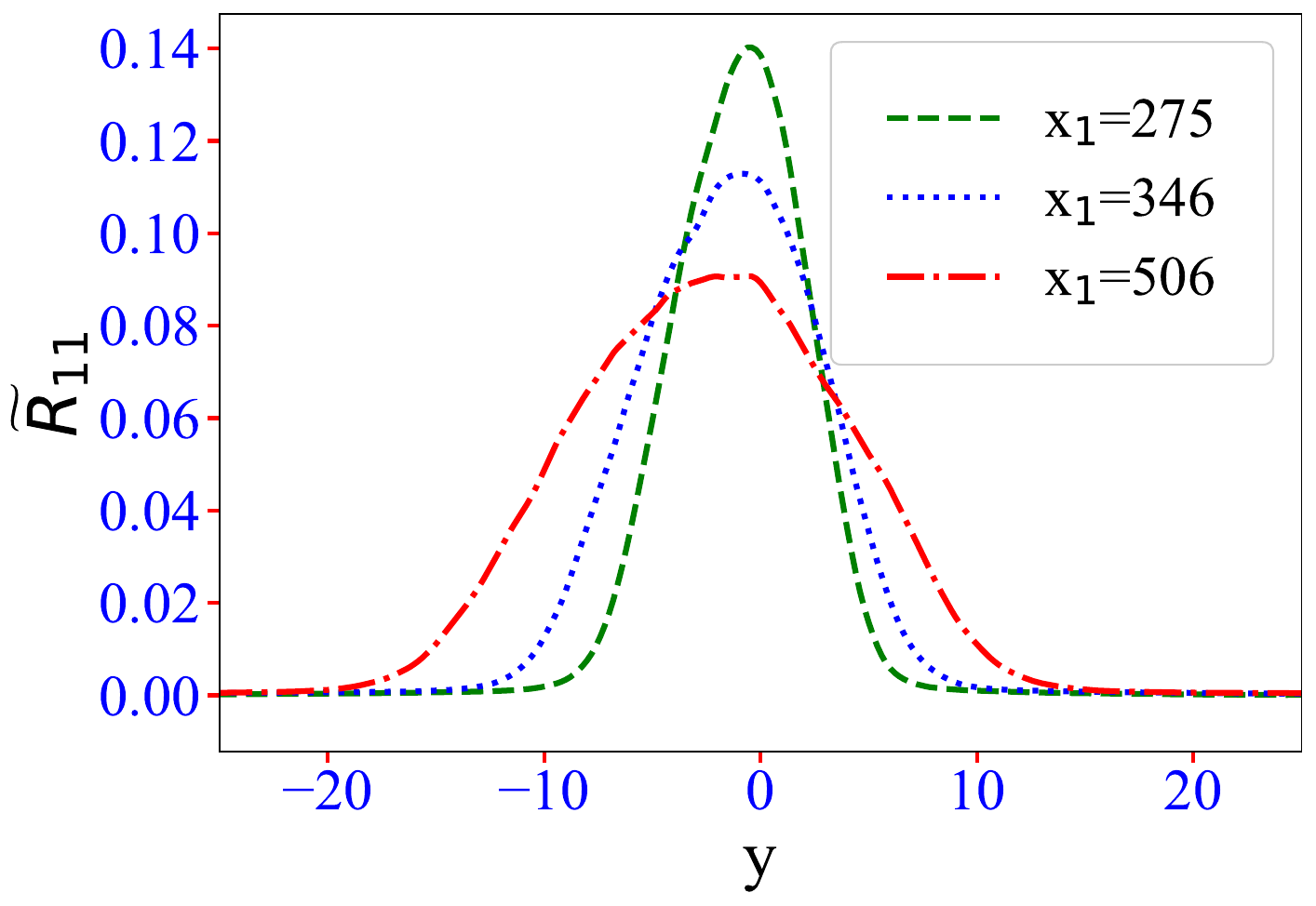}
         \subcaption{}
     \end{subfigure}
     \centering
     \begin{subfigure}[b]{0.40\textwidth}
         \centering
         \includegraphics[width=\textwidth]{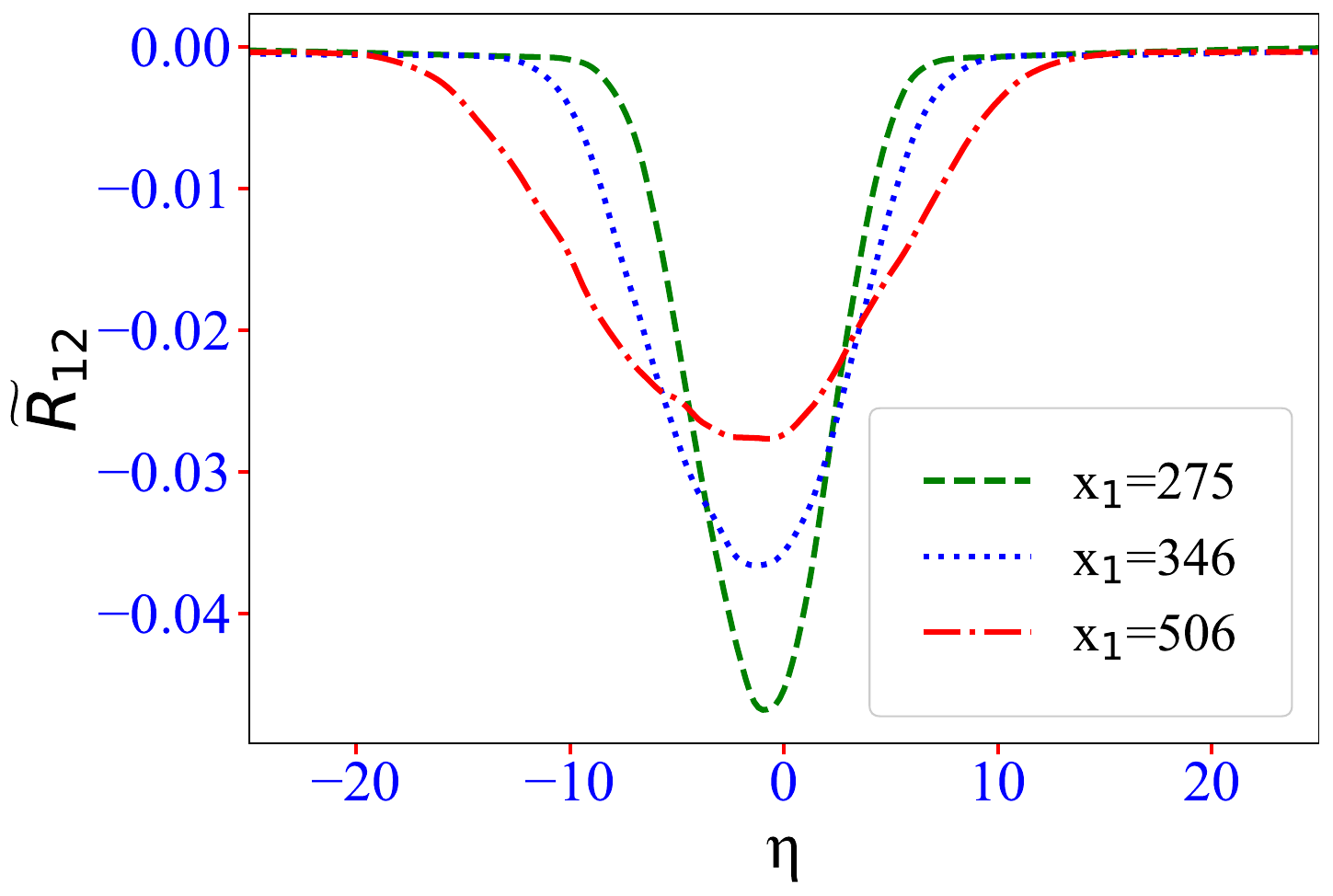}
         \subcaption{}
     \end{subfigure}
     \centering
     \begin{subfigure}[b]{0.39\textwidth}
         \centering
         \includegraphics[width=\textwidth]{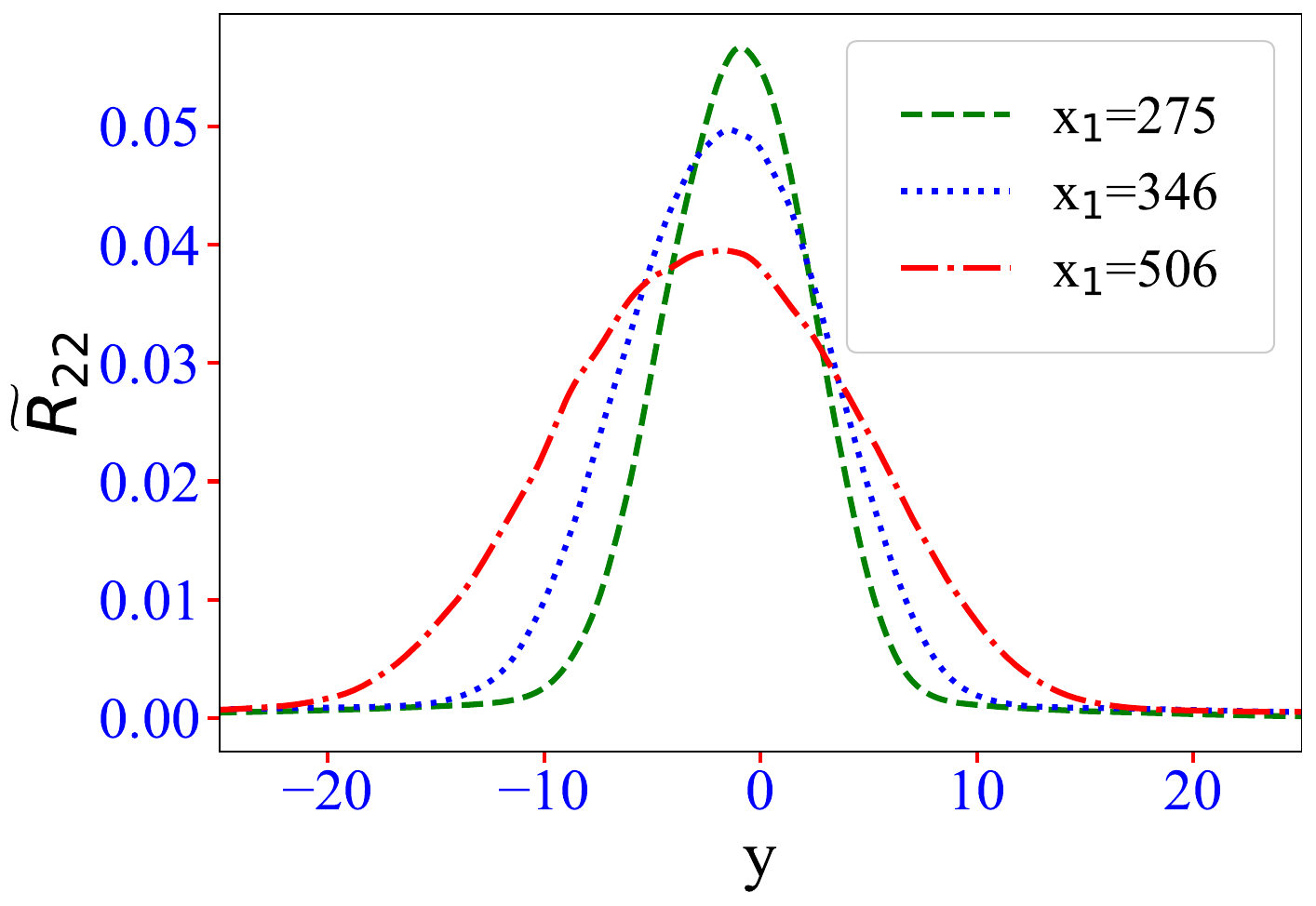}
         \subcaption{}
     \end{subfigure}
     \caption{Streamwise evolution of mean (a) streamwise velocity, (b) density, (c) pressure, (d) transverse/cross-stream velocity, (e) Reynolds normal stress in the streamwise direction, (f) Reynolds shear stress, and (g) Reynolds normal stress in the transverse direction.}
     \label{fig:mean}
\end{figure}

The spatial variations of (time and spanwise) averaged streamwise velocity, density, pressure, transverse velocity, and Reynolds stresses are shown in figure \ref{fig:mean} for the reference case (M$_c$=1.2 and $\lambda$=0.4). The selection of streamwise locations aims to provide a clearer view of profile variations in different flow regions. The transverse spreading of the streamwise velocity in figure \ref{fig:mean} (a) is consistent with those of the density, pressure, and Reynolds stress components shown in figures \ref{fig:mean} (b), (c), (e), (f) and (g). There is a significant density variation within the core of the shear layer as seen in figure \ref{fig:mean} (b), with a drop in the mean density around 21\% from the free stream value. We will connect this density variation with turbulent kinetic energy dissipation and show that its magnitude depends on $M_c$ in section \ref{sec:self-similarity}. There is also a variation of pressure within the core of the shear layer; however, its magnitude is less significant compared to the density variation (approximately 5\% of the free stream pressure). The streamwise evolution of mean transverse velocity is shown in figure \ref{fig:mean} (d). The transverse velocity has received scarce attention in the literature, in part because it is very difficult to accurately compute or measure.  There is a drop in transverse velocity in the center of the shear layer that will be shown to be tied to entrainment. In addition, there is an asymmetric behavior of the transverse velocity between the high- and low-speed sides. This asymmetric behavior is connected to density variation and centerline shifting which will be discussed in section \ref{sec:self-similarity}. Figures \ref{fig:mean} (e), (f), and (g) show the streamwise evolution of the Favre-averaged Reynolds stress components, defined as $\widetilde{R}_{ij}=\frac{<\rho u_i u_j>}{<{\rho}>}$. The Reynolds stress profiles have peak values around the center of the shear layer and go to zero outside the shear layer, as expected.

In the next section, we will discuss the self-similar equations and scalings for different mean and turbulent quantities mentioned above. We will show how these profiles collapse using our suggested scalings in the self-similar coordinate system, and find approximate close-form solutions of the self-similar equations that describe the variations of the transverse velocity and entrainment ratio.

\section{Self-Similarity} \label{sec:self-similarity}

It is generally accepted that, sufficiently far downstream after the inlet transition, the mixing layer undergoes a self-similar development. This phenomenon is characterized by the consistent collapse of statistical quantities in various spatial locations, when appropriately scaled. In self-similar scaling, the partial differential equations governing the turbulence statistics become ordinary differential equations. The present self-similar analysis for spatially developing shear layer configuration is similar to the self-similar analysis performed by \cite{wei22}, \cite{pantano02}, and \cite{livescu20}. In this study, we have generalized all the previous attempts in the literature and, for the first time, we have formulated the self-similar forms of continuity,  streamwise momentum, transverse momentum, and energy equations, incorporating both compressibility and centerline shifts. We show that the $M_c$ and $\lambda$ emerge naturally in the self-similar equations, and therefore we conduct separate investigations on the influences of these two parameters. 
%We suggest self-similar scaling for various mean and turbulent quantities including, for the first time, the density, pressure, temperature, transverse velocity, and dissipation. 

 Since turbulent shear layers are thin and spread slowly in the transverse direction, it is possible to use Prandtl's boundary layer description approach to study them (\citealt{wei22}). Assuming ideal gas behavior, steady-state mean flow conditions, and negligible variations in the spanwise direction, the compressible mean continuity, $x_1$ momentum, $x_2$ momentum, and energy equations after applying the boundary layer approximation are:
\begin{equation}\label{eq:continuity}
    \frac{\partial(\bar{\rho}\tilde{u}_1)}{\partial x_1}+\frac{\partial (\bar{\rho}\tilde{u}_2)}{\partial x_2}=0,
\end{equation}

\begin{equation}\label{eq:xmomentum}
    \bar{\rho} \tilde{u}_1 \frac{\partial\tilde{u}_1}{\partial x_1}+\bar{\rho} \tilde{u}_2 \frac{\partial\tilde{u}_1}{\partial x_2}+\frac{\partial (\bar{\rho}\tilde{R}_{12})}{\partial x_2}=0,
\end{equation}

\begin{equation}\label{ymomentum}
    \bar{\rho} \tilde{u}_1 \frac{\partial\tilde{u}_2}{\partial x_1}+\bar{\rho} \tilde{u}_2 \frac{\partial\tilde{u}_2}{\partial x_2}+\frac{\partial (\bar{\rho}\tilde{R}_{22})}{\partial x_2}=-\frac{\partial\bar{P}}{\partial x_2},
\end{equation}

\begin{equation}\label{eq:energy}
    C_v \frac{\partial(\bar{\rho}\tilde{T}\tilde{u}_1)}{\partial x_1} 
    + C_v \frac{\partial(\bar{\rho}\tilde{T}\tilde{u}_2)}{\partial x_2} 
    +\bar{P} \frac{\partial\tilde{u}_1}{\partial x_1}
    +\bar{P} \frac{\partial\tilde{u}_2}{\partial x_2}
    = \bar{\epsilon}_t-\tilde R_{\Delta(\rho T u)}-\tilde R_{P\Delta u}.
\end{equation}
Here, $\tilde{u}_1$ and $\tilde{u}_2$ are the Favre-averaged mean velocities in the streamwise and transverse directions, respectively. Favre-averaged kinematic Reynolds stress is defined as $\widetilde{R}_{ij}=\frac{<\rho u_i u_j>}{<{\rho}>}$. It is also assumed that the viscous terms are negligible in the momentum equations and heat conduction is negligible in the energy equation. $\bar{\epsilon_t}$ represents the mean dissipation, $\tilde R_{\Delta(\rho T u)}$ represents triple correlation terms $\frac{\partial(\overline{\rho^{\prime} T^{\prime} u_1^{\prime}})}{\partial x_1} $ and $\frac{\partial(\overline{\rho^{\prime} T^{\prime} u_2^{\prime}})}{\partial x_2} $ and $\tilde R_{P\Delta(u)}$  is associated with the pressure fluctuation terms $\overline{P^{\prime} \frac{\partial {u_1}^\prime}{\partial x_1}}$ and $\overline{P^{\prime} \frac{\partial {u_2}^\prime}{\partial x_2}}$. 

The statistics of a spatially developing planar turbulent shear layer depend on the streamwise coordinate, $x_1$, and inhomogeneous direction, $x_2$. Self-similarity implies that by introducing a new variable, $\eta$, these two dependencies can be replaced by a single variable.  $\eta$ is defined here as:
\begin{equation}
 \eta=\frac{x_2-\bar{x}_2}{\delta (x_1)},
 \end{equation}
such that
 \begin{equation}
  \frac{\partial\eta}{\partial x_2}=\frac{1}{\delta},
  \end{equation}
 \begin{equation} 
 \frac{\partial\eta}{\partial x_1}=-\frac{\eta}{\delta}\frac{d\delta}{dx_1}-\frac{1}{\delta}\frac{d\bar{x}_2}{dx_1}.
   \end{equation}
where $\delta$ is a length scale that represents the transverse spreading length of the shear layer. In this study, $\delta$ is considered as the shear layer thickness. $\bar{x}_2$ is the centerline shifting distance from the geometric center of the shear layer defined as the location where the streamwise velocity becomes equal to the average velocity ($U_c$), as shown in figure \ref{fig:Screenshot 2022-10-24 145102}. Due to entrainment by the higher speed stream, the centerline shifts towards the lower speed stream. To account for the changes in the shear layer due to this shifting, the coordinate system is rotated to align with the new centerline.
As explained later, the mean continuity, momentum, and energy equations  exhibit self-similar growth (i.e. can be reduced to ordinary differential equations) if both the growth rate $\frac{d\delta}{dx_1}$ and the centerline shifting rate $\frac{d\bar{x}_2}{dx_1}$ remain constant, and the variables are non-dimensionalized as:
\begin{equation}
\bar{\rho}(x_1,x_2)=\rho_{01}\hat{\rho}(\eta)+\rho_0
\end{equation}
\begin{equation}
\tilde{T}(x_1,x_2)=T_{01}\hat{T}(\eta)+T_0
\end{equation}
\begin{equation}
\bar{P}(x_1,x_2)=P_{01}\hat{P}(\eta)+P_0
\end{equation}
\begin{equation}
\tilde{u}_1(x_1,x_2)=u_{01}\hat{u}_1(\eta)+U_c 
\end{equation}
\begin{equation}
\tilde{u}_2(x_1,x_2)=u_{02}\hat{u}_2(\eta)+u_2^*
\end{equation}
\begin{equation}
\tilde{R}_{11}(x_1,x_2)=R_{011}\hat{R}_{11}(\eta)
\end{equation}
\begin{equation}
\tilde{R}_{12}(x_1,x_2)=R_{012}\hat{R}_{12}(\eta)
\end{equation}
\begin{equation}
\tilde{R}_{22}(x_1,x_2)=R_{022}\hat{R}_{22}(\eta)
\end{equation}
\begin{equation}
\bar{\epsilon}_t(x_1,x_2)=\epsilon_{01}\hat{\epsilon}_t(\eta)
\end{equation}
\begin{equation}
\tilde{R}_{\Delta (\rho T u)}(x_1,x_2)=R_{0\rho T u}\hat{R}_{\Delta (\rho T u)}(\eta)
\end{equation}
\begin{equation}
\tilde{R}_{P\Delta u}(x_1,x_2)=R_{0 P u}\hat{R}_{P\Delta u}(\eta)
\end{equation}.
Here, $\rho_{01}$, $T_{01}$, $P_{01}$, $u_{01}$, $u_{02}$, $R_{011}$, $R_{012}$, $R_{022}$ and $\epsilon_{01}$ are the reference scales of the mean density, temperature, pressure, streamwise velocity, transverse velocity, normal Reynolds stress in the streamwise direction, shear Reynolds stress, normal Reynolds stress in the transverse direction and dissipation, respectively. $R_{0\rho T u}$ is the reference scale for the density - temperature - velocity triple correlation, and $R_{0 P u}$ is the reference scale for the pressure - velocity correlation. $u_2^*$ is the mean transverse velocity at the shear layer centerline in the rotated coordinate system ($\eta$=0). $\rho_0$, $T_0$, and $P_0$ are the initial density, temperature, and pressure, respectively. $U_c$ is the convective or average velocity of the two streams. These reference scales are obtained using the scaling patch approach used by \cite{wei22, wei05, wei052}. This approach has been used for wakes (\citealt{wei222}), planar jets (\citealt{wei21}), turbulent channel flows (\citealt{wei05}), and plumes (\citealt{wei212}).  

\subsection{Self-similar continuity equation}\label{sec:sscon}

Substituting the self-similar variables into equation \ref{eq:continuity}, the mean continuity equation can be formulated as follows:
\begin{equation}\label{eq:con2}
\begin{array}{l}
    -(\eta+\phi) [\psi(M_c)\frac{d(\hat{\rho}\hat{u}_1)}{d\eta}+\frac{\psi(M_c)}{2\lambda} \frac{d\hat{\rho}}{d\eta}+\frac{d\hat{u}_1}{d\eta}]+[\psi(M_c)\frac{u_{02}}{u_{01}\frac{d\delta}{dx_1}}\frac{d(\hat{\rho}\hat{u}_2)}{d\eta}+\psi(M_c)\frac{u_{2}^*}{u_{01}\frac{d\delta}{dx_1}}\frac{d\hat{\rho}}{d\eta} \\
    +\frac{u_{02}}{u_{01}\frac{d\delta}{dx_1}}\frac{d\hat{u}_2}{d\eta}]=0\ ,
\end{array}
\end{equation}
\begin{equation}
 \mbox{where},  \phi=\frac{\frac{d\bar{x}_2}{dx_1}}{\frac{d\delta}{dx_1}}
\end{equation}
Here, $\phi$ is the ratio between the centerline shifting speed and the growth rate of the shear layer (\citealt{wei22}). To make the original continuity equation (\ref{eq:con2}) self-similar, $\phi$ is required to be a constant, which means the shear layer thickness and the centerline distance have vary linearly with streamwise distance in the self-similar zone. The present DNS data suggest that in the far downstream region, both $\frac{d\bar{x_2}}{dx_1}$ and $\frac{d\delta}{dx_1}$ are indeed constant, which makes $\phi$ constant. Note that in the rotated coordinate system, the positive $x_2$ direction points from the low-speed to the high-speed side. Given that the shear layer shifts toward the low-speed side so that $\frac{d\bar{x_2}}{dx_1}$ is negative, it yields that $\phi$ is negative. Experimental data also suggest that $\phi$ is constant and negative (\citealt{wei22,mehta85,chang11,chang12}). The variations of $\phi$ with both $M_c$ and velocity parameter are significant and are discussed in section \ref{sec:asym}. In addition to self-similarity, for equation (\ref{eq:con2}) to retain the transverse velocity terms, these terms need to be of order 1. Therefore, $\tilde{u}_2$ has to scale with $u_{01}\frac{d\delta}{dx_1}$. If the scaling of the mean streamwise velocity is $\Delta U$ (\citealt{livescu20, wei22}), $\Delta U\frac{d\delta}{dx_1}$ is the proper scaling for $\tilde{u}_2$. The scaling for density includes the factor $\psi(M_c)$, which is necessary to collapse the density profiles for different $M_c$ values. This factor will be discussed later. According to the definition of velocity parameter, $\frac{U_c}{u_{01}}$ may also be written as $\frac{1}{2\lambda}$ in equation (\ref{eq:con2}). 

\subsection{Self-similar streamwise momentum equation}\label{eq:sssmomentum}

The mean viscous terms have little effect on the mean momentum at sufficiently high Reynolds numbers far downstream, as they scale with $1/Re$. The major contribution of viscous terms is in the energy equation (\citealt{livescu20}). After applying the self-similar scaling of the mean quantities to the mean $x_1$ momentum equation, the equation becomes:
\;
\begin{equation}\label{eq:xmomentum2}
\begin{array}{l}
    -(\eta+\phi) [2\lambda\psi(M_c)\hat{\rho}\hat{u_1}\frac{d\hat{u_1}}{d\eta}+\psi(M_c)\hat{\rho}\frac{d\hat{u_1}}{d\eta}+2\lambda\hat{u_1}\frac{d\hat{u_1}}{{d\eta}}+\frac{d\hat{u_1}}{d\eta}]+[2\lambda\psi(M_c)\frac{u_{02}}{u_{01}\frac{d\delta}{dx_1}}\hat{\rho}\hat{u_{2}}\frac{d\hat{u_{1}}}{d\eta} \\
    +2\lambda\psi(M_c)\frac{u_{2}^*}{u_{01}\frac{d\delta}{dx_1}}\hat{\rho}\frac{d\hat{u_1}}{d\eta}
    +2\lambda\frac{u_{02}}{u_{01}\frac{d\delta}{dx_1}}\hat{u_2}\frac{d\hat{u_1}}{d\eta}+2\lambda\frac{u_{2}^*}{u_{01}\frac{d\delta}{dx_1}}\frac{d\hat{u_1}}{d\eta}]+\psi(M_c)\frac{R_{012}}{\frac{u_{01}^2\frac{d\delta}{dx_1}}{2\lambda}}\frac{d(\hat{\rho}\hat{R}_{12})}{d\eta}\\
    +\frac{R_{012}}{\frac{u_{01}^2\frac{d\delta}{dx_1}}{2\lambda}}\frac{d(\hat{R}_{12})}{d\eta}=0
     \end{array}
\end{equation}
\;
Since the scalings of $\tilde {u}_1$ and $\tilde {u}_2$ are  $\Delta U$ and $\Delta U\frac{d\delta}{dx_1}$, respectively, the self-similar $x_1$ momentum equation indicates that the scaling for $\tilde R_{12}$ should be [$\frac{1}{2\lambda} \frac{d\delta}{dx_1}$]$(\Delta U)^2$ or $\Delta U U_c \frac{d\delta}{dx_1}$. On the other hand, the self-similar scaling for $\tilde R_{11}$ is $(\Delta U)^2$, which is consistent with the current DNS data and previous studies (\citealt{wei22,livescu20}). It is important to note that in numerous prior investigations of turbulent mixing layers, $\tilde R_{12}$ was also scaled as $(\Delta U)^2$. However, \cite{pantano02,livescu20,wei22} have emphasized the importance of including $\frac{d\delta}{dx_1}$ in the proper scaling of $\tilde R_{12}$.

\subsection{Self-similar transverse momentum equation}\label{eq:sstmomentum}

Upon applying self-similar scalings to the $x_2$ momentum equation, the equation becomes:
\begin{equation}
\begin{array}{l}
    -(\eta+\phi) [\psi(M_c)\hat{\rho}\hat{u_1}\frac{d\hat{u_2}}{d\eta}+\frac{\psi(M_c)}{2\lambda}\hat{\rho}\frac{d\hat{u_2}}{d\eta}+\hat{u_1}\frac{d\hat{u_2}}{{d\eta}}+\frac{1}{2\lambda}\frac{d\hat{u_2}}{d\eta}]+[\psi(M_c)\frac{u_{02}}{u_{01}\frac{d\delta}{dx_1}}\hat{\rho}\hat{u_{2}}\frac{d\hat{u_{2}}}{d\eta} \\
    +\psi(M_c)\frac{u_{2}^*}{u_{01}\frac{d\delta}{dx_1}}\hat{\rho}\frac{d\hat{u_2}}{d\eta} 
    +\frac{u_{02}}{u_{01}\frac{d\delta}{dx_1}}\hat{u_2}\frac{d\hat{u_2}}{d\eta}+\frac{u_{2}^*}{u_{01}\frac{d\delta}{dx_1}}\frac{d\hat{u_2}}{d\eta}]+
    \frac{\psi(M_c)}{(2\lambda)^2}\frac{R_{022}}{\frac{u_{01}u_{02}\frac{d\delta}{dx_1}}{(2\lambda)^2}}\frac{d(\hat{\rho}\hat{R}_{22})}{d\eta} \\
    +\frac{1}{(2\lambda)^2}\frac{R_{022}}{\frac{u_{01}u_{02}\frac{d\delta}{dx_1}}{(2\lambda)^2}}\frac{d(\hat{R}_{22})}{d\eta}
    =\frac{-1}{(2\lambda)^2}\frac{P_{01}}{\frac{\rho_0 u_{01}u_{02}\frac{d\delta}{dx_1}}{(2\lambda)^2}}\frac{d\hat{P}}{d\eta}.
     \end{array}
\end{equation}

The self-similar $x_2$ momentum equation suggests that the proper scaling for $\tilde R_{22}$ is [$\frac{1}{(2\lambda)^2} (\frac{d\delta}{dx_1})^2$]$(\Delta U)^2$ or $ (U_c)^2(\frac{d\delta}{dx_1})^2$ and for $\bar P$ is $\rho_{0}$[$\frac{1}{(2\lambda)^2} (\frac{d\delta}{dx_1})^2$]$(\Delta U)^2$ or $ \rho_{0}(U_c)^2(\frac{d\delta}{dx_1})^2$. Our data collapses in the self-similar zone for both variables, as shown in section \ref{sec:sszone}.

\subsection{Self-similar energy equation}\label{eq:ssenergy}

Upon applying the self-similar scaling, the transformed form of the equation is obtained as follows:
\begin{equation}\label{eq:en2}
\begin{array}{l}
    (-\eta-\phi) [ 2\lambda \frac{d (\hat{P}\hat{u_1})}{d\eta}  + \frac{2\lambda}{\psi^2} \frac{d\hat{u_1}}{d\eta} + \frac{d\hat{P}}{d\eta}] + 2\lambda [\frac{d (\hat{P}\hat{u_2})}{d\eta} +\frac{1}{\psi^2} \frac{d\hat{u_2}}{d\eta} + \frac{{u_2}^*}{u_{02}} \frac{d\hat{P}}{d\eta}]+  \\\\
    2\lambda (-\eta-\phi) (\gamma-1) [\hat{P} \frac{d\hat{u_1}}{d\eta}+ \frac{1}{\psi^2} \frac{d\hat{u_1}}{d\eta}] + 2\lambda (\gamma-1) [\hat{P} \frac{d\hat{u_2}}{d\eta}+ \frac{1}{\psi^2} \frac{d\hat{u_2}}{d\eta}] = \\\\
    4\gamma (\gamma-1) \frac{{M_c}^2}{\psi} \frac{2\lambda \delta [\epsilon_{01}\hat{\epsilon}_t-R_{0\rho Tu}\hat{R}_{\Delta(\rho T u)}-R_{0Pu}\hat{R}_{P\Delta u}]}{\rho_{01}  {u_{01}}^3\frac{d\delta}{dx}}.
    \end{array}
\end{equation}

For the energy equation to be self-similar, the scaling of the mean dissipation should be $\frac{\rho_{01}(\Delta u)^3 \frac{d\delta}{dx1}}{2\lambda \delta}$.  Our observations indicate that, indeed, dissipation exhibits a quasi-self-similar behavior when appropriately normalized (see section \ref{sec:sszone}). Thus, data from the present as the profiles collapse well across different streamwise locations (see figure \ref{fig:selfsimilarC} below). The behavior of mean dissipation at different $M_c$ and $\lambda$ values is described in Appendix \ref{appA}. As an additional requirement for the energy equation to be self-similar, the mean temperature is scaled as $T_{01}$= $T_0$$\psi(M_c)$ which is similar to the mean density scaling.  

\subsection{Summary of self-similar scalings and equations}\label{eq:ssfinal}

All self-similar scalings for the turbulence statistics considered in the present analysis are summarized in table \ref{tab:ssscaling}. For the shear layer with constant global density, the freestream densities are constant and equal to the initial density $\rho_0$. However, in our supersonic flow, there is a significant density variation inside the core of the shear layer and the variation intensifies with increasing compressibility (see the discussion in section \ref{sec:Mc}). Although the mean density profiles collapse well for each $M_c$ case, which is consistent with self-similarity, in order to be able to collapse the mean density profiles for cases with different $M_c$ values, $\rho_{01}$ was written as $\rho_0\psi(M_c)$, where $\psi(M_c)$ is a function of $M_c$ which satisfies the definition of $\hat{\rho}$ at both $M_c$ limits (i.e., in the center of the layer $\hat{\rho}$ goes to 1 as $M_c$ goes to 0 and $\hat{\rho}$ goes to zero as $M_c$ goes to $\infty$) and collapses the density profiles across different $M_c$ values. The details of this scaling are discussed in section~\ref{sec:Mc}. The mean temperature is scaled consistently with the mean density to address the rise of temperature with compressibility inside the core of the shear layer.

\begin{table}
  \centering
  \begin{tabular}{|c|c|}
  \hline
Parameters & ~~Suggested Self-Similar Scaling \\[3pt]
       \hline
       $\bar \rho$ &  $\rho_0 \psi (M_c)$ \\
       \hline
       $\tilde T$ &  $T_0 \psi (M_c)$ \\
       \hline
       $\tilde{u}_{1}$ & $\Delta U$ \\
       \hline
       $\tilde{u}_{2}$ & $\Delta U$ $\frac{d\delta}{dx_1}$ \\
       \hline
       $\tilde{R}_{11}$ & ($\Delta U)^2$ \\
       \hline
       $\tilde{R}_{12}$ & [$\frac{1}{2\lambda} (\frac{d\delta}{dx_1})](\Delta U)^2$ or $\Delta U$ $U_c$ $\frac{d\delta}{dx_1}$\\
       \hline
       $\tilde{R}_{22}$ & [$\frac{1}{2\lambda} (\frac{d\delta}{dx_1})](\Delta U)^2$ or $\Delta U$ $U_c$ $\frac{d\delta}{dx_1}$\\
                & and $[\frac{1}{2\lambda} (\frac{d\delta}{dx_1})]^2$$(\Delta U)^2$ or  $(U_c)^2$ $(\frac{d\delta}{dx_1})^2$\\
                \hline
        $\bar P$ & $\rho_0$[$\frac{1}{2\lambda} (\frac{d\delta}{dx_1})](\Delta U)^2$ or $\rho_0$$\Delta U$ $U_c$ $\frac{d\delta}{dx_1}$\\
                & and $\rho_0$$[\frac{1}{2\lambda} (\frac{d\delta}{dx_1})]^2$$(\Delta U)^2$ or  $\rho_0$$(U_c)^2$ $(\frac{d\delta}{dx_1})^2$\\
                \hline
        $\bar {\epsilon}_t$ & $\rho_0$$\psi(M_c)$[$\frac{1}{2\lambda\delta} (\frac{d\delta}{dx_1})](\Delta U)^3$ or $\rho_0$$\psi(M_c)$[$\frac{1}{\delta} (\frac{d\delta}{dx_1})] U_c (\Delta U)^2$ \\
        \hline
  \end{tabular}
  \caption{Suggested self-similar scaling for different quantities}
  \label{tab:ssscaling}
\end{table}

Using the self-similar scalings shown in table \ref{tab:ssscaling}, the final forms of the self-similar  continuity, $x_1$ momentum, $x_2$ momentum, and energy equations are as follows:
\begin{equation}\label{eq:sscon}
\begin{array}{l}
    -(\eta+\phi) [2 \lambda \frac{d\hat{u_1}}{d\eta} (1+\psi(M_c) \hat{\rho}) +\psi(M_c)\frac{d\hat{\rho}}{d\eta} (1+ 2\lambda \hat{u_1})] + 2 \lambda (1+ \psi(M_c)\hat{\rho}) \frac {d\hat {u_2}}{d\eta}\\
    + 2\lambda\psi(M_c)(\hat{u_2}+\frac{u_2^*}{u_{02}}) \frac{d\hat{\rho}}{d\eta}=0\;,
    \end{array}
\end{equation}
\;
\begin{equation}\label{eq:ssxmom}
\begin{array}{l}
    -(\eta+\phi) [(1+2\lambda \hat{u_1}) (1+\psi(M_c)\hat{\rho}) ] \frac{d\hat{u_1}}{d\eta} +2\lambda(1+\psi(M_c)\hat{\rho})(\hat{u_2}+\frac{u_2^*}{u_{02}})\frac{d\hat{u_1}}{d\eta}\\
    +\frac{d[\hat{R_{12}} (1+\psi(M_c)\hat{\rho})]}{d\eta}=0\;,
    \end{array}
\end{equation}
\;
\begin{equation}\label{eq:ssymom}
\begin{array}{l}
    -(\eta+\phi) [2\lambda(1+2\lambda \hat{u_1}) (1+\psi(M_c)\hat{\rho}) ] \frac{d\hat{u_2}}{d\eta} +(2\lambda)^2 (1+\psi(M_c)\hat{\rho})(\hat{u_2}+\frac{u_2^*}{u_{02}})\frac{d\hat{u_2}}{d\eta}\\
    +\frac{d[\hat{R_{22}} (1+\psi(M_c)\hat{\rho})]}{d\eta}=-\frac{d\hat{P}}{d\eta}\;,
    \end{array}
\end{equation}
\;
\begin{equation}\label{eq:ssenr}
\begin{array}{l}
    -(\eta+\phi) [ 2\lambda \frac{d (\hat{P}\hat{u_1})}{d\eta}  + \frac{2\lambda}{\psi^2} \frac{d\hat{u_1}}{d\eta} + \frac{d\hat{P}}{d\eta}] + 2\lambda [\frac{d (\hat{P}\hat{u_2})}{d\eta} +\frac{1}{\psi^2} \frac{d\hat{u_2}}{d\eta} + \frac{{u_2}^*}{u_{02}} \frac{d\hat{P}}{d\eta}]+  \\\\
    2\lambda (-\eta-\phi) (\gamma-1) [\hat{P} \frac{d\hat{u_1}}{d\eta}+ \frac{1}{\psi^2} \frac{d\hat{u_1}}{d\eta}] + 2\lambda (\gamma-1) [\hat{P} \frac{d\hat{u_2}}{d\eta}+ \frac{1}{\psi^2} \frac{d\hat{u_2}}{d\eta}] = \\\\
    4\gamma (\gamma-1) \frac{{M_c}^2}{\psi} [\hat{\epsilon_t}-(\hat{R}_{\Delta(\rho T u)}+\hat{R}_{P\Delta u})].
    \end{array}
\end{equation}

In equations (\ref{eq:sscon})-(\ref{eq:ssenr}), the convective Mach number ($M_c$) and velocity parameter ($\lambda$) appear as independent parameters. Therefore, in the next section we will study the effects of these two parameters separately. To recover the incompressible limit, one can let $M_c \rightarrow 0$. In this case, the continuity, $x_1$ momentum and $x_2$ momentum equations become:
\begin{equation}\label{eq:conincom}
    -(\eta+\phi) \frac{d\hat{u_1}}{d\eta} + \frac {d\hat {u_2}}{d\eta}=0\;
\end{equation}
\;
\begin{equation}\label{eq:xmomincom}
    -(\eta+\phi) (1+2\lambda \hat{u_1}) \frac{d\hat{u_1}}{d\eta} +2\lambda (\hat{u_2}+\frac{u_2^*}{u_{02}})\frac{d\hat{u_1}}{d\eta}+\frac{d\hat{R}_{12}}{d\eta}=0\;
\end{equation}
\;
\begin{equation}\label{eq:ymomincom}
    -(\eta+\phi) [2\lambda(1+2\lambda \hat{u_1})] \frac{d\hat{u_2}}{d\eta} +(2\lambda)^2 (\hat{u_2}+\frac{u_2^*}{u_{02}})\frac{d\hat{u_2}}{d\eta}+\frac{d\hat{R}_{22}}{d\eta}=-\frac{d\hat{P}}{d\eta}\;
\end{equation}
If equation (\ref{eq:conincom}) is used in equation (\ref{eq:xmomincom}), the incompressible self-similar $x_1$ momentum equation becomes:
\begin{equation}\label{eq:xmomincom2}
    \frac{d\hat{u_2}}{d\eta} -[2\lambda \frac{u_2^*}{u_{02}}]\frac{d\hat{u_1}}{d\eta}+2\lambda(\hat{u_1}\frac{d\hat{u_2}}{d\eta}-\hat{u_2}\frac{d\hat{u_1}}{d\eta})-\frac{d\hat{R}_{12}}{d\eta}=0\;
\end{equation}
\;

Equations (\ref{eq:conincom}), (\ref{eq:xmomincom}), and (\ref{eq:xmomincom2}) are similar to the incompressible self-similar equations derived by \cite{wei22}. After formulating the self-similar equations, the next crucial step is to identify the self-similar region in the DNS data. Employing our suggested scalings, we will then demonstrate how various mean and turbulent quantities, as outlined in section \ref{sec:mean}, collapse within the self-similar region.  

\subsection{Self-similarity of simulated shear layer}\label{sec:sszone}

Achieving self-similar or fully-developed conditions within the computational domain requires a careful set of simulations. The downstream distance to the self-similar region is very sensitive to inlet flow conditions, as well as to both $M_c$ and $\lambda$. We have already discussed inlet perturbations in the streamwise velocity, and here we will address the role of $M_c$ and $\lambda$, as well as the procedure to identify the onset of the self-similar region in the DNS data. The transient region changes for different cases and for different turbulence statistics, so it is very important to set an outline for selecting self-similar zones. In many previous studies, the onset of self-similarity was decided based on the constant shear layer growth rate and collapse of the mean streamwise velocity and Reynolds stress profiles  
(e.g. \cite{mehta90,dutton91}). In the present study, we consider, in addition, the collapse of the mean transverse velocity, density, and pressure profiles. We show that the shear layer growth becomes linear earlier than the collapse of the Reynolds stresses, mean pressure and dissipation. The self-similar region is defined as the downstream region where all turbulence statistics in the self-similar equations (\ref{eq:sscon})-(\ref{eq:ssenr}) reach their self-similar behavior. 
\begin{figure}
     \centerline
     \centering
     \begin{subfigure}[b]{0.38\textwidth}
         \centering
         \includegraphics[width=\textwidth]{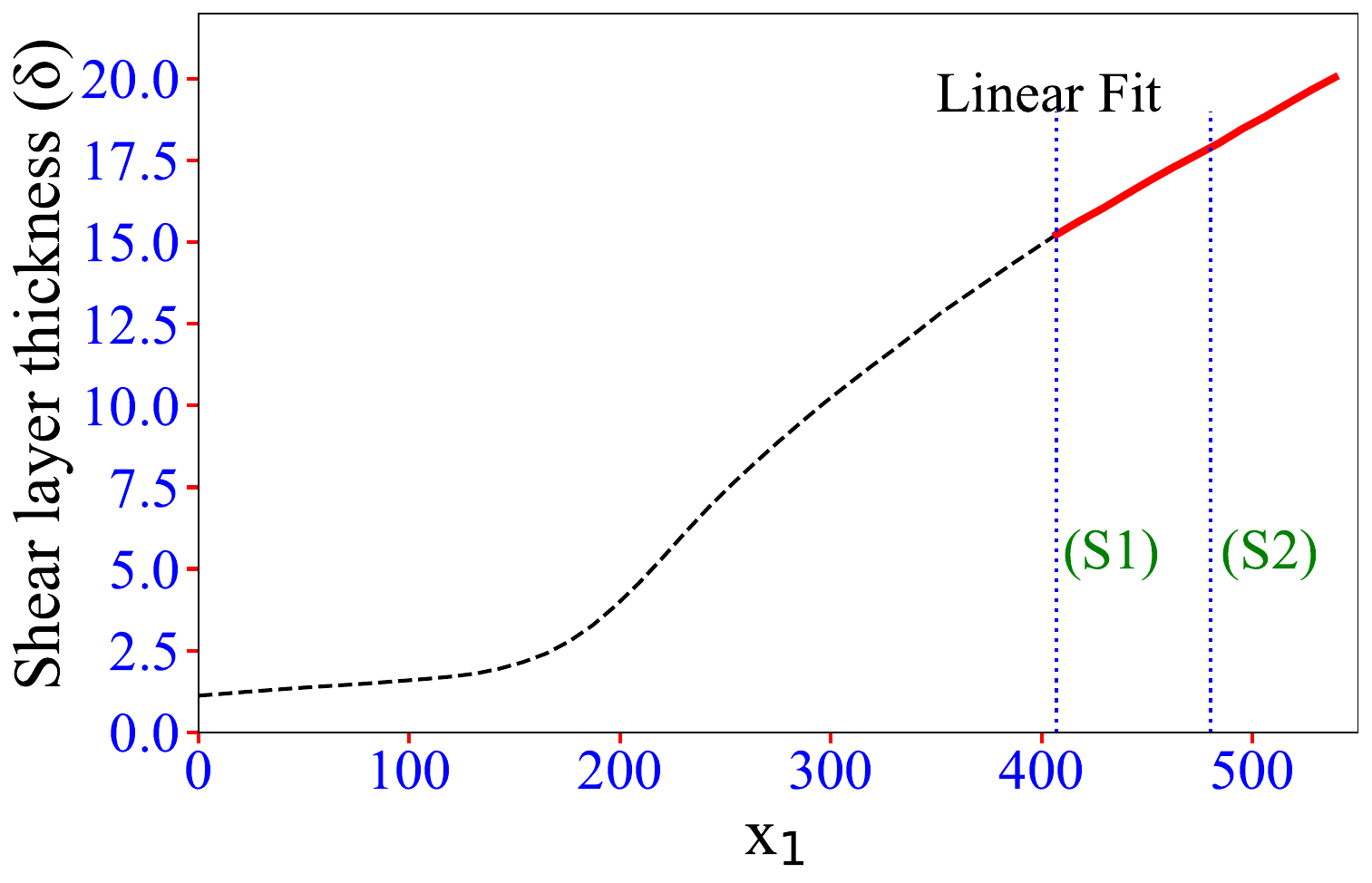}
         \subcaption{}
     \end{subfigure}
     \centering
     \begin{subfigure}[b]{0.38\textwidth}
         \centering
         \includegraphics[width=\textwidth]{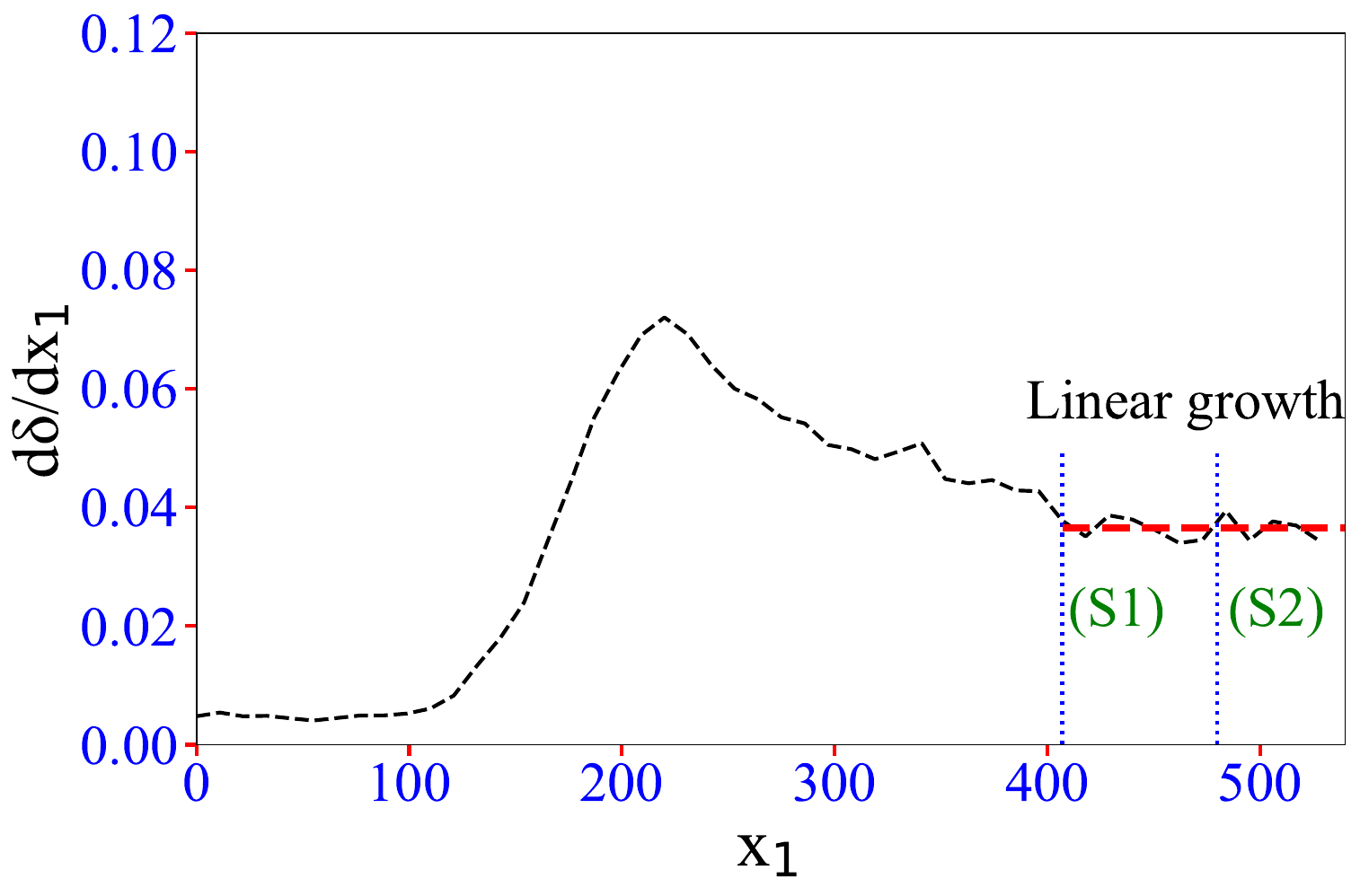}
         \subcaption{}
     \end{subfigure}
     \centering
     \begin{subfigure}[b]{0.38\textwidth}
         \centering
         \includegraphics[width=\textwidth]{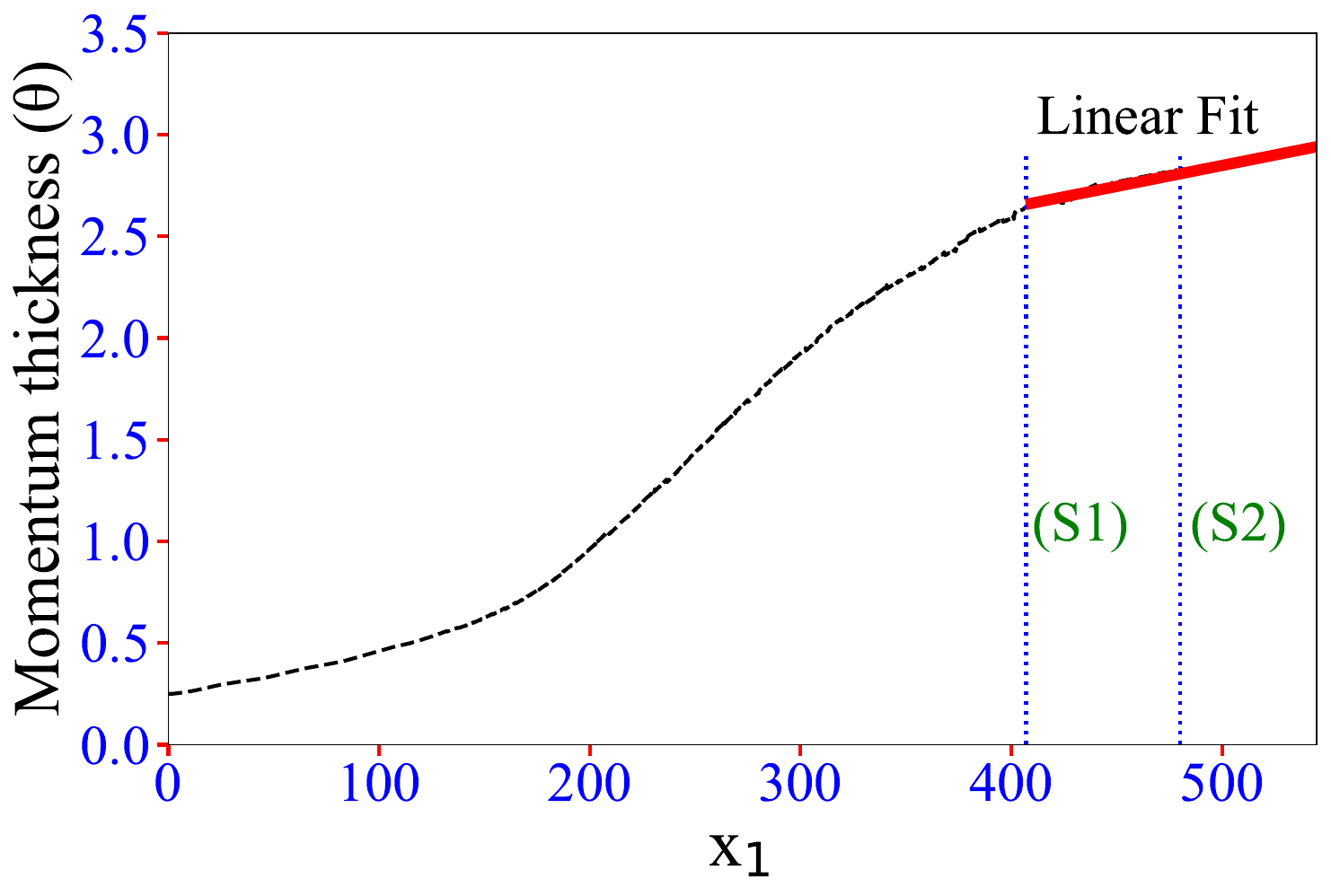}
         \subcaption{}
     \end{subfigure}
     \centering
     \begin{subfigure}[b]{0.40\textwidth}
         \centering
         \includegraphics[width=\textwidth]{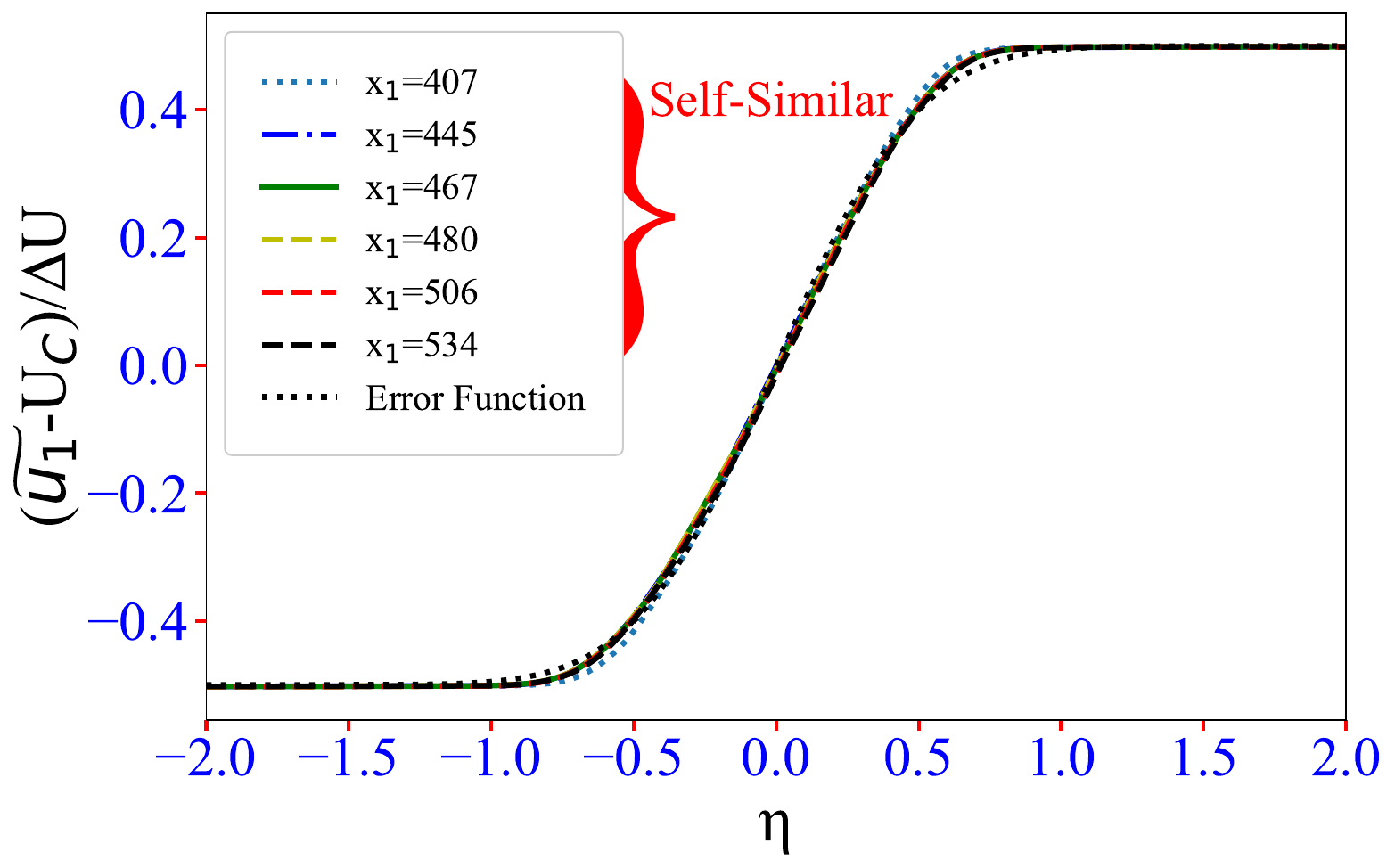}
         \subcaption{}
     \end{subfigure}
     \centering
     \begin{subfigure}[b]{0.39\textwidth}
         \centering
         \includegraphics[width=\textwidth]{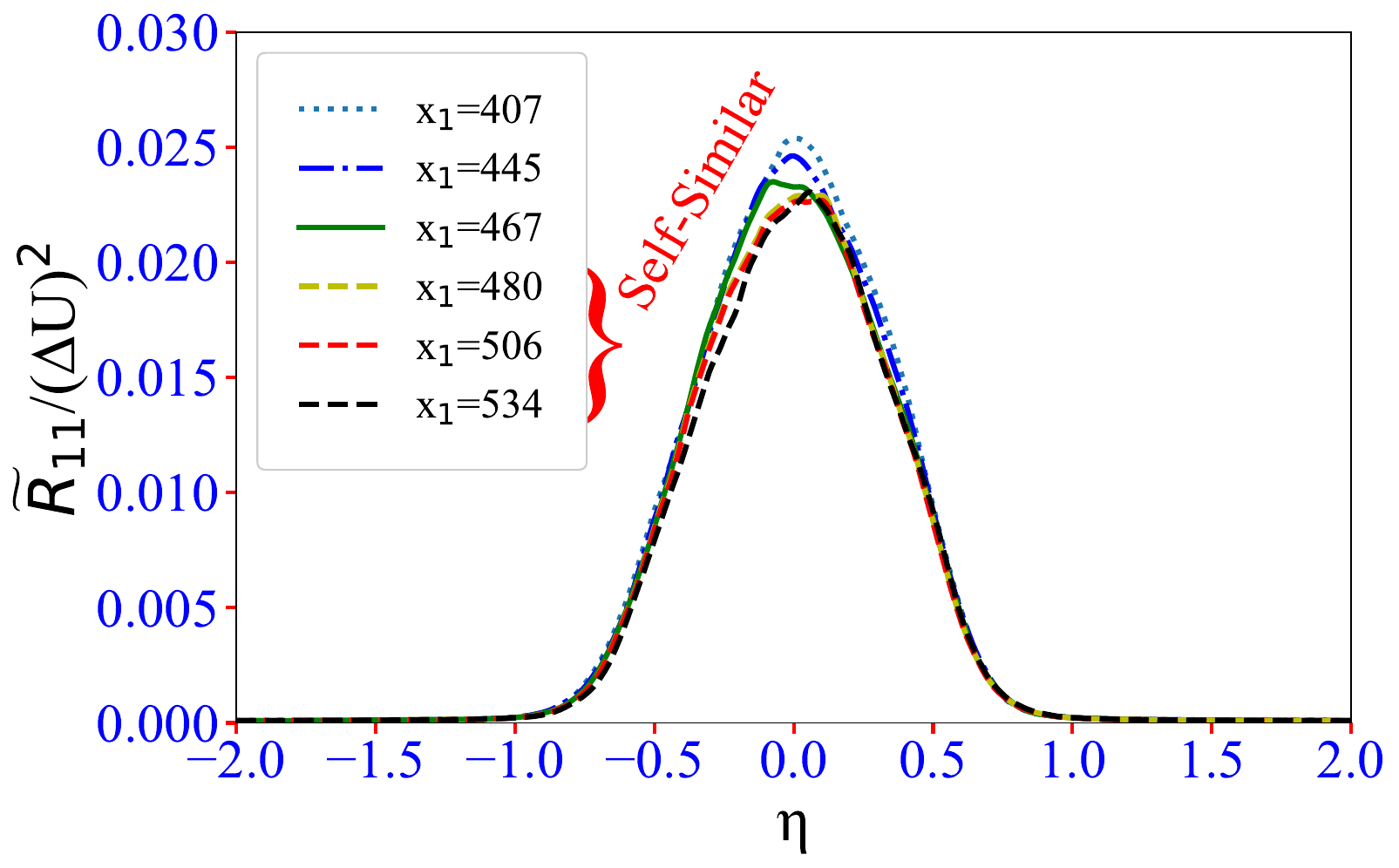}
         \subcaption{}
     \end{subfigure}
     \centering
     \begin{subfigure}[b]{0.39\textwidth}
         \centering
         \includegraphics[width=\textwidth]{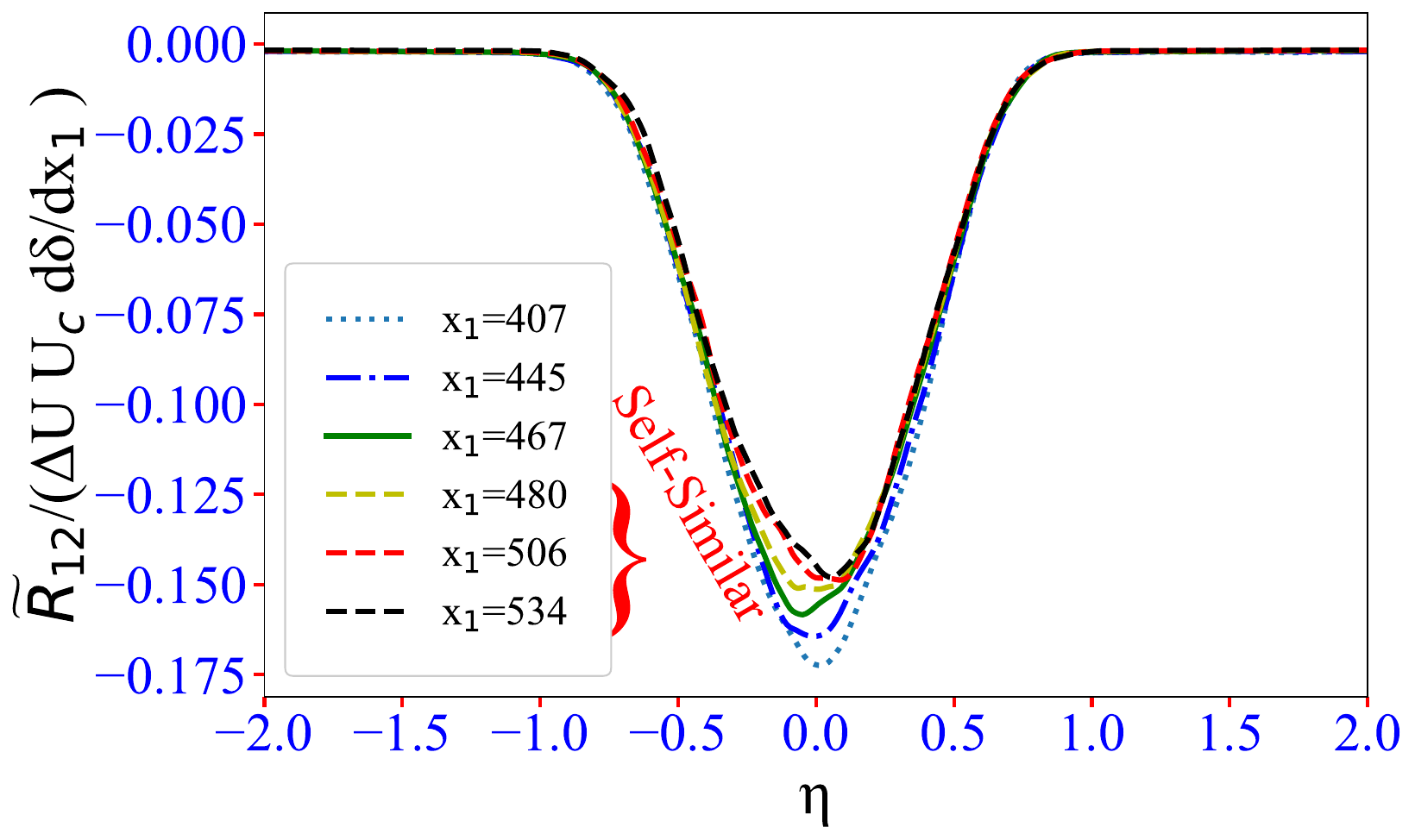}
         \subcaption{}
     \end{subfigure}
     \centering
     \begin{subfigure}[b]{0.39\textwidth}
         \centering
         \includegraphics[width=\textwidth]{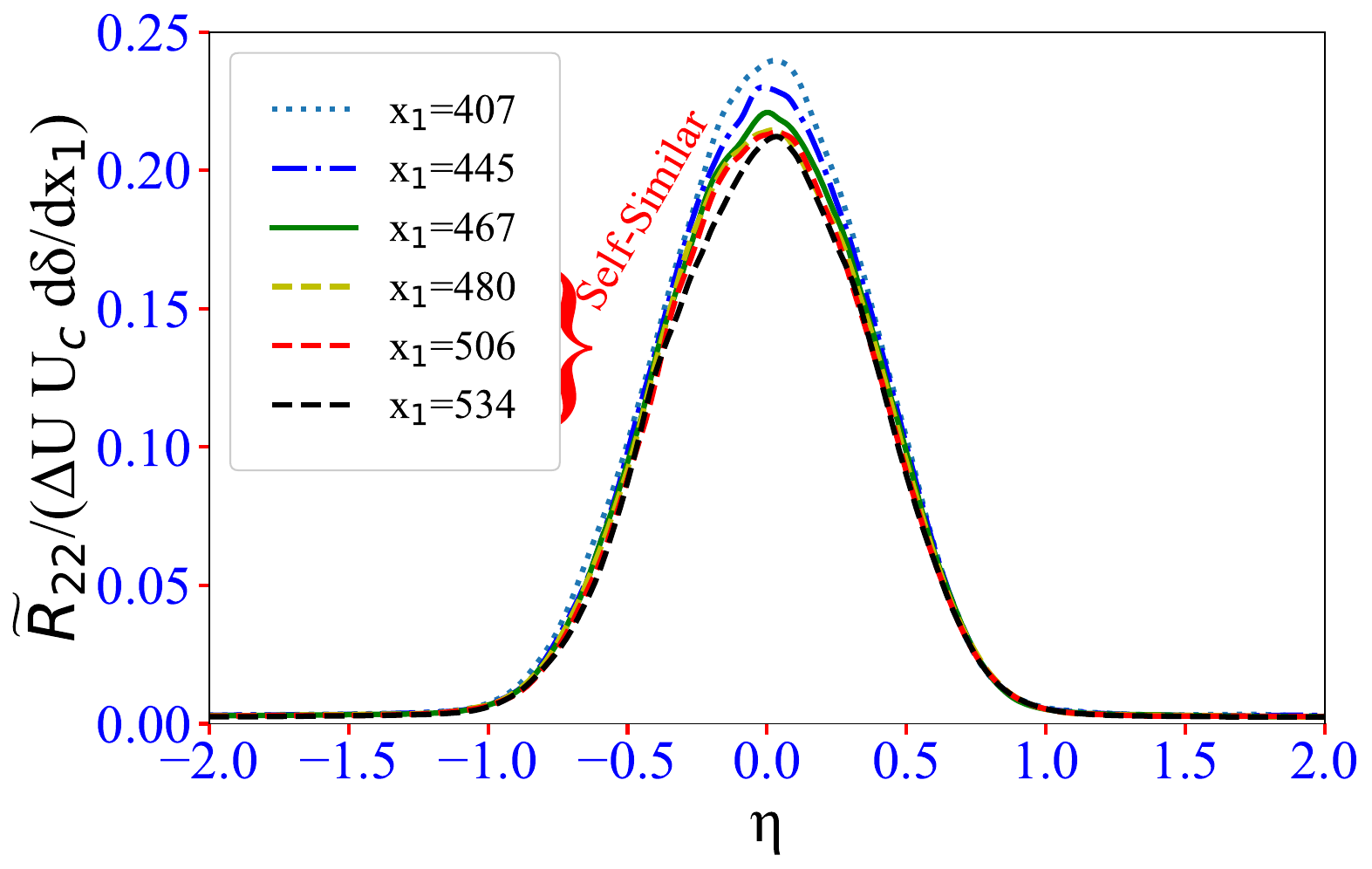}
         \subcaption{}
     \end{subfigure}
     \centering
     \begin{subfigure}[b]{0.37\textwidth}
         \centering
         \includegraphics[width=\textwidth]{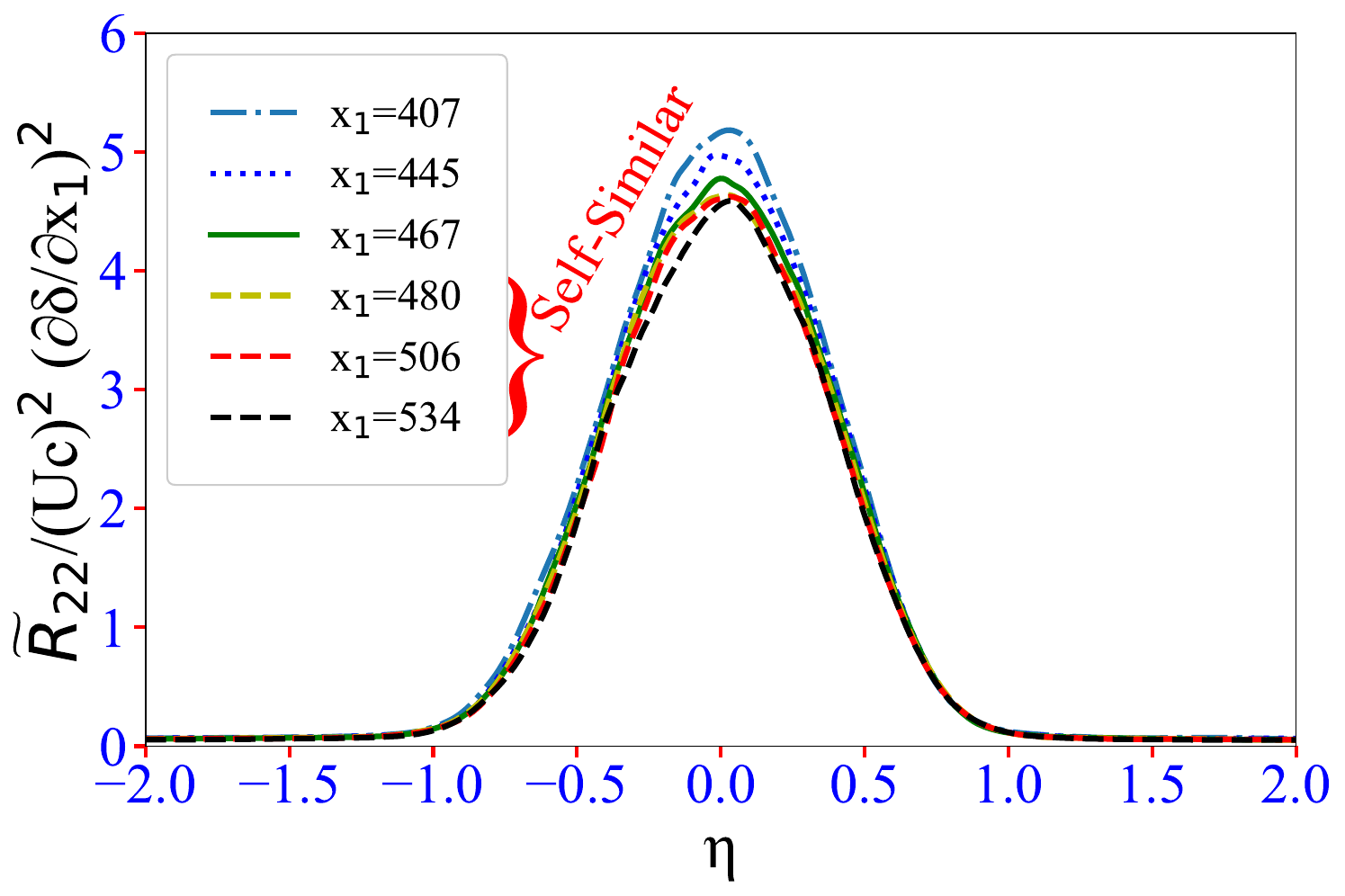}
         \subcaption{}
     \end{subfigure}
     \caption{The evolution of mean and turbulent quantities in self-similar coordinates. Figures (a), (b), and (c) show the linear growth of the shear layer and momentum thickness. Figure (d) shows the collapse of the streamwise velocity profiles and figures (e), (f), (g), and (h) show the collapse of Reynolds stresses in the self-similar zone. Quantities like shear layer thickness, momentum thickness, and streamwise velocity become self-similar around location S1 ($x_1\approx407$) whereas the Reynolds stresses become self-similar at location S2 ($x_1\approx 480$).}
     \label{fig:selfsimilar1}
\end{figure}

\begin{figure}
     \centering
     \begin{subfigure}[b]{0.49\textwidth}
         \centering
         \includegraphics[width=\textwidth]{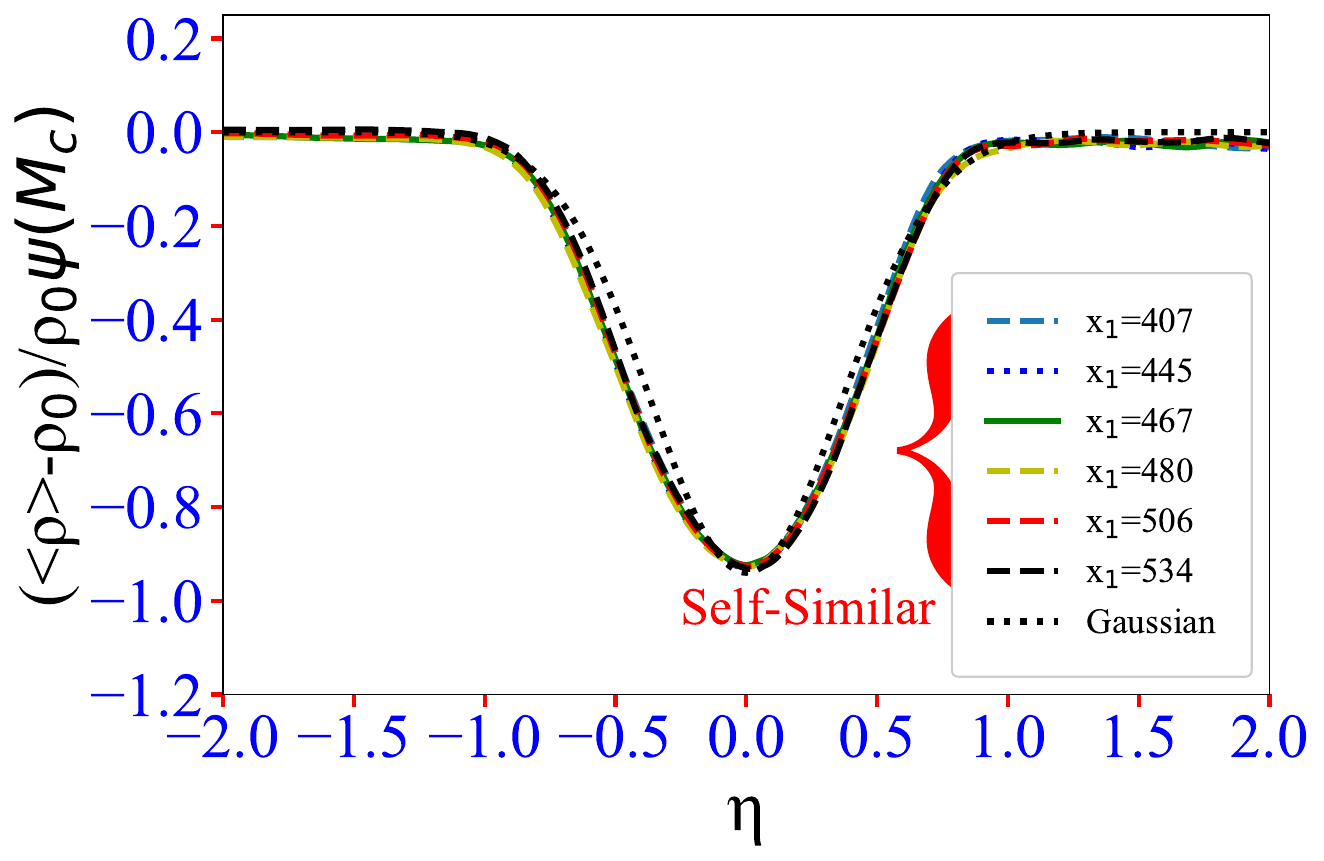}
         \subcaption{}
     \end{subfigure}
      \begin{subfigure}[b]{0.48\textwidth}
         \centering
         \includegraphics[width=\textwidth]{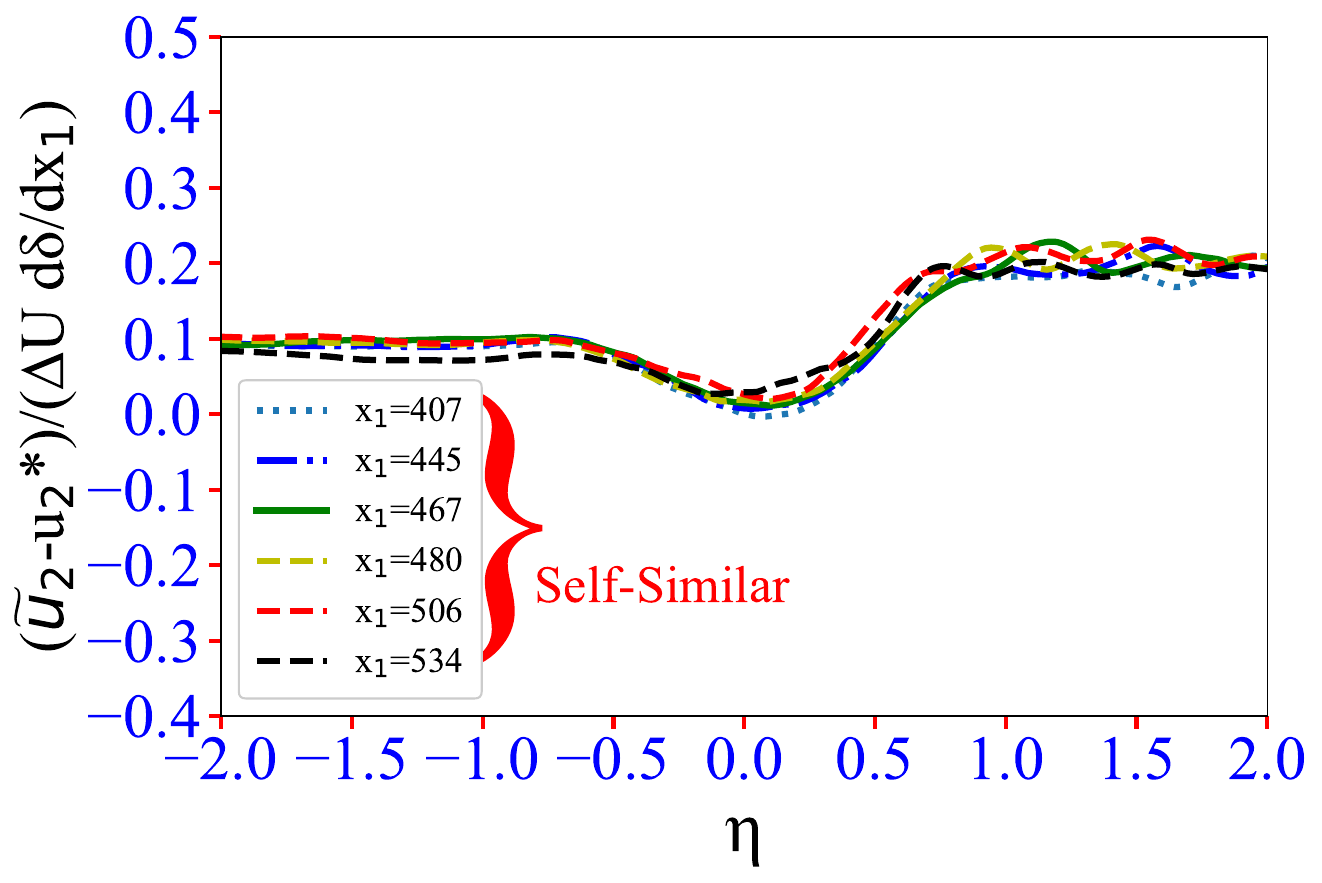}
         \subcaption{}
     \end{subfigure}
      \begin{subfigure}[b]{0.50\textwidth}
         \centering
         \includegraphics[width=\textwidth]{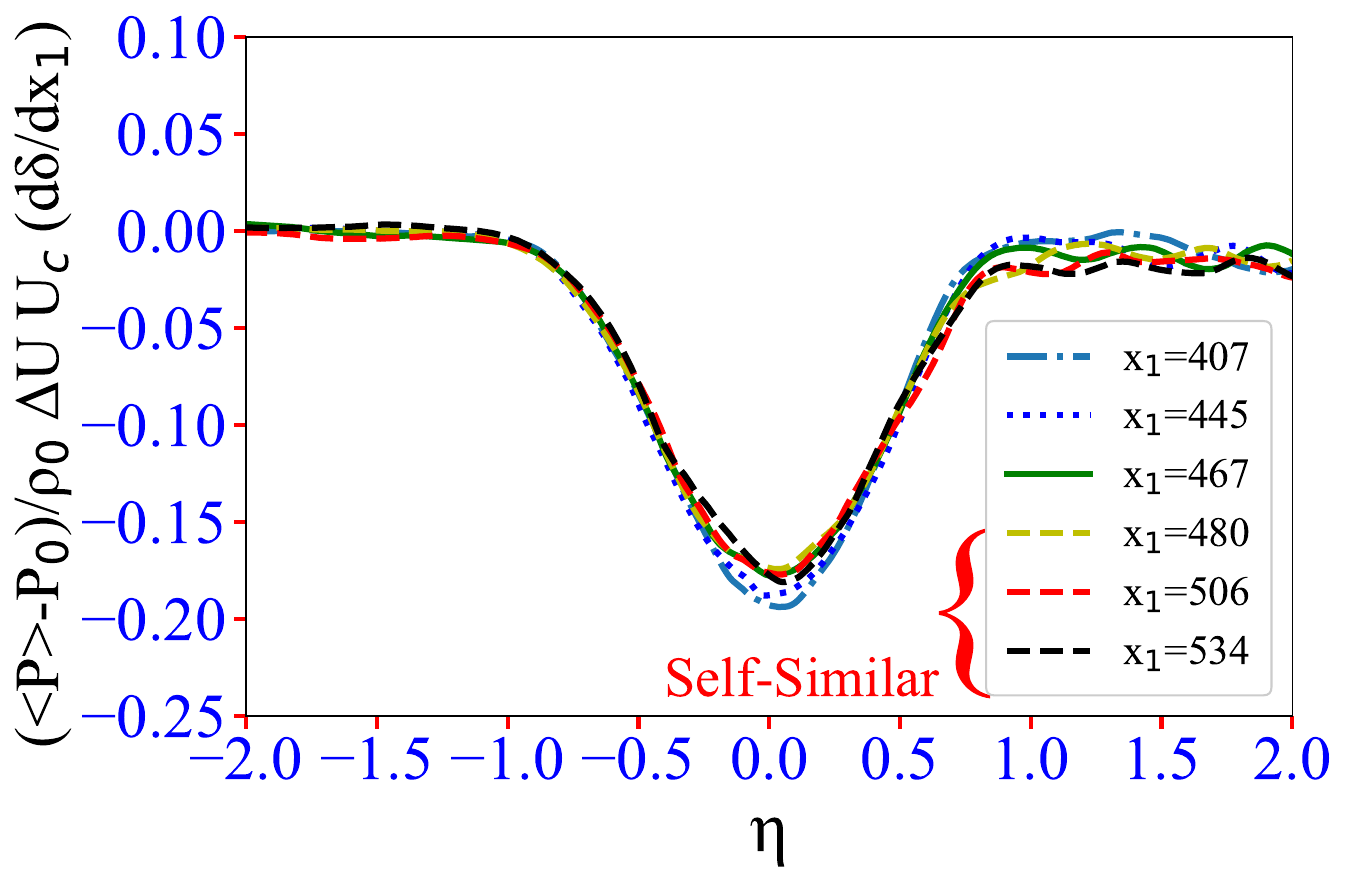}
         \subcaption{}
     \end{subfigure}
      \begin{subfigure}[b]{0.455\textwidth}
         \centering
         \includegraphics[width=\textwidth]{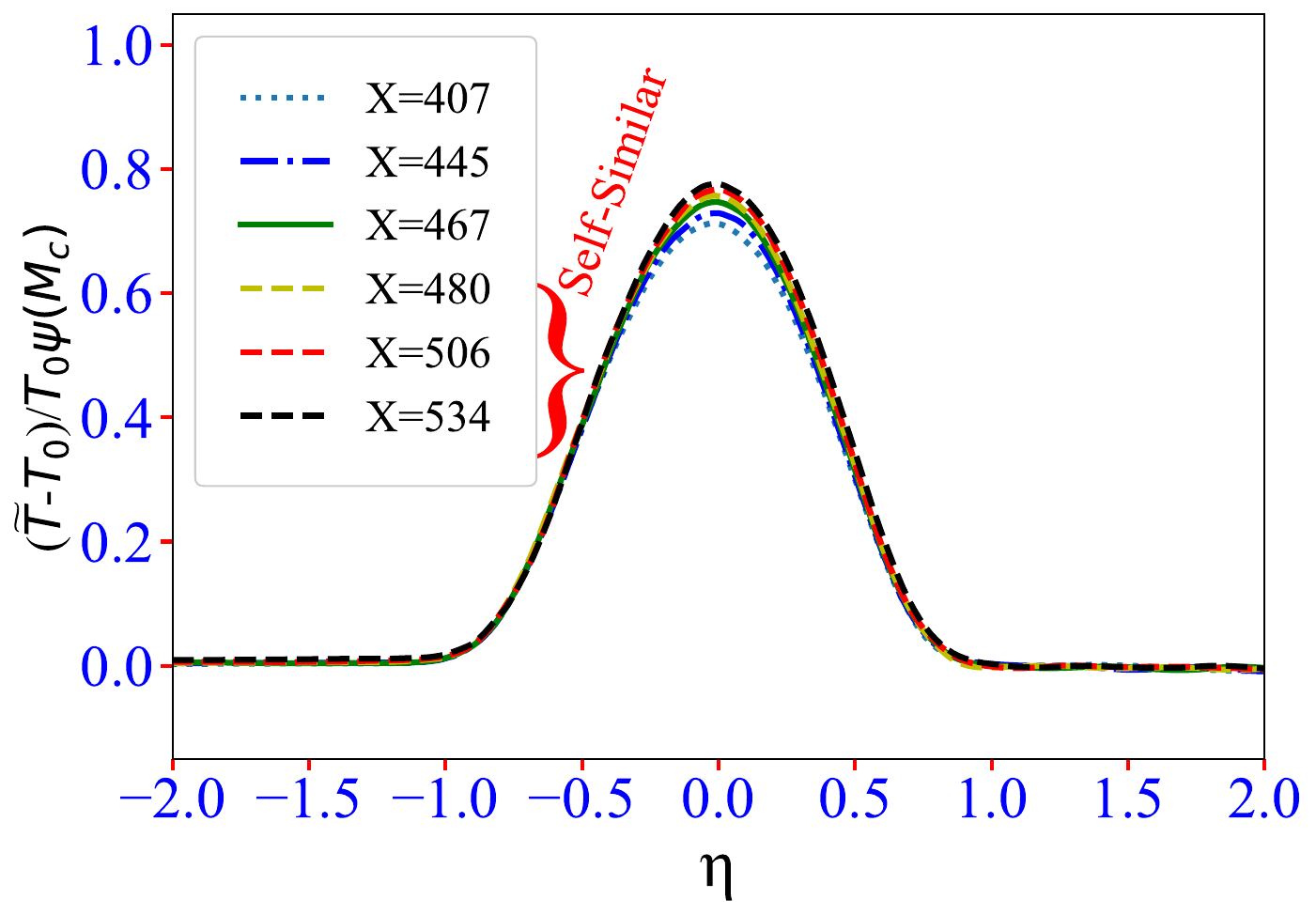}
         \subcaption{}
     \end{subfigure}
     \begin{subfigure}[b]{0.495\textwidth}
         \centering
         \includegraphics[width=\textwidth]{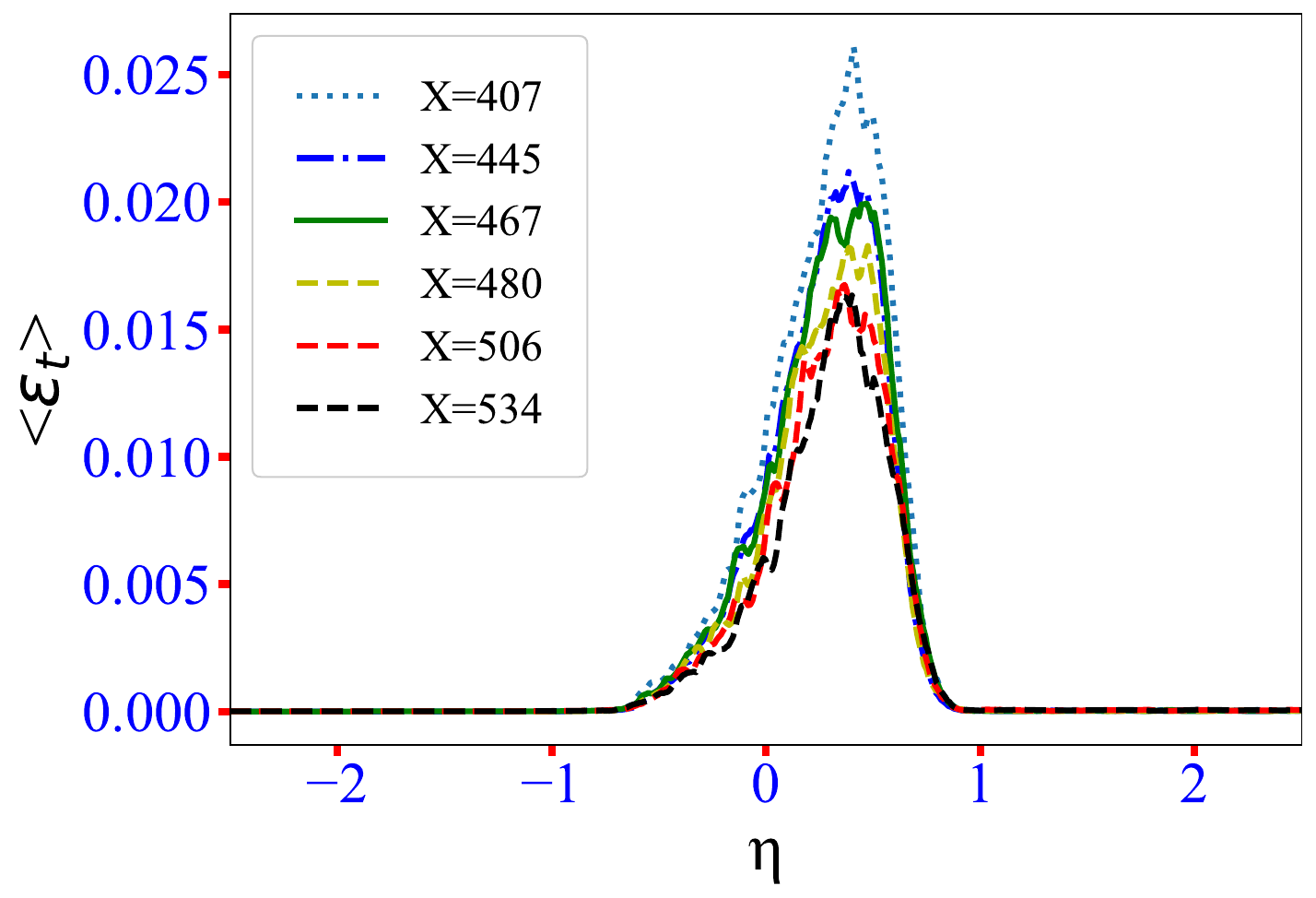}
         \subcaption{}
     \end{subfigure}
     \begin{subfigure}[b]{0.47\textwidth}
         \centering
         \includegraphics[width=\textwidth]{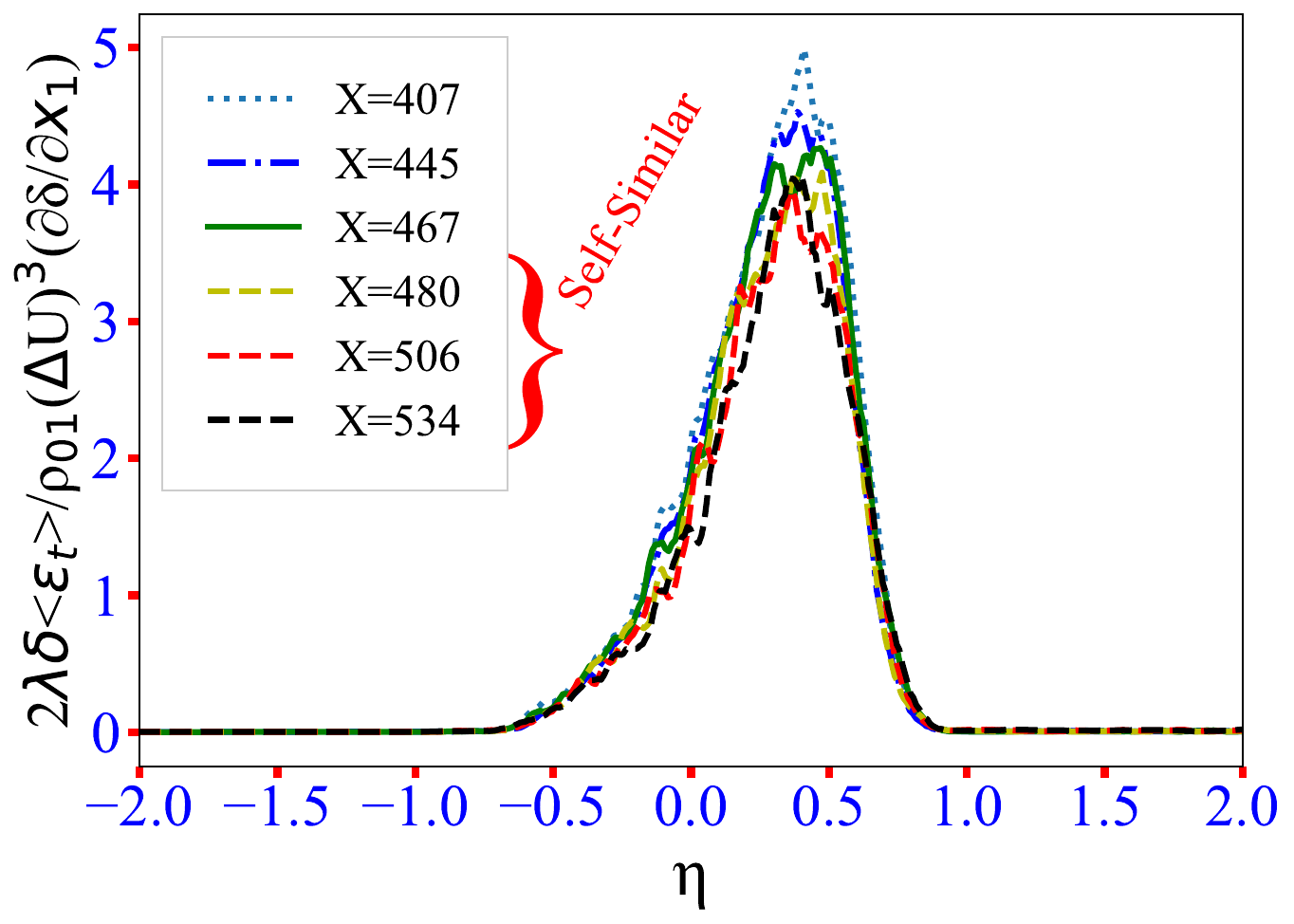}
         \subcaption{}
     \end{subfigure}
     \caption{The collapse of (a) scaled mean density, (b) scaled mean transverse velocity, (c) scaled mean pressure, and (d) scaled mean temperature in self-similar coordinates. Mean density becomes self-similar around $x_1\approx 407$, and mean pressure and temperature become self-similar around $x_1\approx467$. Figure (e) shows the mean dissipation at different streamwise locations and figure (f) shows the collapse of scaled mean dissipation in self-similar coordinates.}
     \label{fig:selfsimilarC}
\end{figure}

Figure~\ref{fig:selfsimilar1} shows the mean and turbulent quantities at different streamwise locations for the reference case. From figure \ref{fig:selfsimilar1} (a), (b), and (c), it can be seen that the shear layer thickness and momentum thickness start to grow linearly after $x_1\approx 407$. Therefore, these two quantities become self-similar around $x_1\approx 407$. The self-similarity condition for mean streamwise velocity is also satisfied for $x_1 \gtrapprox 407$. The scaled mean streamwise velocity can be approximated as an error function ($\hat{u_1}$=$Aerf[B\eta]$) which will be used to derive a closed-form solution for the mean transverse velocity in section \ref{sec:Mc}. The higher-order statistics become self-similarity further downstream. Thus, the Reynolds stresses reach constant peak values for $x_1 \gtrapprox 480$. We show next that $x_1 \gtrapprox 480$ is a reasonable range for achieving self-similar behavior for the rest of the mean quantities appearing in the self-similar equations considered here. 

All cases examined here fall within the highly compressible range of values of $M_c$, and from figure \ref{fig:selfsimilarC} (a), it is seen that density variations are significant within the core of the shear layer. The scaled mean density profiles collapse for $x_1 \gtrapprox 407$. However, the mean density depends on dissipation, as will be explained below, and is very sensitive to the $M_c$ value, so it is challenging to collapse the density profiles over cases with different $M_c$ values. Therefore, we have introduced the function $\psi(M_c)$ in the mean density scaling.  With this scaling, the mean density profiles collapse over different $M_c$ runs and can be approximated as a Gaussian function ($\hat{\rho}$=$ae^{-b\eta^2}$, figure \ref{fig:selfsimilarC} a). A detailed discussion on this approximation will be given in section \ref{sec:Mc}. For the mean transverse velocity, self-similarity is achieved at a similar downstream location, i.e. at $x_1 \gtrapprox 407$ (figure figure \ref{fig:selfsimilarC} b). The self-similarity of scaled mean pressure starts at further downstream locations. For the reference case, the self-similar condition for the mean pressure is achieved at $x_1 \gtrapprox 467$ (figure \ref{fig:selfsimilarC} c). The mean temperature also becomes self-similar around the same location (figure figure \ref{fig:selfsimilarC} d), which is expected as the pressure and temperature are related through the equation of state. Furthermore, after nondimensionalization the mean dissipation attains a self-similar behavior for $x_1 \gtrapprox 480$, similar to the Reynolds stresses (see figures \ref{fig:selfsimilarC} e and f). 

Based on these results, we choose to represent the self-similar range for the base case as $x_1 \gtrapprox 480$. The same criteria are used for all cases in this study. Table~\ref{tab:sszone} summarizes the self-similar ranges for all DNS cases, starting from the first downstream location that satisfies the self-similar criteria for all turbulence statistics considered to the end of the computational domain. 

\begin{table}
  \centering
  \begin{tabular}{|c|c|c|c|}
  \hline
Simulated Cases  & $M_c$   &   $\lambda$ & $x_1/\delta_{0}$ \\[3pt]
       \hline
       A02M12   & 1.2 & 0.2 & 642-714\\
       \hline
        A03M12   & 1.2 & 0.3 & 550-662\\
       \hline
       A04M12  & 1.2 & 0.4 & 428-491\\
       \hline
       A05M12   & 1.2 & 0.5 & 375-446\\
       \hline
       A04M08 & 0.8 & 0.4 & 378-446\\
       \hline
       A04M10 & 1.0 & 0.4 & 403-491\\
       \hline
       A04M14 & 1.4 & 0.4 & 456-535\\
       \hline
       A04M16 & 1.6 & 0.4 & 473-535\\
       \hline
  \end{tabular}
  \caption{The self-similar zones for all the simulated cases}
  \label{tab:sszone}
\end{table}

\subsection{Behavior of scaled transverse velocity at $\eta = \pm \infty$}\label{sec:u2hat1}

As mentioned in section \ref{sec:mean}, the behavior of mean transverse velocity is very different on the high-speed and low-speed sides. In this section, we explain this asymmetric behavior through the integral form of the self-similar continuity equation (\ref{eq:sscon}). Integrating this equation from $\eta=-\infty$ to $\eta=+\infty$ and applying the boundary conditions from table~\ref{tab:bc}, it yields:
\begin{equation}\label{eq:intcon}
    \hat{u}_{2,+\infty}-\hat{u}_{2,-\infty}=-\psi(M_c)\int_{-\infty}^{+\infty}\hat{\rho}\hat{u_1}\,d\eta- \frac{\psi(M_c)}{2\lambda}\int_{-\infty}^{+\infty}\hat{\rho}\,d\eta+\int_{-\infty}^{+\infty}\eta\frac{d\hat{u_1}}{d\eta}d\eta+\phi.
\end{equation}

\begin{table}
  \centering
  \begin{tabular}{|c|c|c|}
  \hline
  & $\hat{\rho}$  & $\hat{u_1}$ \\[3pt]
  \hline
    $\eta={+\infty}$ &   0  &  $\frac{1}{2}$\\
    \hline
    $\eta={-\infty}$ &   0  & -$\frac{1}{2}$\\
    \hline
  \end{tabular}
  \caption{Boundary conditions for the dimensionless density and streamwise velocity}
  \label{tab:bc}
\end{table}

Empirically, it is observed that $\hat{u_1}$ is approximately antisymmetric while and $\hat{\rho}$ is symmetric with respect to the centerline in the rotated coordinate system (i.e. $\eta =0$ line), as discussed in sections \ref{sec:sszone} and \ref{sec:Mc}. Therefore, the 1st and 3rd integrals on the right-hand side of the equation (\ref{eq:intcon}) are nearly zero and the integrated compressible self-similar continuity equation can be approximated by:
\begin{equation}\label{eq:int_con1}
    \hat{u}_{2,+\infty}-\hat{u}_{2,-\infty}=- \frac{\psi(M_c)}{2\lambda}\int_{-\infty}^{+\infty}\hat{\rho}\,d\eta+\phi.
\end{equation}

In the incompressible limit ($M_c \rightarrow 0$, $\psi(M_c) \rightarrow 0$)equation  (\ref{eq:int_con1}) becomes:
\begin{equation}\label{eq:int_con2}
    \hat{u}_{2,+\infty}-\hat{u}_{2,-\infty}=\phi.
\end{equation}

Equation (\ref{eq:int_con2}) is the same as that derived by  \citealt{wei22}. From equation (\ref{eq:int_con1}), it can be seen that for the compressible spatially evolving shear layer, the difference between $\hat{u}_{2,+\infty}$ and $\hat{u}_{2,-\infty}$ is primarily due to the density variation inside the shear layer and centerline shifting. However, in the incompressible limit, since there is no density variation, the difference between $\hat{u}_{2,+\infty}$ and $\hat{u}_{2,-\infty}$ is only due to the centerline shifting.

\subsection{Density variations and dissipation}\label{sec:densityvariation}

In this section, we explore the underlying factors contributing to the density variation inside the shear layers, and through that, we will try to explain the asymmetric behavior of the transverse velocity using the integral forms of both continuity and energy equations. A similar attempt to connect density variations with dissipation was made by \cite{bretonnet2007deflection} for the temporal case. An additional distinction from the analysis performed here lies in the flow considered, as their investigation focused on laminar flow conditions. Finally, \cite{bretonnet2007deflection} assumed constant $\delta$ over some short time scale, which is nevertheless inconsistent with the linear $\delta$ growth in the self-similar region of the shear layer. As shown in section \ref{sec:sszone}, this assumption is not necessary, as the dissipation itself has a self-similar behavior when properly scaled. 

\begin{figure}
     \centering
     \begin{subfigure}{0.55\textwidth}
         \centering
         \includegraphics[width=\textwidth]{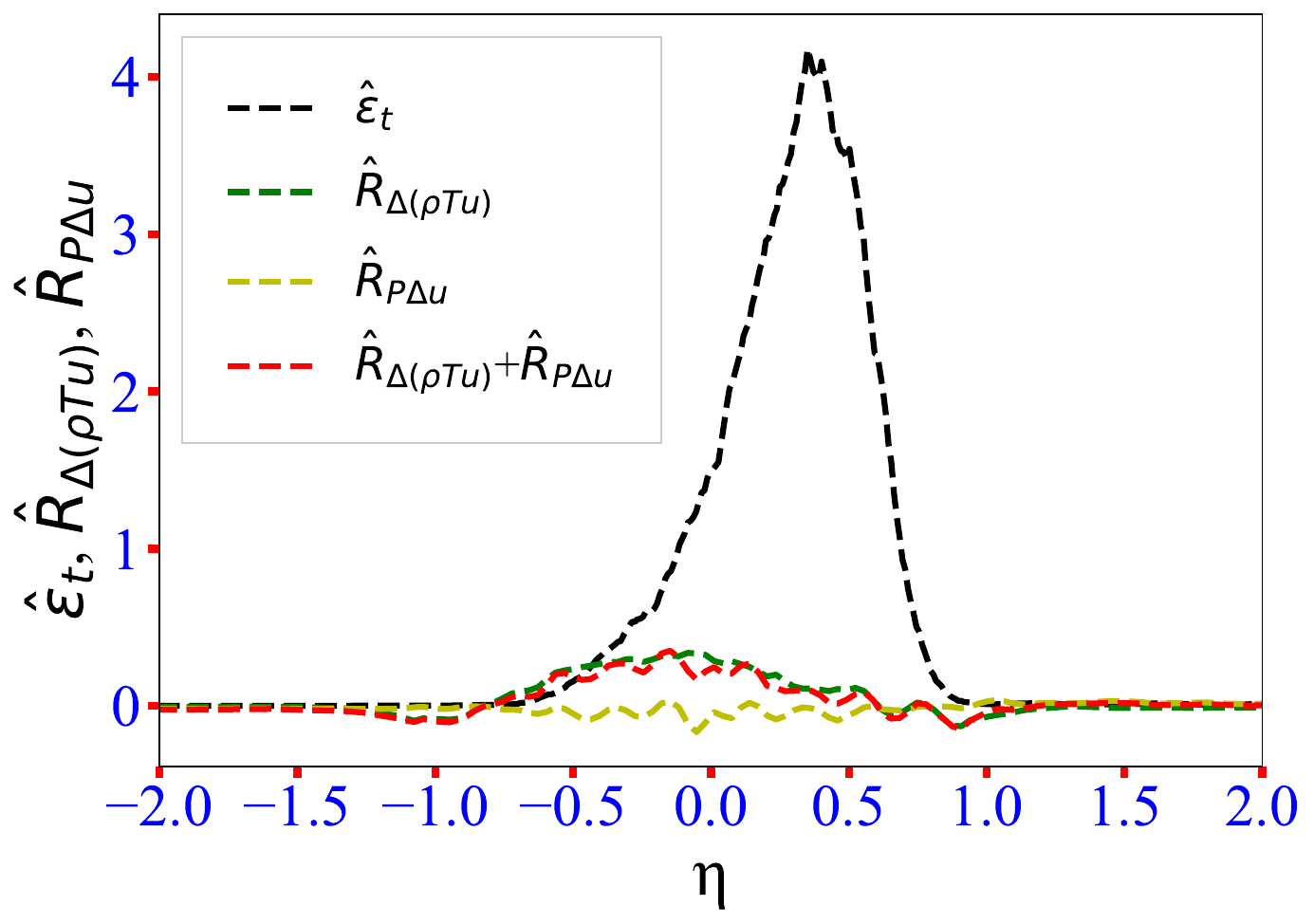}
     \end{subfigure}
     \caption{ $\hat{\epsilon_t}$ compared to $\hat R_{\Delta (\rho T u)}$ and $\hat R_{P \Delta u}$ for the reference case. It is clear from the figure that $\hat{\epsilon_t}$ is much larger than the summation of $\hat R_{\Delta (\rho T u)}$ and $\hat R_{P \Delta u}$.}
     \label{fig:fluc}
\end{figure}

After integrating the energy equation (\ref{eq:ssenr}) from $\eta=-\infty$ to $\eta=+\infty$ and applying the boundary conditions from table \ref{tab:bc}, the resulting integral form of the equation can be expressed as follows:
\begin{equation}\label{eq:dis_2}
    \hat{u}_{2,+\infty}-\hat{u}_{2,-\infty} -\phi = 4(\gamma-1)  \frac{M_c^2 \psi}{2\lambda} \int_{-\infty}^{+\infty} [\hat{\epsilon_t}-(\hat{R}_{\Delta(\rho T u)}+\hat{R}_{P\Delta u})]\,d\eta,
\end{equation}
where the pressure terms were neglected as our DNS data suggest that the variation of mean pressure inside the shear layer is negligible. Furthermore, if equation (\ref{eq:dis_2}) is combined with the integral form of the continuity equation presented in (\ref{eq:int_con1}), the resulting form of the equation can be expressed as follows:
\begin{equation}\label{eq:dis_3}
   -\int_{-\infty}^{+\infty}\hat{\rho}\,d\eta = 4(\gamma-1)M_c^2 \int_{-\infty}^{+\infty} [\hat{\epsilon_t}-(\hat{R}_{\Delta(\rho T u)}+\hat{R}_{P\Delta u})]\,d\eta.
\end{equation}
According to our DNS data, the $\hat{\epsilon_t}$ is much larger compared to the summation of $\hat R_{\Delta (\rho T u)}$ and $\hat R_{P \Delta u}$ (Figure \ref{fig:fluc}) in the equation \ref{eq:dis_3}. Therefore, by neglecting these terms, the resulting form of the equation is:
\begin{equation}\label{eq:dis_4}
   -\int_{-\infty}^{+\infty}\hat{\rho}\,d\eta \approx 4(\gamma-1)M_c^2 \int_{-\infty}^{+\infty} \hat{\epsilon_t}\,d\eta.
\end{equation}

Given the intrinsically positive value of the dissipation term, the right-hand side of the equation (\ref{eq:dis_4}) is always positive. This requires a negative region in the $\hat{\rho}$ variation, confirming the drop in mean density within the shear layer compared to the free stream value.
Physically, the density drop can be attributed to overheating due to the viscous dissipation at the core of the shear layer. This results in a localized maximum in the temperature profile and a corresponding localized minimum in the density profile. \cite{bretonnet2007deflection} also observed a local maximum in the temperature profile at a higher $M_c$ value. Heating due to viscous dissipation is also responsible for the generation of transverse velocity, as shown using the same approximations of equation (\ref{eq:dis_2}) as for equation (\ref{eq:dis_4}): 

\begin{equation}\label{eq:dis_5}
    \hat{u}_{2,+\infty}-\hat{u}_{2,-\infty}  \approx 4(\gamma-1)  \frac{M_c^2 \psi}{2\lambda} \int_{-\infty}^{+\infty} \hat{\epsilon_t}\,d\eta +\phi,
\end{equation}

\subsection{Approximate solution for scaled mean transverse velocity}\label{sec:approxu2}

Next, we will build on our self-similar analysis to find an approximate formula for the transverse velocity. The transverse velocity variation in a planar shear layer has received only scarce attention in the literature, in part due to the difficulty in obtaining accurate experimental data. \cite{wei22} summarized the existing literature for the incompressible case and derived an approximate solution, but the compressible version has not been addressed so far. The analysis presented here is an generalization of the  \cite{wei22} analysis to compressible shear layers. Integrating the self-similar continuity equation (\ref{eq:sscon}) from $-\infty$ to $\eta$ and applying boundary conditions of table~\ref{tab:bc}, one can produce an analytical formula for the scaled mean transverse velocity as a function of $\eta$:
\begin{equation}
\label{equation:u2a}
\begin{array}{l}
    \hat{u}_{2}= 
    \frac{\psi(M_c)}{1+\hat{\rho}\psi(M_c)}[\eta \hat{\rho} \hat{u_{1}} -\int_{-\infty}^{\eta}\hat{\rho}\hat{u_1}\,d\eta +\phi\hat{\rho} \hat{u_{1}}] +\frac{\psi(M_c)}{2\lambda(1+\hat{\rho}\psi(M_c))}[\eta \hat{\rho}-\int_{-\infty}^{\eta}\hat{\rho}\,d\eta+\phi\hat{\rho}] \\\\ + \frac{\phi}{2(1+\hat{\rho}\psi(M_c))}[1+2\hat{u_{1}}] + \frac{1}{1+\hat{\rho}\psi(M_c)}\int_{-\infty}^{\eta}\eta\frac{d\hat{u_1}}{d\eta}\,d\eta-\frac{\psi(M_c)}{1+\hat{\rho}\psi(M_c)} \frac{u_2^*}{u_{02}}\hat{\rho}+\frac{\hat{u}_{2,-\infty}}{1+\hat{\rho}\psi(M_c)}.\;
    \end{array}
\end{equation}
This equation depends on $u_2^*$, which is the transverse velocity at the location where the mean streamwise velocity is zero, namely $\eta=0$. Using formula (\ref{equation:u2a}) to evaluate $u_2^*$ and following some simplifications outlined in Appendix \ref{appc}, the $\hat{u}_{2}$ solution becomes:
\begin{equation}
\begin{array}{l}
    \hat{u}_{2}= \phi\hat{u_{1}} +\frac{1}{1+\hat{\rho}\psi(M_c)}[\psi(M_c)\int_{-\infty}^{\eta} \eta \frac{d(\hat{\rho}\hat{u_1})}{d\eta}\,d\eta+
    \frac{\psi(M_c)}{2\lambda}\int_{-\infty}^{\eta} \eta \frac{d\hat{\rho}}{d\eta}\,d\eta+ \int_{-\infty}^{\eta}\eta\frac{d\hat{u_1}}{d\eta}\,d\eta]\\\\
    -\frac{1}{1+\hat{\rho}_0\psi(M_c)} [\psi(M_c)\int_{-\infty}^{0} \eta \frac{d(\hat{\rho}\hat{u_1})}{d\eta}\,d\eta+
    \frac{\psi(M_c)}{2\lambda}\int_{-\infty}^{0} \eta \frac{d\hat{\rho}}{d\eta}\,d\eta+ \int_{-\infty}^{0}\eta\frac{d\hat{u_1}}{d\eta}\,d\eta]\\\\
     +[\frac{1}{1+\hat{\rho}\psi(M_c)} -\frac{1}{1+\hat{\rho}_0\psi(M_c)}] (\psi(M_c)\frac{\lambda-1}{2\lambda}\int_{-\infty}^{+\infty}\hat{\rho}\hat{u_1}\,d\eta+\frac{\psi(M_c)}{4\lambda}\int_{-\infty}^{+\infty}\hat{\rho}\,d\eta\\\\
    +\frac{1-\lambda}{2\lambda}\int_{-\infty}^{+\infty}\eta\frac{d\hat{u_1}}{d\eta}\,d\eta
    -\psi(M_c)\int_{-\infty}^{+\infty}\hat{\rho}\hat{u_1}\hat{u_1}\,d\eta+2\int_{-\infty}^{+\infty}\eta\hat{u_1}\frac{d\hat{u_1}}{d\eta}d\eta+\phi \int_{-\infty}^{+\infty}\hat{u_1}\frac{d\hat{u_1}}{d\eta}\,d\eta).
    \end{array}
\end{equation}

This formula can be further simplified by using approximations of the mean streamwise velocity and density profiles. We use an error function profile to approximate $\hat{u}_1$ as $\hat{u}_1=A erf(B\eta)$ and a Guassian function to approximate $\hat{\rho}$ as $\hat{\rho}=a e^{-B^2\eta^2}$ as outlined in Appendix \ref{appB}. The  numerical fit described in section \ref{sec:Mc} suggests the values for the constants $A$ and $a$ as 0.5 and -1.0, respectively. Then the approximate closed-form solution for $\hat{u_2}$ becomes: 
\begin{equation}\label{eq:finalu2hat1}
\begin{array}{l}
\hat{u}_{2} =  \frac{\phi}{2}erf(B\eta)-\frac{\eta e^{-B^2\eta^2}\psi(M_c)[1+\lambda erf(B\eta)]}{2\lambda(1-\psi(M_c) e^{-B^2\eta^2})}
+\frac{1}{1-\psi(M_c)}[\frac{\psi(M_c)\sqrt{\pi}(1+\lambda)}{8B\lambda}\\
+\frac{1}{2\sqrt{\pi}B}]-\frac{1}{1-\psi(M_c)e^{-B^2\eta^2}}[\frac{\psi(M_c)\sqrt{\pi}[1+erf(B\eta)]}{8B}(\frac{1}{\lambda}+\frac{1}{2}[erf(B\eta)-1])+ \frac{e^{-B^2\eta^2}}{2\sqrt{\pi}B}]\\
+ [\frac{1}{1-\psi(M_c)e^{-B^2\eta^2}}-\frac{1}{1-\psi(M_c)}][\frac{-\psi(M_c)(3-\lambda)}{12\lambda B}+\frac{1}{\sqrt{2\pi}B}]
\end{array}
\end{equation}

Formula (\ref{eq:finalu2hat1}) represents the compressible form of the approximate solution of the scaled transverse velocity to the self-similar equations and serves as the basis for investigating the characteristics of the scaled mean transverse velocity and obtaining further approximate solutions for the entrainment ratio. The predictions based on formula (\ref{eq:finalu2hat1}) are compared with the current DNS results across different $M_c$ and $\lambda$ values (sections \ref{sec:Mc} and \ref{sec:At}, respectively). In the incompressible limit, $\psi (M_c)$ converges to zero and the incompressible form of the approximate self-similar transverse velocity solution becomes:
\begin{equation}\label{eq:u2incom1}
\begin{split}
\hat{u}_{2} = & \frac{\phi}{2} erf(B\eta) + \frac{1}{2\sqrt{\pi} B}-\frac{1}{2\sqrt{\pi}B}e^{-B^2\eta^2}\\
\end{split}
\end{equation}

Equation (\ref{eq:u2incom1}) is similar to the scaled mean transverse velocity derived by \cite{wei22} for the incompressible constant density case. 

\section{Role of convective Mach number and velocity parameter on the flow evolution}\label{ss2} 

In the previous section, we formulated self-similar forms of continuity, $x_1$ momentum, $x_2$ momentum, and energy equations, incorporating both compressibility and centerline shifts, and showed that they depend on two distinct parameters that characterize the flow, namely $M_c$ and $\lambda$. In this section, we consider the role of these two parameters on the evolution of the flow. In addition to the usual metrics such as shear layer thickness and mean streamwise velocity and density, we will also examine the transverse velocity dependence on $M_c$ and $\lambda$, which was only scarcely studied in the literature. In addition, the role of $\lambda$ has been much less studied compared to that of $M_c$ in previous studies.

\subsection{Convective Mach Number}\label{sec:Mc}

\begin{figure}
     \centering
     \begin{subfigure}{0.55\textwidth}
         \centering
         \includegraphics[width=\textwidth]{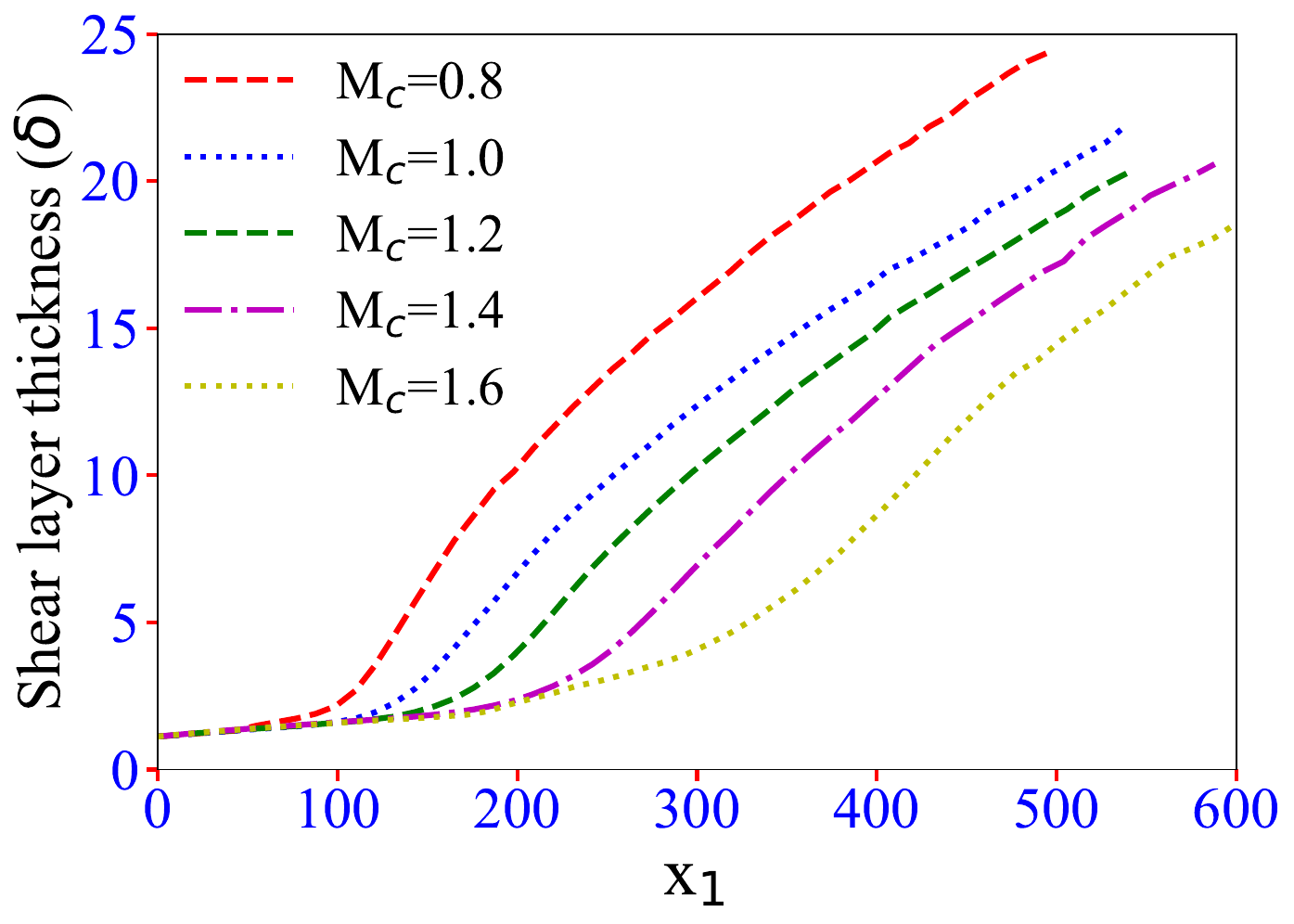}
     \end{subfigure}
     \caption{The evolution of shear layer thickness at different $M_c$}
     \label{fig:deltaMc}
\end{figure}

The streamwise variations of shear layer thickness, shown in figure~\ref{fig:deltaMc} for all simulated $M_c$ values, indicate three distinct regions. The first region is referred to as the initial slow development region where the three-dimensional instability starts to grow. The second region is the transition region or the middle higher development region where the large-scale vortex structure starts to rapidly grow. The multiplicity of ring-like vortices in the transition region leads to a much faster growth rate of the mixing layer. The third region with a slower growth rate is the self-preserving or self-similar region. The linear growth rate predicted by the self-similar analysis is achieved in this region (\citealt{zhang19}). Regardless of which thickness definition is employed, the shear layer growth rate decreases as the $M_c$ increases. Figure~\ref{fig:deltaMc} shows that the streamwise location where the shear layer begins to rapidly grow is different for different $M_c$ values with delayed growth at higher $M_c$. %For M$_c$=0.8, the layer starts to grow around $x_1$=100 whereas, for M$_c$=1.6, the layer starts to grow around $x_1$=250. 
As explained above, the domain size was extended for the higher $M_c$ simulations to ensure that the self-similar region is well captured.

In numerous previous studies on shear layers, the growth rate of the shear layer was found to consistently decrease with increasing $M_c$. The reduced growth rate can be related to the decrease of turbulent production as the compressibility increases. Turbulent production is, in turn, reduced as a result of the decrease in normalized pressure-strain terms (\cite{vreman1996compressible}), which inhibits energy transfer from streamwise to cross-stream fluctuations. The decreased production term in the turbulent kinetic energy equation is also responsible for the reduction of turbulent kinetic energy within the shear layer (\citealt{pantano02}). In the more idealized configuration of homogeneous shear flow, analytical solutions can be obtained showing the direct connection between the pressure mode, the dilatational velocity in the transverse direction and the reduction in the turbulence production term with the distortion Mach number (\citet{livescu2004small}).

\begin{figure}
     \centering
     \begin{subfigure}{0.75\textwidth}
         \centering
         \includegraphics[width=\textwidth]{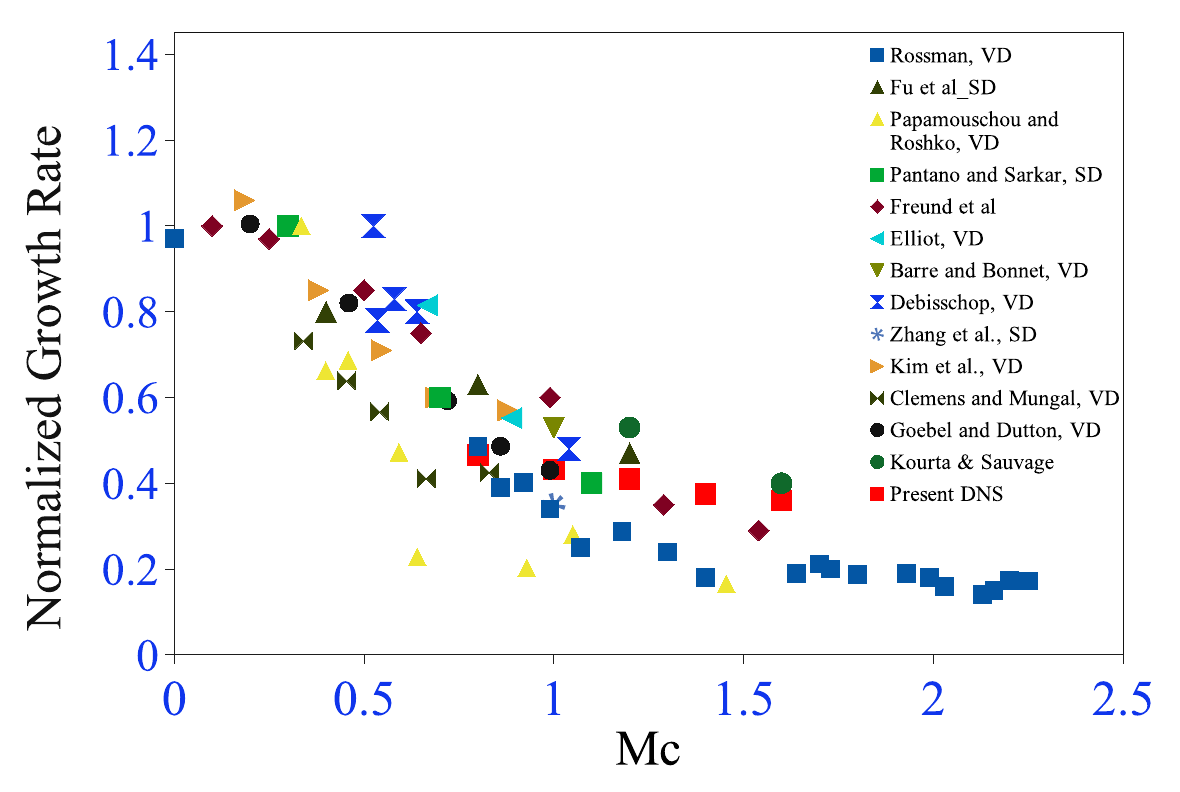}
     \end{subfigure}
     \caption{Normalized growth rate at different $M_c$ comparing current DNS results to previous literature (\cite{kim20}, \cite{zhang19}, \cite{rashko88}, \cite{pantano02}, \cite{elliot90}, \cite{rossmann02},\cite{mungal92}, \cite{fu2006numerical}, \cite{lele02}, \cite{barre1997influence}, \cite{debisschop1993mean}, \cite{goebel1990velocity}, \cite{kourta2002computation}).}
     \label{fig:ngr}
\end{figure}

The growth rates obtained in the present DNS study exhibit a similar decreasing trend with compressibility, as seen in figure \ref{fig:deltaMc}. To compare with other studies, the growth rates derived from current simulations were normalized by the corresponding incompressible growth rate established in previous works \cite{wei22,abramovich84,sabin65,pope2000turbulent,dimotakis1991turbulent} and the results are shown in figure~\ref{fig:ngr}. 
All studies show a clear reduction in the normalized growth rate with increasing compressibility, and the normalized growth rates reported for the present study are in line with the literature. However, figure~\ref{fig:ngr} also shows that the data are scattered significantly. There are a number of factors behind this scattering. First, researchers use various definitions for mixing layer thickness, which we believe is the main reason for the scatter in the collected results (\citealt{kim20}). For example, the growth rate is defined differently depending on the diagnostic techniques used in each experiment and  depending on the definition of the shear layer width using, e.g. the visual thickness with 1\% or 10\% cut-off levels, the vorticity thickness or momentum growth thickness, which generate different results (\citealt{kim20}). Here, we used the shear layer growth rate calculated from the 10\% velocity difference compared to the free stream velocity.  \cite{abramovich84}  and \cite{sabin65} suggested a linear relationship between the shear layer growth rate and $\lambda$. The proportionality constant of the relation suggested by \cite{abramovich84} is twice the spreading rate defined by \cite{pope2000turbulent}. We used the value of the spreading rate reported by \cite{dimotakis1991turbulent} to calculate the corresponding incompressible growth rate.
However, different definitions of the shear layer width do not fully explain the scatter in the results shown in figure~\ref{fig:ngr}. As we discuss in the next section, the growth rate also depends on $\lambda$, besides the Mach number and density ratio of the free streams.

%Our analysis suggests that parameters other than velocity ratio and density ratio are involved in inhibiting the shear layer growth (\citealt{kim20}). 

\begin{figure}
     \centering
     \begin{subfigure}[b]{0.43\textwidth}
         \centering
         \includegraphics[width=\textwidth]{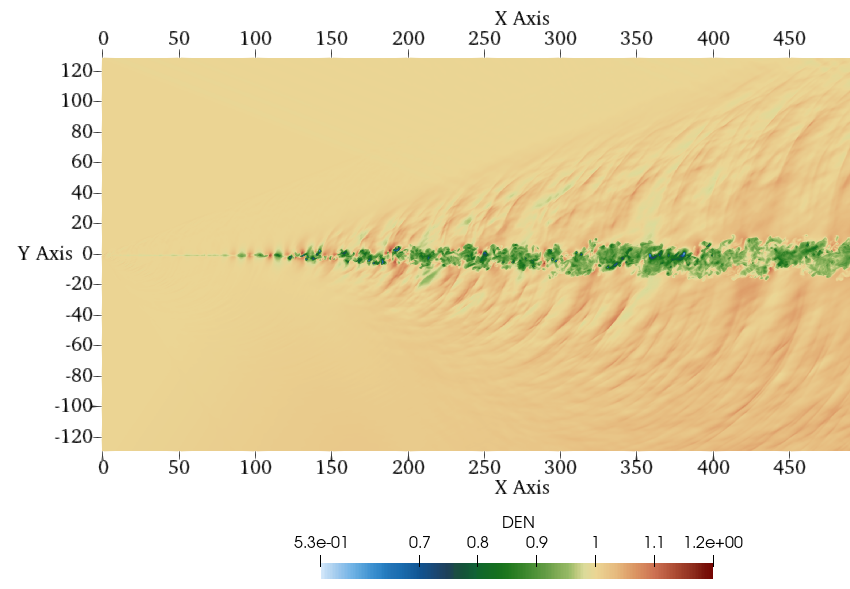}
         \subcaption{}
     \end{subfigure}
     \hfill
     \begin{subfigure}[b]{0.56\textwidth}
         \centering
         \includegraphics[width=\textwidth]{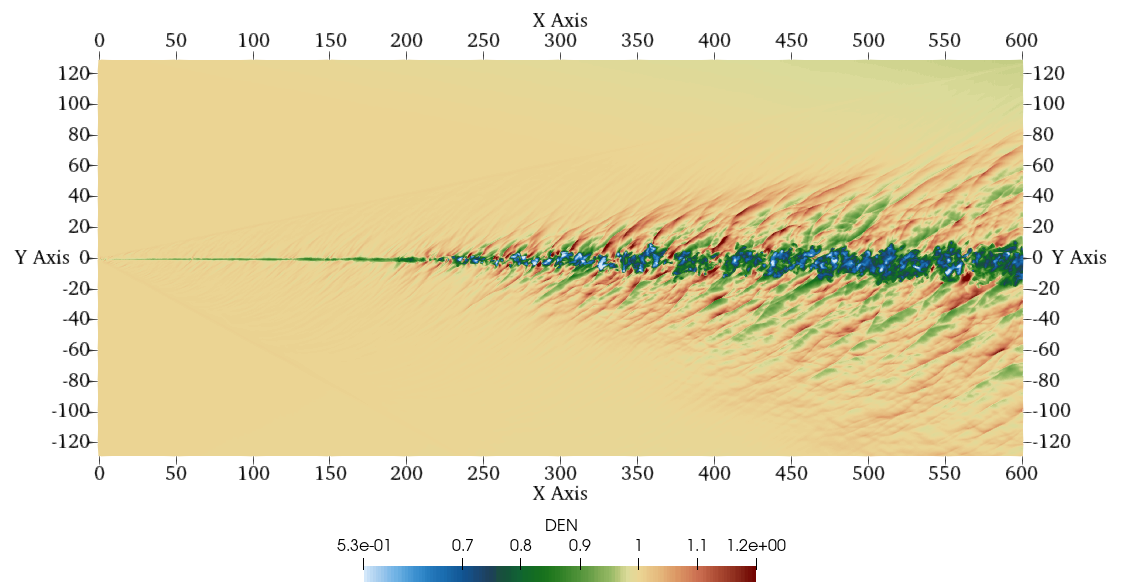}
         \subcaption{}
     \end{subfigure}
     \caption{Color contour ($x_1$–$x_2$ plane) of instantaneous density for a) M$_c$= 0.8 and b) M$_c$= 1.6}
     \label{fig:contourMc}
\end{figure}

Another way to illustrate the reduction in shear layer growth rate with $M_c$ is to consider the contours of various flow variables. Instantaneous color contours of the density for $M_c=0.8$ and $M_c=1.6$  in the $x_1-x_2$ plane at the mid-spanwise location and the same instant in time are shown in figure~\ref{fig:contourMc}. Figure~\ref{fig:contourMc}(a) indicates the presence of dominant large-scale structures at comparatively lower $M_c$ (\citealt{mungal92}, \citealt{kim20}). On the other hand, figure~\ref{fig:contourMc}(b) shows that at higher $M_c$, the shear layer contains much smaller eddies and the structures are less coherent. Moreover, the contours indicate that the shear layer begins to develop most upstream for the smallest convective Mach number, $M_c=0.8$.

\begin{figure}
     \centering
     \begin{subfigure}{0.48\textwidth}
         \centering
         \includegraphics[width=\textwidth]{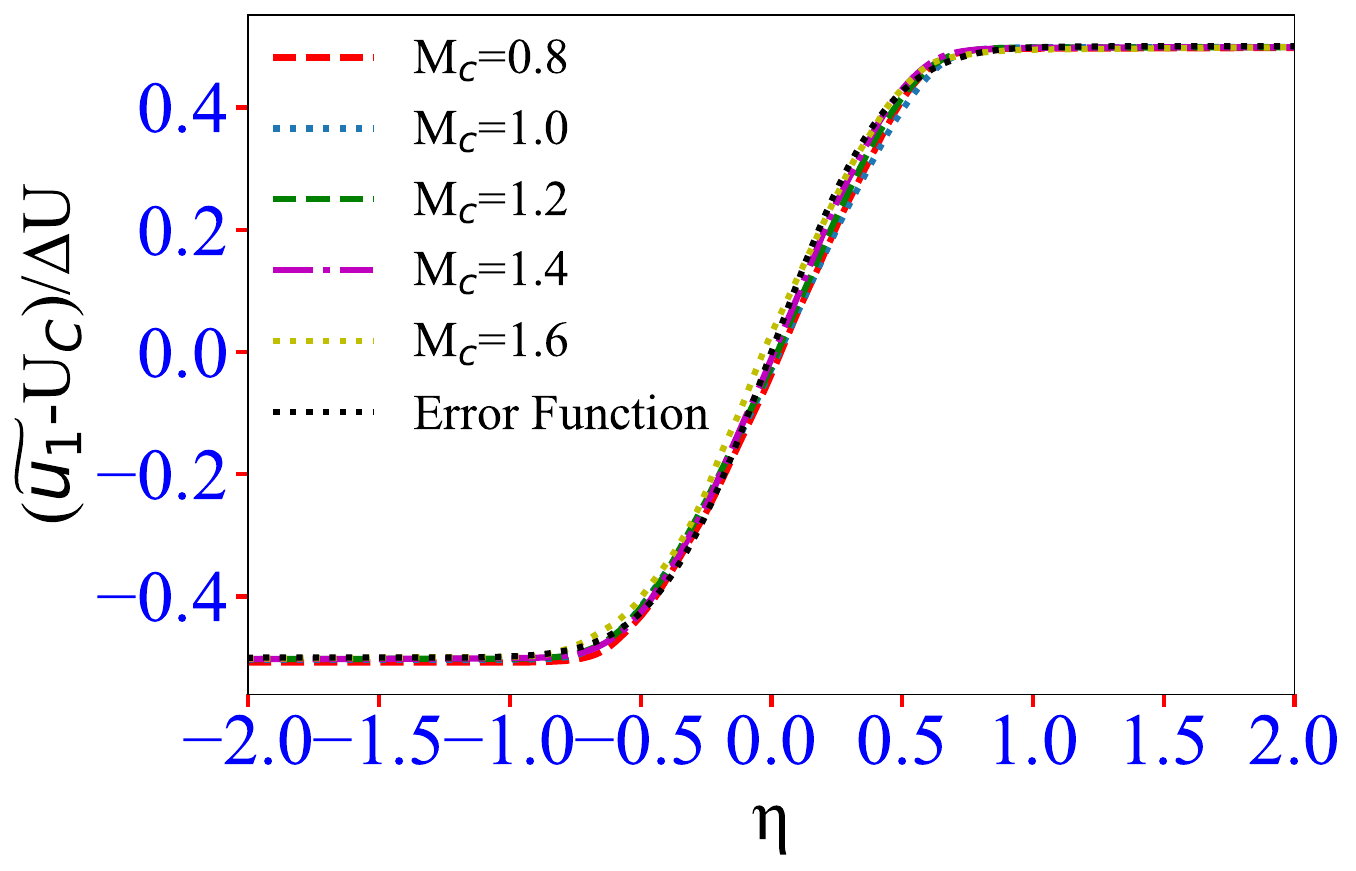}
         \subcaption{}
     \end{subfigure}
     \begin{subfigure}{0.50\textwidth}
         \centering
         \includegraphics[width=\textwidth]{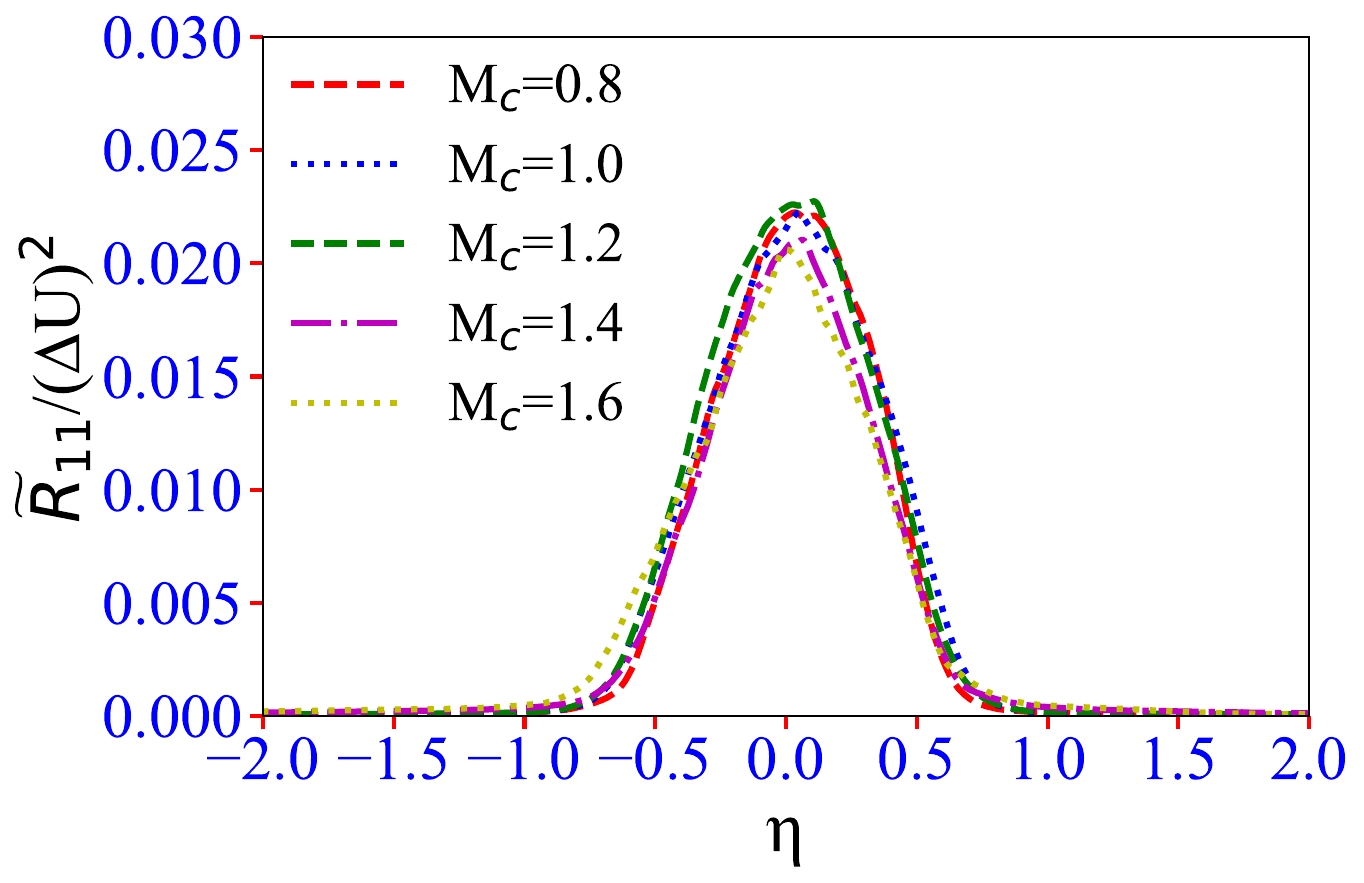}
         \subcaption{}
     \end{subfigure}
     \begin{subfigure}{0.47\textwidth}
         \centering
         \includegraphics[width=\textwidth]{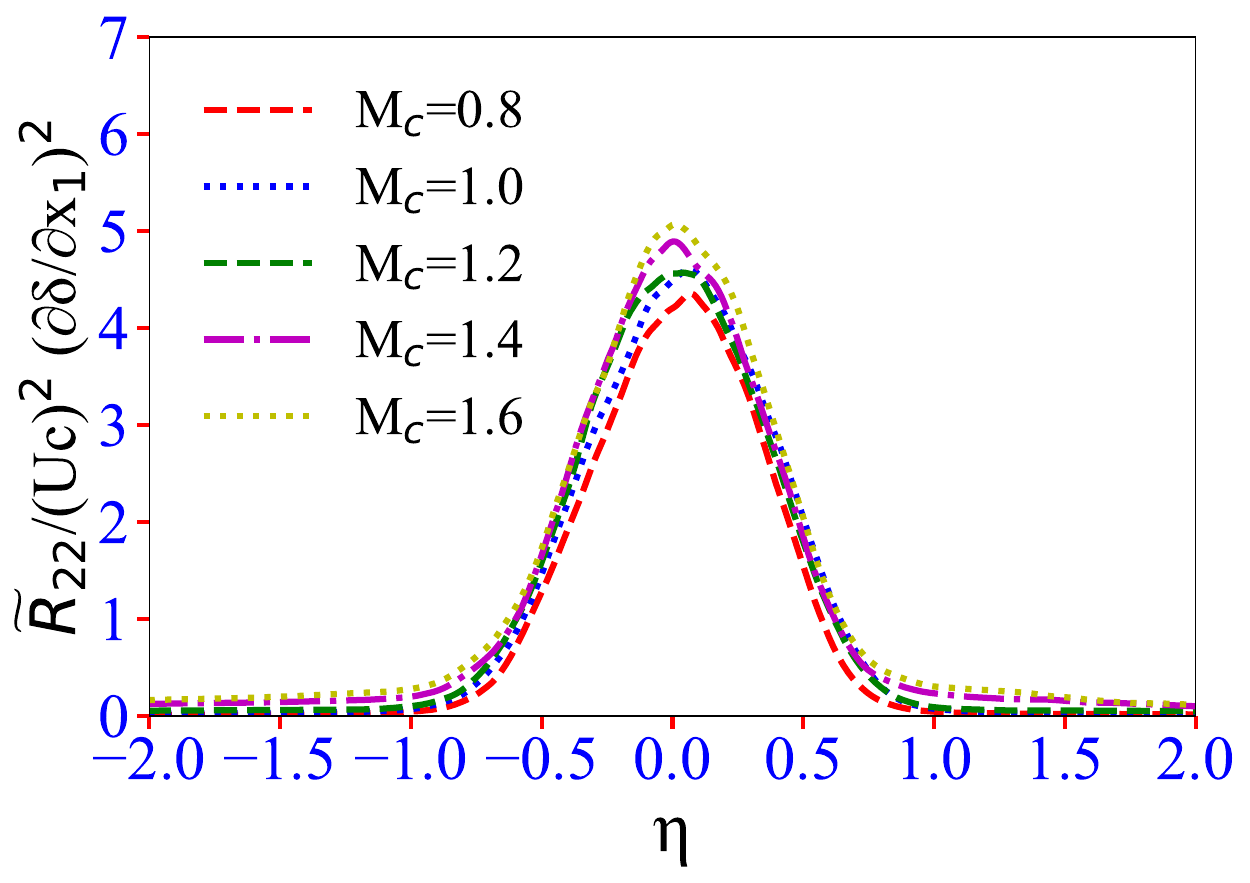}
         \subcaption{}
     \end{subfigure}
     \begin{subfigure}{0.49\textwidth}
         \centering
         \includegraphics[width=\textwidth]{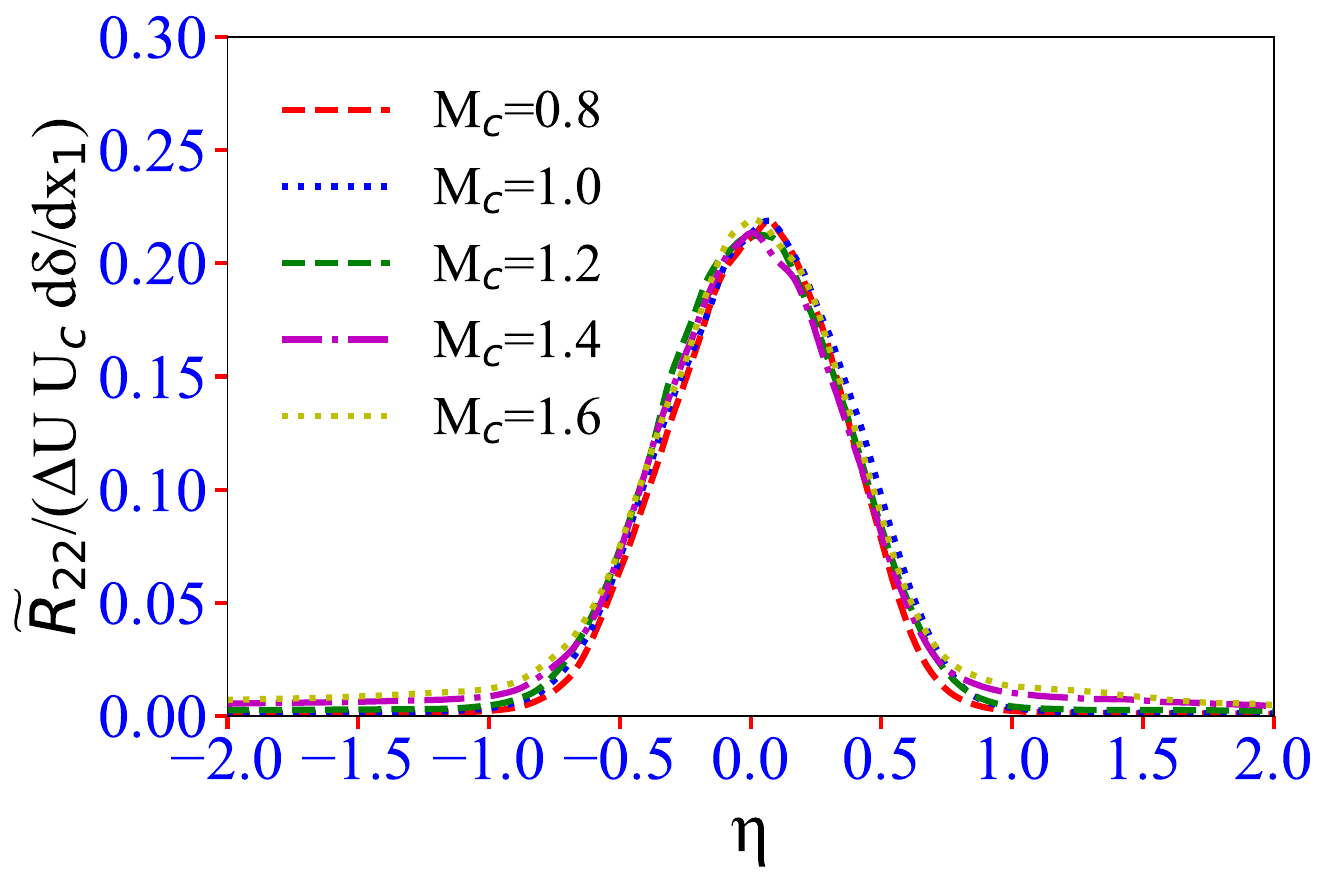}
         \subcaption{}
     \end{subfigure}
     \centering
     \begin{subfigure}{0.50\textwidth}
         \centering
         \includegraphics[width=\textwidth]{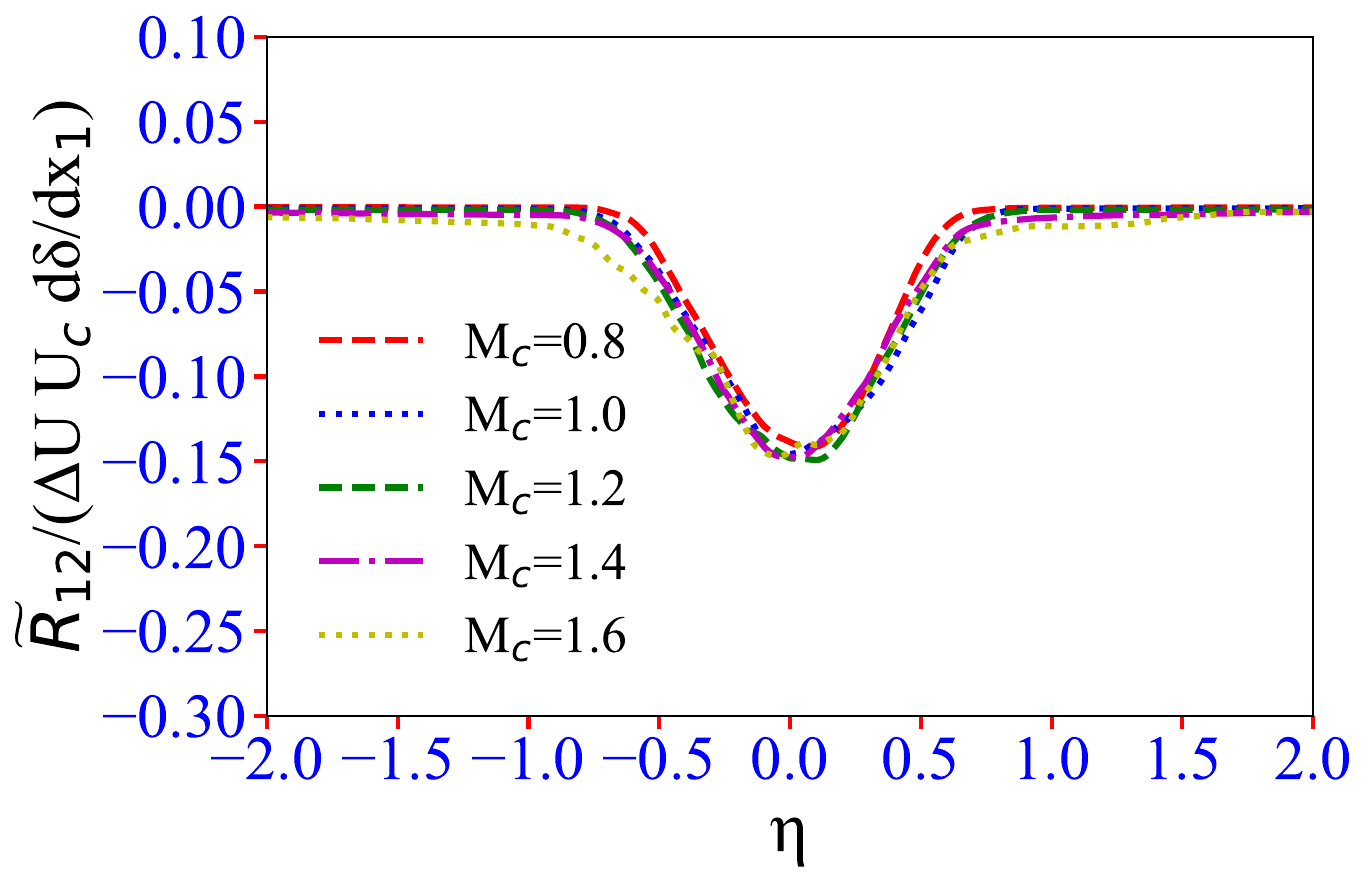}
         \subcaption{}
     \end{subfigure}
     \caption{The collapse of (a) mean streamwise velocity (b) $\widetilde{R}_{11}$ (c) and (d) $\widetilde{R}_{22}$ in two different scalings and (e) $\widetilde{R}_{12}$ in self-similar coordinates for all the $M_c$ cases}
     \label{fig:scalingMc1}
\end{figure}

\begin{figure}
     \centering
     \begin{subfigure}{0.49\textwidth}
         \centering
         \includegraphics[width=\textwidth]{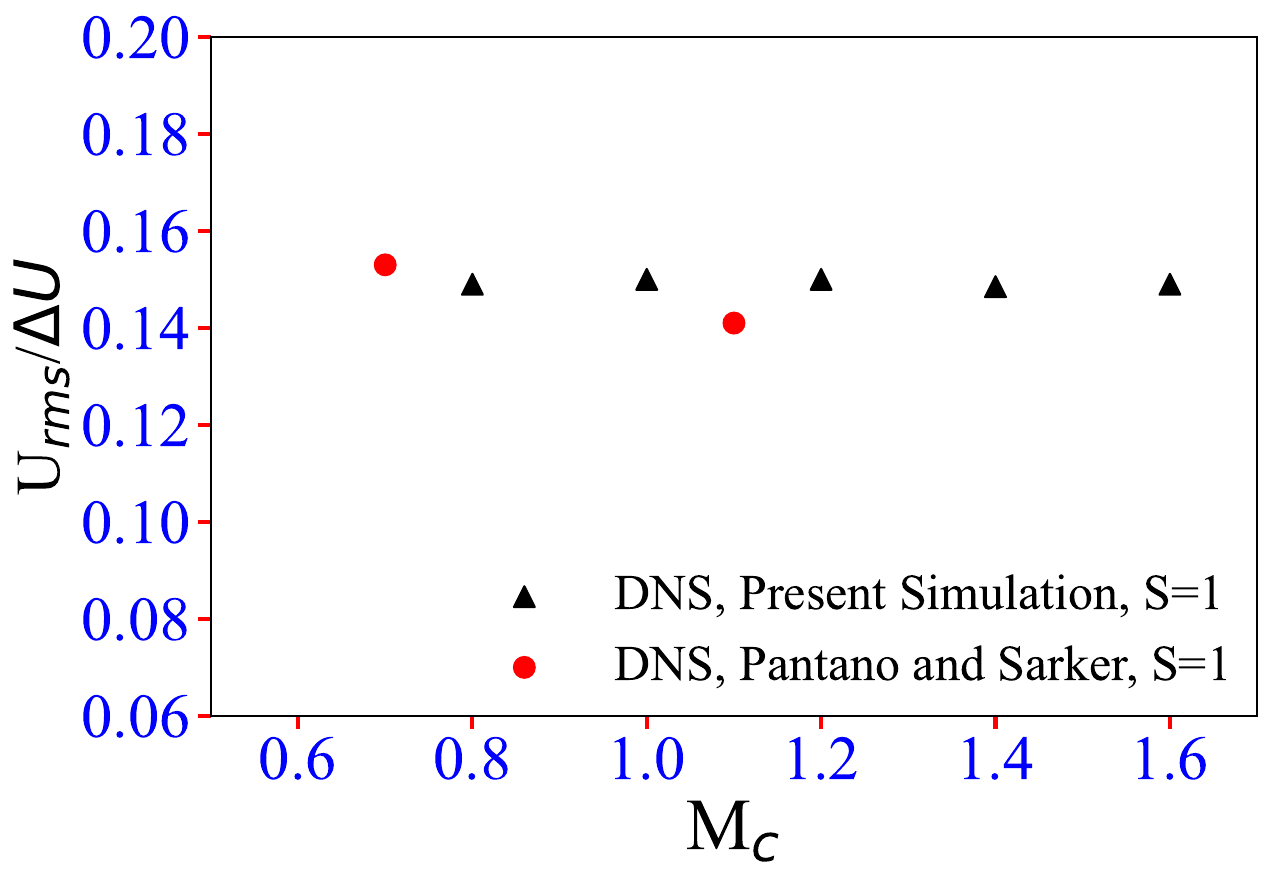}
         \subcaption{}
     \end{subfigure}
     \begin{subfigure}{0.49\textwidth}
         \centering
         \includegraphics[width=\textwidth]{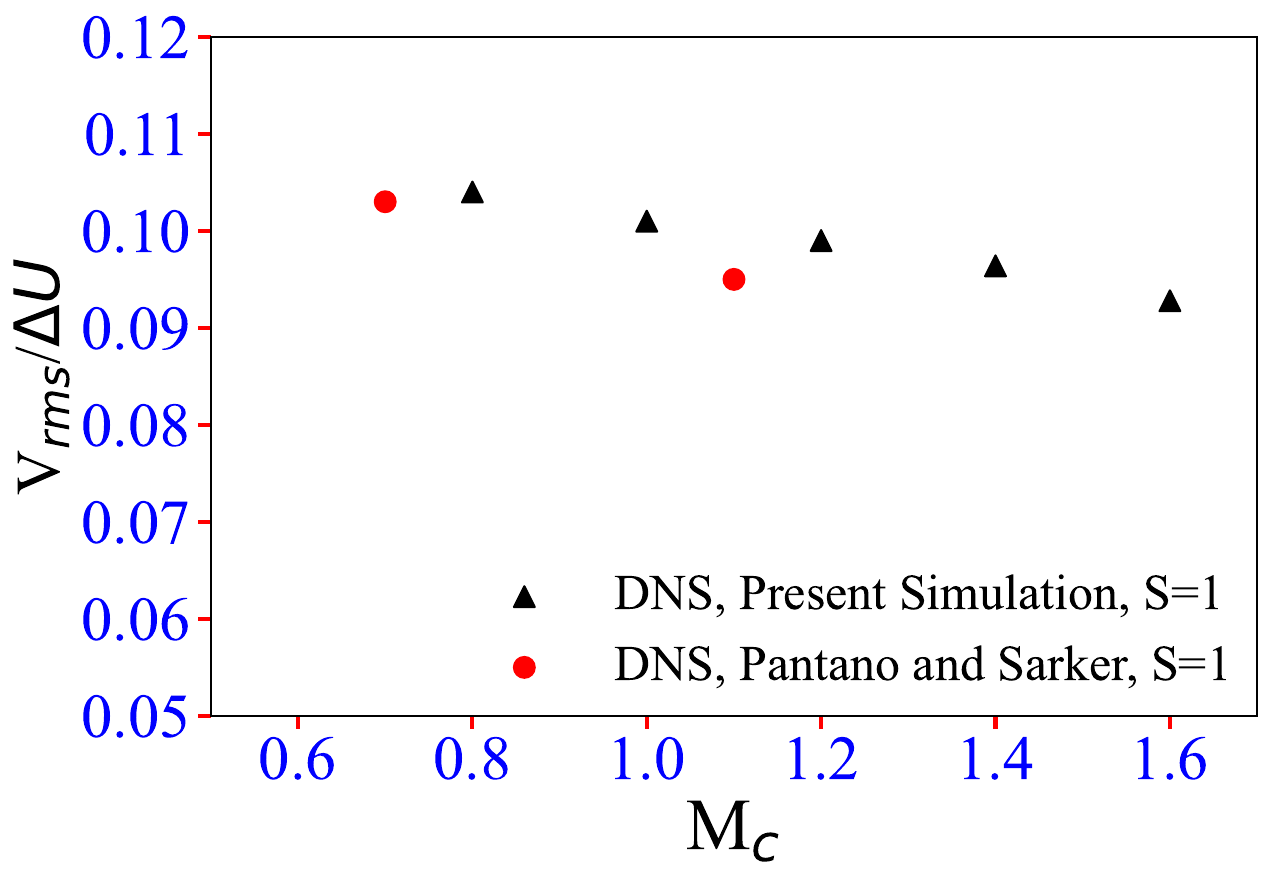}
         \subcaption{}
     \end{subfigure}
     \begin{subfigure}{0.49\textwidth}
         \centering
         \includegraphics[width=\textwidth]{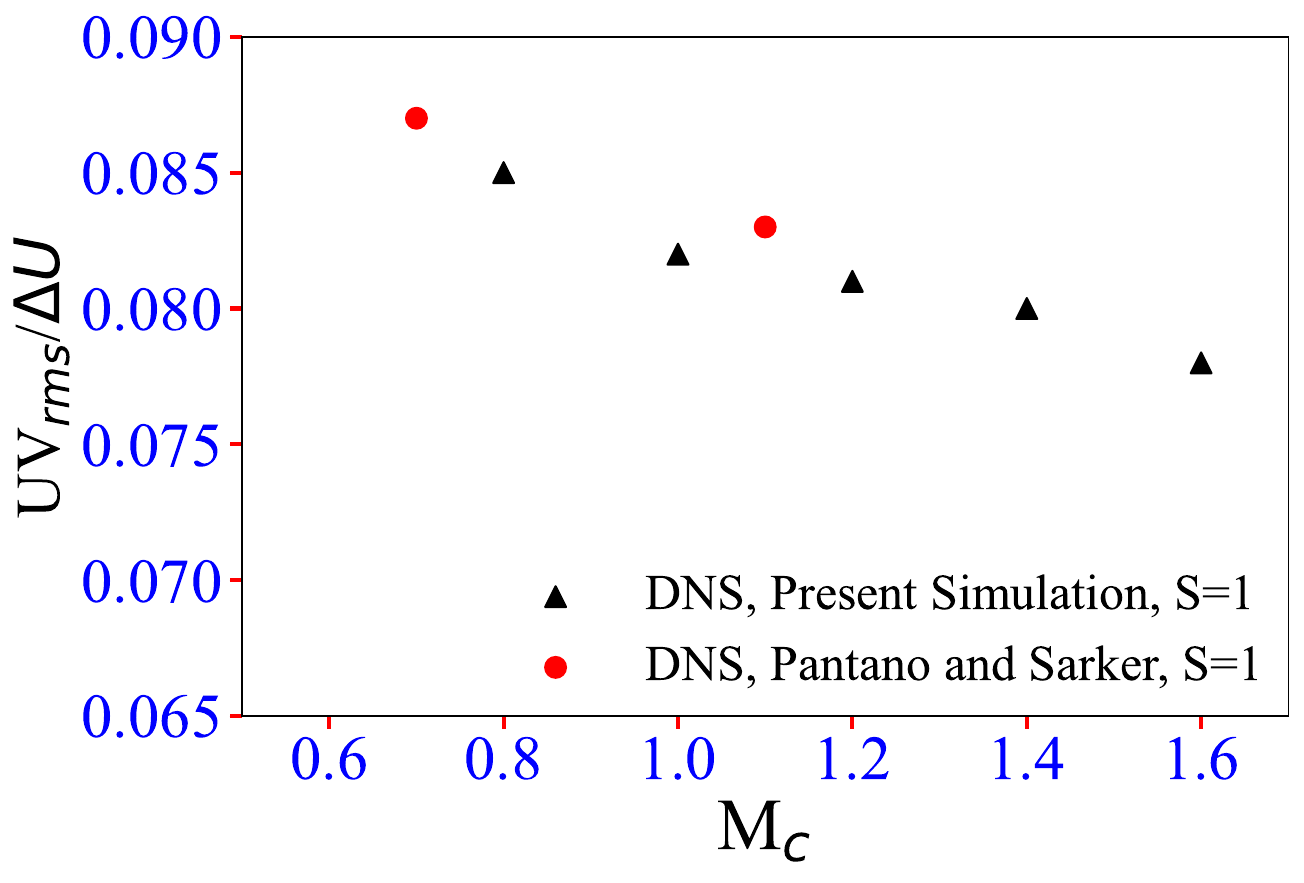}
         \subcaption{}
     \end{subfigure}
     \caption{Normalized turbulent intensities (a) $U_{rms}$/$\Delta U$=$\sqrt{\widetilde{R}_{11}}$/$\Delta U$  (b) $V_{rms}$/$\Delta U$=$\sqrt{\widetilde{R}_{22}}$/$\Delta U$ (c) $UV_{rms}$/$\Delta U$=$\sqrt{\widetilde{R}_{12}}$/$\Delta U$ from previous literature. \cite{pantano02}  performed DNS of constant density (density ratio, s=1) temporal shear layer. Though the present simulations were not performed exactly under the same conditions, they agree with the previous literature.}
     \label{fig:lit1}
\end{figure}

The consistent trend of decreasing growth rate with increasing $M_c$ is evident, yet all mean and turbulent quantities maintain self-similarity across different $M_c$ values, as demonstrated in below in figures \ref{fig:scalingMc1} and \ref{fig:scalingMc2}. The results imply that, although the growth rate of a shear layer is significantly affected by the compressibility, once the mean and turbulent quantities become fully developed and properly scaled, their normalized profiles look almost identical for all $M_c$s.

The profiles of the mean streamwise velocity and Reynolds stresses are very different at different $M_c$ values but in the self-similar zone, they all collapse when we apply our suggested self-similar scaling mentioned in the table~\ref{tab:ssscaling}. This is observed in Figure~\ref{fig:scalingMc1}, where the self-similar behavior of scaled mean streamwise velocity and Reynolds stress components is shown.  For $\widetilde{R}_{22}$, both scalings presented in table~\ref{tab:ssscaling} worked reasonably well. However, the scaling similar to that of $\widetilde{R}_{12}$ seems to work better. As expected, the Reynolds stress profiles exhibit the peak value near the center of the shear layer and go to zero outside the shear layer. The smoothness of the profiles indicates that all the cases are statistically well converged when sufficient time and space averaging is performed. 

Experimental data suggests that the scaled mean streamwise velocity profile attains a sigmoid shape. Fitting an error function to the self-similar streamwise velocity profile is the most common approximation (\citealt{wei22}). \cite{abramovich84} and \cite{ovidio} found Schlichting's formula convenient to use for the scaled mean streamwise velocity profile. %Fitting a hyperbolic tangent profile can be another option. We tried all the available options in the literature to approximate our scaled mean streamwise velocity profile and we found the error function to be the most suitable approximation. 
We also found the error function to be a suitable approximation and convenient for calculating the derivatives and integrals in the derivations. In this study, we use a similar form of error function as \cite{pope2000turbulent} and \cite{wei22}:
\begin{equation}
    \hat{u_1}=Aerf(B\eta),
\end{equation}
where $A=\frac{1}{2}$ and $B=2.05$ match our data well, as shown in figure~\ref{fig:scalingMc1} (a). The value of $B$ found here is slightly higher than $B=1.812$ used by \cite{wei22} to match the incompressible case data. 
%Since B=1.812 and B=2.05 gave reasonably close results for all the analyses presented here, we used B=2.05 in this study.  

Compressibility effects on higher-order turbulent statistics are also examined and compared with those in the literature in figure~\ref{fig:lit1}.  Figure \ref{fig:lit1} indicates that the peak of $\widetilde{R}_{11}$ in the self-similar zone is not affected by compressibility, while the peaks of $\widetilde{R}_{12}$ and $\widetilde{R}_{22}$ are reduced with increasing $M_c$, with $\widetilde{R}_{11} < \widetilde{R}_{33} < \widetilde{R}_{22}$. Thus, the mean flow is highly three-dimensional and the results align with the findings of \cite{pantano02}, \cite{zhang19}, \cite{dutton93}, and \cite{kim20}.The decreasing trend of $\widetilde{R}_{12}$ and $\widetilde{R}_{22}$ with increasing $M_c$ has been observed in previous studies. However, the trend of $\widetilde{R}_{11}$ has not been universally consistent (\citealt{kim20}). The results of the present study are consistent with those of \cite{dutton90}, \cite{dutton91} and \cite{kim20}. 
%in that the peak of $\widetilde{R}_{11}$ is almost unaffected by the level of compressibility while the peak of $\widetilde{R}_{12}$ and $\widetilde{R}_{22}$ decreases with increasing convective Mach numbers. 
The physical explanation for $\widetilde{R}_{11}$ not being affected by the convective Mach number is given by \cite{kim20} who states that as compressibility increases, the large-scale turbulent structures become elongated in the streamwise direction compared to the transverse and spanwise directions, and the streamwise velocity fluctuations are preserved and remain of the same magnitude as in the incompressible case. \cite{dutton90} explained that the reduction in transverse turbulence intensity results from the decrease in pressure-transverse velocity correlations, which redistribute turbulent kinetic energy from the streamwise to the transverse direction. 

\begin{figure}
     \centering
     \begin{subfigure}{0.48\textwidth}
         \centering
         \includegraphics[width=\textwidth]{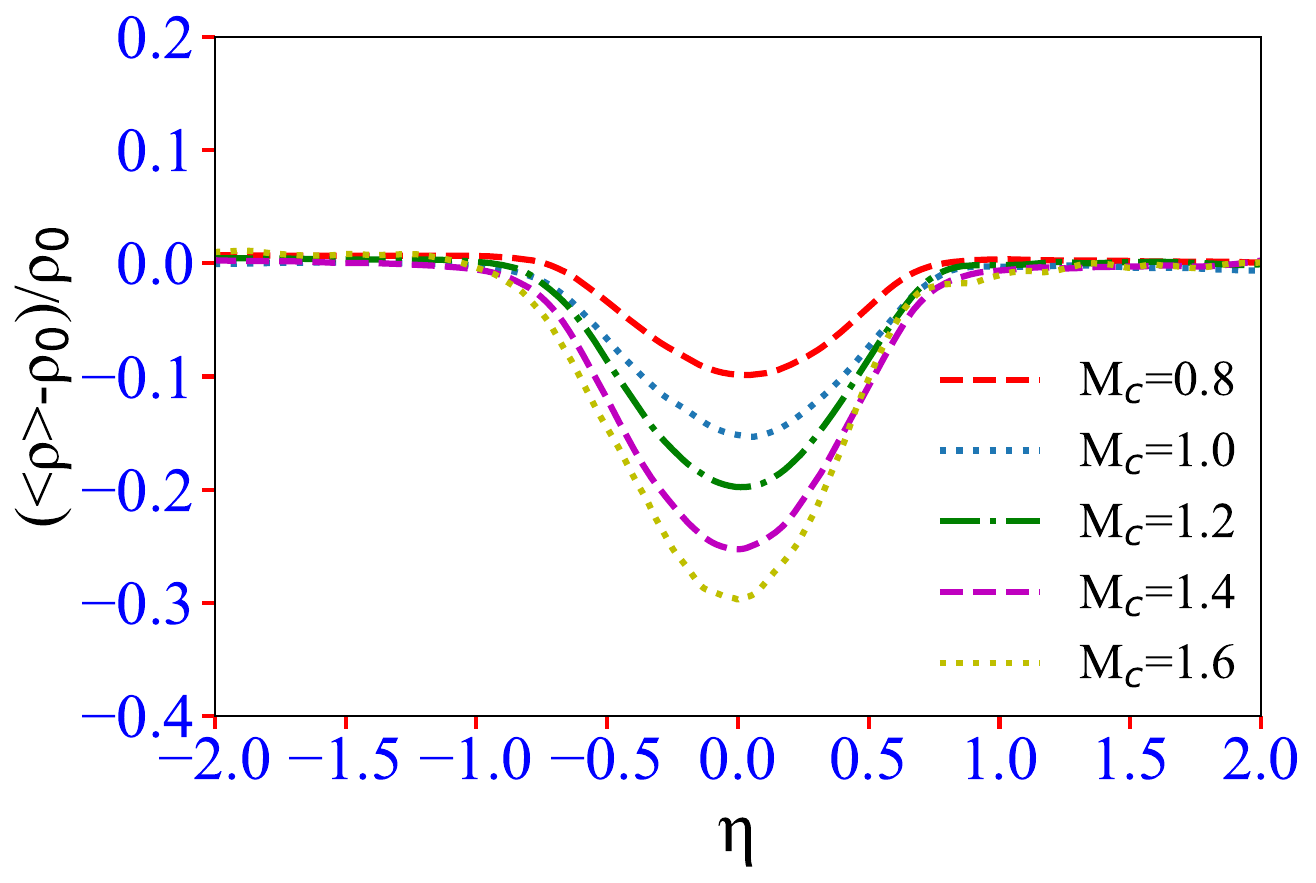}
         \subcaption{}
     \end{subfigure}
     \begin{subfigure}{0.48\textwidth}
         \centering
         \includegraphics[width=\textwidth]{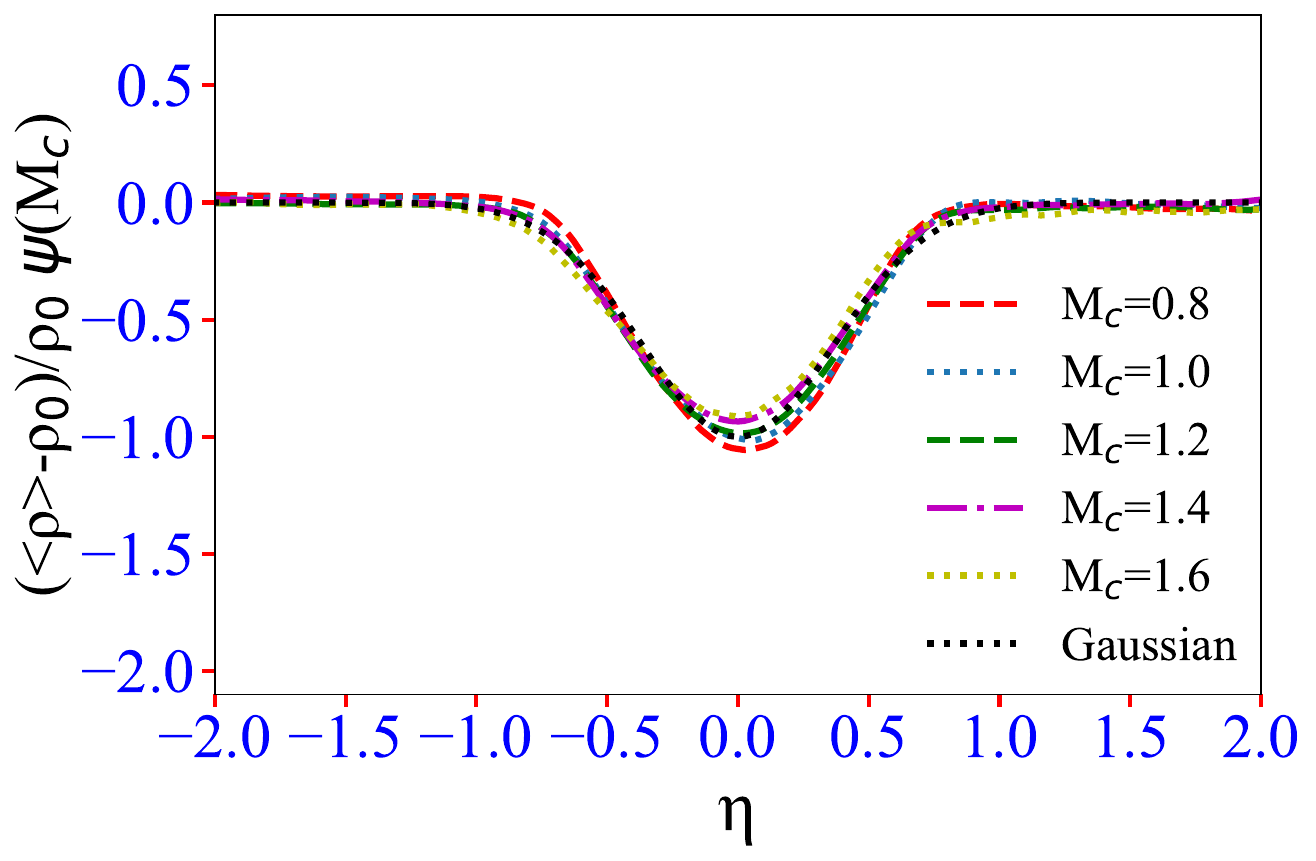}
         \subcaption{}
     \end{subfigure}
     \begin{subfigure}{0.48\textwidth}
         \centering
         \includegraphics[width=\textwidth]{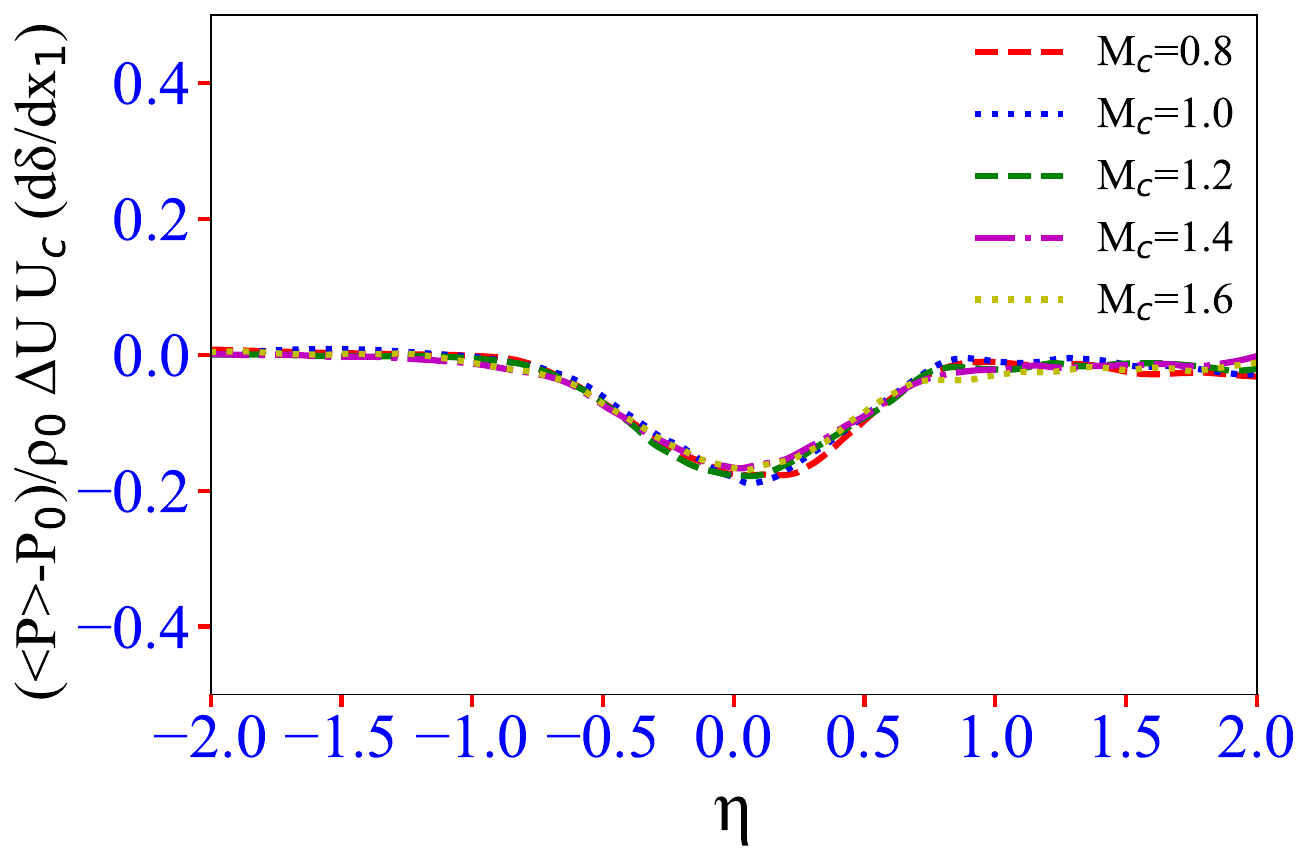}
         \subcaption{}
     \end{subfigure}
     \caption{The variation of mean density and mean pressure in the self-similar coordinate at different $M_c$ values. (a) Density profiles do not collapse using $\rho_{01}$= $\rho_0$ scaling  (b) Collapse of density profile using $\rho_{01}$= $\rho_0$ $\psi$(M$_c$) scaling (c) Collapse of pressure at different $M_c$ values.}
     \label{fig:scalingMc2}
\end{figure}

\begin{figure}
     \centering
     \begin{subfigure}{0.6\textwidth}
         \centering
         \includegraphics[width=\textwidth]{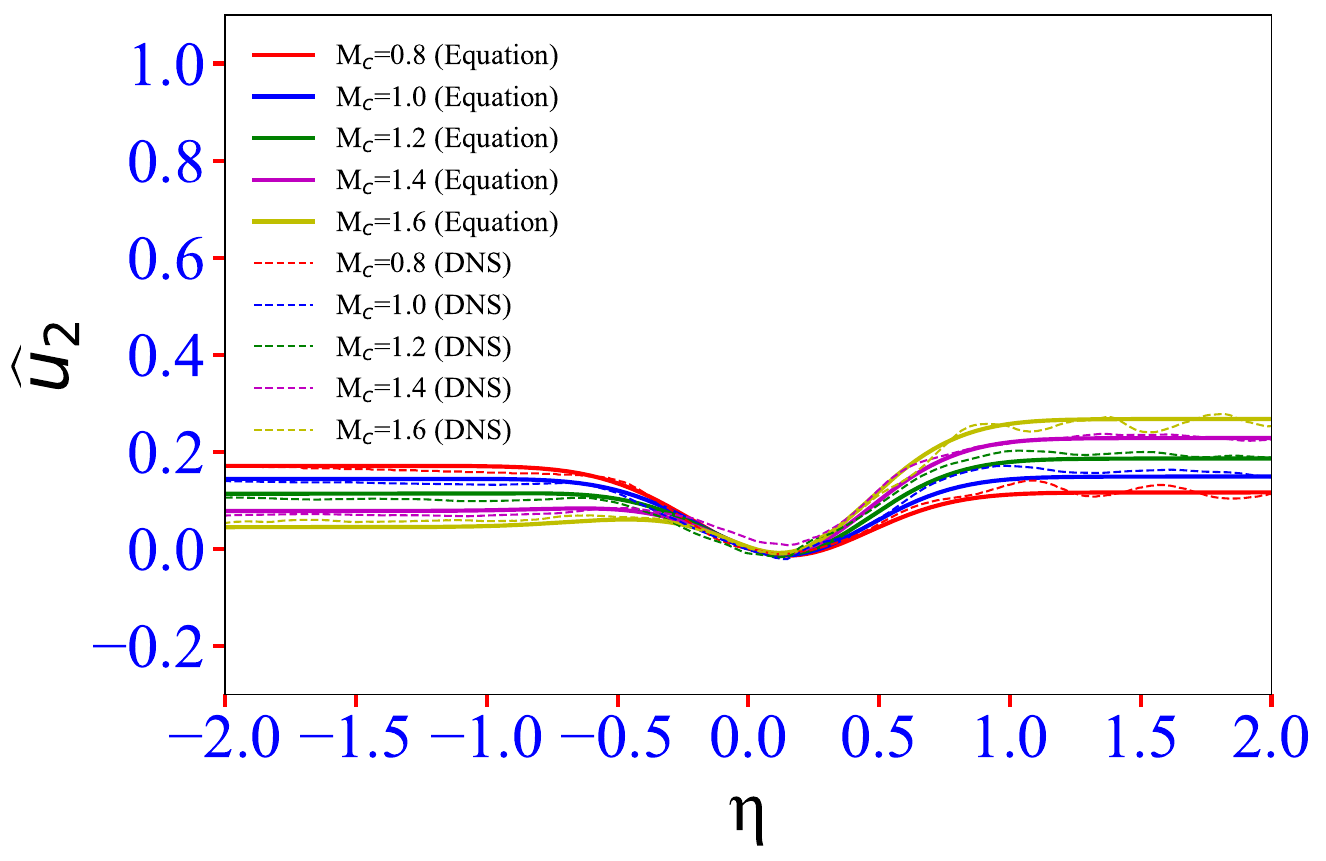}
     \end{subfigure}
     \caption{The scaled mean transverse velocity in the self-similar coordinate compared with profiles from our analytical equation at different $M_c$ values. The analytical solution is reasonably close to the DNS results.}
     \label{fig:vMc1}
\end{figure}

Mean density profiles at different $M_c$ values in the self-similar coordinate are shown in Figure~\ref{fig:scalingMc2} (a). It is observed that there is a significant density variation within the shear layer zone for all cases considered. The lowest density values are near the center of the shear layer. For the lower convective Mach number of $M_c=0.8$, there is a density drop of about $8.2\%$ inside the shear layer. When the Mach number increases to $M_c=1.6$, the density drop increases to approximately $28.2\%$ from the free stream value. As we explained earlier, dissipation is mostly responsible for the density drop inside the shear layer. 

To collapse the density profiles for different $M_c$ cases, it was suggested during the derivation of self-similar equations (see section \ref{eq:ssfinal}) a scaling of type $\rho_{01}$=$\rho_{0} \psi(M_c)$, where the function $\psi(M_c)$ 
should provide the correct behavior for the peak of $\hat{\rho}$ under both limiting conditions ($M_c\rightarrow  0$ and $M_c \rightarrow \infty$). Here, we suggest a Gaussian function as ($\psi(M_c)$=1-$e^{-0.17 M_c^2}$) that collapses our data well (Figure~\ref{fig:scalingMc2} (b)) and reaches the limiting values $0$ and $1$. We found this function by curve-fitting the peak of $\hat{\rho}$ as a function of the $M_c$. %However, we are not suggesting that this is the only function one can use. Any function that satisfies the limits and collapses the data can be used. 
In summary, for fitting the density profiles, we suggest a Gaussian function of the following form:
\begin{equation}
    \hat{\rho}=a e^{-b\eta^2},
\end{equation}
with the parameters $a=-1.0$ and $b=4.20$ the function $\psi(M_c)=1-e^{-0.17 M_c^2}$ fitted here. %However, the magnitude of the coefficient $a$ depends on the function $\psi(M_c)$. If any function, other than the function used in this study, is taken as $\psi(M_c)$ to collapse the self-similar density profiles, the magnitude of coefficient $a$ should vary accordingly. 

For the scaled mean pressure, we found a good collapse of the data using our suggested self-similar scaling (Figure \ref{fig:scalingMc2} (c)). Finally, figure \ref{fig:vMc1} compares the scaled mean transverse velocity profile in the self-similar coordinate with our suggested approximate equation at different $M_c$ values. The profiles from the DNS results are very close to the approximate equation proposed.

\subsection{Velocity Parameter}\label{sec:At}

As explained above and suggested in previous studies (\citealt{mehta91,ragab88,wei22}) the shear layer growth rate is significantly affected by the velocity parameter, $\lambda$, but no comprehensive study was carried out for the compressible case, where the effects of $\lambda$ could be separated from those of $M_c$ and density ratio of the free streams. In order to fill in the gaps in the literature, in this section the effects of $\lambda$ on the growth rate, mean, and turbulence statistics of the shear layer are considered. To decouple the effect of $\lambda$ from  $M_c$, we have changed the velocity ratio (and the velocity difference between the free streams) while keeping the convective Mach number the same at $M_c=1.2$. 

\begin{figure}
     \centering
     \begin{subfigure}{0.6\textwidth}
         \centering
         \includegraphics[width=\textwidth]{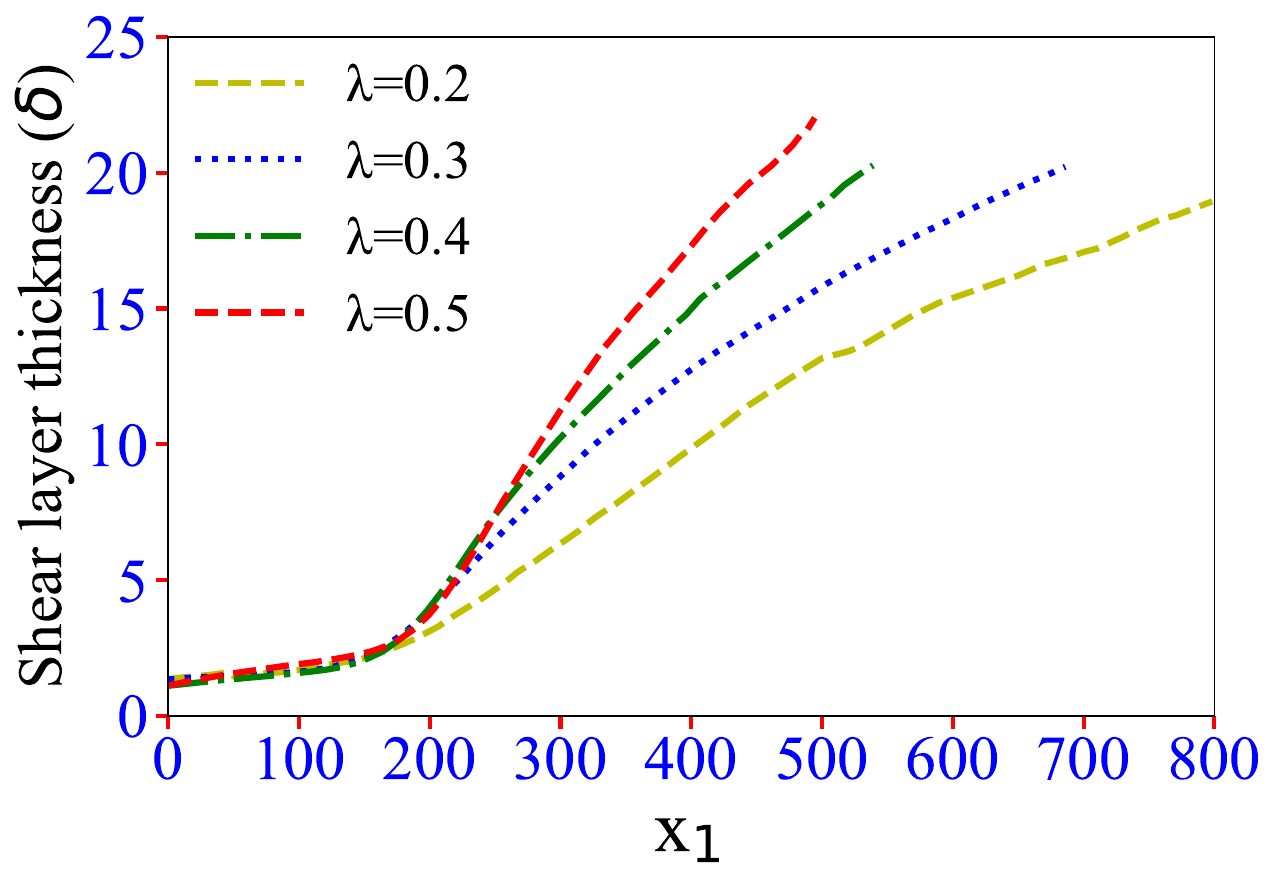}
     \end{subfigure}
     \caption{Shear layer thickness at different $\lambda$ values}
     \label{fig:deltaAt}
\end{figure}

\begin{figure}
     \centering
     \begin{subfigure}[b]{0.85\textwidth}
         \centering
         \includegraphics[width=\textwidth]{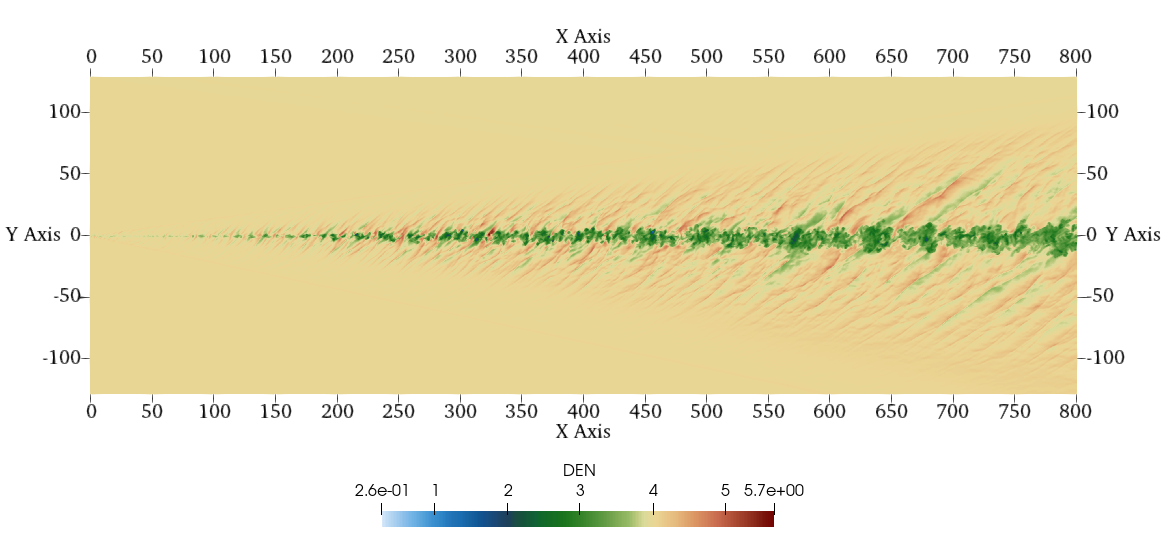}
         \subcaption{}
     \end{subfigure}
     \begin{subfigure}[b]{0.6\textwidth}
         \centering
         \includegraphics[width=\textwidth]{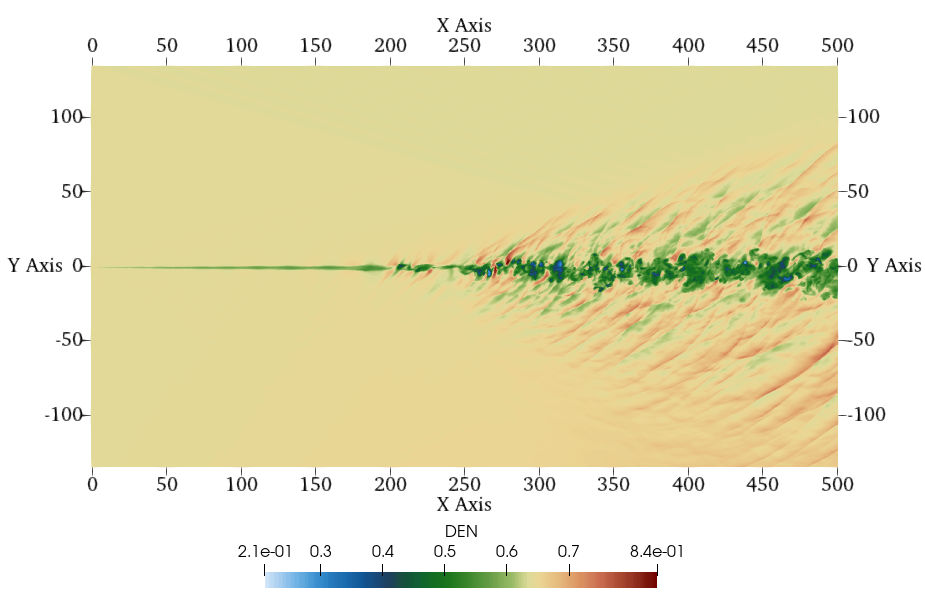}
         \subcaption{}
     \end{subfigure}
     \caption{Color contour ($x_1$-$x_2$ plot) of instantaneous density for a) $\lambda$=0.2 and b) $\lambda$=0.5}
     \label{fig:contourAt}
\end{figure}

The evolution of the shear layer thickness for all $\lambda$ values is shown in figure~\ref{fig:deltaAt}. 
Regardless of the specific thickness definition, the shear layer growth rate increases with increasing $\lambda$. 
%As mentioned in section \ref{sec:Mc}, the shear layer thickness has three distinct regions: initial slow development region, middle higher development region, and self-similar region. %The linear growth rate is observed in the self-similar zone after the initial transition (\citealt{zhang19}). 
Moreover, figure~\ref{fig:deltaAt} suggests that the distance required to reach the self-similar linear growth depends on $\lambda$, since all other parameters in the simulations were kept constant. 
For low values of $\lambda$ ($\lambda=0.2$ and $\lambda=0.3$), the downstream distance to the self-similar region increases considerably compared to the higher values of $\lambda$ considered here, and therefore the computational domain was extended accordingly. This is further illustrated in figure~\ref{fig:contourAt}, which shows the instantaneous contours of density for $\lambda=0.2$ and $\lambda=0.5$ in the $x_1$-$x_2$ plane at the mid-spanwise location. Larger-scale structures are observed from the contour plots at $\lambda$=0.5, which is consistent with the larger growth rate.

\begin{figure}
     \centering
     \begin{subfigure}[b]{0.48\textwidth}
         \centering
         \includegraphics[width=\textwidth]
         {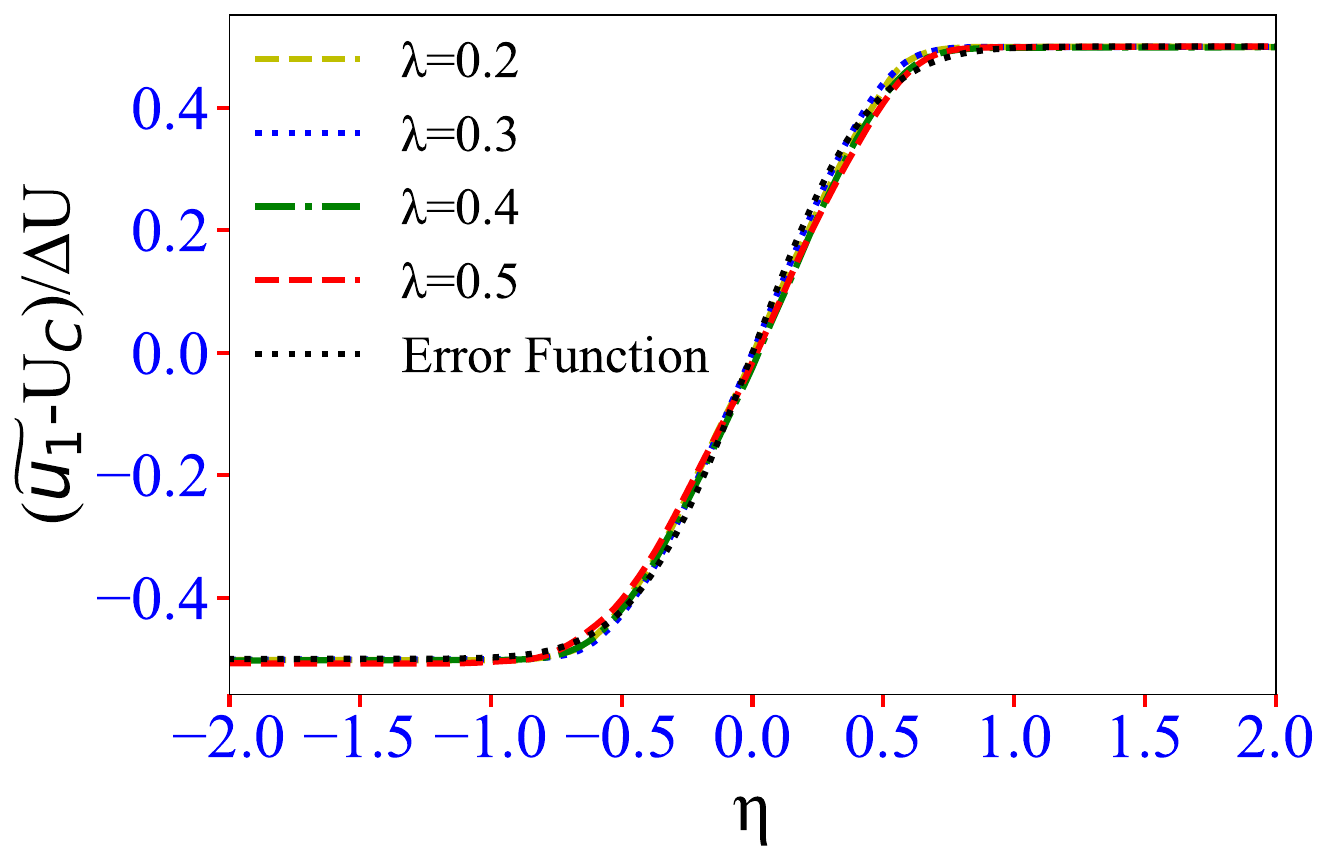}
         \subcaption{}
     \end{subfigure}
     \begin{subfigure}[b]{0.5\textwidth}
         \centering
         \includegraphics[width=\textwidth]
         {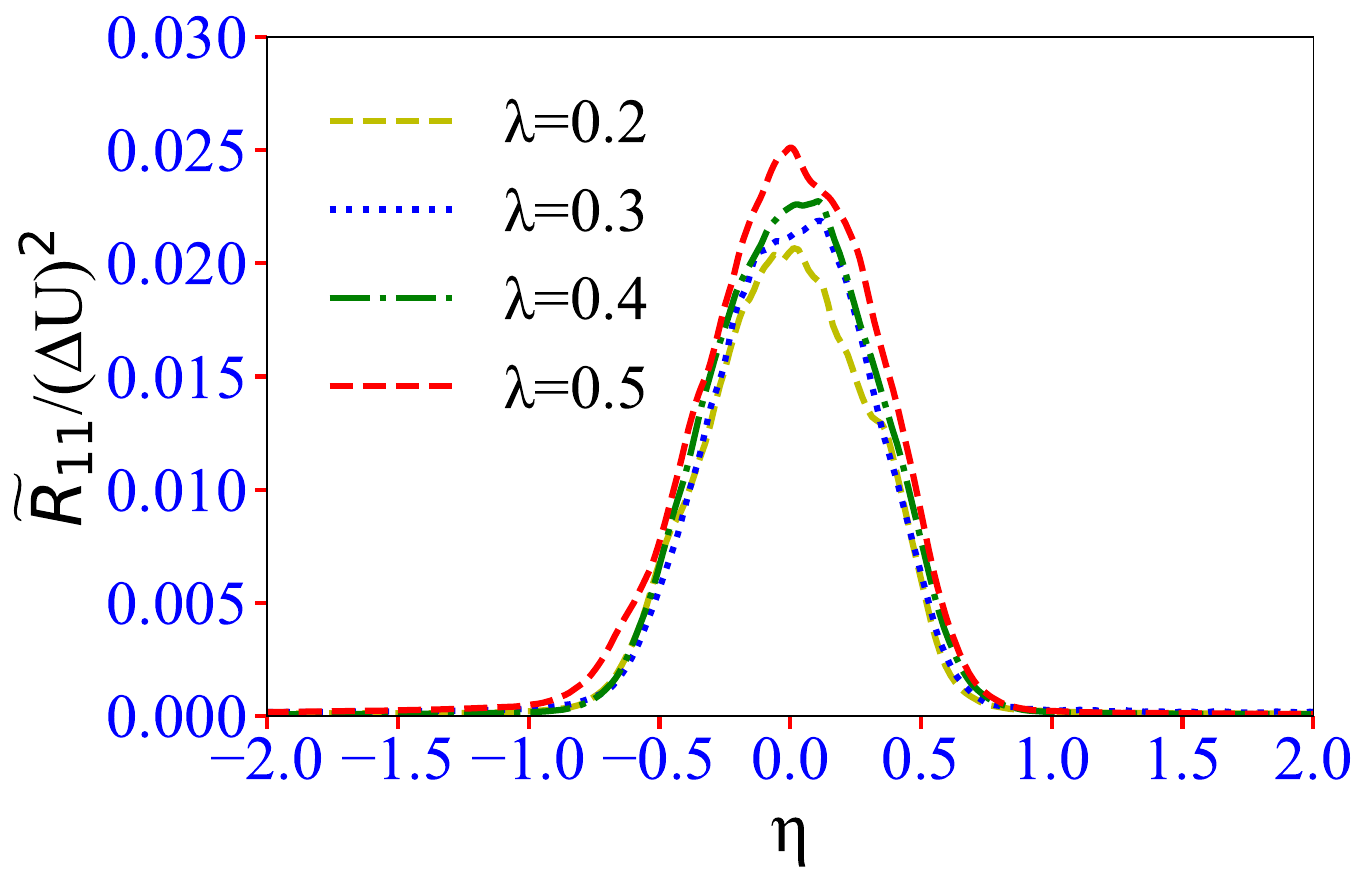}
         \subcaption{}
     \end{subfigure}
    \begin{subfigure}[b]{0.48\textwidth}
         \centering
         \includegraphics[width=\textwidth]
         {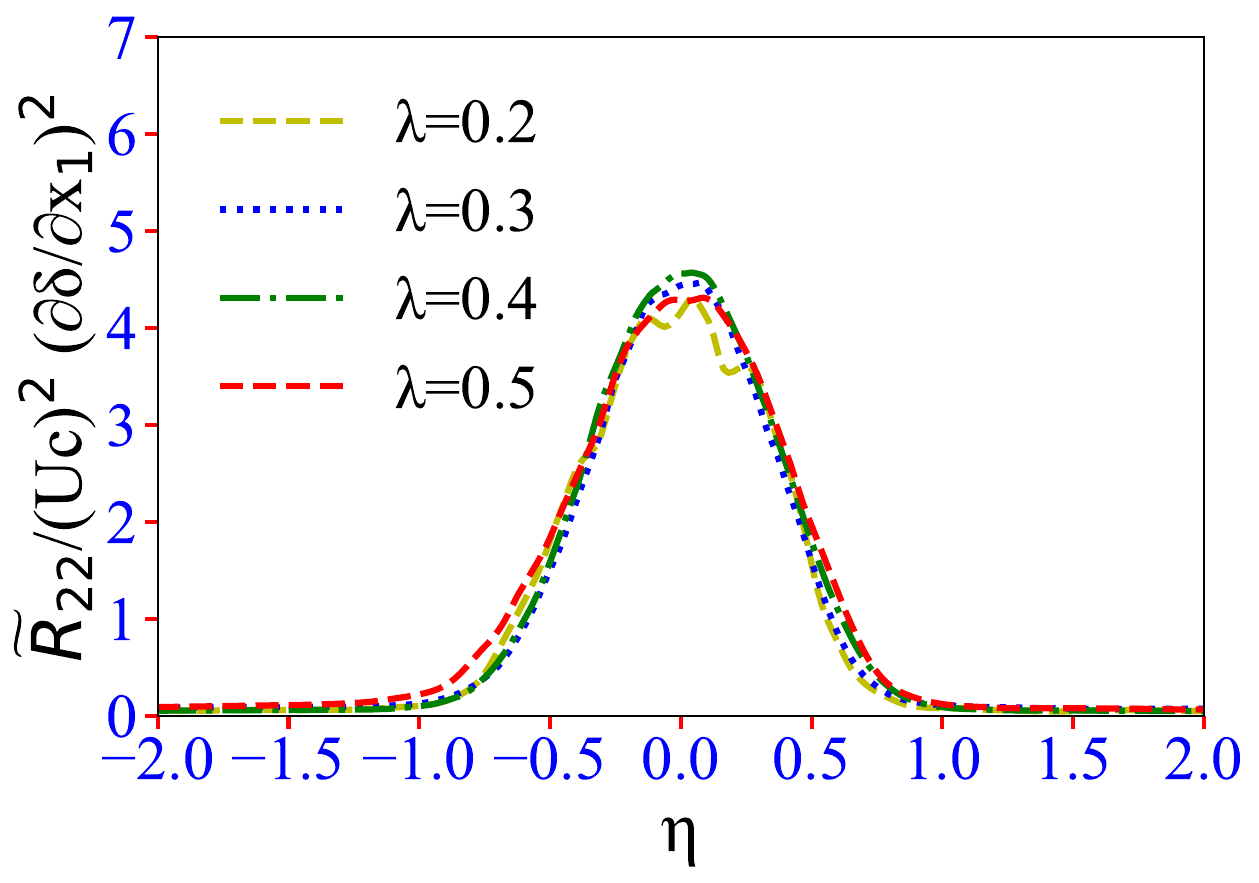}
         \subcaption{}
     \end{subfigure}
%     \begin{subfigure}[b]{0.5\textwidth}
%         \centering
%         \includegraphics[width=\textwidth]
%         {Figures/4lam}
%         \subcaption{}
%     \end{subfigure}
     \begin{subfigure}[b]{0.5\textwidth}
         \centering
         \includegraphics[width=\textwidth]{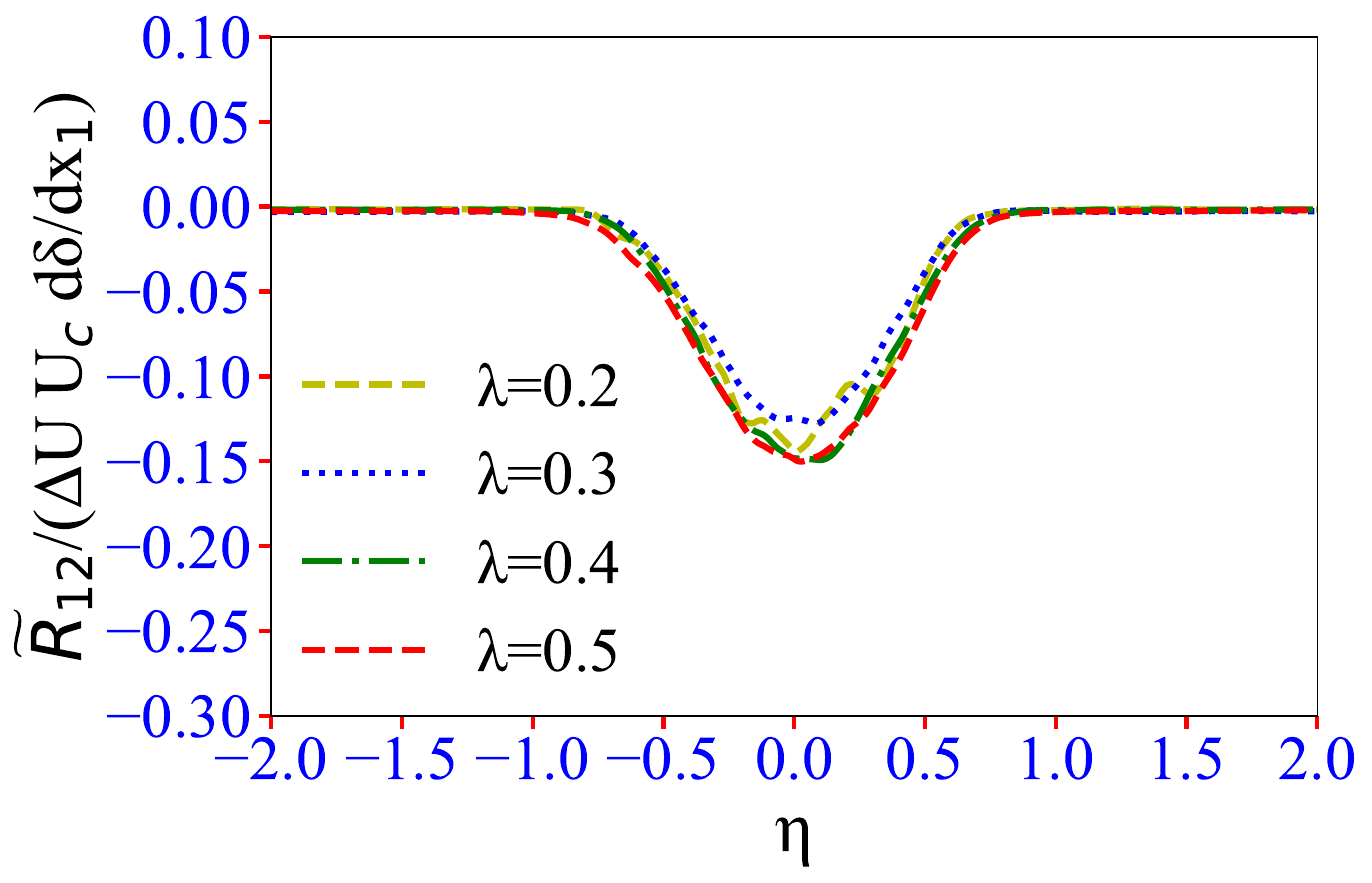}
         \subcaption{}
     \end{subfigure}
     \centering
     \caption{The collapse of scaled (a) mean streamwise velocity, (b) $\widetilde{R}_{11}$, (c) $\widetilde{R}_{22}$ and (d) $\widetilde{R}_{12}$ for different $\lambda$ values in the self-similar coordinate}
     \label{fig:scalingAt1}
\end{figure}

\begin{figure}
     \centering
     \begin{subfigure}{0.48\textwidth}
         \centering
         \includegraphics[width=\textwidth]{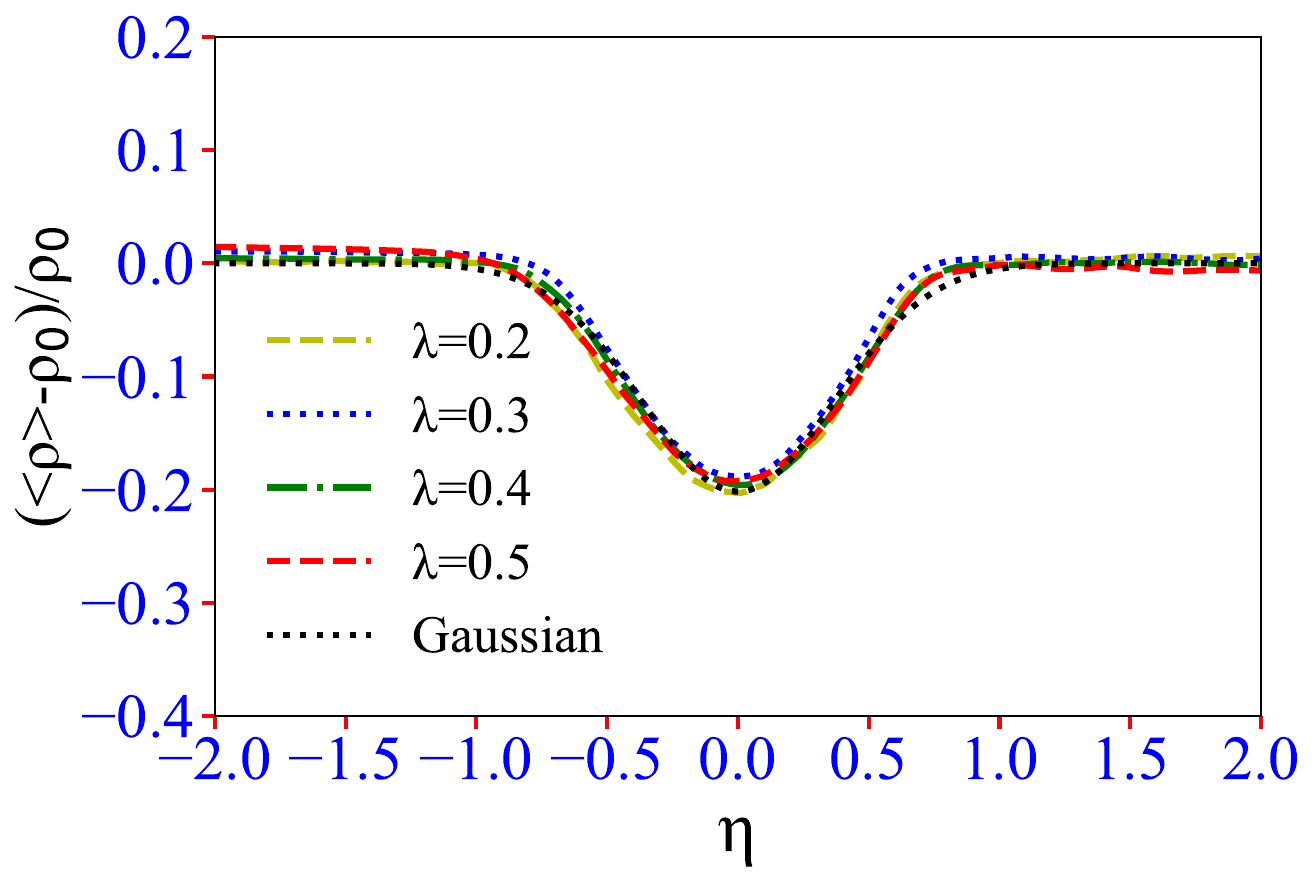}
         \subcaption{}
     \end{subfigure}
%     \begin{subfigure}{0.48\textwidth}
%         \centering
%         \includegraphics[width=\textwidth]{Figures/7lam}
%         \subcaption{}
%     \end{subfigure}
     \begin{subfigure}{0.48\textwidth}
         \centering
         \includegraphics[width=\textwidth]{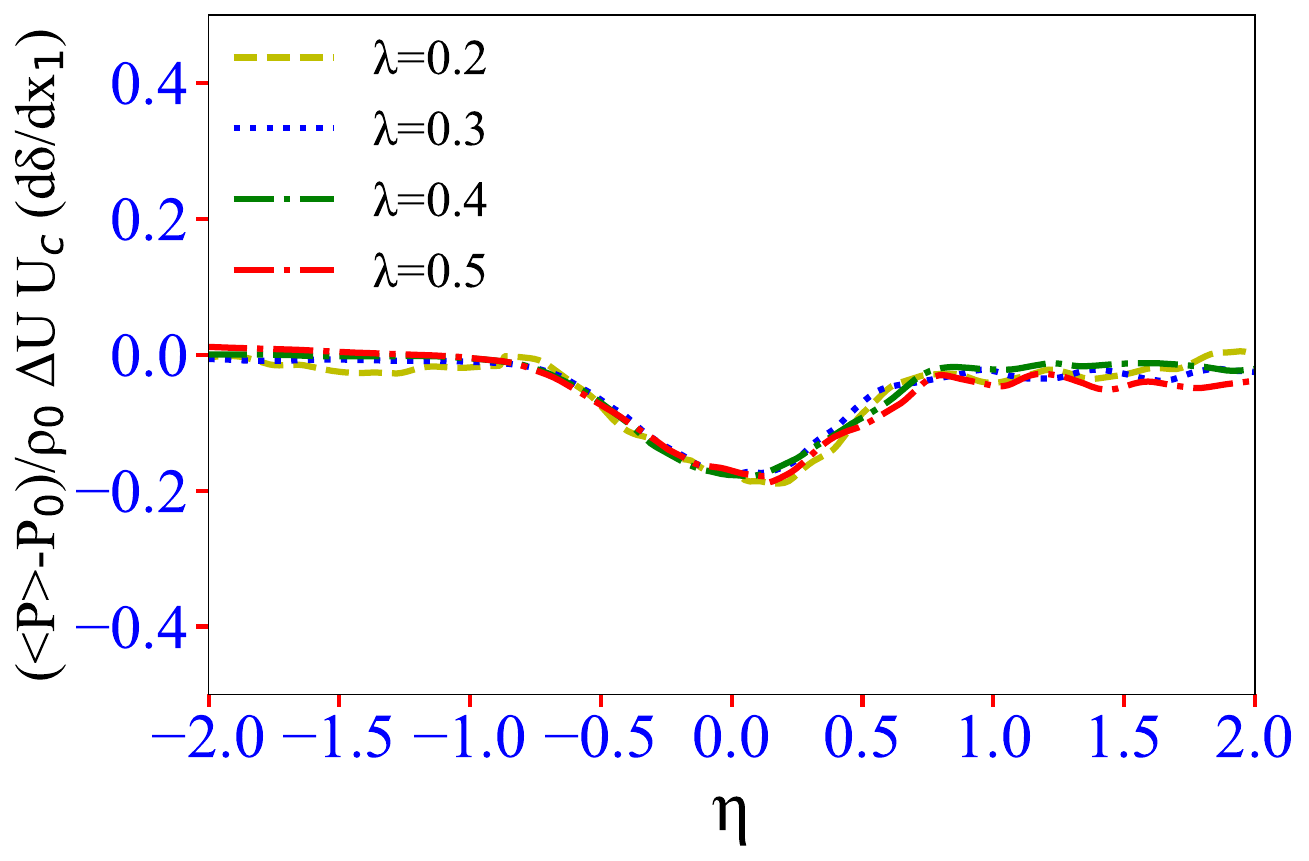}
         \subcaption{}
     \end{subfigure}
     \caption{The variation of mean density and mean pressure in the self-similar coordinate at different $\lambda$ values. (a) Collapse of scaled mean density using $\rho_{01}$= $\rho_0$ scaling 
     (b) Collapse of scaled mean pressure at different $\lambda$ values.}
     \label{fig:scalingAt2}
\end{figure}

\begin{figure}
     \centering
     \begin{subfigure}{0.6\textwidth}
         \centering
         \includegraphics[width=\textwidth]{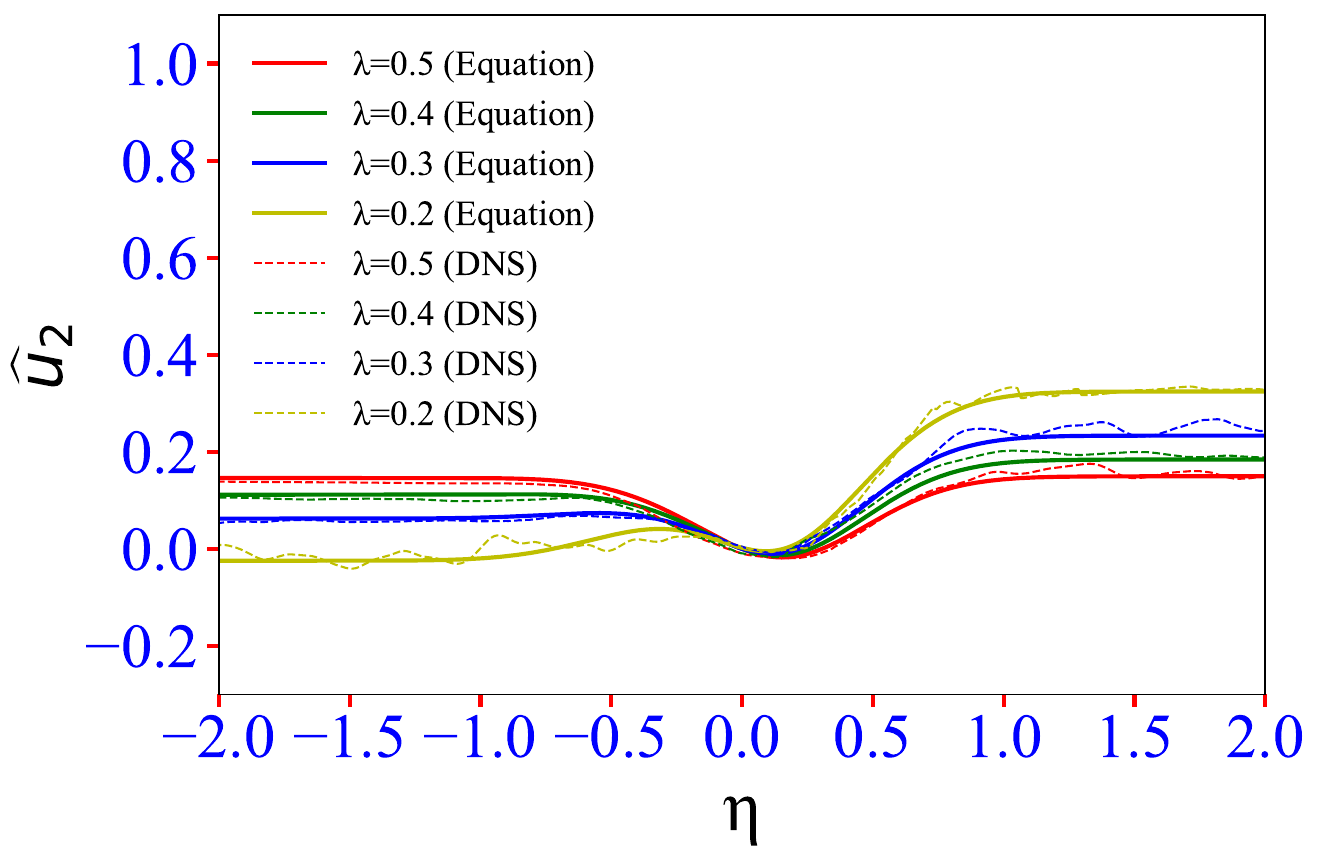}
     \end{subfigure}
     \caption{The scaled mean transverse velocity compared with the profiles obtained from our approximate equation in the self-similar coordinate at different $\lambda$ values. The approximate solution closely follow the DNS results.}
     \label{fig:vAt1}
\end{figure}

Figure~\ref{fig:scalingAt1} (a) shows that the streamwise velocity collapses well into a single profile in the self-similar coordinate for the different $\lambda$ values considered here. To be consistent, the same error function was used to fit the scaled mean streamwise velocity at different $\lambda$ values. In several previous studies, an alternative definition was suggested to collapse the streamwise mean velocity profiles in the self-similar zone. For example, \cite{kim20} subtracted the velocity of the low-speed side instead of the average velocity from the mean streamwise velocity to collapse their profiles. Our scaled mean streamwise velocity profiles also collapse using this alternative scaling (not shown). Similarly to the scaled mean streamwise velocity, the scaled Reynolds stress profiles also collapse in the self-similar zone using our suggested self-similar scalings. In particular, $\widetilde{R}_{22}$ collapses better into a single profile based on the proper scaling of $\tilde{u}_2$. %convective Mach numbers cases, $\widetilde{R}_{22}$  if we use the scaling similar to $\widetilde{R}_{12}$ as we can see from figure \ref{fig:scalingAt1} (c) and (d). 
As anticipated, all Reynolds stress components reach their peaks near the center of the shear layer and gradually diminish to zero moving away from the center of the shear layer. The smoothness of the profiles further indicates that all the cases are statistically well converged and sufficient time and space averaging was considered. 

Figure~\ref{fig:scalingAt2} (a) shows the collapsed profile of scaled mean density at different $\lambda$ values in our self-similar coordinates. The unscaled mean densities are significantly different at different $\lambda$ values, but they collapse well using our suggested self-similar scaling. Unlike the results at different $M_c$ values, the drop in density is not sensitive to the velocity parameter and the $\rho_{01}=\rho_{0}\psi(M_c)$ scaling is not shown here.
Figure~\ref{fig:scalingAt2} (b) shows the collapse of the mean pressure in the self-similar coordinate at different $\lambda$ values. There is a drop in the mean pressure around the core of the shear layer and, similar to the mean density, our self-similar scaling collapses the $\lambda$ variation quite well. 

In figure \ref{fig:vAt1}, the scaled mean transverse velocity profiles in the self-similar coordinate are compared with the profiles obtained from our suggested analytical approximation at different $\lambda$ values. It is seen that the analytical expressions approximate well the DNS variations over the $\lambda$ values examined here.

\section{Entrainment and Asymmetric Behaviour}\label{sec:asym}

\begin{figure} 
     \centering
     \begin{subfigure}{0.48\textwidth}
         \centering
         \includegraphics[width=\textwidth]{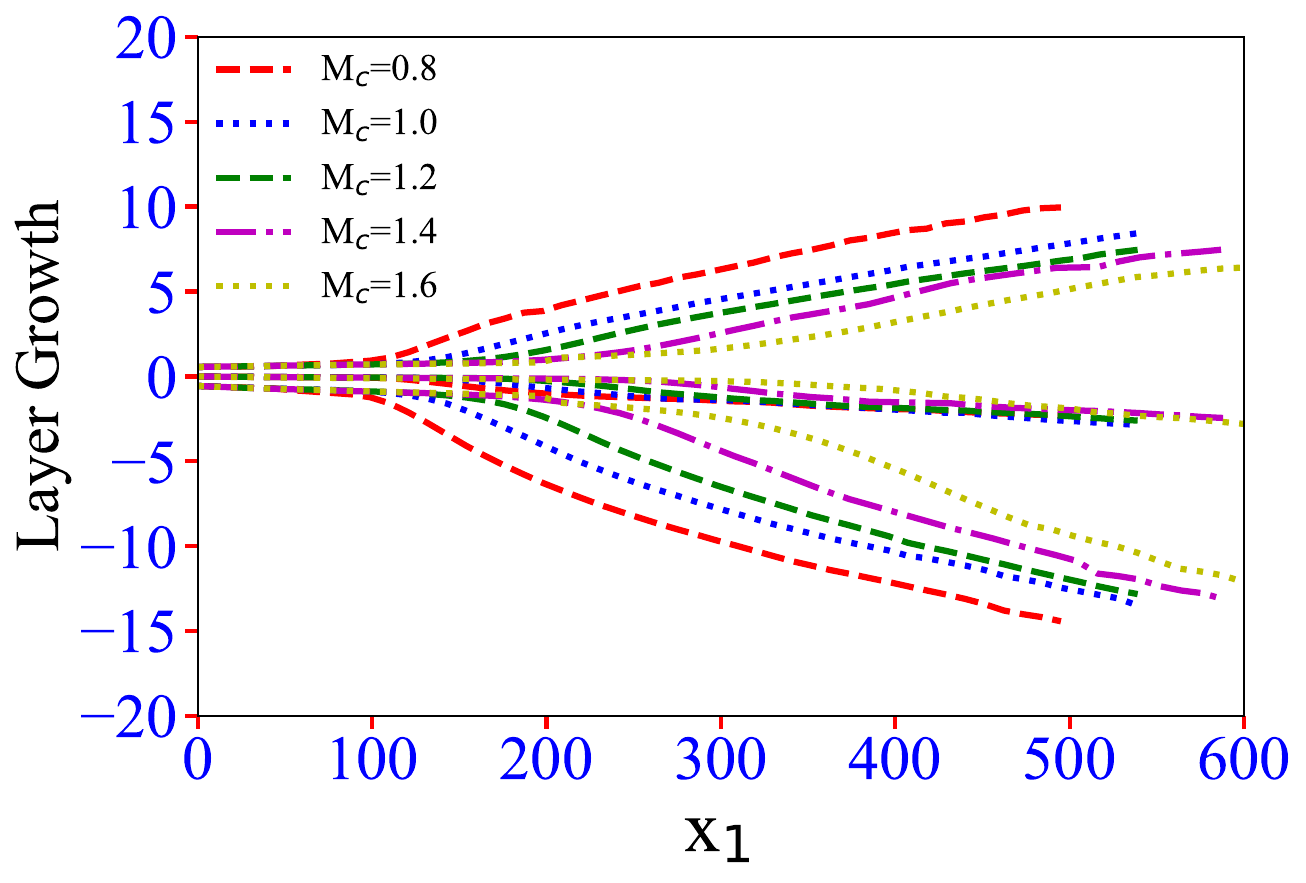}
         \subcaption{}
     \end{subfigure}
     \begin{subfigure}{0.48\textwidth}
         \centering
         \includegraphics[width=\textwidth]{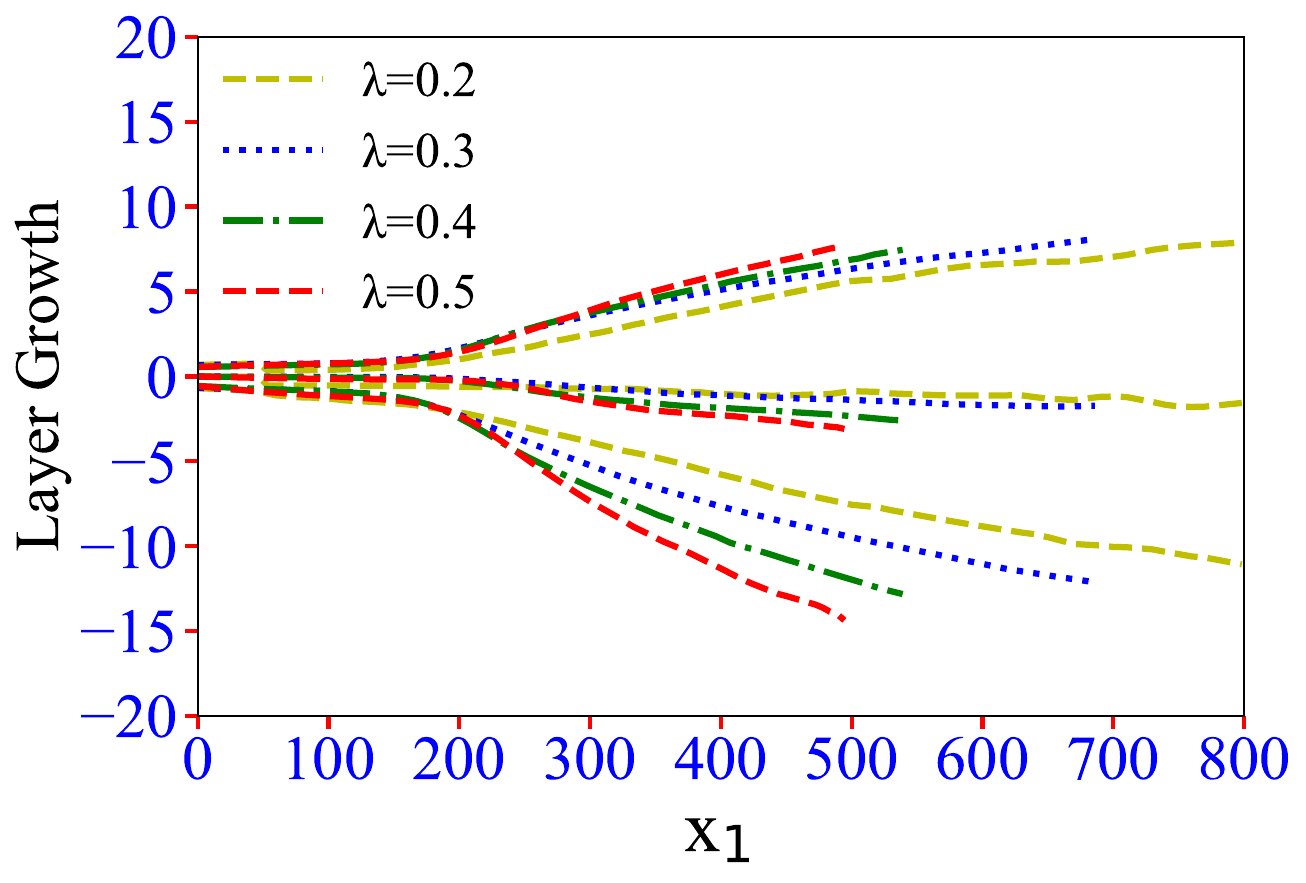}
         \subcaption{}
     \end{subfigure}
     \begin{subfigure}{0.48\textwidth}
         \centering
         \includegraphics[width=\textwidth]{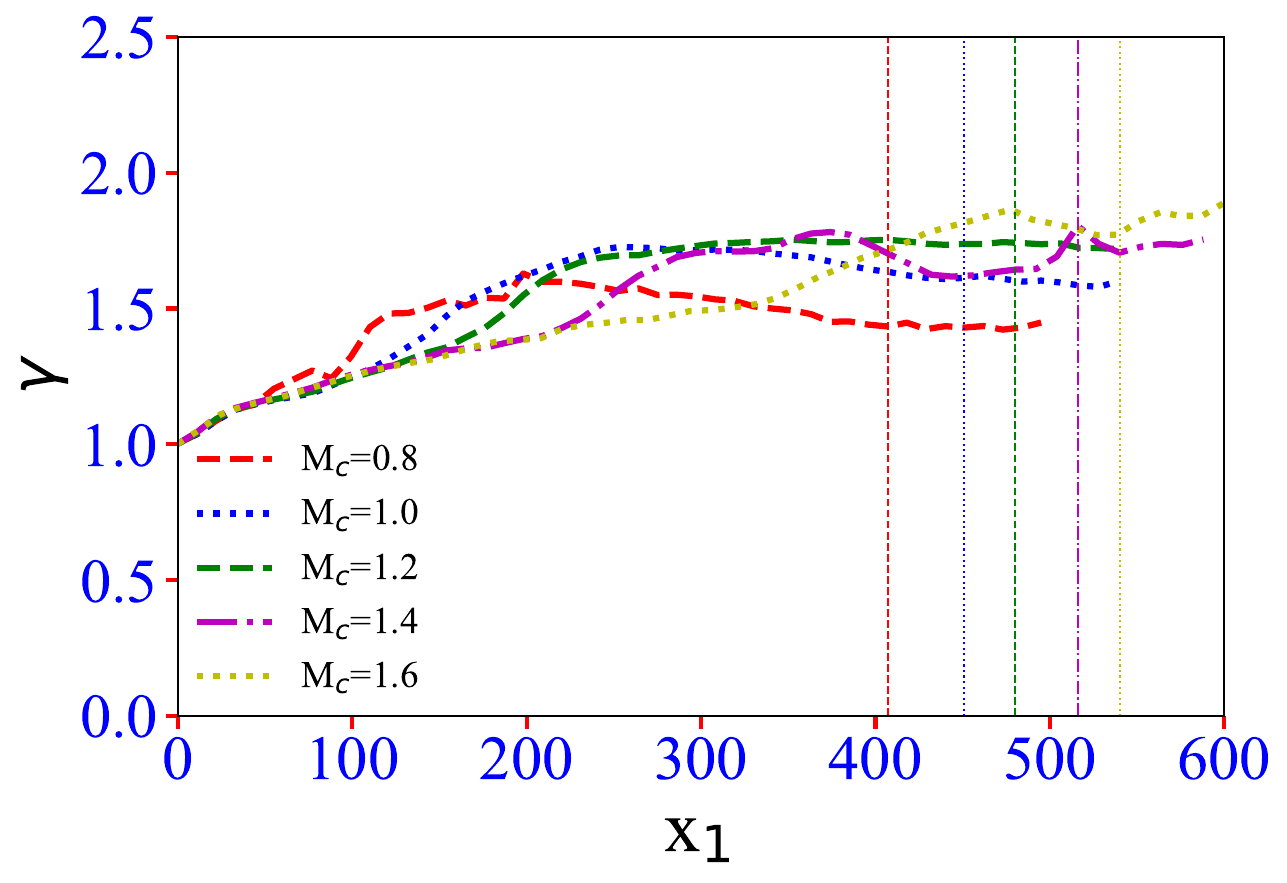}
         \subcaption{}
     \end{subfigure}
     \begin{subfigure}{0.48\textwidth}
         \centering
         \includegraphics[width=\textwidth]{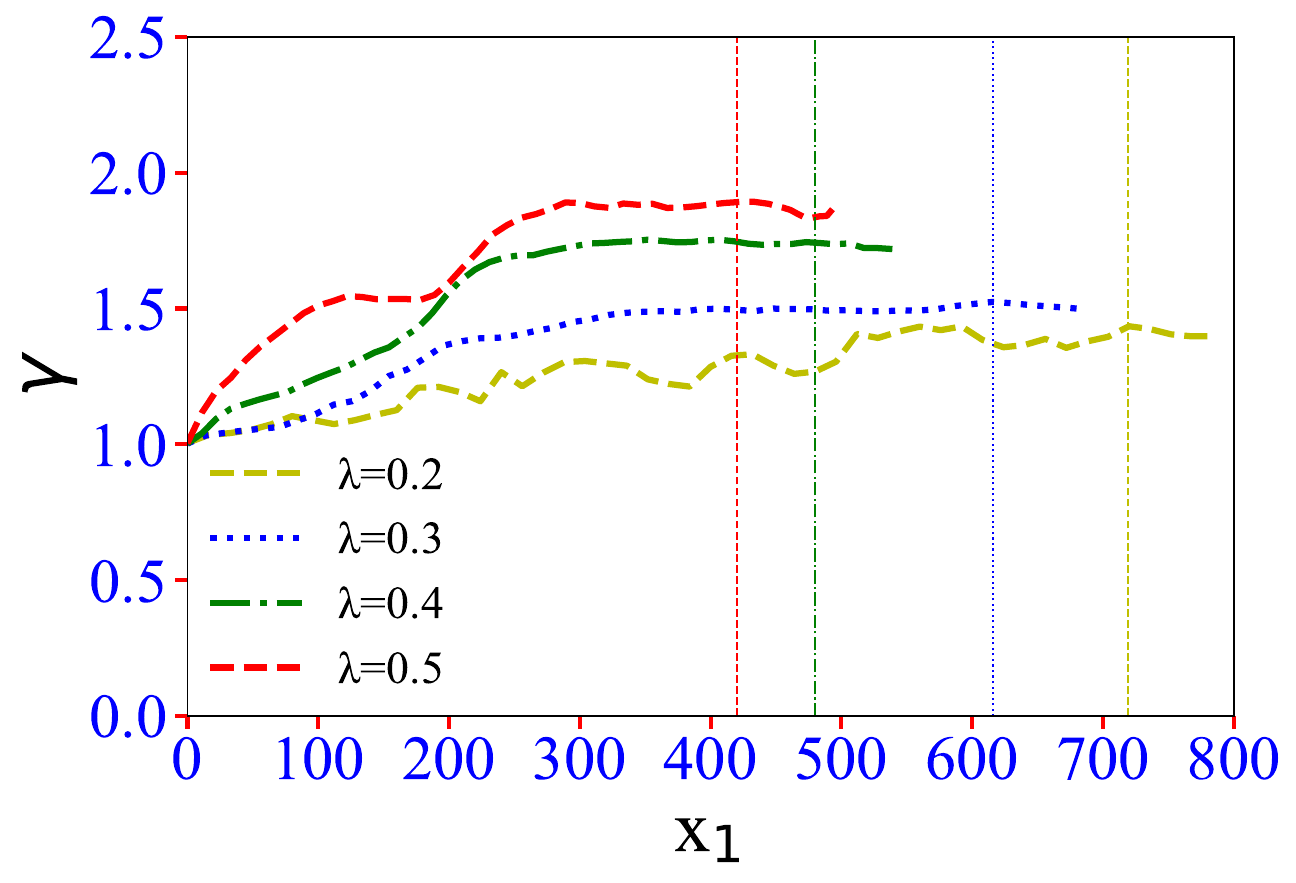}
         \subcaption{}
     \end{subfigure}
     \caption{The growth of the shear layer denoted by the positions where the mean axial velocity reaches $90\%$ of the free stream velocity and the centerline position (where the mean axial velocity is zero) at different (a) $M_c$ (b) v$\lambda$ values. The edges are defined following asymmetry parameter at different (c) $M_c$ and (d) $\lambda$ values. The vertical lines indicate the locations where a self-similar zone starts for a particular case. Figures (c) and (d) show that the asymmetry parameter becomes constant in the self-similar zones.}
     \label{fig:asym}
\end{figure}

The disparity in shear layer growth on the high- and low-speed sides at different $M_c$ and $\lambda$ values is another interesting topic to investigate. The growth of the layer is different on the two sides of the layer, as seen in figures~\ref{fig:asym} (a) and (b), with faster growth on the low-speed side as $M_c$ or $\lambda$ increases. 
Additionally, the centerline of the shear layer shifts towards the low-speed side, which makes $\frac{d\bar{x}_2}{dx_1}$ always negative, where $\bar{x}_2$ is the centerline shifting distance discussed in the section~\ref{sec:self-similarity}. %In the transition region, the shear layer growth rate is higher, leading to an increased rate of centerline shifting towards the low-speed side. 
In the self-similar zone, the centerline position varies linearly and $\frac{d\bar{x}_2}{dx_1}$ becomes constant, confirming the assumption discussed in section~\ref{sec:self-similarity} to make the equations self-similar. 

To quantify the asymmetric behavior of the shear layer, an asymmetry parameter ($\gamma$) is introduced, which is defined as the ratio of the thickness of the shear layer on the low-speed side and the thickness of the shear layer on the high-speed side. 
\begin{equation}
   \gamma=\frac{\delta_{low-speed}}{\delta_{high-speed}}
\end{equation}
\;

The thicknesses are measured from the geometric centerline ($x_2$=0 line in figure~\ref{fig:Screenshot 2022-10-24 145102}) of the shear layer to the position where the mean axial velocity reaches $90\%$ of the free stream velocity. 
Figures~\ref{fig:asym} (c) and (d), and  tables~\ref{tab:asym1} and ~\ref{tab:asym2} show that in the self-similar region $\gamma$ becomes almost constant. The value of $\gamma$ is always greater than unity, as the shear layer grows faster on the low-speed side and increases with $M_c$ and $\lambda$.

\begin{table}
  \centering
  \begin{tabular}{|c|c|c|}
  \hline
$M_c$   &   $\lambda$ & $\gamma$ \\[3pt]
        \hline
         0.8   & 0.4 & 1.43\\
         \hline
         1.0   & 0.4 & 1.60\\
         \hline
         1.2   & 0.4 & 1.73\\
         \hline
         1.4   & 0.4 & 1.74\\
         \hline
         1.6   & 0.4 & 1.83\\
         \hline
  \end{tabular}
\caption{Asymmetry parameter at different $M_c$ values for $\lambda=0.4$.}
  \label{tab:asym1}
\end{table}

\begin{table}
  \centering
  \begin{tabular}{|c|c|c|}
  \hline
$\lambda$   &   $M_c$ & $\gamma$ \\[3pt]
\hline
         0.2   & 1.2 & 1.38\\
         \hline
         0.3   & 1.2 & 1.50\\
         \hline
         0.4   & 1.2 & 1.73\\
         \hline
         0.5   & 1.2 & 1.86\\
         \hline
  \end{tabular}
  \caption{Asymmetry parameter at different $\lambda$ for $M_c=1.2$.}
  \label{tab:asym2}
\end{table}

\begin{figure}
     \centering
     \begin{subfigure}{0.7\textwidth}
         \centering
         \includegraphics[width=\textwidth]{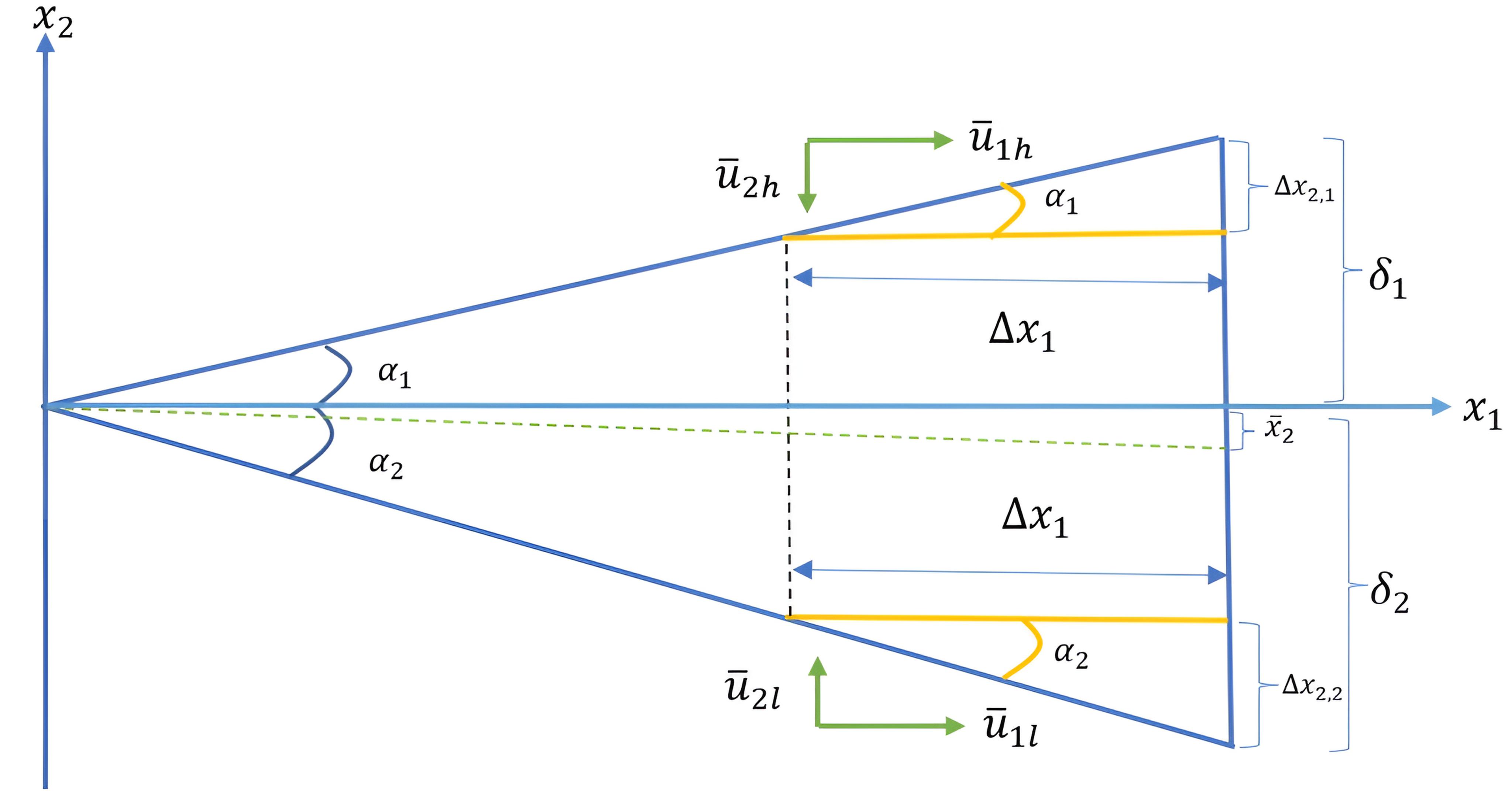}
     \end{subfigure}
     \caption{Coordinate system for entrainment ratio analysis. Subscript h denotes the high-speed side and subscript l denotes the low-speed side. }
     \label{fig:evgeo}
\end{figure}

The shear layer grows by entraining fluid from the free stream, through two main mechanisms, (i) engulfment and (ii) nibbling (\citealt{kim2020compressibility,dimotakis_entrainment}). In engulfment, pockets of non-turbulent fluids are trapped in the turbulent zone whereas, in the nibbling, free-stream fluids become part of the mixing layer due to viscous diffusion by small-scale eddies. 

As previously noted, in the spatially developing shear layer, the centerline shifts towards the low-speed side, and due to this change in orientation, the shear layer entrains varying amounts of fluid from the free streams (\citealt{ovidio}, \citealt{dimotakis_entrainment}). The entrainment ratio ($E_v$) serves as an effective means to describe this particular phenomenon. When the shear layer exhibits self-similar growth, the volumetric entrainment from the two-speed sides can be calculated from the magnitude of the free stream vector and the angle of the edge of the shear layer (\citealt{ovidio}). The theoretical estimation of the entrainment ratio presented in this study is motivated by the analytical solution developed by \cite{abramovich84} for the classical problem of the turbulent mixing layer of the two infinite streams. Figure~\ref{fig:evgeo} illustrates the coordinate system of the shear layer used for the entrainment analysis presented in the study. The edges of the shear layer are oriented at angles $\alpha_1$ and $\alpha_2$ with respect to the high- and low-speed streams, respectively. As the shear layer shifts towards the low-speed side, $\alpha_2$ is greater than $\alpha_1$. $\bar{x_2}$ is the centerline shifting, $\bar{u}_{1h}$ and $\bar{u}_{1l}$ are the mean streamwise velocities, $\bar{u}_{2h}$ and $\bar{u}_{2l}$ are the mean transverse velocities and $\delta_1$ and $\delta_2$ are the shear layer thicknesses for high-speed and low-speed sides respectively. If a small segment of the shear layer in the self-similar zone is considered, from the continuity equation, the entrainment would be the integral of mass flow within the mixing layer through any plane normal to $x_1$ axis. Therefore, the entrainment on the high-speed and low-speed sides are: 
\begin{equation}
V_{01}= (\Bar{u}_{2h} + \Bar{u}_{1h} \tan{\alpha_1}) \Delta x_1 \\
\end{equation}
\begin{equation}
V_{02} =(\Bar{u}_{2l} + \Bar{u}_{1l} \tan{\alpha_2})\Delta x_1 \\
\end{equation}
The entrainment ratio $E_v$ can then be written as:
\begin{equation}\label{eq:ev1}
E_v = \frac{V_{01}}{V_{02}}=\frac{\Bar{u}_{2h} + \Bar{u}_{1h} \tan{\alpha_1}}{\Bar{u}_{2l} + \Bar{u}_{1l} \tan{\alpha_2}} \\
\end{equation}

Geometrically, 
\begin{equation}
\delta_1 + \delta_2= \delta \\
\end{equation}
\begin{equation}
\delta_1 - \delta_2= \bar{x}_2 \\
\end{equation}
Therefore,
\begin{equation}
\begin{split}
& \frac{d\delta_1}{dx}= \frac{1}{2} (\frac{d\delta}{dx} + \frac{d\bar{x}_2}{dx})  \\
& \frac{d\delta_2}{dx}= \frac{1}{2} (\frac{d\delta}{dx} - \frac{d\bar{x}_2}{dx})
\end{split}
\end{equation}
 
Furthermore, from Figure~\ref{fig:evgeo}:
\begin{equation}
\begin{split}
& \tan{\alpha_1}= \frac{d\delta_1}{dx}= \frac{1}{2} (\frac{d\delta}{dx} + \frac{d\bar{x}_2}{dx})  \\
& \tan{\alpha_2}= \frac{d\delta_2}{dx}= \frac{1}{2} (\frac{d\delta}{dx} - \frac{d\bar{x}_2}{dx})
\end{split}
\end{equation}

After replacing the tangents, equation (\ref{eq:ev1}) becomes:
\begin{equation}\label{eq:ev2}
E_v =\frac{\Bar{u}_{2h} + \frac{1}{2} \Bar{u}_{1h} (\frac{d\delta}{dx} + \frac{d\bar{x}_2}{dx})}{\Bar{u}_{2l} + \frac{1}{2} \Bar{u}_{1l}  (\frac{d\delta}{dx} - \frac{d\bar{x}_2}{dx})} \\
\end{equation}

The alternative form of equation (\ref{eq:ev2}) can be written using the definition of $\phi$:
\begin{equation}\label{eq:ev3}
E_v =\frac{\Bar{u}_{2h} + \frac{1}{2} \Bar{u}_{1h}  \frac{d\delta}{dx} (1+\phi)} {\Bar{u}_{2l} + \frac{1}{2} \Bar{u}_{1l}  \frac{d\delta}{dx} (1 - \phi)} \\
\end{equation}

If the variables in equation (\ref{eq:ev3}) are replaced by our self-similar variables discussed in section~\ref{sec:self-similarity}, the equation becomes: 
\begin{equation}\label{eq:ev4}
E_v = \frac{V_{01}}{V_{02}}= \frac{\hat{u}_{2h} + \frac{u_2*}{u_{02}}+ \frac{1}{2} (\hat{u}_{1h} + \frac{1}{2\lambda}) (1+\phi)} {\hat{u}_{2l} + \frac{u_2*}{u_{02}}+ \frac{1}{2} (\hat{u}_{1l} + \frac{1}{2\lambda}) (1-\phi)}, \\
\end{equation}
where:
\begin{equation}\label{eq:ev5}
V_{01} =\Delta U \frac{d\delta}{dx} [\hat{u}_{2h} + \frac{u_2*}{u_{02}}]+ \frac{1}{2}\Delta U (\hat{u}_{1h} + \frac{1}{2\lambda})\frac{d\delta}{dx} (1+\phi) \\
\end{equation}
and
\begin{equation}\label{eq:ev6}
V_{02} =\Delta U \frac{d\delta}{dx} [\hat{u}_{2l} + \frac{u_2*}{u_{02}}]+ \frac{1}{2}\Delta U (\hat{u}_{1l} + \frac{1}{2\lambda})\frac{d\delta}{dx} (1-\phi).\\
\end{equation}

By replacing ($\hat{u}_2$+$\frac{u_2*}{u_{02}}$) in equation (\ref{eq:ev5}) from our $\hat{u}_2$ derivation, $V_{01}$ for compressible shear layer becomes: 
\begin{equation}
\begin{split}
V_{01} = [& \frac{1}{1+\hat{\rho}\psi(M_c)} [\psi(M_c)\hat{\rho}(\eta+\phi)+\phi]\hat{u}_{1}+\frac{\psi(M_c)\hat{\rho}(\eta+\phi)}{2\lambda[1+\hat{\rho}\psi(M_c)]} + \\
& \frac{\psi(M_c)}{1+\hat{\rho}\psi(M_c)} [-\int_{-\infty}^{\eta}\hat{\rho}\hat{u_1}\,d\eta -\frac{1}{2\lambda} \int_{-\infty}^{\eta}\hat{\rho}\,d\eta] +\frac{\phi}{2\lambda[1+\hat{\rho}\psi(M_c)]} \\
& + \frac{1}{1+\hat{\rho}\psi(M_c)} \int_{-\infty}^{\eta} \eta \frac{d\hat{u_1}}{d\eta}\,d\eta + \frac{\psi(M_c)}{1+\hat{\rho}\psi(M_c)} \frac{\lambda-1}{2\lambda} \int_{-\infty}^{+\infty}\hat{\rho}\hat{u_1}\,d\eta \\
& + \frac{\psi(M_c)}{4\lambda [1+\hat{\rho}\psi(M_c)]}\int_{-\infty}^{+\infty}\hat{\rho}\,d\eta + \frac{1-\lambda}{2\lambda [1+\hat{\rho}\psi(M_c)]} \int_{-\infty}^{+\infty} \eta \frac{d\hat{u_1}}{d\eta}\,d\eta \\
& - \frac{\psi(M_c)}{1+\hat{\rho}\psi(M_c)} \int_{-\infty}^{+\infty}\hat{\rho}\hat{u_1}\hat{u_1}\,d\eta + \frac{2}{1+\hat{\rho}\psi(M_c)}\int_{-\infty}^{+\infty}\eta\hat{u_1}\frac{d\hat{u_1}}{d\eta}d\eta \\
& + \frac{\phi}{1+\hat{\rho}\psi(M_c)} \int_{-\infty}^{+\infty}\hat{u_1}\frac{d\hat{u_1}}{d\eta}\,d\eta] \Delta{U} \frac{d\delta}{dx} + \frac{1}{2}\Delta{U} (\hat{u}_{1}+\frac{1}{2\lambda})\frac{d\delta}{dx} (1+\phi). \\
\end{split}
\end{equation}

Replacing the values of integrals from the appendix gives: 
\begin{equation}
\begin{split}
V_{01} = [& \frac{erf(B\eta)}{2[1+a e^{-B^2\eta^2}\psi(M_c)]}[\psi(M_c)ae^{-B^2\eta^2}(\eta+\phi)+\phi] + \frac{\psi(M_c)ae^{-B^2\eta^2}(\eta+\phi)}{2\lambda[1+ae^{-B^2\eta^2}\psi(M_c)]}  \\
& - \frac{\psi(M_c)a\sqrt{\pi}[1+erf(B\eta)]}{4B[1+ae^{-B^2\eta^2}\psi(M_c)]} [\frac{1}{\lambda}+\frac{1}{2}erf(B\eta)-\frac{1}{2}] +\frac{1}{1+ae^{-B^2\eta^2}\psi(M_c)} \\
& [\frac{\phi}{2\lambda}-\frac{1}{2\sqrt{\pi}B}(e^{-B^2\eta^2}-\sqrt{2})] + \frac{\psi(M_c)\sqrt{\pi}a(3-\lambda)}{12\lambda B[1+ae^{-B^2\eta^2}\psi(M_c)]}]\Delta{U}\frac{d\delta}{dx}  \\
& + \frac{1}{2}\Delta{U} (\frac{1}{2}erf(B\eta)+\frac{1}{2\lambda})\frac{d\delta}{dx} (1+\phi). \\
\end{split}
\end{equation}

Similarly, $V_{02}$ is: 
\begin{equation}
\begin{split}
V_{02} = [& \frac{erf(B\eta)}{2[1+a e^{-B^2\eta^2}\psi(M_c)]}[\psi(M_c)ae^{-B^2\eta^2}(\eta+\phi)+\phi] + \frac{\psi(M_c)ae^{-B^2\eta^2}(\eta+\phi)}{2\lambda[1+ae^{-B^2\eta^2}\psi(M_c)]}  \\
& - \frac{\psi(M_c)a\sqrt{\pi}[1+erf(B\eta)]}{4B[1+ae^{-B^2\eta^2}\psi(M_c)]} [\frac{1}{\lambda}+\frac{1}{2}erf(B\eta)-\frac{1}{2}] +\frac{1}{1+ae^{-B^2\eta^2}\psi(M_c)} \\
& [\frac{\phi}{2\lambda}-\frac{1}{2\sqrt{\pi}B}(e^{-B^2\eta^2}-\sqrt{2})] + \frac{\psi(M_c)\sqrt{\pi}a(3-\lambda)}{12\lambda B[1+ae^{-B^2\eta^2}\psi(M_c)]}]\Delta{U}\frac{d\delta}{dx}  \\
& + \frac{1}{2}\Delta{U} (\frac{1}{2}erf(B\eta)+\frac{1}{2\lambda})\frac{d\delta}{dx} (1-\phi). \\
\end{split}
\end{equation}

In the incompressible limit, when $M_c$ goes to 0, $\psi(M_c)$ goes to 0. Therefore, for the incompressible shear layer, $V_{01}$ and $V_{02}$ are:
\begin{equation}
V_{01} = [\frac{\phi}{2}erf(B\eta)+\frac{\phi}{2\lambda}-\frac{1}{2\sqrt{\pi}B}e^{-B^2\eta^2}+\frac{1}{\sqrt{2\pi}B}]\Delta{U}\frac{d\delta}{dx}+ \frac{1}{2}\Delta{U}(\frac{1}{2}erf(B\eta)+\frac{1}{2\lambda})\frac{d\delta}{dx} (1+\phi) \\  
\end{equation}
and
\begin{equation}
V_{02} = [\frac{\phi}{2}erf(B\eta)+\frac{\phi}{2\lambda}-\frac{1}{2\sqrt{\pi}B}e^{-B^2\eta^2}+\frac{1}{\sqrt{2\pi}B}]\Delta{U}\frac{d\delta}{dx}+ \frac{1}{2}\Delta{U}(\frac{1}{2}erf(B\eta)+\frac{1}{2\lambda})\frac{d\delta}{dx} (1-\phi). \\  
\end{equation}

In order to evaluate the integrals, a specific $\eta$ location should be chosen. To avoid the arbitrariness in the choice of the shear layer edge location, it assumed that at the edges of the shear layer $erf(B\eta)=\pm 1$, and $e^{-B^2\eta^2}=0$. With this choice, for the fully compressible case, $E_v$ becomes: 
\begin{equation}\label{eq:ev8}
E_{v,comp} = \frac{\frac{4\lambda}{\sqrt{2\pi}B}+ (1+\lambda)(1+3\phi)+\frac{\psi (M_c)\sqrt{\pi}}{B}[1+\frac{\lambda}{3}]}{\frac{4\lambda}{\sqrt{2\pi}B}+ (1-\lambda)(1+\phi)+\frac{\psi (M_c)\sqrt{\pi}}{B}[-1+\frac{\lambda}{3}]}. \\  
\end{equation}

The incompressible limit of $E_v$ is:
\begin{equation}\label{eq:ev7}
E_{v,incomp} = \frac{\frac{4\lambda}{\sqrt{2\pi}B}+ (1+\lambda)(1+3\phi)}{\frac{4\lambda}{\sqrt{2\pi}B}+ (1-\lambda)(1+\phi)}. \\  
\end{equation}

For consistency, in the limit of $\lambda$ going to 0, $\phi$ goes to 0 and $E_v$ becomes 1 for both incompressible and compressible cases. For compressible flow at $\lambda$=0, (by definition) $M_c=$ 0 and therefore $\psi(M_c)=$ 0 which eventually makes $E_v$=1. 

The value of $B=1.812$ is used by \cite{wei22} to approximate the profile of the mean axial velocity for the case of the incompressible shear layer. Our data suggest that $B=2.05$ fits the DNS mean axial velocity profiles well for the compressible shear layer case. 
These values are reasonably close and we will use $B=2.05$ for the rest of this analysis. 

\begin{figure} 
     \centering
     \begin{subfigure}{0.49\textwidth}
         \centering
         \includegraphics[width=\textwidth]{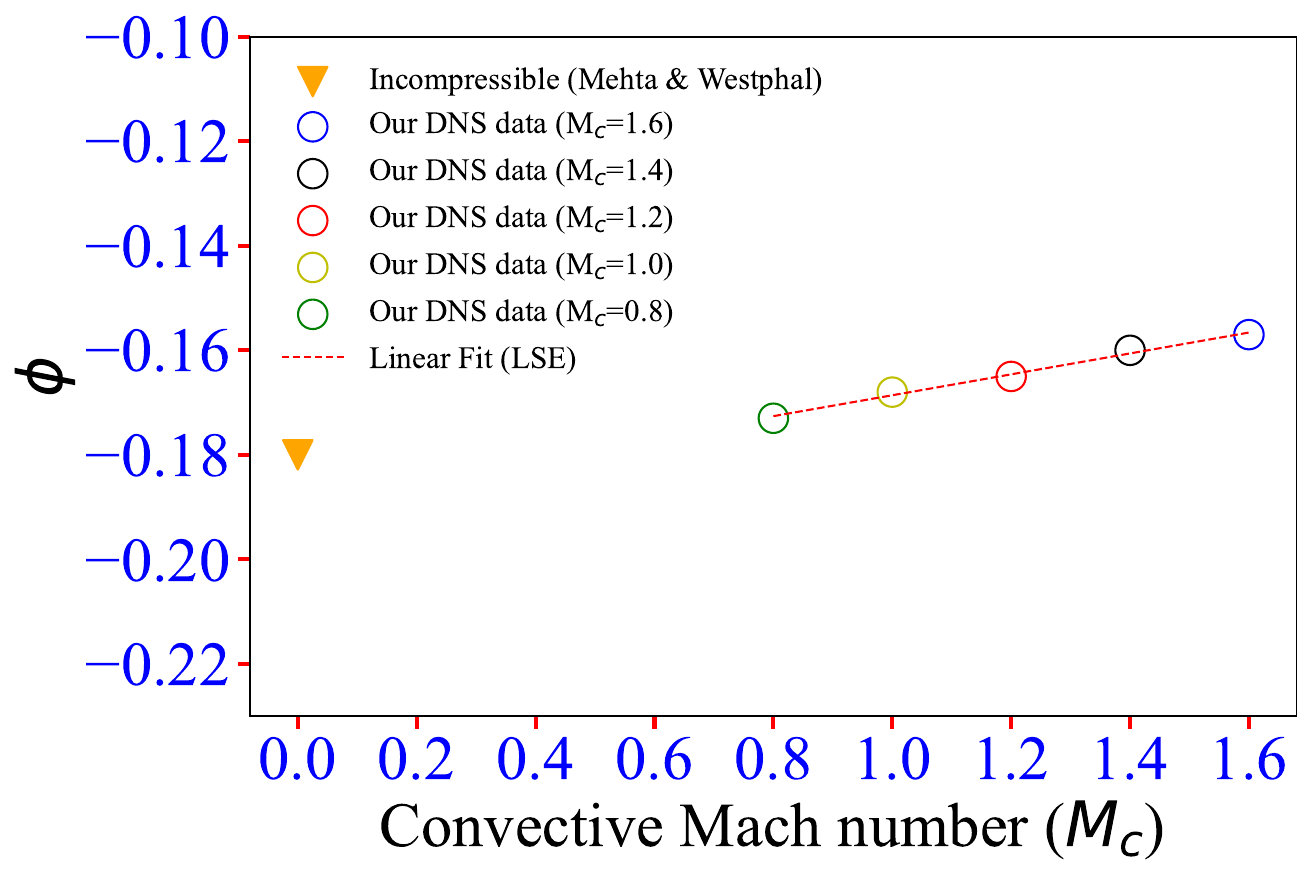}
         \subcaption{}
     \end{subfigure}
     \begin{subfigure}{0.49\textwidth}
         \centering
         \includegraphics[width=\textwidth]{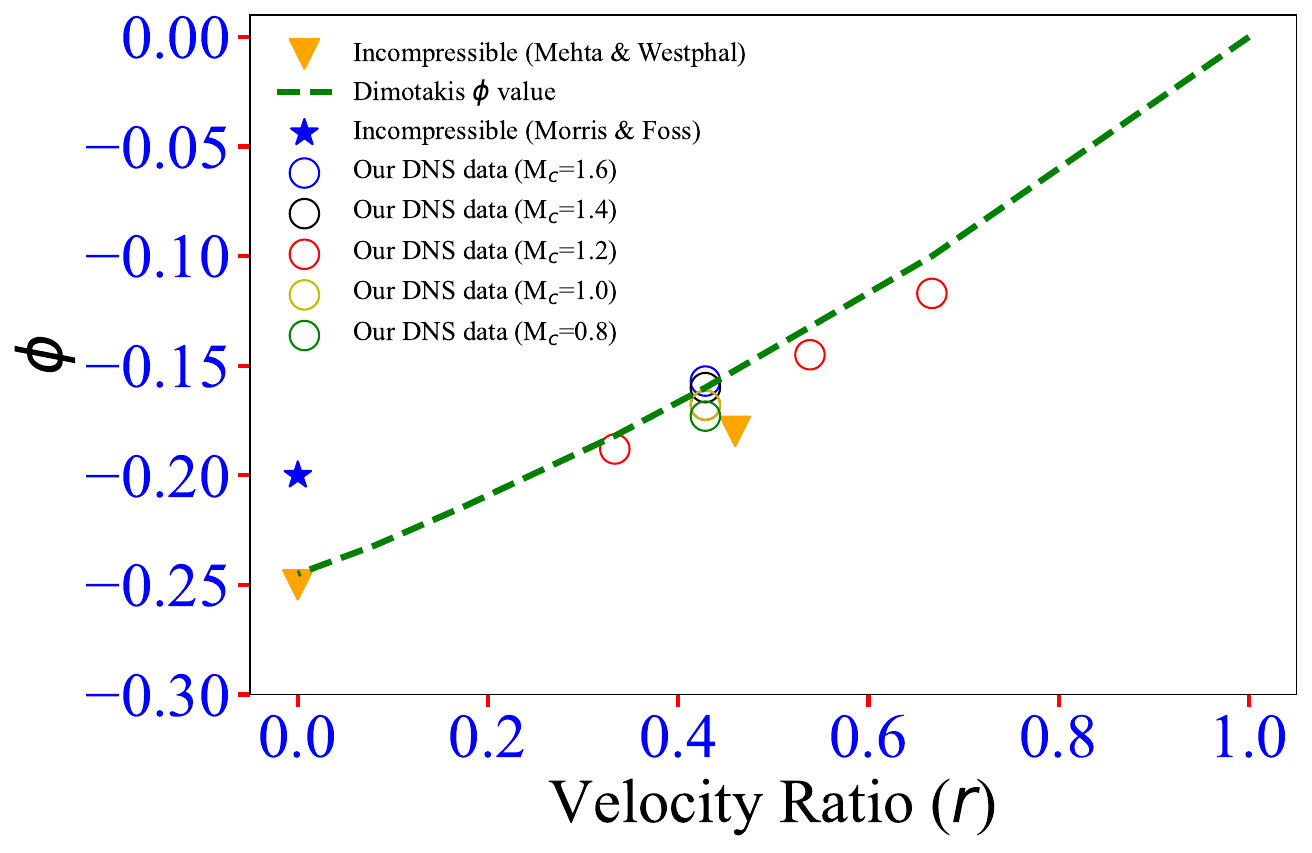}
         \subcaption{}
     \end{subfigure}
     \begin{subfigure}{0.48\textwidth}
         \centering
         \includegraphics[width=\textwidth]{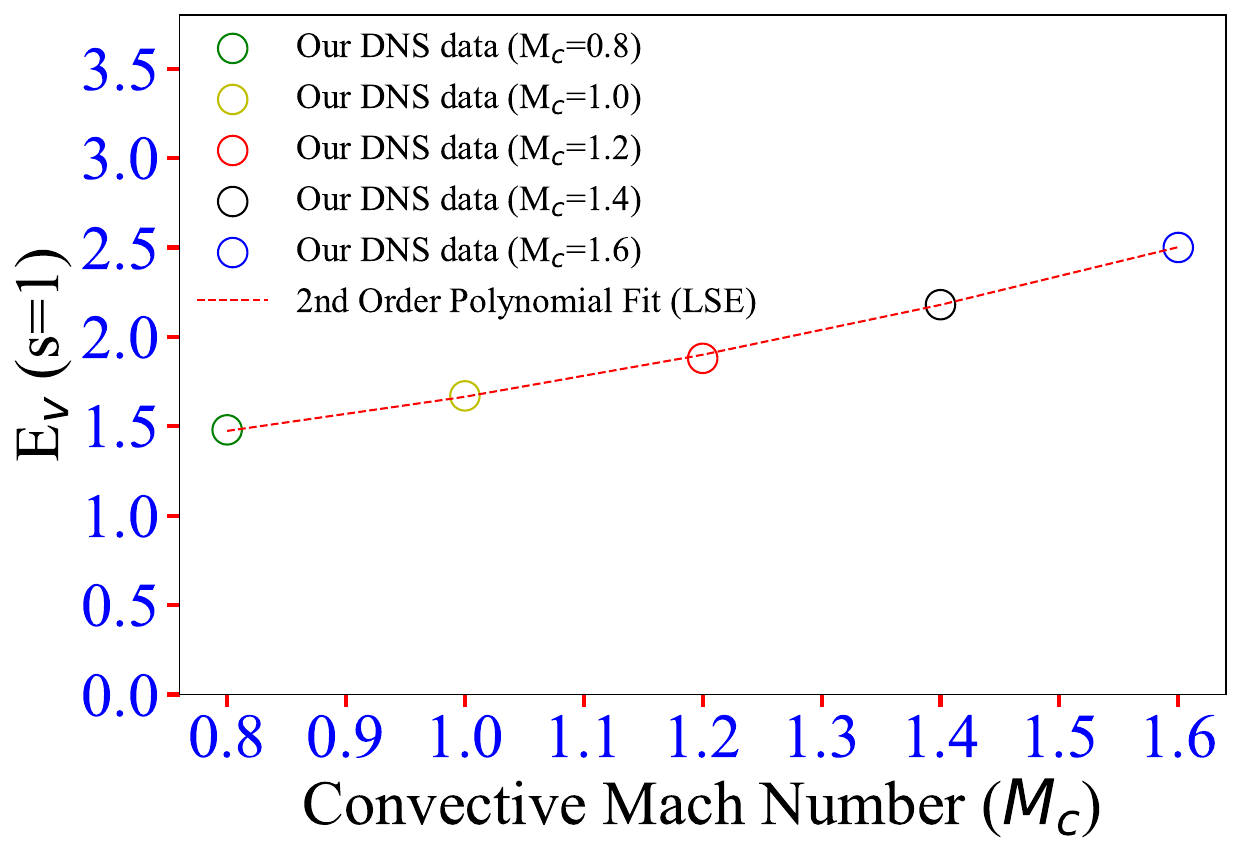}
         \subcaption{}
     \end{subfigure}
     \begin{subfigure}{0.48\textwidth}
         \centering
         \includegraphics[width=\textwidth]{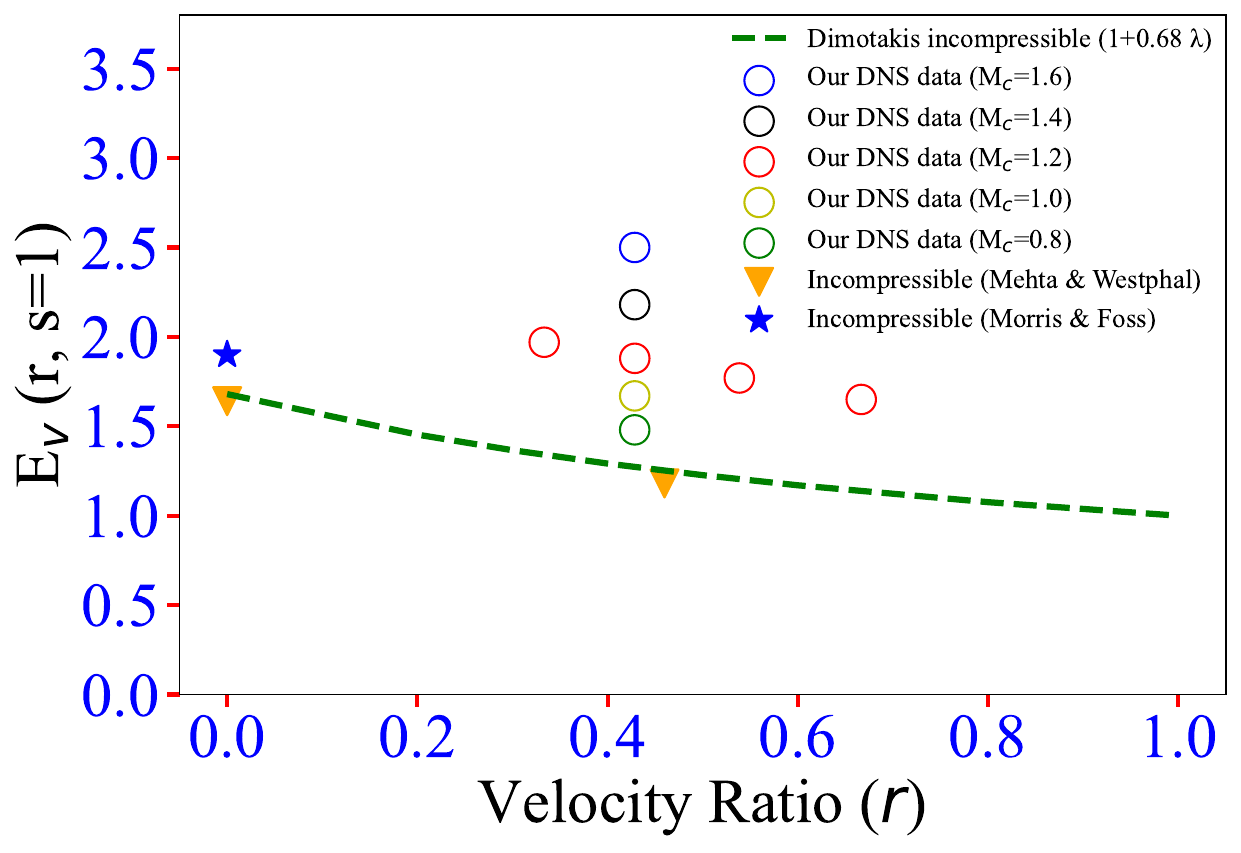}
         \subcaption{}
     \end{subfigure}
     \caption{(a) Variation of $\phi$ with $M_c$ at $\lambda=0.4$. The incompressible value at $M_c=0$ is taken from \cite{mehta85} at $\lambda=0.37$. In the range of higher $M_c$ values, $\phi$ varies linearly with $M_c$. The equation of the linear fit using Least Square Error (LSE) is $\phi=0.02 M_c-0.1886$. %However, it is recommended not to use this formula outside this $M_c$ range.
     (b) Incompressible and compressible $\phi$ values at different $M_c$ and velocity ratios. The incompressible data at $r= 0$ and $0.46$ have been taken from the experiments of \cite{mehta85} and \cite{foss}. The compressible $\phi$ values have been taken from present simulations. The dashed line has been extracted from \cite{dimotakis_entrainment} $Ev=1+0.68\lambda$ formula. (c) Variation of $E_v$ with $M_c$. A second order polynomial ($E_v=0.55M_c^2-0.035M_c+1.15$) using the Least Square Error has been fitted to our $E_v$ vs. $M_c$ data. (d) Variation $E_v$ with $r$ using our incompressible and compressible $E_v$ equations. When using the incompressible $\phi$ values from \cite{mehta85} and \cite{foss} in our derived incompressible $E_v$ equation, the $E_v$ values closely align with Dimotakis' curve \cite{dimotakis_entrainment}.}
     \label{fig:evincom}
\end{figure}

Equations (\ref{eq:ev7}) and (\ref{eq:ev8}) indicate that the incompressible $E_v$ is a function of $\lambda$ and $\phi$ while the compressible $E_v$ is influenced by $\lambda$, $\phi$ and $M_c$. The value of $\phi$ depends on $M_c$, $r$, and the density ratio between the two free streams, $s$. Since $s$ remains unity in this paper, the variation of $\phi$ is considered based on $M_c$ and $r$. In the literature, the variation of $\phi$ with $M_c$ has not been reported. Our DNS data (Figure \ref{fig:evincom} a) indicate a linear correlation between $\phi$ and $M_c$ for the range of $M_c$ studied here. 
 The relation between $\phi$ and $\lambda$ or velocity ratio ($r$) is less clear. In the literature, the variation of $\phi$ with $r$ has been scarcely reported, especially at high-velocity ratios (\citealt{wei22}). $\phi$ from our DNS data is shown and compared with the incompressible $\phi$ values reported by \cite{wei22}, \cite{mehta85} and \cite{foss} in figure \ref{fig:evincom} (b). 
 Due to the scarcity of data points of $\phi$ for different $\lambda$ values, it is difficult to establish a clear relationship between $\phi$ and $r$.  

Figure \ref{fig:evincom} (c) shows that a quadratic fit for $E_v$ captures well the variation $M_c$ from our data. In figure \ref{fig:evincom} (d), the $E_v$ values resulting from using the available $\phi$ values in the corresponding incompressible and compressible formulas are compared against the $E_v$=$1+0.68\lambda$ relation suggested by \cite{dimotakis_entrainment}. Employing the $\phi$ values from \cite{mehta85} in our incompressible $E_v$ equation yields results that closely align with the linear $\lambda$ fit from \cite{dimotakis_entrainment}. The $E_v$ value of \cite{foss} is found to be slightly higher than this fit, most likely due to the imposition of cross-stream entrainment velocity in the experiment. 
%Using the extrapolated $\phi$ at $M_c$=0 from our DNS data, the incompressible $E_v$ equation gives the value of $E_v$ that is reasonably close to \cite{dimotakis_entrainment}. 
As $M_c$ increases, the value of $E_v$ increases, consistent with the higher asymmetry of the layer.

To further examine the variation of $E_v$ with $M_c$ and $\lambda$, the values of the total entrainment and entrainment in each stream are shown in tables~\ref{tab:entrain1} and \ref{tab:entrain2}. The total entrainment and the entrainment in individual streams also increase with $\lambda$, similar to the entrainment ratio and consistent with previous studies (\citealt{dimotakis_entrainment,shahadat2024entrainment}). For a fixed density ratio of 7, \cite{brown1975entrainment} showed a similar trend of entrainment in each stream with $\lambda$. On the other hand, table~\ref{tab:entrain2} shows that the total entrainment decreases while the entrainment ratio increases with increasing $M_c$, which is connected to the reduced growth rate. The excess entrainment on the high-speed side with increasing $M_c$ is responsible for this higher entrainment ratio (\citealt{shahadat2024entrainment}).

\begin{table}
  \centering
  \begin{tabular}{|c|c|c|c|c|}
  \hline
 $\lambda$   & $V_{01}$ & $V_{02}$  & $V_{01}$+$V_{02}$  & $V_{01}$/$V_{02}$ \\
 \hline
    0.2   & 0.0273 & 0.0165 & 0.0438 & 1.65 \\
    \hline
    0.3   & 0.0382 & 0.0215 & 0.0597 & 1.776 \\
    \hline
    0.4   & 0.0572 & 0.0302 & 0.0874 & 1.896 \\
    \hline
    0.5   & 0.0758 & 0.0382 & 0.114 & 1.98 \\
    \hline
  \end{tabular}
  \caption{Variation of total entrainment and entrainment ratio with $\lambda$. $M_c=1.2$ for all cases.}
  \label{tab:entrain1}
\end{table}

\begin{table}
  \centering
  \begin{tabular}{|c|c|c|c|c|}
   \hline
$M_c$   &   $V_{01}$ & $V_{02}$  & $V_{01}$+$V_{02}$  & $V_{01}$/$V_{02}$ \\
\hline
    0.8   & 0.0546 & 0.0369 & 0.0915 & 1.48 \\
    \hline
    1.0   & 0.0551 & 0.0329 & 0.088 & 1.67 \\
    \hline
    1.2   & 0.0572 & 0.0302 & 0.0874 & 1.896\\
    \hline
    1.4   & 0.0575 & 0.0263 & 0.0838 & 2.186 \\
    \hline
    1.6   & 0.0578 & 0.0228 & 0.0806 & 2.535 \\
    \hline
  \end{tabular}
  \caption{Variation of total entrainment and entrainment ratio with $M_c$. $\lambda=0.4$ for all cases. }
  \label{tab:entrain2}
\end{table}

\section{Concluding Remarks}\label{sec:conclusion}
Direct Numerical Simulation (DNS) of a spatially developing supersonic turbulent shear layer is conducted for a detailed study of the shear layer growth, statistical behavior, self-similarity, flow asymmetry, and entrainment. %Fully compressible dimensionless mass, momentum, energy, scalar, and equation of state are solved with high-order numerical methods. 
The convective Mach number, $M_c$, and velocity parameter, $\lambda$, appear naturally in our self-similar equations for the compressible single fluid case and, therefore are considered as the important parameters affecting the evolution, structure, and growth rate of the shear layer. To systematically study these parameters, five different $M_c$ values were chosen keeping the $\lambda$ fixed and four different $\lambda$ values were chosen keeping the $M_c$ fixed. A wide range of grid sizes and domain sizes were considered in convergence tests to establish the accuracy of the numerical results. Statistical analyses confirm the robustness of the numerical simulations, indicating no sensitivity to grid resolution, domain size and that adequate time and space averaging were implemented. 

Attaining self-similarity is very important to establish the scaling behavior, especially given the comparative framework involving multiple simulation cases. Given that prior studies have employed diverse criteria for defining self-similarity, the present work systematically describes the criteria used for identifying self-similar zones. Our analysis reveals that different flow parameters attain self-similarity at different streamwise locations. For instance, the mean streamwise velocity, shear layer thickness, and momentum thickness demonstrate self-similarity at comparatively earlier streamwise locations, while Reynolds stress components achieve self-similar behavior further downstream. %It would be misleading if the self-similar zone is selected based on the self-similarity of the mean streamwise velocity, shear layer thickness, and momentum thickness because, at that particular location, Reynolds stress components might not be still self-similar. 
Therefore, the self-similar zone should be selected based on all of the quantities of interest achieving self-similarity. In addition, the layer growth occurs at different rates for different case studies. Therefore, to ensure that self-similarity is captured for different cases, the domain sizes in the streamwise direction were extended accordingly.

%Another interesting fact is that the shear layer starts to grow at different streamwise locations across different case studies. Moreover, the evolution, growth, and structure of the shear layer are also very different. Therefore, to ensure self-similarity for different cases, the domain sizes in the streamwise direction were extended accordingly. Once we found the self-similar zone, additional quantities such as scaled mean density, pressure, temperature, transverse velocity, and dissipation also achieve self-similarity. 

In the present investigation, we extend upon prior research by employing the fully compressible forms of self-similar continuity, streamwise momentum, and transverse momentum and energy equations. For the first time, compressibility, centerline shifting, and $\phi$ which is the ratio of the centerline shifting speed and growth rate of the shear layer are employed in the self-similar analysis. We show that $M_c$ and $\lambda$ emerge naturally in self-similar equations, and therefore we conducted separate investigations on the influences of these two parameters. Furthermore, based on the compressible self-similar solutions, we derive, for the first time, closed-form analytical relations for different mean and turbulent quantities including the density, pressure, temperature, Reynolds normal stress in the transverse direction, transverse velocity, and dissipation. Except for scaled mean transverse velocity, all mean and turbulent quantities exhibit a collapse over different $M_c$ and $\lambda$ values using our proposed self-similar scalings. The scaled mean transverse velocity is also self-similar, but it is difficult to find a universal scaling to collapse its variation across different $M_c$ and $\lambda$ values because it behaves very differently on the high-speed and low-speed sides.

%The collapsed normalized streamwise velocity profiles can be approximated as an error function. 
%The suggested self-similar scaling for the mean transverse velocity is $\Delta U$$\frac{d\delta}{dx_1}$. For Reynolds shear stress, we proposed a mixed scaling dependent upon the average velocity $U_c$. We suggested two different scalings for the transverse Reynolds stress component but the scaling similar to Reynolds shear stress worked the best for present DNS data. Mean pressure is scaled in a similar manner to transverse Reynolds stress but incorporates a factor of $\rho_0$. The scalings of mean density and temperature are not straightforward and a function $\psi(M_c$) was incorporated to collapse the profiles at different convective Mach numbers. The mean dissipation can be self-similar if normalized properly. The same function $\psi(M_c$) is used in the normalization of mean dissipation.
%There is significant density variation inside the shear layer and the drop of density increases with compressibility. We investigated this density variation within the shear layer, linking this variation to dissipation effects as revealed by our analysis of self-similar energy equation. It is shown that 

The mean velocity profiles are used to determine the growth rate of the compressible shear layer. Consistent with prior observations, we found that the growth rate decreases with increasing compressibility. The peak transverse Reynolds stress and Reynolds shear stress decrease with increasing compressibility. Conversely, the peak streamwise Reynolds stress remains unaffected by changes in compressibility. A detailed validation against previous literature has been performed and the turbulence intensities and growth rates are found to agree well. 

It is observed that there is considerable variation in mean density within the shear layer and density reduction becomes more pronounced as compressibility increases. This density variation is linked to the dissipation effects as revealed by our analysis of the self-similar energy equation. Thus, due to viscous dissipation, there is an overheating at the core of the shear layer. This phenomenon results in a localized maximum in the temperature profile and a corresponding localized minimum in the density profile. Given the challenges in collapsing density profiles across various $M_c$ values, we introduced a novel scaling approach to successfully collapse these profiles. Unlike its dependence on $M_c$, the mean density is not sensitive to $\lambda$ variation.

The results also show significant disparities in the growth of the shear layer on the high- and low-speed sides. Due to the momentum difference between the high- and low-speed sides, the centerline (position of the zero mean axial velocity) shifts towards the low-speed side and $\phi$ is always negative. This brings asymmetry to the spatially developing shear layer and makes the growth of the shear layer very different on the high-speed and low-speed sides. We have introduced an asymmetry parameter $\gamma$ to quantify this disparity. Within the self-similar zone, $\gamma$ approaches a constant value and exhibits an increase with increasing $M_c$ and $\lambda$. For the $M_c$ range studied here ($M_c=0.8-1.6$), $\phi$ varies linearly with $M_c$. More research is required to see the trend of the $\phi$ variation at lower $M_c$ values. The absolute value of $\phi$ increases with increasing $\lambda$. However, due to limited data availability, it is difficult to establish a definitive trend between $\lambda$ and $\phi$. 

Using our approximate equation for the transverse velocity, we have derived a general expression for the entrainment ratio, $E_v$. Employing the $\phi$ values from the literature in our incompressible $E_v$ equation yields results closely aligned with the model proposed in \cite{dimotakis_entrainment}. The entrainment ratio was found to increase with $M_c$ and $\lambda$, favoring excess entrainment on the high-speed side.

This research contributes to the fundamental understanding of supersonic shear layers, providing technical advancements that are directly translatable into improved designs and operational efficiencies in air-breathing propulsion, jet noise reduction, missile aerodynamics, and high-speed mixing systems. Thus, the findings of our spatially developing compressible shear layer analysis, and especially the closed-form analytical relations derived here, can facilitate the testing of experimental and numerical data and have direct applications in advanced high-speed flow technologies. In air-breathing propulsion systems, such as scramjets, understanding the shear layer growth rate and its sensitivity to compressibility is essential to optimize fuel-air mixing, which directly impacts combustion efficiency and overall engine performance. The self-similar scalings for mean and turbulent quantities, as well as the entrainment ratio, between different values of $M_c$ and $\lambda$ allow quick predictions of flow behavior in these engines. 

\begin{bmhead}[Acknowledgments.]
This work was supported by the Los Alamos National Laboratory (LANL). LANL is operated by Triad National Security, LLC under Contract No. 89233218CNA000001 with the U.S. Department of Energy/National Nuclear Security Administration. DL acknowledges support from the ASC Mix \& Burn and OES programs at LANL. Z. Li gratefully acknowledges support from the U.S. Department of Education under Award No. P116H240026. The authors also acknowledge the High Performance Computing Center at Michigan State University for providing computational resources.
\end{bmhead}

\begin{bmhead}[Declaration of interests.]
The authors report no conflict of interest.
\end{bmhead}

\begin{appen}

\section{The scaling for the mean dissipation}\label{appA}

%\appendix  % Start appendix mode
%\renewcommand{\thefigure}{A.\arabic{figure}}  % This will label figures as A1, A2, etc.
%\renewcommand{\thesubfigure}{(\alph{subfigure})}  % %This will label subfigures as (a), (b), etc.

\begin{figure}
     \centering
     \begin{subfigure}{0.49\textwidth}
         \centering
         \includegraphics[width=\textwidth]{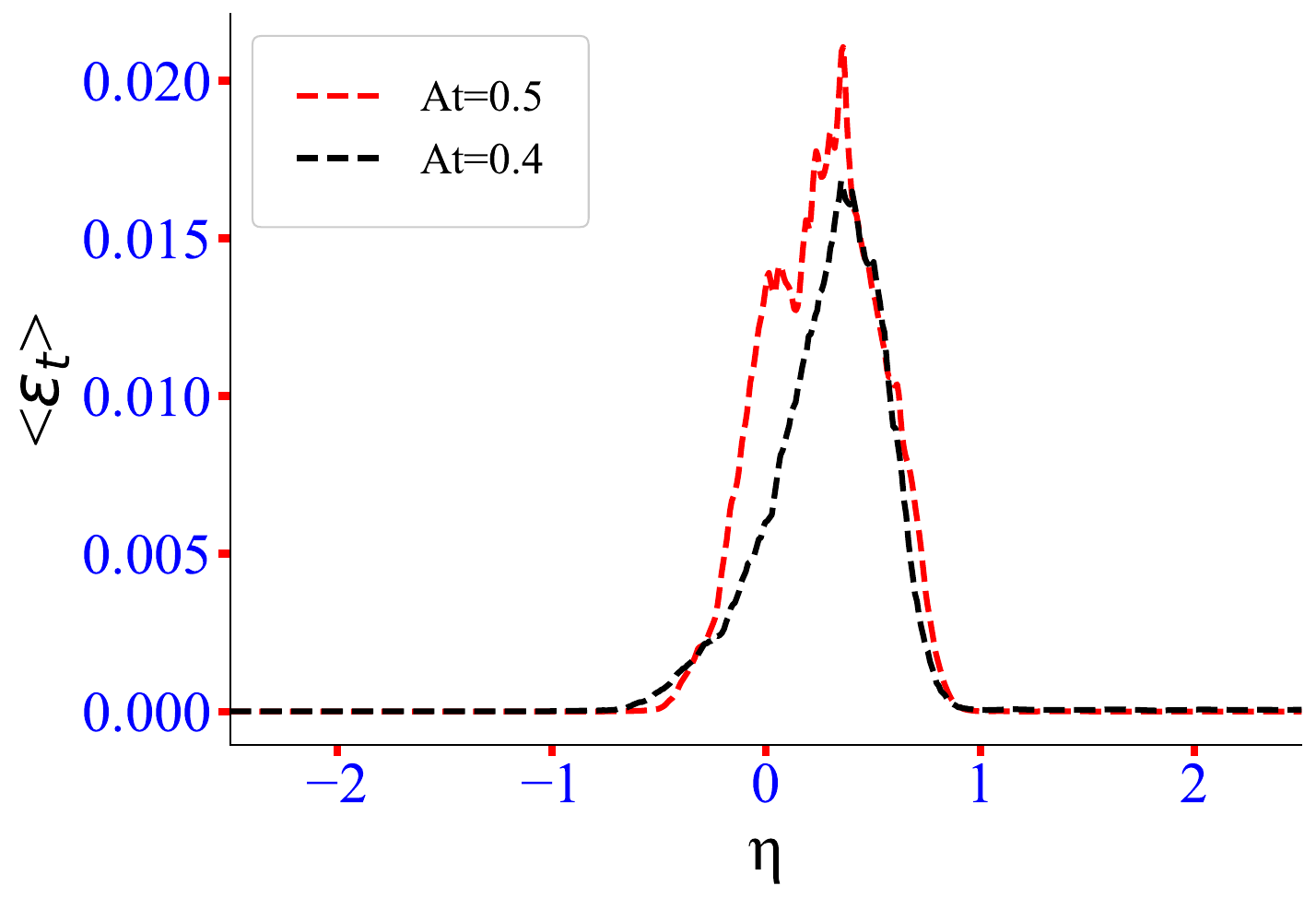}
         \subcaption{}
     \end{subfigure}
     \hfill
     \centering
     \begin{subfigure}{0.47\textwidth}
         \centering
         \includegraphics[width=\textwidth]{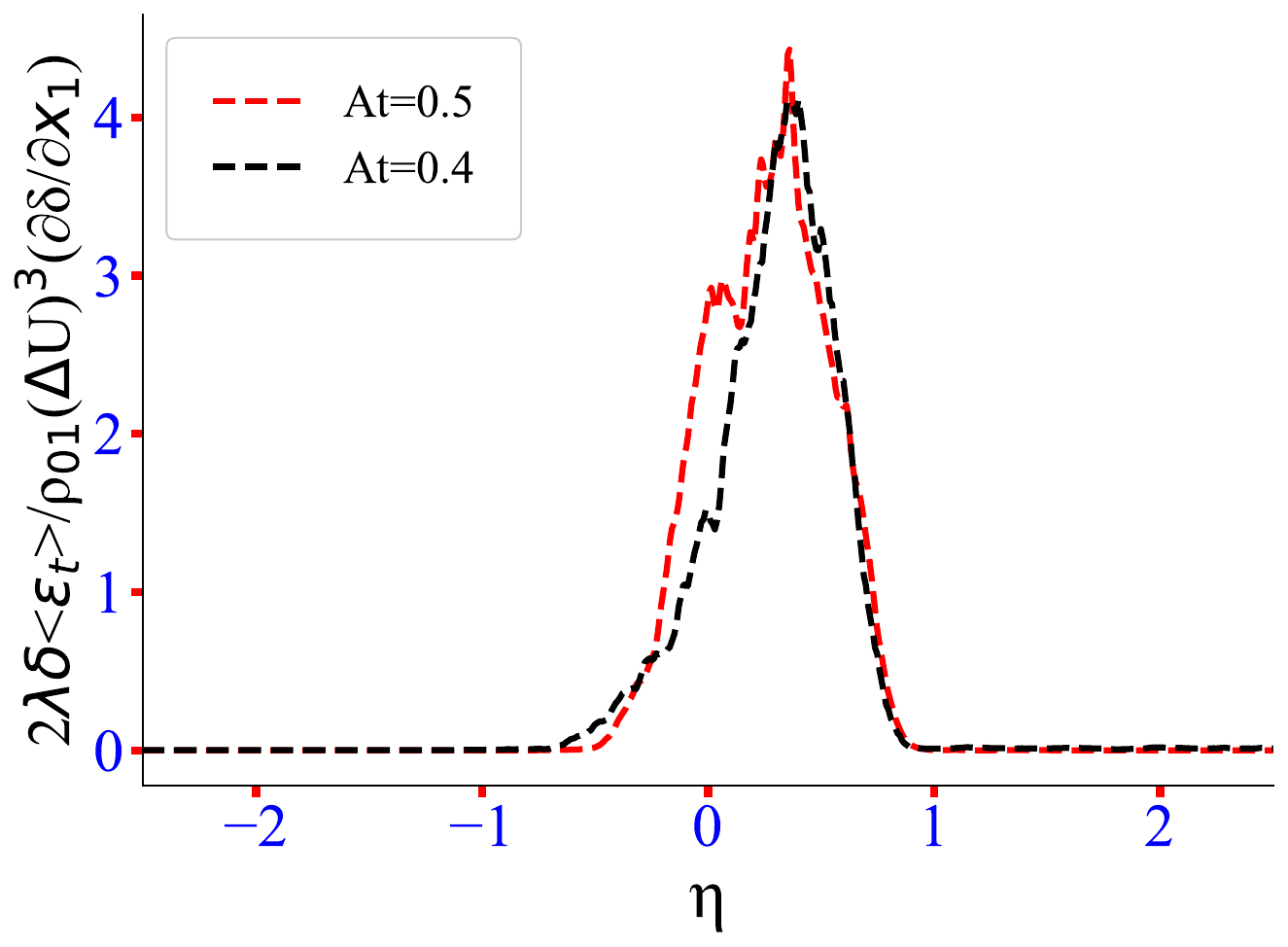}
         \subcaption{}
     \end{subfigure}
     \hfill
     \centering
     \begin{subfigure}{0.49\textwidth}
         \centering
         \includegraphics[width=\textwidth]{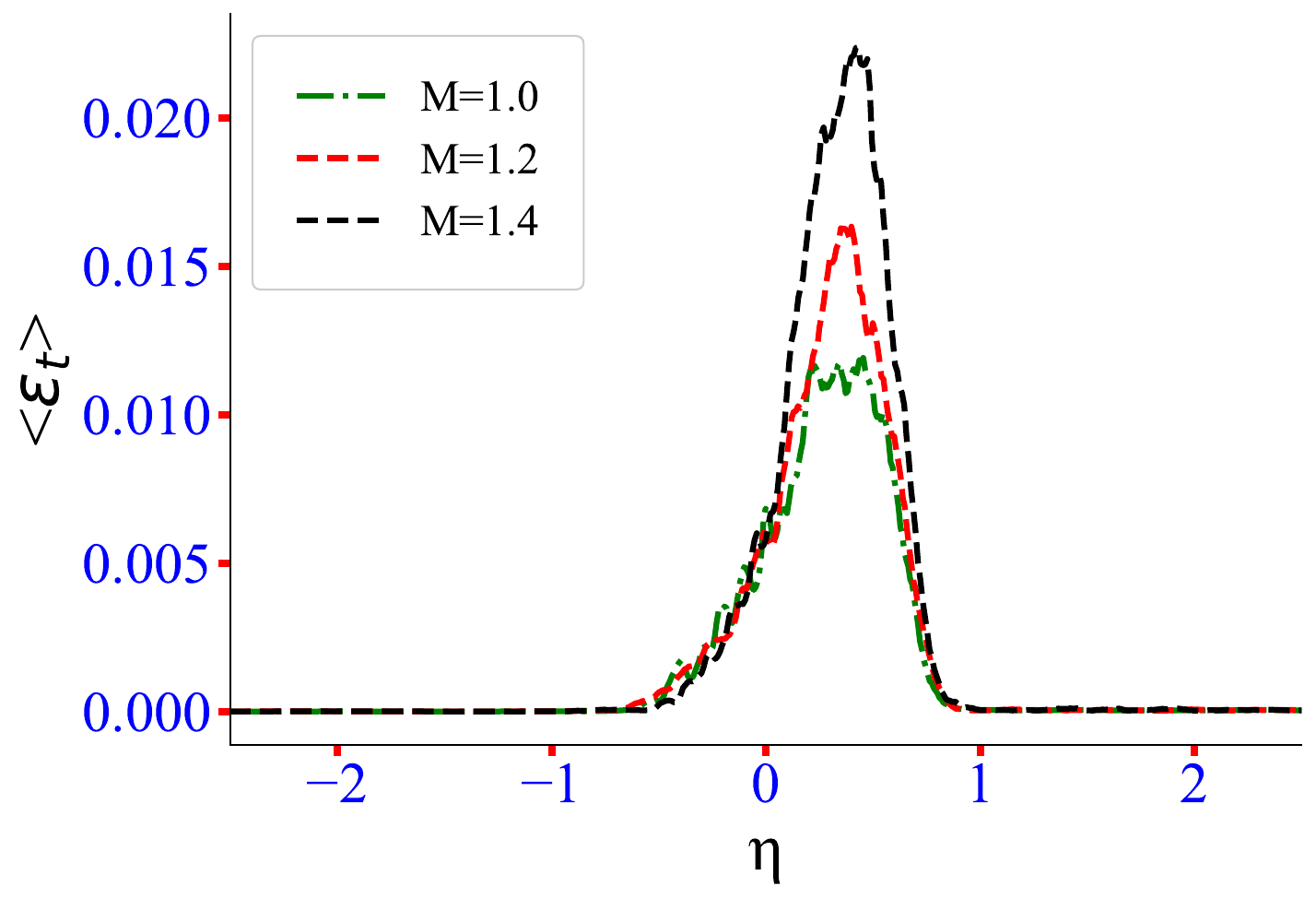}
         \subcaption{}
     \end{subfigure}
     \hfill
     \centering
     \begin{subfigure}{0.46\textwidth}
         \centering
         \includegraphics[width=\textwidth]{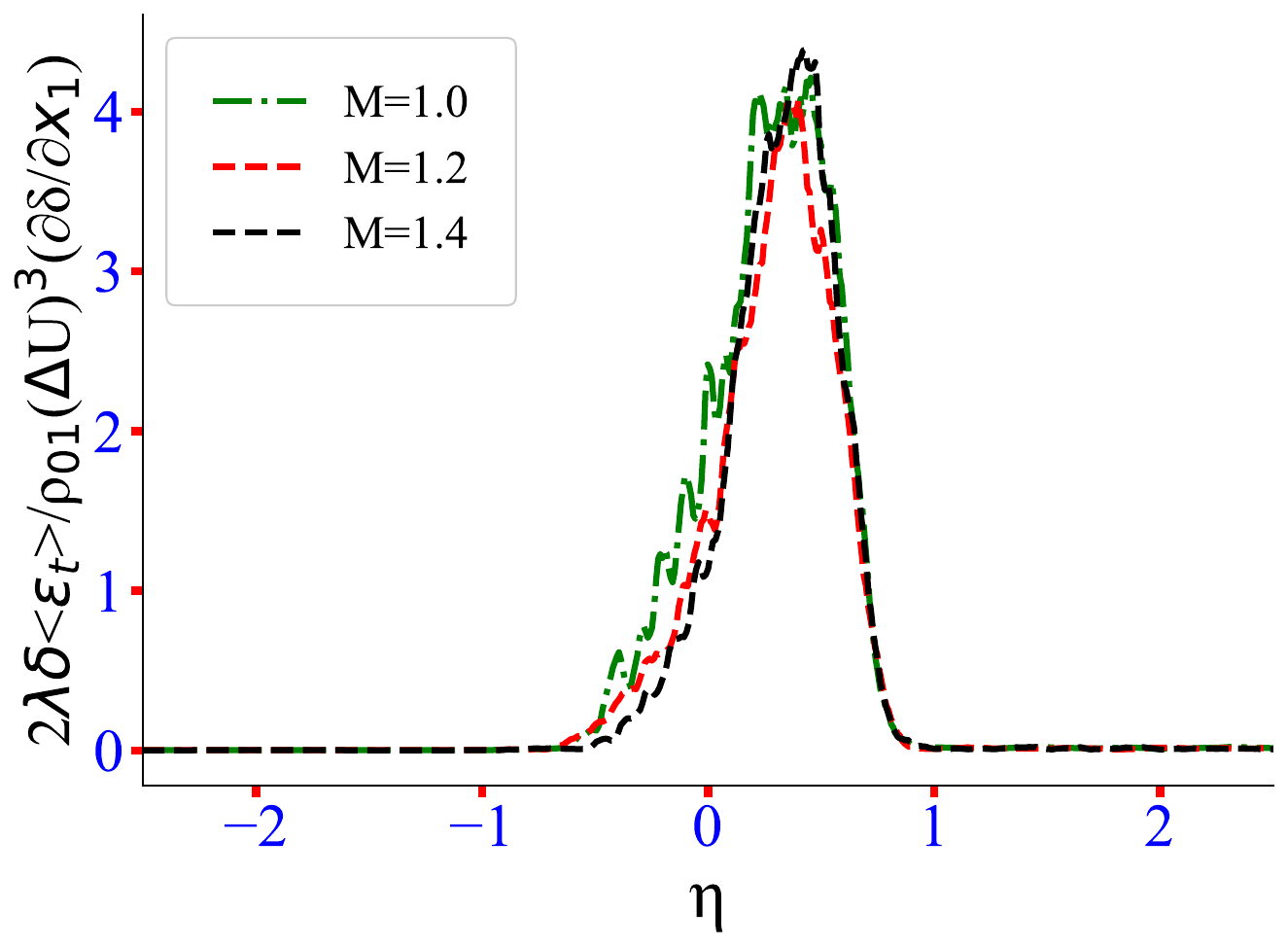}
         \subcaption{}
     \end{subfigure}
     \hfill
     \caption{ (a) The variation in mean dissipation at different $\lambda$ in the self-similar zones. (b) The collapse of mean dissipation at different $\lambda$ using our suggested scaling. (c) The variation of mean dissipation at different $M_c$.  (d) The collapse of mean dissipation at different $M_c$ using our suggested scaling  }
     \label{fig:diss}
\end{figure}

In this section, the behavior of mean dissipation is presented at different $M_c$ and $\lambda$ values. The mean dissipation shows a quasi-self-similar behavior as discussed in section \ref{sec:self-similarity} if normalized using our self-similar scaling, as seen in figure \ref{fig:diss}. It is important to note that, since the dissipation produces the density variation inside the shear layer, the same function $\psi(M_c)$ is required to collapse the dissipation profiles at different $M_c$ values. 

\section{Integrals involving self-similar scaled transverse velocity}\label{appB}

We have already demonstrated that the mean streamwise velocity, when normalized in self-similar coordinates for a range of convective Mach numbers and velocity parameters, exhibits a profile that can be approximated by the following error function: 
\begin{equation}
\hat{u_1}=Aerf(B\eta)\\  
\end{equation}

On the other hand, the collapsed density profiles can be approximated by the following Gaussian function:
\begin{equation}
\hat{\rho}=ae^{-b\eta^2}\\  
\end{equation}

The following integrals were used in the derivation of the approximate equation for $\hat{u}_2$.

\begin{equation}
\int_{-\infty}^{+\infty}\hat{\rho}\hat{u_1}\,d\eta= 0\\  
\end{equation}

\begin{equation}
\int_{-\infty}^{+\infty}\eta\frac{d\hat{u_1}}{d\eta}\,d\eta= 0\\  
\end{equation}

\begin{equation}
\int_{-\infty}^{+\infty}\hat{u_1}\frac{d\hat{u_1}}{d\eta}\,d\eta= 0\\  
\end{equation}

The rest of the integrals were calculated based on the assumption that $b$=$B^2$, which simplifies the equations. This assumption is consistent with the data. Based on this assumption, the integrals reduce to: 
\begin{equation}
\int_{-\infty}^{\eta}\hat{\rho}\,d\eta= \int_{-\infty}^{\eta} ae^{-B^2\eta^2}\,d\eta = \frac{a\sqrt{\pi}(1+erf[B\eta])}{2B}\\  
\end{equation}

\begin{equation}
\int_{-\infty}^{0}\hat{\rho}\,d\eta= \int_{-\infty}^{\eta} ae^{-B^2\eta^2}\,d\eta = \frac{a\sqrt{\pi}}{2B}\\  
\end{equation}

\begin{equation}
\int_{-\infty}^{+\infty}\hat{\rho}\,d\eta= \int_{-\infty}^{\eta} ae^{-B^2\eta^2}\,d\eta = \frac{a\sqrt{\pi}}{B}\\  
\end{equation}

\begin{equation}
\int_{-\infty}^{\eta}\eta\frac{d\hat{u_1}}{d\eta}\,d\eta =\frac{-A}{\sqrt{\pi}B}e^{-B^2\eta^2}\\  
\end{equation}

\begin{equation}
\int_{-\infty}^{0}\eta\frac{d\hat{u_1}}{d\eta}\,d\eta =\frac{-A}{\sqrt{\pi}B}\\  
\end{equation}

\begin{equation}
\int_{-\infty}^{0}\hat{\rho}\hat{u_1}\,d\eta= \int_{-\infty}^{0} ae^{-B^2\eta^2}Aerf(B\eta)\,d\eta  =\frac{-aA}{\sqrt{\pi}B} \frac{\pi}{4}\\  
\end{equation}

\begin{equation}
\int_{-\infty}^{\eta}\hat{\rho}\hat{u_1}\,d\eta= \int_{-\infty}^{\eta} ae^{-B^2\eta^2}Aerf(B\eta)\,d\eta  =\frac{aA\sqrt{\pi}}{4B} [(erf[B\eta])^2-1]\\  
\end{equation}

\begin{equation}
\int_{-\infty}^{+\infty}\eta\hat{u_1}\frac{d\hat{u_1}}{d\eta}d\eta= \int_{-\infty}^{+\infty} Aerf(B\eta)Aerf(B\eta)\,d\eta  =\frac{\sqrt{2}A^2}{\sqrt{\pi}B}\\  
\end{equation}

\begin{equation}
\int_{-\infty}^{+\infty}\hat{\rho}\hat{u_1}\hat{u_1}\,d\eta= \int_{-\infty}^{+\infty} ae^{-B^2\eta^2} [Aerf(B\eta)]^2\,d\eta  =\frac{a\sqrt{\pi}A^2}{3B}\\  
\end{equation}

\section{Scaled transverse velocity for a compressible shear layer }\label{appc}

In this section, we discuss the steps to derive the final compressible version of the equation $\hat{u}_2$. The analysis presented here is an extension of \cite{wei22} to the compressible case. Integrating the compressible self-similar $x_1$ momentum equation from $-\infty$ to $+\infty$ we get:
\begin{equation}
\begin{array}{l}
    \lambda\hat{u}_{2,+\infty}+\lambda\hat{u}_{2,-\infty}=2\lambda\psi(M_c)\int_{-\infty}^{+\infty}\eta\hat{\rho}\hat{u_1}\frac{d\hat{u_1}}{d\eta}\,d\eta+\psi(M_c)\int_{-\infty}^{+\infty}\eta\hat{\rho}\frac{d\hat{u_1}}{d\eta}\,d\eta +2\lambda\int_{-\infty}^{+\infty}\eta\hat{u_1}\frac{d\hat{u_1}}{d\eta}d\eta\\\\
    +\int_{-\infty}^{+\infty}\eta\frac{d\hat{u_1}}{d\eta}d\eta
    +2\lambda\phi\psi(M_c)\int_{-\infty}^{+\infty}\hat{\rho}\hat{u_1}\frac{d\hat{u_1}}{d\eta}\,d\eta+\phi\psi(M_c)\int_{-\infty}^{+\infty}\hat{\rho}\frac{d\hat{u_1}}{d\eta}\,d\eta+\phi\\\\
    -2\lambda\psi(M_c)\int_{-\infty}^{+\infty}\hat{\rho}\hat{u_2}\frac{d\hat{u_1}}{d\eta}\,d\eta-2\lambda\psi(M_c)\frac{u_2*}{u_{02}}\int_{-\infty}^{+\infty}\hat{\rho}\frac{d\hat{u_1}}{d\eta}\,d\eta
    +2\lambda\int_{-\infty}^{+\infty}\hat{u_1}\frac{d\hat{u_2}}{d\eta}\,d\eta\ -2\lambda\frac{u_2*}{u_{02}}\;
    \end{array}
\end{equation}

Applying the integral form of the continuity equation and replacing $\hat{u}_{2,-\infty}$ in the previous equation, we get the following.
\begin{equation}
\begin{array}{l}
    \hat{u}_{2}[1+\hat{\rho}\psi(M_c)]= 
    \psi(M_c)[\eta \hat{\rho} \hat{u}_{1} -\int_{-\infty}^{\eta}\hat{\rho}\hat{u_1}\,d\eta +\phi\hat{\rho} \hat{u_{1}}] +\frac{\psi(M_c)}{2\lambda}[\eta \hat{\rho}-\int_{-\infty}^{\eta}\hat{\rho}\,d\eta+\phi\hat{\rho}] \\\\ 
    + \frac{\phi}{2}[1+2\hat{u_{1}}] + \int_{-\infty}^{\eta}\eta\frac{d\hat{u_1}}{d\eta}\,d\eta- \psi(M_c)\frac{u_2^*}{u_{02}}\hat{\rho}+ \frac{\psi(M_c)}{2}\int_{-\infty}^{+\infty}\hat{\rho}\hat{u_1}\,d\eta+\frac{\psi(M_c)}{4\lambda}\int_{-\infty}^{+\infty}\hat{\rho}\,d\eta\\\\
    -\frac{1}{2}\int_{-\infty}^{+\infty}\eta\frac{d\hat{u_1}}{d\eta}\,d\eta
    -\frac{\phi}{2}+\psi(M_c)\int_{-\infty}^{+\infty}\eta\hat{\rho}\hat{u_1}\frac{d\hat{u_1}}{d\eta}\,d\eta+\frac{\psi(M_c)}{2\lambda}\int_{-\infty}^{+\infty}\eta\hat{\rho}\frac{d\hat{u_1}}{d\eta}\,d\eta+\int_{-\infty}^{+\infty}\eta\hat{u_1}\frac{d\hat{u_1}}{d\eta}d\eta\\\\
    +\frac{1}{2\lambda}\int_{-\infty}^{+\infty}\eta\frac{d\hat{u_1}}{d\eta}d\eta
    +\phi\psi(M_c)\int_{-\infty}^{+\infty}\hat{\rho}\hat{u_1}\frac{d\hat{u_1}}{d\eta}\,d\eta+\frac{\phi\psi(M_c)}{2\lambda}\int_{-\infty}^{+\infty}\hat{\rho}\frac{d\hat{u_1}}{d\eta}\,d\eta+\frac{\phi}{2\lambda}\\\\
    -\psi(M_c)\int_{-\infty}^{+\infty}\hat{\rho}\hat{u_2}\frac{d\hat{u_1}}{d\eta}\,d\eta-\psi(M_c)\frac{u_2*}{u_{02}}\int_{-\infty}^{+\infty}\hat{\rho}\frac{d\hat{u_1}}{d\eta}\,d\eta+\int_{-\infty}^{+\infty}\hat{u_1}\frac{d\hat{u_2}}{d\eta}\,d\eta\ -\frac{u_2*}{u_{02}}
    \end{array}
\end{equation}

By applying the self-similar continuity equation in the previous equation, we get the following:
\begin{equation}
\begin{array}{l}
    \hat{u}_{2}[1+\hat{\rho}\psi(M_c)]= 
    \psi(M_c)[\eta \hat{\rho} \hat{u}_{1} -\int_{-\infty}^{\eta}\hat{\rho}\hat{u_1}\,d\eta +\phi\hat{\rho} \hat{u_{1}}] +\frac{\psi(M_c)}{2\lambda}[\eta \hat{\rho}-\int_{-\infty}^{\eta}\hat{\rho}\,d\eta+\phi\hat{\rho}] \\\\ 
    + \frac{\phi}{2}[1+2\hat{u_{1}}] + \int_{-\infty}^{\eta}\eta\frac{d\hat{u_1}}{d\eta}\,d\eta+ \frac{\psi(M_c)}{2}\int_{-\infty}^{+\infty}\hat{\rho}\hat{u_1}\,d\eta+\frac{\psi(M_c)}{4\lambda}\int_{-\infty}^{+\infty}\hat{\rho}\,d\eta-\frac{1}{2}\int_{-\infty}^{+\infty}\eta\frac{d\hat{u_1}}{d\eta}\,d\eta\\\\
    -\frac{\phi}{2}+\psi(M_c)\int_{-\infty}^{+\infty}\eta\hat{\rho}\hat{u_1}\frac{d\hat{u_1}}{d\eta}\,d\eta+\frac{\psi(M_c)}{2\lambda}\int_{-\infty}^{+\infty}\eta\hat{\rho}\frac{d\hat{u_1}}{d\eta}\,d\eta+\int_{-\infty}^{+\infty}\eta\hat{u_1}\frac{d\hat{u_1}}{d\eta}d\eta+\frac{1}{2\lambda}\int_{-\infty}^{+\infty}\eta\frac{d\hat{u_1}}{d\eta}d\eta\\\\
    +\phi\psi(M_c)\int_{-\infty}^{+\infty}\hat{\rho}\hat{u_1}\frac{d\hat{u_1}}{d\eta}\,d\eta+\frac{\phi\psi(M_c)}{2\lambda}\int_{-\infty}^{+\infty}\hat{\rho}\frac{d\hat{u_1}}{d\eta}\,d\eta+\frac{\phi}{2\lambda}+\psi(M_c)\int_{-\infty}^{+\infty}\eta\hat{u_1}\frac{d(\hat{\rho}\hat{u_1})}{d\eta}d\eta\\\\
    +\frac{\psi(M_c)}{2\lambda}\int_{-\infty}^{+\infty}\eta\hat{u_1}\frac{d\hat{\rho}}{d\eta}d\eta+\int_{-\infty}^{+\infty}\eta\hat{u_1}\frac{d\hat{u_1}}{d\eta}d\eta+\psi(M_c)\phi\int_{-\infty}^{+\infty}\hat{u_1}\frac{d(\hat{\rho}\hat{u_1})}{d\eta}d\eta\\\\
    +\frac{\phi \psi(M_c)}{2\lambda}\int_{-\infty}^{+\infty}\hat{u_1}\frac{d\hat{\rho}}{d\eta}d\eta+\phi\int_{-\infty}^{+\infty}\hat{u_1}\frac{d\hat{u_1}}{d\eta}d\eta
    - \frac{u_2^*}{u_{02}}(1+\hat{\rho}\psi(M_c))
    \end{array}
\end{equation}

Applying the boundary conditions and integration by parts, the equation becomes:
\begin{equation}
\begin{array}{l}
    \hat{u}_{2}(1+\hat{\rho}\psi(M_c))= 
    \psi(M_c)[\eta \hat{\rho} \hat{u_{1}} -\int_{-\infty}^{\eta}\hat{\rho}\hat{u_1}\,d\eta +\phi\hat{\rho} \hat{u_{1}}] +\frac{\psi(M_c)}{2\lambda}[\eta \hat{\rho}-\int_{-\infty}^{\eta}\hat{\rho}\,d\eta+\phi\hat{\rho}] \\\\ 
    + \frac{\phi}{2}[1+2\hat{u_{1}}] + \int_{-\infty}^{\eta}\eta\frac{d\hat{u_1}}{d\eta}\,d\eta+\psi(M_c)\frac{\lambda-1}{2\lambda}\int_{-\infty}^{+\infty}\hat{\rho}\hat{u_1}\,d\eta+\frac{\psi(M_c)}{4\lambda}\int_{-\infty}^{+\infty}\hat{\rho}\,d\eta+\frac{1-\lambda}{2\lambda}\int_{-\infty}^{+\infty}\eta\frac{d\hat{u_1}}{d\eta}\,d\eta\\\\
    -\frac{\lambda-1}{2\lambda}\phi-\psi(M_c)\int_{-\infty}^{+\infty}\hat{\rho}\hat{u_1}\hat{u_1}\,d\eta+2\int_{-\infty}^{+\infty}\eta\hat{u_1}\frac{d\hat{u_1}}{d\eta}d\eta+\phi \int_{-\infty}^{+\infty}\hat{u_1}\frac{d\hat{u_1}}{d\eta}\,d\eta
    - \frac{u_2^*}{u_{02}}[1+\hat{\rho}\psi(M_c)]
    \end{array}
\end{equation}

At $\eta$=0, $\hat{u}_{1}$=0 and $\hat{u}_{2}$=0 according to the definition. Therefore,
\begin{equation}
\begin{array}{l}
    \frac{u_2^*}{u_{02}}= \frac{\psi(M_c)}{1+\hat{\rho_0}\psi(M_c)}
    [-\int_{-\infty}^{0}\hat{\rho}\hat{u_1}\,d\eta] +\frac{\psi(M_c)}{2\lambda\ (1+\hat{\rho_0}\psi(M_c))}[-\int_{-\infty}^{0}\hat{\rho}\,d\eta+\phi\hat{\rho}_0]
    + \frac{\phi}{2(1+\hat{\rho_0}\psi(M_c))}\\\\ 
    +\frac{1}{1+\hat{\rho_0}\psi(M_c)}\int_{-\infty}^{0}\eta\frac{d\hat{u_1}}{d\eta}\,d\eta+\psi(M_c)\frac{\lambda-1}{2\lambda (1+\hat{\rho_0}\psi(M_c))}\int_{-\infty}^{+\infty}\hat{\rho}\hat{u_1}\,d\eta+\frac{\psi(M_c)}{4\lambda(1+\hat{\rho_0}\psi(M_c))}\int_{-\infty}^{+\infty}\hat{\rho}\,d\eta\\\\
    +\frac{1-\lambda}{2\lambda(1+\hat{\rho_0}\psi(M_c))}\int_{-\infty}^{+\infty}\eta\frac{d\hat{u_1}}{d\eta}\,d\eta
    -\frac{\lambda-1}{2\lambda\ (1+\hat{\rho_0}\psi(M_c))}\phi-\frac{\psi(M_c)}{1+\hat{\rho_0}\psi(M_c)}\int_{-\infty}^{+\infty}\hat{\rho}\hat{u_1}\hat{u_1}\,d\eta\\\\
    +\frac{2}{1+\hat{\rho_0}\psi(M_c)}\int_{-\infty}^{+\infty}\eta\hat{u_1}\frac{d\hat{u_1}}{d\eta}d\eta+\frac{\phi}{1+\hat{\rho_0}\psi(M_c)}\int_{-\infty}^{+\infty}\hat{u_1}\frac{d\hat{u_1}}{d\eta}\,d\eta\;
    \end{array}
\end{equation}

After $\frac{u_2^*}{u_{02}}$ is replaced in equation (3.19), we get the final form of $\hat{u}_2$ equation:

\begin{equation}
\begin{array}{l}
    \hat{u}_{2}= \phi\hat{u_{1}} +\frac{1}{1+\hat{\rho}\psi(M_c)}[\psi(M_c)\int_{-\infty}^{\eta} \eta \frac{d(\hat{\rho}\hat{u_1})}{d\eta}\,d\eta+
    \frac{\psi(M_c)}{2\lambda}\int_{-\infty}^{\eta} \eta \frac{d\hat{\rho}}{d\eta}\,d\eta+ \int_{-\infty}^{\eta}\eta\frac{d\hat{u_1}}{d\eta}\,d\eta]\\\\
    -\frac{1}{1+\hat{\rho}_0\psi(M_c)} [\psi(M_c)\int_{-\infty}^{0} \eta \frac{d(\hat{\rho}\hat{u_1})}{d\eta}\,d\eta+
    \frac{\psi(M_c)}{2\lambda}\int_{-\infty}^{0} \eta \frac{d\hat{\rho}}{d\eta}\,d\eta+ \int_{-\infty}^{0}\eta\frac{d\hat{u_1}}{d\eta}\,d\eta]\\\\
     +[\frac{1}{1+\hat{\rho}\psi(M_c)} -\frac{1}{1+\hat{\rho}_0\psi(M_c)}] (\psi(M_c)\frac{\lambda-1}{2\lambda}\int_{-\infty}^{+\infty}\hat{\rho}\hat{u_1}\,d\eta+\frac{\psi(M_c)}{4\lambda}\int_{-\infty}^{+\infty}\hat{\rho}\,d\eta\\\\
    +\frac{1-\lambda}{2\lambda}\int_{-\infty}^{+\infty}\eta\frac{d\hat{u_1}}{d\eta}\,d\eta
    -\psi(M_c)\int_{-\infty}^{+\infty}\hat{\rho}\hat{u_1}\hat{u_1}\,d\eta+2\int_{-\infty}^{+\infty}\eta\hat{u_1}\frac{d\hat{u_1}}{d\eta}d\eta+\phi \int_{-\infty}^{+\infty}\hat{u_1}\frac{d\hat{u_1}}{d\eta}\,d\eta)
    \end{array}
\end{equation}

We have already discussed that $\hat{u}_1$ is an anti-symmetric function (error function) and $\hat{\rho}$ is a symmetric function (Gaussian). Applying the functional form of $\hat{u}_1$=$0.5 erf(B\eta)$ and $\hat{\rho}$=-$e^{-B^2\eta^2}$ and the value of the integrals discussed in Appendix \ref{appB}, the final analytical form of the approximate $\hat{u_2}$ equation is: 
\begin{equation}\label{eq:finalu2hat}
\begin{array}{l}
\hat{u}_{2} =  \frac{\phi}{2}erf(B\eta)-\frac{\eta e^{-B^2\eta^2}\psi(M_c)[1+\lambda erf(B\eta)]}{2\lambda(1-\psi(M_c) e^{-B^2\eta^2})}
+\frac{1}{1-\psi(M_c)}[\frac{\psi(M_c)\sqrt{\pi}(1+\lambda)}{8B\lambda}\\
+\frac{1}{2\sqrt{\pi}B}]-\frac{1}{1-\psi(M_c)e^{-B^2\eta^2}}[\frac{\psi(M_c)\sqrt{\pi}[1+erf(B\eta)]}{8B}(\frac{1}{\lambda}+\frac{1}{2}[erf(B\eta)-1])+ \frac{e^{-B^2\eta^2}}{2\sqrt{\pi}B}]\\
+ [\frac{1}{1-\psi(M_c)e^{-B^2\eta^2}}-\frac{1}{1-\psi(M_c)}][\frac{-\psi(M_c)(3-\lambda)}{12\lambda B}+\frac{1}{\sqrt{2\pi}B}]
\end{array}
\end{equation}

Equation (\ref{eq:finalu2hat}) represents the compressible form of the approximate self-similar scaled transverse velocity equation. Subsequently, this equation serves as the basis for investigating the characteristics of the scaled mean transverse velocity, and its outcomes were compared with current DNS findings across varying convective Mach numbers and velocity parameters. In the incompressible limit, $\psi (M_c)$ converges to 0 and the incompressible form of the approximate self-similar transverse velocity equation becomes: 
\begin{equation}\label{eq:u2incom}
\begin{split}
\hat{u}_{2} = & \frac{\phi}{2} erf(B\eta) + \frac{1}{2\sqrt{\pi} B}-\frac{1}{2\sqrt{\pi}B}e^{-B^2\eta^2}\\
\end{split}
\end{equation}

Equation (\ref{eq:u2incom}) is similar to the scaled mean transverse velocity derived by \cite{wei22} for the incompressible case.

\end{appen}\clearpage

\bibliographystyle{jfm}
\bibliography{mainJFM1}

%\begin{thebibliography}{}
%\expandafter\ifx\csname natexlab\endcsname\relax
%\def\natexlab#1{#1}\fi
%\expandafter\ifx\csname selectlanguage\endcsname\relax
%\def\selectlanguage#1{\relax}\fi

%\end{thebibliography}

%% End of file `jfm.bib'.

\end{document}